# ARISING INFORMATION REGULARITIES IN AN OBSERVER

**Vladimir S. Lerner, USA,** lernervs@gmail.com

This information approach develops Wheeler's concept of Bit as Observer-Participator leading to formal description of information process from probabilistic observation, emerging microprocess, time, space, entanglement, qubit, encoding bits, evolving macroprocess, and extends to Observer information geometrical structure. They emerge in observing interactive process where each yes-no inter-action $\downarrow\uparrow$ impulse models information Bit. Multiple interactions are random composing random process which covers both the bit and information process of the interacting bits.
Uncovering the bits and information process through observation of the random process is aim of an information observer.
In the considered formal model of the observation, the impulse observation runs axiomatic probabilities of random field linking Kolmogorov 0-1 law and Markov process probabilities. The field connects sets of possible and actual events with their probabilities. Field energy covers actual events. This triad specifies the observation. The Markov process models the arising observer's process, collecting observation which changes its probabilities measure similar to sequence of apriori-aposteriori Bayes probabilities. These objective probabilities, being immanent part of the process, virtually observe and measure the Markov correlation connecting states-events, discretely change the entropy of correlation generating probabilistic impulses. Each such impulse virtually cuts the observing entropy-uncertainty hidden in the cutting correlation. The cutting entropy decreases the initial Markov entropy and increases the entropy of cutting impulse.
Such multiple interactions minimize uncertainty of the Markov process and maximize entropy of each following observing impulse. That runs minimax principle for each observing impulse along the Markov interactive impulses.
When observing probability approaches 1, the impulse cutting entropy automatically converts to information.
The merging impulse curves and rotates the impulse interactive yes-no conjugated entropies in the microprocess. The entropies entanglement starts within impulse time interval before its space forms and the ends with beginning the space during reversible relative time interval of $0.015625\pi$ part of the impulse invariant measure $\pi$. Since the entanglement has no space measure, the entangled states can be everywhere in a space. Virtual observer emerges from the probabilistic model. The opposite curvature, enclosing entropy of the interacting impulses, lowers potential energy of an external process that converts entropy to bit of the interacting process, which memorizes the bit by delivering the Landauer energy.
Sequential interactive cuts along the process integrate the cutoff hidden information in information macroprocess with irreversible time course. Each memorized information binds the reversible microprocess within impulse with the irreversible information macroprocess along the multi-dimensional process. The impulse observation consecutively converts entropy to information in emerging information observer, conveying information causality, certain logic, and complexity. Multiple interacting Bits self-organize information process encoding information causality, logic and complexity. Trajectory of observation process carries wave function both probabilistic and certain, which self-builds structural information macrounits-triplets. Macrounits logically self-organize information networks IN encoding the units in geometrical structures enclosing triplet code. Multiple IN binds their ending triplets which encloses observer information cognition and intelligence. The observer cognition assembles common units through multiple attraction and resonances at forming IN triplet hierarchy which accept only units that recognizes each IN node. Maximal number of accepted triplet levels in multiple IN measures the observer maximum comparative information intelligence. The intelligent observer recognizes and encodes digital images in message transmission, being self-reflective enables understanding the message meaning. The cognitive logic self-controls the process encoding the intelligence in double helix coding structure. Integrating the process entropy in the entropy functional and the Bits in information path integral' measures formalize the variation problem in the minimax law which determines all regularities of the processes. Solving the problem, mathematically describes the micro-macro processes, the IN, and invariant conditions of observer's self-organization and self-replication. These functional regularities create united information mechanism whose integral logic self-operates, transforming interacting uncertainties to physical reality-matter, human information and designing AI.
Both information and information process emerge as phenomena of natural interaction. Each specific field triad generates information process creating its observer. The described information equations finalize the main results, validate them numerically, and present information models of many interactive physical processes.
*Keywords:* impulse probabilistic observation; cutting correlation; minimax information law; wave function; micro-macro processes, integral information measure; causal logic; time arrow, encoding, triplet code, variation equations,cooperative information dynamics; hierarchical network; objective and subjective observers;self-forming cognition and intellect; AI observer; applications.

**Introduction**

Observers are everywhere from communicating people, animals, different species to any interacting subjects, accepting, transforming and exchanging information.

Up to now, their common information regularities, emergence, differentiation, and evolution have not been studied by united information approach.

The tasks are information regularities describing each observer information path from observation to encoding images.

This study shows how an information observer with regularities emerges from observing random process at conversion of its uncertainty to certainty-information, creating information dynamics, information network with its logic and code generating an information observer which evolve toward cognition and intellect.

It allows information modeling many physical processes, including human brain which works with information processes.

Even though multiple physical studies [1-9] reveal information nature of the analyzed physical processes in an observer, until A. Wheeler's theory [10-14] of information-theoretic origin of an observer ('it from bit'), the observer has studied mostly from physical point of view.

The questions still are: How this bit appears and how does information acquire physical properties?

The information processes of an observer, its information structure and regularities have not been adequately understood.

Weller has included the observer in wave function [14] according to standard paradigm: Quantum Mechanics is Natural.

Since information initially originates in quantum process with conjugated probabilities, its study should focus not on physics of particles but on its information-theoretical essence of observing process tracing their probabilistic interactions. .

Important scientific task is revealing information nature of various interactive processes, including multiple physical interactions, human observations, Human-machine communications, biological, social, economic, other interactive systems, integrated in information observer.

However both information and information process have not scientifically conclusive definitions, neither implicit origin.

Elementary unit of information-Bit is a Yes-No single interactive action-impulse; series of the interactive bits may produce information process. Thus, interaction originates information.

Since interactions are fundamental phenomena building structure of Universe, information is its natural phenomenon.

Multiple interactions are random composing random process which covers the interacting bits. Uncovering the bits and information process through observation of the random process is the aim creating the information observer.

What is fundamental source of random interacting impulses, which potentially cover the information bit and their sequence in information process?

How to measure the rapprochement of the observing process uncertainty to certainty and extract a certain physical bit?

How to rebuild the information process from multiple sequence of observing bits hidden in interactive random process?

How the observing information creates the information observer with regularities?

These are fundamental questions, essential for Science and Applications, arise at observing multiple events and processes in physics, biology, cognition, cosmos, economy, sociology, experimental science, learning, reading, acquiring knowledge, examining investigations, playing games, dealing with human activities; for example, in human interactive communications (discussions) searching facts (truths).

That primary includes understanding notion and nature of observation searching information by interacting impulse. This defines the observation as a random process of interactive impulses. Otherwise, the interactive observations are source of randomness which model series of random impulses-a random process–uncertain or imaginary before measuring.

*In observing interactive processes, the yes-no interactive actions, converting observed uncertainty to observing certainty, creates information Bits. Or, certain yes-no interaction is natural phenomenon creating unit of information.*

This paper extends and review results [26, 27, 39, and 40] analyzing the emergence and rise the observer's *regularities*.

 (Numbering formulas in each section starts with (1..);the citations numbers use the first number of the section (1.1..),… , (4.1..)).

# I. The Initial Points, Results, and Principles of the Approach

## 1.1. Random interactions, probability field, Kolmogorov 0-1 law, Markov process, and link to Bayes probabilities.

Interactions are natural fundamental phenomena of multiple events in common environment of Universe which is also source of interactions. Interactions build Nature and reality is only the interactions. .

Elementary natural inter-action consists of action and reaction, which represents symbols: yes-no of $\downarrow\uparrow$ actions modeling a Bit. How does the bit and their logical sequence originate?

We show that probabilistic interactions create information, physical processes, and an observer of this information.

The interactions can be an observing probability trace of interacting objects, particles, others up to human being.

The aim is the formal principles and methodology explaining the procedure of emerging interacting observer self-creating information. The aim derives from unifying the different interactions independent of origin, and focusing on observation as the interactive observer. The observing interactive random process of multiple interactions generates the observer, which evaluates the probabilities measured with equivalent entropy.

Natural (real) interactions converts this entropy to information.

*The initial points*

Axiom

Multiple interactive processes build Universe independently of their origin and reality is only the emerging interactions, which unify natural and communications' interactions through sequence of interactive impulses yes-no or no-yes actions.

Corollaries and concepts

1. Each real (certain) inter-action is opposite yes-no action modeling elementary Bit of information or a discrete impulse.
2. The multiple interactions are random and represent a random process of the interacting impulses in a surrounding random field. The random process and its states (events) are formally considered independent of specific substances.
3. To uncover a real Bit and/or multiple Bits from random process its observation requires.
4. Observation is formal acts emerging from changing a probability of interactive events of random impulse with probability of yes (no) action to probability of no (yes) action.

The random process virtually observes a discrete yes-no axiomatic probabilities of the random field.

For example, changing Bayes a priori probability to a posteriori probability in sequence of the interacting yes-no or no-yes actions of the random impulses arising within an observable random process emerging from the formal probability field.

5. An observer (virtual) emerges with beginning of observing each interactive no-yes actions in sequence of the random impulses arising within the random observable process. Such observer is a virtual receptor of the observing process.

For example, the observer emerges with observing each posteriori Bayes probability collecting all of them in the sequence of random interactive impulses within the initial random observable process.

The probabilities of both observable and observing processes are immanent parts of the field that virtually links them and links the observing process and its observer.

6. These processes relative entropy measures uncertainty of random events between the impulse yes-no probabilities, or uncertain multiple impulses of the uncertain Bits.

(The entropy is formal measure of the process' probabilities logarithmic function).

7. Each discrete probability of the yes (no) or no (yes) action virtually cuts the entropy of these random events, which decreases entropy of the observable process.
8. The sequential cutoff (erasure) the entropies minimize uncertainty of the observed impulse up to finally revealing a certain impulse in the observation.

9. Reaching the certainty requires the interactive process' probability approaching one when the real cut applies. That exposes Bit emerging from the observing impulse interactions as unit of information, certainty in an information process.

10. Information observer emerges with beginning of observing multiple information impulses creating information process.

11. Considering interaction as phenomena of Nature defines information as phenomena emerging from interactions.

Essence of methodology implementing the Axiom and Corollaries

1. Describing the random sequence of the interacting impulses yes-no or no-yes actions by a sequence of Kolmogorov'0-1 law events' objective probabilities measure for the random impulses in a formal random field.

2. Definition of this axiomatic Kolmogorov field formally connects the sets of possible events, the sets of actual events, and their probability function. This triad specifies the occurrence of specific events and starts each sequence in a probabilistic observation. (Probability measures only multiple events, or the probabilistic observation measures multiple inter-acting events in a random process where each sequence of observation begins).

Manifold of the specific events provides manifold of the observable sequences in observable processes.

3. Describing the observing process by changing the probabilities of the impulses within the observable process under the sequence of interacting Kolmogorov'0-1 law probabilities.

4. Describing the observation of changing the probabilities in the observing process as the emerging observer.

5. Modeling the observable random process by a Markov process chain of probabilistic impulses emerging from the random field.

6. Modeling the observing process of changing the probabilities in the observable process by sequence of the interactive random impulses within the Markov process.

10. Modeling the observation of the emerging observer by Markov process probabilities, where act of observation consists of changing probabilities of the Markov chain to probabilities of the Markov diffusion. That binds the chain by Markov diffusion correlations, which encloses the act of observation in form of hidden information. .

11. Sequence of the probabilistic 0-1 (No-Yes) impulses, acting on the observable process, initiates Bayes probabilities within the Markov diffusion process, which self-observe the Markov process. These objective probabilities link the Kolmogorov law's axiomatic probabilities with the Bayesian probabilities, bringing discrete Bayes probabilities with observing Yes-No random impulses which build an observing process of potential observer. The interacting probabilistic 0-1 impulses of the observable process initiate the probabilistic impulses of the observing process starting the potential observer as a reaction on the observation. Particular objective Bayes probability observes specific set of events, whose correlation holds the entropy measure.

12. The discrete impulse probabilities, virtually cutting the observable correlation, allow virtually observe entropy-uncertainty hidden in the cutting correlation binding the states in the observable random process.

13. Cutting the entropy when the observing impulse probability approaches one reveals the information Bit of the certain impulse which really cuts the impulse entropy.

14. Observing the information Bit under the cutting impulse uncovers the informative events of observing process.

15. Multiple interacting Bits self-organize the information process which creates Information Observer.

*Finally, both information and information process emerge as phenomena of natural interactions while each of its specific random field triad generates an observer depending on the observation (interactive) process.*

*Initial formalism. Probabilistic model.*

Multiple interactive actions are random events-variables $\omega$ in a surrounding random probability field.

The probability field defines mathematical triad [19]: $\Phi = (\Omega, F, P)$, where $\Omega$ is sets of all possible $\omega$, $F$ is Borel's $\sigma$-algebra subsets from sets $\Omega$, and probability $P$ is a non-negative function of the sets, defined on $F$ at condition $P(\Omega) = 1$.

This triad formally connects the sets of possible events, the sets of actual events, and their probability function.

*Example.*

If the experiment consists of events: one flip of a fair coin, then the result is either heads $H$ or tails $T$, (or none of them), called *tail event*s. Then $\Omega = (H,T)$, and the $\sigma$-algebra $F = 2^\Omega$ contains $2^2 = 4$ these tail events with the probability measures $P(\bullet) = 0, P(H) = 0.5, P(T) = 0.5, P(H,T) = 1$. For three dimensional spaces, from the probability field emerges potential observer with eight possible probability measures in form of random cube.

In the infinite sequence of random variables $\omega$, distributed in the field, a discrete $P(\bullet) = 0$ can happen among $P(\omega)$. •

Let $F$ be the $\sigma$-algebra generated by $\omega$, and $F_o$ be the $\sigma$-algebra generated by the sequence of *mutually independent* variables $\varpi$ and its function $f(\varpi)$.

Then, according to Kolmogorov 0-1 law, "the conditional probability $P_{\bar{\omega}}[f(\varpi) = 0]$ of the relation $f(\varpi) = 0$ remain, when the first $n$ variables (of $\varpi$) are known, equals to the absolute probability $P_{\bar{\omega}}[f(\varpi) = 0] = 0$ or $P_{\bar{\omega}}[f(\varpi) = 0] = 1$ for every $n$. The assumptions of the law are fulfilled if random variables $\varpi$ are mutually independent and if the value of function $f(\varpi)$ remain unchanged when only a finite variable change "[19.p.69].

The probability of the independent events is an equivalent to the probability in Big Number law[19, p.69].

Let also have in the field a Markov process chain [19] $X_t = X(\omega)$ of the event $\omega = (x,t)$ including states $x$ and their time moment $t$. The Markov process is $n$-dimensional and all dimensions start in probability field with different triads associated with local random frequencies of the initial events $\omega$ attribute. Generally $n \to \infty$.

Since the act of observation consists of changing probabilities of the Markov chain to probabilities of the Markov diffusion, let us consider a formal Markov diffusion process.

The Markov diffusion process trajectories are defined on $n$-dimensional probability distribution $P_n = P_n[X(\omega)]$ with transitional probabilities $P(s,\tilde{x},t,B)$, $\sigma$-algebra $F(s,t)$ created by events $\{\tilde{x}(\tau) \in B\}$ at $s \leq \tau \leq t$, and conditional probability distributions $P_{s,x}$ on $F(s,t)$, where transitional probability is equivalent to the conditional.

Transformation of this probability measures relation [22]:

$$\tilde{P}_{s,x}(d\omega) = p(\omega) P_{s,x}(d\omega) \tag{1.0}$$

on trajectories of Markov process $(\tilde{x}_t, P_{s,x})$ which holds distributions $\tilde{P}_{s,x} = \tilde{P}_{s,x}(A)$ on extensive $\sigma$-algebra $F(s,\infty)$ with density measure:

$$p(\omega) = \frac{\tilde{P}_{s,x}(d\omega)}{P_{s,x}(d\omega)} \tag{1.1}$$

Applying the definition of conditional entropy [23] to mathematical expectation of logarithmic probability *functional density measure* (0.1), we introduce *entropy functional on trajectories of Markov diffusion process* [24]:

$$S = E\{-\ln[p(\omega)]\} = \int_{\tilde{x}(t) \in B} -\ln[p(\omega)] P_{s,x}(d\omega), \tag{1.2}$$

where $E = E_{x,s,\tilde{x}_t}$ is conditional mathematical expectation, taken along the process trajectories $\tilde{x}_t$ at a varied $(\tilde{x},s)$ (by analogy with M.Kac [25]).

Markov diffusion process describes drift function $a = a(t,x)$ and function of diffusion $\sigma = \sigma(t,x)$ which define additive functional [22,26-29];

$$\varphi_s^T = 1/2 \int_s^T a(t,\tilde{x})^T (2b(t,\tilde{x}))^{-1} a(t,\tilde{x}_t) dt + \int_s^T \sigma(t,\tilde{x})^{-1} a(t,\tilde{x}) d\xi(t), 2b(t,\tilde{x}) = \sigma(t,\tilde{x})\sigma^T(t,\tilde{x}) > 0 \ ). \tag{1.3}$$

Functional (1.3) describes transformation of the Markov processes' random time traversing the process trajectory.

Additive functional (1.3) measures probability density (1.1) and can control $\varphi_s^t(\omega)$ and vice versa:

$$p(\omega) = \exp\{-\varphi_s^t(\omega)\}, \text{ or } \varphi_s^t(\omega) = -\ln p(\omega). \tag{1.4}$$

For a diffusion process $\varsigma_t$, transforming density (1.4) with aid of additive functional (1.3), transitional probabilities satisfying relation (1.0) is

$$\tilde{P}(s,\varsigma,t,B) = \int_{\tilde{x}(t)\in B} \exp\{-\varphi_s^t(\omega)\} P_{s,x}(d\omega). \tag{1.5}$$

Applying the definition of entropy *functional* (1.2) to process $\tilde{x}_t$ regarding process $\varsigma_t$, we get the entropy functional measure expressed via additive functional $\varphi_s^t(\omega)$ on the trajectories of the diffusion processes:

$$S[\tilde{x}_t / \varsigma_t] = E[\varphi_s^t(\omega)]. \tag{1.6}$$

Minimum of this functional, depending on the additive functional, measures closeness the above distributions:

$$\min_{\varphi_s^t} S[\tilde{x}_t / \varsigma_t] = S^o. \tag{1.7}$$

Let the transformed process be

$$\varsigma_t = \int_s^t \sigma(v,\xi_v) d\xi_v \tag{1.8}$$

having the same diffusion as the initial process $\tilde{x}_t$, but the zero drift.

Process $\varsigma_t$ is a transformed version of process $\tilde{x}_t$ whose transition probability satisfies (1.1), and transformed probability $\tilde{P}_{s,x}$ for this process evaluates the Feller kernel measure [30, 31]. Since transformed process $\varsigma_t$ (1.8) has the same diffusion matrix but zero drift, the right part of additive functional in (1.3) satisfies

$$E[\int_s^T (\sigma(t,\tilde{x}))^{-1} a(t,\tilde{x}) d\xi(t)] = 0. \tag{1.9}$$

It brings the integral measure of the entropy functional expressed via parameters of Ito stochastic equation [34] in form:

$$\Delta S[\tilde{x}_t]|_s^T = 1/2 E_{s,x}\{\int_s^T a^u(t,\tilde{x}_t)^T (2b(t,\tilde{x}_t))^{-1} a^u(t,\tilde{x}_t) dt\} = \int_{\tilde{x}(t)\in B} -\ln[p(\omega)] P_{s,x}(d\omega) = -E_{s,x}[\ln p(\omega)]. \tag{1.10}$$

Formulas (1.2-1.5), (1.6), (1.7), (1.8) and (1.10) are in [24] with related citations and references.

The entropy functional (EF) in forms (1.6, 1.10) is an *information indicator* of distinction between the probability measures of processes $\tilde{x}_t$ and $\varsigma_t$; it measures a *quantity of information* of process $\tilde{x}_t$ regarding process $\varsigma_t$.

At EF (1.10)=0, the observing process' Bayes impulses indistinct from the noise.

Since the time, following from the zero EF, is also equal to zero, the observation has *not* begun.

The right side of (1.10) is the EF equivalent formula, expressed via probability density $p(\omega)$ of random events $\omega$, integrated with the probability measure $P_{s,x}(d\omega)$ along the process trajectories $\tilde{x}(t) \in B$, which are defined at the set $B$.

*Formula (1.10) directly connects the probabilities, defining the EF, and the process functions drift and diffusion without necessary to the implement probabilities measurement for the given process.*

For the processes with the equivalent probability measures, quantity (1.10) is zero, and it is positive for the process' nonequivalent measures. Variation problem (1.7) for the EF functional was solved in [32].

Mathematical expectation (1.10) on the *process' trajectories,* conditional to transformed probability measure of Feller kernel $\tilde{P}_{s,x}$, defines an invariant at Markov transformations along trajectories.

The invariant measures the Radon-Nikodym probability density measure (1.1) where both $P_{s,x}$ and $\tilde{P}_{s,x}$ are defined.

Thus, integral (1.10) is Markov process' $\tilde{x}_t$ functional entropy measure $\tilde{x}_t$ conditional to the kernel probability measure.

Entropy measure in (1.2) is conditional to any transformed Markov diffusion process, not necessary satisfying (1.10). Measuring conditional entropy (1.10) relatively to diffusion process $\varsigma_t$ is practically usable, since $\varsigma_t$ models standard perturbations in controllable systems as a typical "white noise".

Each dimensional probability is local for each random ensemble as a part of the $n$-dimensional process ensemble.

The field of events, having probabilities $P_{\tilde{\omega}} = \in 0,1$, may interact with Markov process probability $P_{s,x}(\omega)$ satisfying relation for joint probability of independent events:

$P = P_{\tilde{\omega}} \times P_{s,x}(d\omega), P_{\tilde{\omega}} = \in 0,1$.

From that follows the sequence of probabilities: at $P_{\tilde{\omega}} = 0, P = P_{s,x}(d\omega) = 0$ at $P_{\tilde{\omega}} = 1, P = P_{s,x}(d\omega)$.

Probability $P_{s,x}(d\omega) = 1$ of 0-1, acting via the additive functional in (1.5), switches both the initial transitional probability $P(s, \tilde{x}, t, B)$ to transitional probability $\overline{P}(\overline{s}, \overline{\tilde{x}}, t, B)$ and density $p(\omega)$. The density, defines via to the process additive functional (1.4), links to drift function in (1.3). Transitional probability $\overline{P}(\overline{s}, \overline{\tilde{x}}, t, B)$ changes related Markov probability $\overline{P}_{s,x}(\omega_\alpha)$ at $(\overline{s}, \overline{\tilde{x}}) = \omega_\alpha$, which, changes its drift functions: $a^\omega = a(\omega) \xrightarrow{\overline{P}_{s,x}(\omega_\alpha)} \overline{a}^\omega = a(\omega_\alpha)$.

That switches Markovian movement from its current drift $a^\omega = a(\omega)$ to Markovian movement under another drift $\overline{a}^\omega = a(\omega_\alpha)$. The random sequence 0-1-0-1-0-0-1-…of the probabilistic actions $\downarrow\uparrow$ affects probabilities $\overline{P}_{s,x}(d\omega), \overline{P}_{s,x}(d\omega_\alpha)$ which through Markov $p(\omega)$ randomly switch the process movement accordingly.

An example is *like* an infinite sequence of coin-tosses tail events, hitting a table that randomly moves it at each interaction with the table.

*The Bayes' probabilities rules.*

For each $i,k$ random event $A_i, B_k$ along the observing events, each conditional a priori probability $P(A_i / B_k)$ follows conditional a posteriori probability:
$P(B_k / A_i) = P(A_i / B_k) / P(A_i)$.

Substituting $P(A_i / B_k) = P(A_i \cup B_k) / P(B_k)$ we get
$P(B_k / A_i) = P(A_i \cup B_k) / P(A_i) P(B_k)$,
where average probability of expecting events along the observing events is:

$$P(B_k) = \sum_{i=1}^{n} P(B_k / A_i) P(A_i).  \qquad (1.11)$$

Ratio a priory to a posteriori probabilities:
$P(A_i / B_k) / P(B_k / A_i) = P(A_i)$

determines current observing probability $P(A_i)$ which may include some observation of previous event $A_{i-1}, A_{i-2}, ..., $.

3. For each $i,k$ random event $A_i, B_k$ along the observing events, each conditional a priori probability $P(A_i / B_k)$ follows conditional a posteriori probability:

$P(B_k / A_i) = P(A_i / B_k) / P(A_i)$.

Substituting $P(A_i / B_k) = P(A_i \cup B_k) / P(B_k)$ we get

$P(B_k / A_i) = P(A_i \cup B_k) / P(A_i) P(B_k)$,

where average probability of expecting events along the observing events is:

$$P(B_k) = \sum_{i=1}^{n} P(B_k / A_i) P(A_i) . \qquad (1.11)$$

Ratio a priory to a posteriori probabilities:

$P(A_i / B_k) / P(B_k / A_i) = P(A_i)$

determines current observing probability $P(A_i)$ which may include some observation of previous event $A_{i-1}, A_{i-2},...,$.

*Applying the Byes formulas and the link to Kolmogorov law for the Markov diffusion process.*

Defining a *priori probability* $P_{s,x}^a(d\omega) = P_{s,x}(d\omega)$ and a *posteriori* $P_{s,x}^p(d\omega)$ probabilities by Bayes rule, we *link* the Kolmogorov 0-1 and Bayes probabilities through the Markov diffusion process.

The switch of Markov drifts $a^\omega \downarrow \bar{a}^\omega$ models No-Yes actions 0-1, while the switch of the drifts $\bar{a}^\omega \uparrow a^\omega$ models Yes-No actions 1-0 of these impulses.

Both impulses, acting on Markov process, change it's a priori probability to a posteriori probability. The impulse $\downarrow\uparrow$ replicates a random bit which the 0-1 probabilities cover, while the switches reveal the interactive link of the Kolmogorov' 0-1 to the Bayes probabilities *within* the Markov diffusion process.

The Markov process probabilities $P_{s,x}^a(d\omega_\alpha)$ and $P_{s,x}^p(d\omega_\alpha)$ correlate under its drift function and diffusion. Increasing $\bar{P}(\bar{s}, \bar{\tilde{x}}, t, B)$ raises probability density $p(d\omega_\alpha) = P_{s,x}^p(d\omega_\alpha) / P_{s,x}^a(d\omega_\alpha)$ which grows the correlations and increases closeness of these probabilities to density $p(\omega)$, finally to $p(d\omega_\alpha) \to 1$.

Rising $p(d\omega_\alpha)$ grows a posteriori probability of revealing actual drift $\bar{a}^\omega = a(\omega_\alpha)$ moving the process. That increases probability of impulse $\downarrow\uparrow$ up to certain Bit acting really on $\bar{a}^\omega = a(\omega_\alpha)$ which processing real movement.

Each considered probability is abstract axiomatic Kolmogorov's probability. It predicts probability measurement on the experiment whose probability distributions, tested by relative frequencies of occurrences of events, satisfy condition of symmetry of the equal probable events [20]. In the theory of randomness, each events' probability is *virtual*, or, at every instance, prescribed to such imaginary event, many of its potential probabilities might occur simultaneously. However, an explicit probability describes a physical possibility of some of them. The multiple random actions describe the probability distributions on the observing sequence of specific set of events, which formally define the observing triple in the probability field. Processing interactions $\varpi$ along $X_t$ possesses a common time course in the field, which for each field ensemble the fractions of random events, being a part of the time in the common time course.

We assume the random field conceals a randomly distributed energy which the events hold.

The probability field of the concealed energy is timeless, reversible, symmetrical, and scalar.

Random interaction, disturbing the field, randomly reveals the energy, which the interaction or its measurement acquires from the field.

**1.2. The notion of observation, observing process, virtual observer, and uncertainty.**

The objective Kolmogorov 0-1 probabilities quantify the idealized (virtual) impulses whose No-Yes actions represent an act of a virtual observation of Markov a priori probability $P_{s,x}^a(d\omega) = P_{s,x}(d\omega)$ shifting to Markov a posteriori probability $P_{s,x}^p(d\bar{\omega})$ during the impulse finite time interval $\delta\tau$.

Each observation measures a probability of possible events for a potential observer.

Thus, the linking a *priori* and *a posteriori* Bayes probabilities of the impulses virtually observe the Markov diffusion process on its trajectory which is moving under the drift.

The transitional probabilities on the trajectories of the Markov diffusion process, transit a multiple virtual observations along a *virtual observing process* $\bar{x}_t$.

The objective probabilities measure the virtual observation probabilities, which along the observing process *measure* the observable process of a potential *(virtual) observer.*

Each such interactive No-Yes actions provide a step-down action (0) and step up action (1) within the impulse.

Let us substantiate both the probabilistic measurement and the measure of the virtual observer.

For each $i$ random events $A_i$ of the observing process, its current a priori probability $\bar{P}_{s,x}(A_i)$ includes previous observed probabilities tracking back to $(n-1)$ transitional probability.

For the Markov process with the $i$-events' transitional probability $P_i(s,\tilde{x},t,B)$, a generalized (integral) a priori probability, which integrates all accessible a priori probabilities has form

$$\bar{P}_{s,x}(A_i) = \int_{s,t,\tilde{x},B,} \prod_{i=1}^{n-1} P_i(s,\tilde{x},t,B), \qquad (2.1)$$

where $\bar{P}_{s,x}(A_i) = P_{s,x}^a(d\omega) = P_{s,x}(d\omega)$ and generalized distribution (2.1) is defined on $\sigma$-algebra events $A_i = A_i(s,\tilde{x},t)$ [29]. Each of these equal probabilities is defined on the same Markovian random states-events and are *observing* in the Markovian process $\bar{x}_t$. The link automatically includes No-Yes actions in the emerging Markov observing process.

Ratio of a posteriori to priori Bayes probabilities defines a posterior probability density

$$\bar{p}(\omega) = \frac{P_{s,x}^p(d\omega)}{P_{s,x}^a(d\omega)} = \frac{P_{s,x}^p(d\omega)}{\bar{P}_{s,x}(A_i)} \ . \qquad (2.2)$$

Each following observation updates the posteriori density $\bar{p}(d\omega)$ as well as the previous observations.

Substituting $\tilde{P}_{s,x}(d\omega) = P_{s,x}(d\omega)p(d\omega)$ to $P_{s,x}^p(d\omega) = \bar{p}(d\omega)P_{s,x}(d\omega)$ leads to

$$P_{s,x}^p(d\omega) = \tilde{P}_{s,x}(d\omega)\bar{p}(d\omega)/p(d\omega) \qquad (2.3)$$

at $\bar{p}(d\omega)/p(d\omega) = P_{s,x}^p(d\omega)/\tilde{P}_{s,x}(d\omega)$. $\qquad (2.4)$

In relations (2.4), all probabilities including $P_{s,x}^p(d\omega)$ are defined directly on the same Markov process.

It allows updating both priori and posteriori probabilities during its observation of the moving observing process.

However, Markov process defines probabilities of two time-steps: one head and one down, which in formula (2.2) limits number of each current observation $i$ by $i=1$ and $i=2$, or $i,i-1$, or $i,k$ in (1.11), which holds two current impulses.

Each previous observations decreases a current priori probability $\bar{P}_{s,x}(A_i)$ in (2.1) at each fixed $i$, since multiplication of the probabilities each less than 1, decreases the total multiplied probability in (2.1) and $\bar{P}_{s,x}(A_i)$. Integration each of these small process' increments did not change the decreasing. That increases $\bar{p}(d\omega)$ updating it through a prior observation.

The posterior observation also updates the posteriori density $\bar{p}(d\omega)$, growing with rising $P_{s,x}^p(d\omega)$.

*Example.* Suppose a process stops at each $P=0$ and the process starts at each $P=1$ and let at step $i=1$ have

$$\bar{P}_{s,x}(A_{i=1}) = P_{s,x}^a(d\omega) = P_{s,x}(d\omega), P_{s,x}(d\omega) \xrightarrow{P=0} \bar{P}_{s,x}(d\omega) = 0.5, p(\omega) = 0.8, \tilde{P}_{s,x}(d\omega) = P_{s,x}^p(d\omega) = 0.4.$$

At step $i=2$, it holds $P_{s,x}(d\omega) \xrightarrow{P=1} \bar{P}_{s,x}(d\omega_\alpha) = P_{s,x}^a(d\omega_\alpha) = 0.55, \bar{p}(\omega_\alpha) = 0.81, \tilde{\bar{P}}_{s,x}(d\omega_\alpha) = P_{s,x}^p(d\omega_\alpha) = 0.4455,$ where $P_{s,x}^a(d\omega_\alpha)$ updates $\bar{P}_{s,x}(A_{i+1})$ which includes $\bar{P}_{s,x}(A_{i=1}) = 0.5$.

That changes the density in (2.4). Integral (2.1) from discrete $i=1$ to $i=2$ we approximate by sum of multiplications: $\bar{P}_{s,x}(A_{i+1}) = P_{s,x}^a(d\omega) \times \bar{P}_{s,x}(d\omega_\alpha) + P_{s,x}^a(d\omega) \times \bar{P}_{s,x}(d\omega_\alpha)$ which brings $\bar{P}_{s,x}(A_{i+1}) = 0.5 \times 0.55 + 0.5 \times 0.55 = 0.55$, and
$$\bar{p}(\omega_\alpha) = P_{s,x}^p(d\omega_\alpha) / \bar{P}_{s,x}(A_{i+1}) = 0.4455/0.55 = 0.81. \tag{2.5}$$

Here probability $P_{s,x}^p(d\omega_\alpha)$ is updating the probability of Markovian drift $a(\omega_\alpha)$ as well as $P_{s,x}^a(d\omega_\alpha)$ is doing it. •

Results [41] prove that cutting the impulse correlation increases its observing Markov probability density and runs the next impulse cutting it off. Each cutoff analytically runs Dirac delta-action or Kronicker 0-1 action.

Let define Bayes entropy on the observing process in form
$$S_B = E_{s,x}[-\ln \bar{p}(\omega)] \tag{2.6}$$
where $E_{s,x}$ is the conditional mathematical expectation taking along this process.

Applying to (2.6) formula (2.2) brings
$$S_B = -\int_{\bar{x}_t \in B} \ln \bar{p}(\omega) P_{s,x}^a(d\omega). \tag{2.7}$$

This formula also concurs with Bayes (conditional) probability in form (1.11) applied to (2.6).

Entropy $S_B$ measures uncertainty of the observing process; $S_B$ also measures a virtual observer uncertainty, which is observing that process.

Maximal uncertainty measures non-correlating a priori-a posteriori probabilities, when their connection approaches zero. Such theoretical uncertainty has infinite entropy measures, whose conditional entropy and time do not exist.

The finite uncertainty measure has non-zero correlating finite a priori–a posteriori probabilities with a finite time interval and the following finite conditional entropy.

Example of such finite uncertainty' process is "white noise". Using it allows measuring the observing process' uncertainty relative to uncertainty-entropy of the white noise (Sec1.1).

Measuring conditional entropy (1.10) relatively to diffusion process $\varsigma_t$ is usable, since average $\varsigma_t$ models zero observing process beginning potential observer. When a Yes(1)–No(2) probable impulse affects this process, the impulse of observing process starts a potential observer as reaction on the observation.

**1.3. Certainty, Information, Cutting Entropy**

If an elementary Kronicker' impulse increases each observing Bayes a posteriori probability, it concurrently increases probability of each virtual impulse (up to real impulse with posteriori probability 1) and decreases the related uncertainty.

Information, as notion of certainty–opposed to uncertainty, is measuring a reduction of the observing uncertainty toward maximal posteriori probability 1, which, we assume, evaluates an observing probabilistic fact.

Each observing impulse cuts entropy of an equivalent impulse of the initial Markov process. The cutting entropy decreases the Markov impulse entropy and increases the entropy of observing impulse. Such multiple interactions minimizes uncertainty of the initial Markov process and maximizes entropy of each following observing impulse.

The observing impulse accumulates maximum entropy when the Markov impulse entropy reach minimum.

Or the Markov impulse acquires maximum probability when the observing impulse reach maximum uncertainty.

Measuring maximum Markov transitional probability indicates the rapprochement of the observing impulse to a moment of cutting its maximal uncertainty, which extracts the certain observing bit.

(When the potential observer obtains bit through an interacting copy within the observer inner process, it excludes measuring uncertainty of the observing external impulse.)

Finally erasure of collected entropy–uncertainty during the impulse observation restores the impulses carrying certain Bits. To rebuild a certain physical bit, the injection of Landauer's energy requires for the erasure. When the Markov impulse' probability reach maximum, its interactive action opens access of this energy, which memorizes the physical bit.

Since such bit becomes asymmetrical, the multiplicity of such bits memorizes specific logic of each observing process.

### 1.3.1. The virtual observer probing impulses with frequencies. Certain Information Observer.

Since $\bar{p}(d\omega)$ increases directly from the growing impulse observations, $S_B$ tends to decreases compared to $S$.

Therefore, impulse observation minimizes uncertainty, from which automatically emerges *principle minimum* entropy:

$$\min_{n-1 \leftarrow i} S_{Bi} = S_{Bo} . \qquad (3.1)$$

At $\bar{p}(d\omega) > p(d\omega)$, $S_B < S$ and this principle, applied to $S$:

$$\min_{a^\omega \to a^\varpi} S \to S_B , \qquad (3.2)$$

generates $S_B$ and therefore the virtual observer. Removing uncertainty generates a certain information Observer.

*The impulse probabilistic description generalizes both a potential random interaction, covering a Bit, and multiple interactive impulse actions running a random process, which unify the Bit's common information origin. During the interactive impulse observations emerge reduction of uncertainty that minimizes the growing impulse observation.*

Moreover, as shown in [41], [44], the impulse observation leads to max-min principle which increases probability reducing each following uncertainty.

Let's specify how a virtual observer, applying the impulses, decreases uncertainty $S_B$, rising to information observer.

To observe, virtual observer applies to the observing process a sequence probing impulses with a frequency $f_m^i$.

The virtual observer, sending the impulse probes sequence with this frequency, generates its observable process, which provides experimental probability test of each axiomatic Kolmogorov probability in the observing process [45, 71].

Using this frequency probes, the virtual observer also provides an experimental probability test of each axiomatic Kolmogorov probability in the observable process.

Probability of random events (inter-actions) defines ratio of a favorable event numbers to total number of events. This ratio measures frequency of the multiple events. When this ratio grows, satisfying symmetry, or equal possibility for the experimental frequencies, independence of the events, and Big Number Law, defining frequency stability, then formulas below counts probability of the random events for that experimental frequency.

Specifically with occurrence in multiple tests $m$-events frequencies $f_m^i$ in $n$-sequence they approach experimental probability

$$E[f_m^i]_{n \to \infty} = P_m^i, \qquad (3.3)$$

where mathematical expectation for these frequencies theoretically satisfies to Big Number Law (which is fulfilling also for probability $P_{\bar{\omega}}[f(\varpi) = 0]$).

Hence the multiple interactions frequencies produce the random events' probabilities, while a sequence of the independent random events, with conditional probabilities depending only on previous event, forms a random process- Markov chain [19]. Existence of trajectories of stochastic process satisfy limitation [20, p.44].

Multiple frequencies, associated with repeating probing impulses, emerge in observation, while their resonances intensify the following a posteriory probability grows and the appearance of a locally stable virtual observer.

The impulse observations replicate frequencies in the observable process, testing the objective probabilities during the observations.

The virtual impulses link the observable process probabilities with the objective probabilities in the observing process which consequently leads to connection of the process correlations, conditional probabilities, conditional entropies, and time intervals of correlations. Both processes describe the Markov process model.

The probing impulses growing frequencies not only initiate probabilities' growth, but also deliver related rotating speed ($W=2\pi f$) to each action, which curves them.

The linking Kolmogorov law and Bayes probabilities, the virtual Dirac-Kronicker impulse representation, and random observation moving the impulses minimize the observing uncertainty comparing with random statistical observation.

### 1.3.2. Evaluation of the cutting process EF fractions by an impulse control

Let us define control on the space $KC(\Delta, U)$ of a piece-wise continuous step functions $u_t$ at $t \in \Delta$:

$$u_- \stackrel{def}{=} \lim_{t \to \tau_k - o} u(t, \tilde{x}_{\tau_k}), \quad u_+ \stackrel{def}{=} \lim_{t \to \tau_k + o} u(t, \tilde{x}_{\tau_k}) \tag{2.1}$$

which are differentiable on the set

$$\Delta^o = \Delta \setminus \{\tau_k\}_{k=1}^m, \quad k = 1, ..., m, \tag{2.1a}$$

and applied on diffusion process $\tilde{x}_t$ from moment $\tau_{k-o}$ to $\tau_k$, and then from moment $\tau_k$ to $\tau_{k+o}$, implementing the process' transformations $\tilde{x}_t(\tau_{k-o}) \to \varsigma_t(\tau_k) \to \tilde{x}_t(\tau_{k+o})$; $n$-dimensional process holds $m$ such transformations.

At a vicinity of moment $\tau_k$, between the jump of control $u_-$ and the jump of control $u_+$, we consider a control *impulse*

$$\delta u_\pm(\tau_k) = u_-(\tau_{k-o}) + u_+(\tau_{k+o}). \tag{2.2}$$

The following statement evaluates the EF information contributions at such transformations.

<u>Proposition 1.</u>

Entropy functional (1.1.10) at the switching moments $t = \tau_k$ of control (2.2) takes values

$$\Delta S[\tilde{x}_t(\delta u_\pm(\tau_k)] = 1/2, \tag{2.3}$$

and at locality of $t = \tau_k$: at $\tau_{k-o} \to \tau_k$ and $\tau_k \to \tau_{k+o}$, produced by each of the impulse control's step functions in (2.1), is estimated by

$$\Delta S[\tilde{x}_t(u_-(\tau_k)] = 1/4, \; u_- = u_-(\tau_k), \; \tau_{k-o} \to \tau_k \tag{2.3a}$$

and

$$\Delta S[\tilde{x}_t(u_+(\tau_k)] = 1/4, u_+ = u_+(\tau_k), \tau_k \to \tau_{k+o}. \tag{2.3b}$$

*Proof.* The jump of control function $u_-$ in (2.1) from moment $\tau_{k-o}$ to $\tau_k$, acting on the diffusion process, might cut off this process after moment $\tau_{k-o} \to \tau_k$.

The cutoff diffusion process has the same drift vector and the diffusion matrix as the initial diffusion process.

The additive functional for this cut off has the form [29]:

$$\varphi_s^{t-} = \begin{cases} 0, t \leq \tau_{k-o} \\ \infty, t > \tau_k \end{cases}. \tag{2.4}$$

The jump of the control function $u_+$ (2.1) from $\tau_k$ to $\tau_{k+o}$ might cut off the diffusion process *after* moment $\tau_k \to \tau_{k+o}$ with the related additive functional

$$\varphi_s^{t+} = \begin{cases} \infty, t > \tau_k \\ 0, t \leq \tau_{k+o} \end{cases}. \tag{2.5}$$

For the control impulse (2.2), the additive functional at a vicinity of $t = \tau_k$ acquires the form of an *impulse function*

$$\varphi_s^{t-} + \varphi_s^{t+} = \delta\varphi_s^{\mp}, \qquad (2.6)$$

which summarizes (2.3) and (2.4).

Entropy functional (1.10) following from (2.4-2.5) takes values

$$\Delta S[\tilde{x}_t(u_-(t \leq \tau_{k-o}; t > \tau_k))] = E[\varphi_s^{t-}] = \begin{cases} 0, t \leq \tau_{k-o} \\ \infty, t > \tau_k \end{cases}, \qquad (2.7a)$$

$$\Delta S[\tilde{x}_t(u_+(t > \tau_k; t \leq \tau_{k+o}))] = E[\varphi_s^{t+}] = \begin{cases} \infty, t > \tau_k \\ 0, t \leq \tau_{k+o} \end{cases}, \qquad (2.7b)$$

from 0 to $\infty$ acquiring *absolute maximum* at $t > \tau_k$, and back from $\infty$ to 0, acquiring *absolute minimum* at $\tau_{k-o}$ and $\tau_{k+o}$.

The multiplicative functional [34], related to (2.4-2.5), are:

$$p_s^{t-} = \begin{cases} 0, t \leq \tau_{k-o} \\ 1, t > \tau_k \end{cases}, \quad p_s^{t+} = \begin{cases} 1, t > \tau_k \\ 0, t \leq \tau_{k+o} \end{cases}. \qquad (2.8)$$

Control impulse (2.2) provides an impulse probability density in form of multiplicative functional

$$\delta p_s^{\mp} = p_s^{t-} p_s^{t+}, \qquad (2.9)$$

where $\delta p_s^{\mp}$ holds $\delta[\tau_k]$-function, which determines probabilities

$$\tilde{P}_{s,x}(d\omega) = 0 \text{ at } t \leq \tau_{k-o}, t \leq \tau_{k+o}, \text{ and } \tilde{P}_{s,x}(d\omega) = P_{s,x}(d\omega) \text{ at } t > \tau_k. \qquad (2.9a)$$

For the cutoff diffusion process, transitional probability (at $t \leq \tau_{k-o}$ and $t \leq \tau_{k+o}$) turns to zero, and states $\tilde{x}(\tau_k - o), \tilde{x}(\tau_k + o)$ become independent, while their mutual time correlations *are dissolved*:

$$r_{\tau_k-o, \tau_k+o} = E[\tilde{x}(\tau_k - o), \tilde{x}(\tau_k + o)] \to 0. \qquad (2.10)$$

Entropy increment $\Delta S[\tilde{x}_t(\delta u_\pm(\tau_k))]$ of additive functional $\delta\varphi_s^{\mp}$ (2.5), produced within or at a border of control impulse (2.2), defines equality

$$E[\varphi_s^{t-} + \varphi_s^{t+}] = E[\delta\varphi_s^{\mp}] = \int_{\tau_{k-o}}^{\tau_{k+o}} \delta\varphi_s^{\mp}(\omega) P_\delta(d\omega), \qquad (2.11)$$

where $P_\delta(d\omega)$ is a probability evaluation of the impulse $\delta\varphi_s^{\mp}$.

Integral of symmetric $\delta$-function $\delta\varphi_s^{\mp}$ between the above time intervals on the border is

$$E[\delta\varphi_s^{\mp}] = 1/2 P_\delta(\tau_k) \text{ at } \tau_k = \tau_{k-o}, \text{ or } \tau_k = \tau_{k+o}. \qquad (2.12)$$

The impulse, produced by deterministic controls (2.2) for each process dimension, is random with probability

$$P_{\delta c}(\tau_k) = 1, k = 1, ..., m \qquad (2.13)$$

at each $\tau_k$-locality.

This probability holds a jump-diffusion transition probability in (2.12) (according to [35]), which is conserved during the jump. For each jump, condition (2.4) leads to $a^u(t, \tilde{x}_t) \to \infty, or \sigma(t, \tilde{x}) \to 0$, while the second one contradicts (1.1.9).

*Therefore, each jump increases Markov speed up to infinitely within a finite impulse.*

From (2.11)-(2.13) it follows estimation of the EF increment under impulse control (2.2) applying at $t = \tau_k$ in form

$$\Delta S[\tilde{x}_t(\delta u_\pm(\tau_k))] = E[\delta\varphi_s^{\mp}] = 1/2 \qquad (2.14)$$

which proves (2.3), while delta impulse $\delta\varphi_s^{\mp} \to \infty$ brings absolute maximum to (2.14) within each $k$ cutoff impulse.

Symmetrical entropy contributions (2.6) at a vicinity of $t = \tau_k$:

$$E[\varphi_s^{t-}] = \Delta S[\tilde{x}_t(u_-(t \leq \tau_{k-o}; t > \tau_k))] \qquad (2.15a)$$

$$E[\varphi_s^{t+}] = \Delta S[\tilde{x}_t(u_+(t > \tau_k; t \leq \tau_{k+o}))] \qquad (2.15b)$$

estimate relations

$$\Delta S[\tilde{x}_t(u_-(t \leq \tau_{k-o}; t > \tau_k))] = 1/4, u_- = u_-(\tau_k), \tau_k \to \tau_{k-o}, \quad (2.16a)$$

$$\Delta S[\tilde{x}_t(u_+(t > \tau_k; t \leq \tau_{k+o}))] = 1/4, u_+ = u_+(t > \tau_k), \tau_k \to \tau_{k+o}, \quad (2.16b)$$

which proves (2.3a,b).

Entropy functional (1.1.10), defined through Radon-Nikodym's probability density measure (1.1.3), holds all properties of the considered cutoff controllable process, where both $P_{s,x}$ and $\tilde{P}_{s,x}$ are defined.

Thus, cutting correlations (2.10) extracts entropy of hidden process information which directly measures each $\delta$-cutoff:

$$\Delta I_k[\tilde{x}_t(\delta u(\tau_k))] = \Delta S[\tilde{x}_t(\delta u_\pm(\tau_k))] = 1/2. \quad (2.17)$$

The known information measures do not provide such measuring.

According the definition of entropy functional (1.1.5), it is measured in natural $\ln$ where each its Nat equals $\log_2 e \cong 1.44 bits$; therefore, it does not using Shannon entropy measure. •

Corollaries

From the Proposition it follows that:

(a)-Stepwise control function $u_- = u_-(\tau_k)$, implementing transformation $\tilde{x}_t(\tau_{k-o}) \to \varsigma_t(\tau_k)$, converts the EF from its minimum at $t \leq \tau_{k-o}$ (2.16a) to maximum at $\tau_{k-o} \to \tau_k$ (2.17);

(b)-Stepwise control function $u_+ = u_+(\tau_k)$, implementing transformation $\varsigma_t(\tau_k) \to \tilde{x}_t(\tau_{k+o})$, converts the EF from its maximum at $\tau_{k-o} \to \tau_k$ (2.17) to minimum at $\tau_k \to \tau_{k+o}$ (2.16b);

(c)-Impulse control function $\delta u^\mp_{\tau_k}$, implementing transformations $\tilde{x}_t(\tau_{k-o}) \to \varsigma_t(\tau_k) \to \tilde{x}_t(\tau_{k+o})$, switches the EF from its minimum to maximum and back from maximum to minimum, while the maximum of the entropy functional at a vicinity of $t = \tau_k$ allows the impulse control to deliver *maximal amount* of information (2.17) from these transformations;

(d)-The dissolving (cutting) correlation between the process' cutoff points (2.10) cuts *functional connections* at these discrete points, which border the Feller kernel measure [31];

(e)-The relation of that measure to additive functional in form (1.1.7) allows evaluating the *kernel's information* by the EF (1.1.5). The jump action (2.2) on Markov process, associated with "killing its drift", selects the Feller measure of the kernel [36, 37], while the cutoff of EF *provides information measure* of the Feller kernel (2.17);

(f)-Stepwise control $u_- = u_-(\tau_k)$, transferring the EF from $\tau_{k-o} \to \tau_k$, maximizes by moment $\tau_k$ the minimal information increment (brought at $t \to \tau_{k-o}$), implementing condition

$$\max_{\tau_k} \min_{\tau_{k-o}} \Delta I_k[\tilde{x}_t(\delta u(\tau_k))]; \quad (2.17a)$$

(g)-Stepwise control $u_+ = u_+(\tau_k)$, transferring the EF from $\tau_k \to \tau_{k+o}$, kills the additive functional at stopping moment $\tau_{k+o}$ minimizing the maximal information increment by the end of this transformation, implementing condition

$$\min_{\tau_{k+o}} \max_{\tau_k} \Delta I_k[\tilde{x}_t(\delta u(\tau_k))]. \quad (2.17b)$$

Such transformation associates with killing the Markovian process at the rate of increment of related additive functional $d\varphi_s^{ti}/\varphi_s^{ti}$ for each single dimension $i$ [34].

Control $u_+ = u_+(\tau_k)$ transfers the rate of killed Markov process to a process with probability (2.13), which is conserved during the jump, and starts a maximal probable (non-random) process with the eigenvalue of diffusion operator [38].

That process balances the killing at the same rate [39].

The step-wise controls, acting on the multi-dimensional diffusion process dimensions, sequentially stops and starts the process, evaluating information of multiple functional, and changes probabilities 0-1 of each impulse.

The dissolved element of the correlation matrix at these moments provides independence of the cutting off fractions, leading to orthogonality of their correlation matrix •.

*Markov transitional probabilities (1.2) acting on the additive functional with drift function $a^u = a(x,t,u)$ ( which is an equivalent of random $a^\omega = a(\omega)$ ), transform in to drift function $a^{\bar{\omega}} = a(\bar{\omega})$ (Sec.1.1) and to equivalent EF measure under the No-Yes actions. Thus, the jump of Markov probabilities densities, cutting the additive functional, runs the controls.*

### 1.3.3. The impulse action on the entropy integral

**1.3.3.1. The regular entropy integral functional**

The EF integrant in (1.1.10) is *partially observable* via measuring only covariation function on the process' trajectories. For a single dimensional Eq. (1.1.6) with drift function $a = c\tilde{x}(t)$ at given nonrandom function $c = c(t)$ and diffusion $\sigma = \sigma(t)$ entropy functional (1.1.10) acquires form

$$S[\tilde{x}_t / \varsigma_t] = 1/2 \int_s^T E[c^2(t)\tilde{x}^2(t)\sigma^{-2}(t)]dt, \quad (3.1.1)$$

from which, at $\sigma(t)$ and nonrandom function $c(t)$, we get

$$S[\tilde{x}_t / \varsigma_t] = 1/2 \int_s^T [c^2(t)\sigma^{-2}(t)E_{s,x}[x^2(t)]dt = 1/2 \int_s^T c^2[2b(t)]^{-1}r_s dt, \quad (3.1.2)$$

where for the diffusion process, the following relations hold true:

$$2b(t) = \sigma(t)^2 = dr/dt = \dot{r}_t, E_{s,x}[x^2(t)] = r_s. \quad (3.1.3)$$

This allows *identifying* the EF on observed Markov process $\tilde{x}_t = \tilde{x}(t)$ by measuring above correlation functions, applying positive function $u(t) = c^2(t)$, and *representing* functional (1.1.10) through the regular integral of non-random functions

$$A(s,t) = [2b(t)]^{-1}r_s = \dot{r}_t^{-1}r_s \quad (3.1.4a) \quad \text{and} \quad u(t) = c^2(t) \quad (3.1.4b)$$

in form

$$S[\tilde{x}_t / \varsigma_t] = 1/2 \int_s^T u(t)A(t,s)dt. \quad (3.1.4)$$

The *n*-dimensional form of functional integrant (3.1.4) follows directly from the related *n*-dimensional covariations (1.1.3), dispersion matrix, and applying *n*-dimensional function $u(t)$.

At given nonrandom function $u(t)$, regular integral, (3.1.4) measures the entropy functional of Markov process at the probability transformation (1.1.2) with additive functional (1.1.7), where the integrant averages the additive functional.

<u>Proposition 3.1.</u>

Integral (3.1.4), satisfying variation condition (1.15) at linear function $c^2(t) = u \bullet t = c^2 \bullet t$, forms

$$S[\tilde{x}_t / \varsigma_t] = 1/2 \int_s^T u(t)o(t)dt, \quad (3.1.5)$$

where the extreme of function (3.1.4a) holds minimum

$$A(t, s_k^{+o}) = o(s)b_k(s_k^{+o})/b_k(t) = o(t), \quad (3.1.5a)$$

which decreases with growing time $t = s_k^{+o} + o(t)$ at $t \to T$ and fixed both $b_k(s_k^{+o})$ and

$$o(s) = A(s,s). \quad (3.1.5b)$$

Since satisfaction of this variation condition includes transitive transformation of a current distribution to that of Feller kernel, $b_k(t)$ is transition dispersion at this transformation, which is growing with the time of the transformation.

<u>Proposition 3.2.</u>

Entropy integral (3.1.15) under impulse control $c^2(t, \tau_k) = \delta u_t(t - \tau_k)$ takes the following information values:

(a)-at a switching impulse middle locality $t = \tau_k$:
$$S[\tilde{x}_t / \varsigma_t]_{t=\tau_k} = 1/2 \, \text{Nats}, \tag{3.1.6}$$

(b)- at switching impulse left locality $t = \tau_k^{-o}$:
$$S[\tilde{x}_t / \varsigma_t]_{t=\tau_k^{-o}} = 1/4 \, Nats, \tag{3.1.6a}$$

(c)-and at switching impulse right locality $t = \tau_k^{+o}$:
$$S[\tilde{x}_t / \varsigma_t]_{t=\tau_k^{+}} = 1/4 \, Nats. \tag{3.1.6b}$$

*Proof.* Applying delta-function $c^2(t, \tau_k) = \delta u_t(t - \tau_k)$ to integral
$$\Delta S[\tilde{x}_t / \varsigma_t]\big|_{\tau_k^{-o}}^{\tau_k^{+o}} = 1/2 \int_{\tau_k^{-}}^{\tau_k^{+}} \delta u_t(t - \tau_k) o(t) dt, \tau_k^{-o} < \tau_k < \tau_k^{+o}, \tag{3.1.7}$$

determines functions [40,p.678-681]:
$$\Delta S[\tilde{x}_t / \varsigma_t]\big|_{t=\tau_k^{-o}}^{t=\tau_k^{+o}} = \begin{cases} 0, t < \tau_k^{-o} \\ 1/4 o(\tau_k^{-o}), t = \tau_k^{-o} \\ 1/4 o(\tau_k^{+o}), t = \tau_k^{+o} \\ 1/2 o(\tau_k), t = \tau_k, \tau_k^{-o} < \tau_k < \tau_k^{+o} \end{cases} \tag{3.1.8}$$

Or such cutoff brings amount of entropy integral $\Delta S[\tilde{x}_t / \varsigma_t]_{t=\tau_k} = 1/2 o(\tau_k) = 1/2 \, \text{Nats}$, while on borders of interval $o(\tau_k)$ the integral amounts are $\Delta S[\tilde{x}_t / \varsigma_t]_{t=\tau_k^{-o}} = 1/4 o(\tau_k^{-o}) Nats$ and $\Delta S[\tilde{x}_t / \varsigma_t]_{t=\tau_k^{+o}} = 1/4 o(\tau_k^{+o}) Nats$ accordingly. •

The results concur with (2.2.3, 2.2.3a,b). Total delivering entropy is $3/4 Nats$ and to right border is transferring $1/4 Nats$.
Sum of the instances for the cutting interval:
$$\sum_{t=\tau_k^{-o}}^{t=\tau_k^{+o}} \Delta S[\tilde{x}_t / \varsigma_t]_t = 1/4 o(\tau_k^{-o}) + 1/2 o(\tau_k) + 1/4 o(\tau_k^{+o}) = o_k \tag{3.1.9}$$

evaluates constant-invariant $1 Nat$ fraction of the cutoff EF on this interval, which the interval encloses.
The invariant cutting fractions follow from variation condition (1.7) imposed on (3.1.4), which leads to (3.1.9).
The impulse 0-1 probability has minimax probability and related minimax entropy measure: $\Delta S_{P=0} \to -\infty, \Delta S_{P=0} = 0$

that is additive at each virtual–probabilistic EF cuts which integrate the virtual observer with entropy measure $S_B$.

### 1.4. Information path functional
*Information path functional* (IPF) defines distributed actions of multi-dimensional delta-function on entropy functional (1.1.10) through the additive functional for all dimensions:
$$I_{pm} = \delta_m \{S[\tilde{x}_t / \varsigma_t]|\} = 1/2 E\{\int_s^T \delta_m [a(t,\tilde{x})^T (2b(t,\tilde{x}))^{-1} a(t,\tilde{x}) dt)]\} \tag{4.4.1}$$

which determines sum (4.4.2) of information measures $\Delta I_k[\tilde{x}_t(\delta u(\tau_k))]$ along the path on cutting process intervals (3.3.1),(3.3.6). In a limit it leads to
$$I_p = \lim_{m \to \infty} \sum_{k=1}^{m} \Delta I_k[\tilde{x}_t(\delta u(\tau_k))]. \tag{4.4.2}$$

Formal definition (4.4.1) allows the IPF representation by Furies integral [40] leading to the frequency analysis with Furies series.
. The IPF is the sum of *extracted* information which approaches theoretical measure (1.1.10):

$$I_p = \lim_{m \to \infty} I_{mo} \mid_s^T = \lim_{m \to \infty} S_{mo} \mid_s^T \to S[\tilde{x}_t / \varsigma_t]_s^T, \qquad (4.4.3)$$

if all finite time intervals $t_1 - s = o_1, t_2 - t_1 = o_2, ..., t_{k-1} - t_k = o_k, ..., t_m - t_{m-1} = o_m$, at $t_m = T$ satisfy condition

$$(T - s) = \lim_{m \to \infty} \sum_{t=s,m}^{t=T} o_m(t). \qquad (4.4.4)$$

Since according to (3.3.6) each cutting interval (4.4.6) encloses invariant information measure, the limited $I_p$ limits the initially undefined upper time $T$ of the EF integral. It also brings direct connection $I_p$ and $T - s$.

Therefore, at infinite sequence of the time intervals, this sequence has limit

$$\lim_{m \to \infty} o(t_m) \to 0, \qquad (4.4.5)$$

and sum of such sequence is finite [40, p.130, 4.8-1].

Realization (4.4.1), (4.4.4), (4.4.5) requires applying the impulse controls at each instant $(\tilde{x}, s), (\tilde{x}, s + o(s)), ...$ along the process trajectories of conditional math expectation (1.1.10).

However for any *finite* number $m$ of these instants, the *integral* process information, composed from the information, measured for the process' fractions, is not complete.

*The $I_p$ properties*:

1. The IPF measures information of the cutting process's interstate connections hidden by the states correlations, which are not covered by traditional Shannon information measure.

2. Since each cutting $\Delta S_k[\tilde{x}_t(\delta u(\tau_k))] = \Delta I_k[\tilde{x}_t(\delta u(\tau_k))]$ maximizes the cutting information on the path intervals, $I_p$ measures a total (integral) maximal information on this path.

The cutting control provides equal maxmin-minimax information contributions

$$\max_{\tau_k} \min_{\tau_{k-o}} \Delta I_k[\tilde{x}_t(\delta u(\tau_k))] = \min_{\tau_{k+1}} \max_{\tau_k} \Delta I_k[\tilde{x}_t(\delta u(\tau_k))] \qquad (4.4.6)$$

on each path $t_{k-1} \to (\tau_{k-o} \to \tau_k \to \tau_{k+o}) \to t_k$ from cutting $t_{k-1}$ to following cut $t_k$ (*Corollaries a-c).*

3. If each $k$-cutoff "kills" the $m$ dimensional process' correlation at moment $\tau_{k+o}$, then at $m = n$, relations (4.4.1-4.4.6) require infinite process dimensions.

4. At $m = n \to \infty, o_k = t_k - t_{k-1} \to \tau_k$, the process time

$$(T - s) = \lim_{n \to \infty} \sum_{t=s,m}^{t=T} \tau_k(t) \qquad (4.4.7)$$

approaches the summary of the discrete time intervals cutting all correlations on the path.

5. Sequential cuts transform the IPF discrete information contributions from each maximum through minimum to next maximal information contributions

$$\max_{\tau_k} \Delta I_k[\tilde{x}_t(\delta u(\tau_k))] \to \min_{\tau_{k+o}} \Delta I_k[\tilde{x}_t(\delta u(\tau_k))] \to \max_{\tau_k+1} \Delta I_k[\tilde{x}_t(\delta u(\tau_{k+1}))], \qquad (4.4.8)$$

where each next maximum decreases at the cutoff moments

$$\max_{\tau_k+1} \Delta I_k[\tilde{x}_t(\delta u(\tau_{k+1}))] < \max_{\tau_k} \Delta I_k[\tilde{x}_t(\delta u(\tau_k))]. \qquad (4.4.9)$$

Each Dirac delta-function preserves its cutting information (3.3.6).

The information contribution by final interval $o_m$ at its inner ending moment $\tau_{m+o}$, according to (2.2.9), (4.4.8a), satisfies

$$\min_{\tau_{m+o}} \Delta I_m[\tilde{x}_t(\delta u(\tau_m))] \to 0, \qquad (4.4.10)$$

which limits sum (4.4.3) at $m = n \to \infty$.

6. Since EF functional $S[\tilde{x}_t / \varsigma_t]_s^T$ limits growth of $S_{mo} \mid_s^T$ in (4.4.3), it limits the IPF in (4.4.7), (4.4.9); hence, the IPF approaches the EF functional during time (4.4.7) at unlimited increase of the process dimensions.

7. Because upper time $T$ in both the EF integral and IPF functional is limited by (4.4.3),(4.4.4) and (4.4.10), the entropy integral converges in the path functional, and both of them are restricted at the unlimited dimension number.

8. For any of these limitations, EF measure, taken along the process trajectories for time $(T-s)$, limits maximum of total process information, while IPF extracts maximum of the process hidden information during the same time and brings more information than Shannon traditional information measure for multiple states of the process.

9. Information density of cutting impulse
$$I_{ko_k} = I_k[\tilde{x}_t(\delta u(\tau_k))]/\tau_k \quad (4.4.11)$$
grows to absolute maximum at
$$I_{ko_n} \to \infty. \quad (4.4.12)$$
The time transition to each following Nat decreases (satisfying the variation condition) since that Nat integrates all preceding Nats concentrating the integral information in the final IPF Nat, which Feller kernel absorbs.

The IPF formally defines the distributed actions of multi-dimensional delta-function on the EF via the multi-dimensional additive functional (1.1.3), which leads to analytical solution and representation by the Furies series.

Final a finite integral information approaches that generated by the last impulse during the final finite $\tau_{k=n}$, which is Kronicker'impulse taking values 0 and 1 and preserving measure (3.3.6).

This probability measure has applied for the impulse probes on an observable random process, which holds opposite Yes-No probabilities – as the unit of probability impulse step-function [41] preserving the max-min.

Comments. Number $M$ of equal probable possibilities determines Hartley's quantity of information $H = \ln M$ measured on Nats, which for the impulse $M=2$ holds $H = \ln 2 Nat$. The impulse information measured in bits is $I = 1/\ln 2 \ln M = \lg Mbit = 1bit$. The correlation cutting by the impulse brings information $0.75 Nat$ from which $\delta S_u \cong 0.0568 Nat$ delivers the impulse controls. Since each cutoff brings invariant $1 Nat$ (3.3.6), the difference $(1-0.7) Nat \cong 0.3 Nat$ presents "free information" for each cutting impulse.

The IPF integral information evaluates maximal speed of enclosing information (4.4.14) in the finite impulse time interval; the instant of the impulse time that cuts correlation's hidden information equals to Bit = ln 2.

All integrated information enfolds the Feller kernel whose time and energy evaluate results [36].

Minimal physical time interval limits the light time interval $\delta t_\tau \cong 1.33 \times 10^{-15}$ sec defined by the light wavelength $\delta l_m \cong 4 \times 10^{-7} m$. That allows us estimate maximal information density (4.4.14) for 1 bit:
$$I_{ko_k} \cong \ln 2 / 1.33 \times 10^{-15} \cong 5.2116 \times 10^{+15} Nat/s. \quad (4.4.13)$$
Or for each invariant impulse $1 Nat$, the maximal density estimates
$$I_{ko_{k1}} \cong 1/1.33 \times 10^{-15} \cong 7.5188 \times 10^{+15} Nat/s. \quad (4.4.14)$$
Variety of the impulse $\downarrow\uparrow$ physical interactions unites the impulse information model, which EF-IPF integrate.

**1.5. The hidden entropy and origin of information. The EF-IPF measure comparison with other entropy measures.**

The link (Secs.1.1-1.2) automatically includes the random No-Yes impulses into the Markov process.

The notion of information *originates* in the probabilistic entropy' hidden correlation whose cut-erasure produces physical information Bit without necessity of any physical particles.

The EF presents a potential (virtual) information functional of the Markov process until the applied impulse control, carrying the impulse cutoff entropy contributions, transforms it to informational path functional (IPF) [43-45].

Markov random process becomes a source of each information contribution, whose entropy increment of cutting random states delivers information hidden between these states' correlation.

The correlation holds the observation' temporal memory.

The cutting function's finite restriction determines the discrete impulse's step-up and step-down actions within impulses interval $\delta_k = \tau_k^{+o} - \tau_k^{-o}$, which capture entropy hidden between impulses on starting instance $\tau_k^{-o}$, cut it and transfer to ending instance $\tau_k^{+o}$ where the cutting entropy converts to the equivalent physical information and memorizes it.

Here, time moment $\tau_k^{+o}$ holds both information and its memory of cutting correlation. The following cutoff interval $\Delta_t \to o(t)$ after the cut delivers new hidden process' information, and so on between each cut along the multi-dimensional process for each impulse. *The cutoff directly generates transitional probabilities densities* $p_s^{t-}(0,1) \to p_s^{t+}(1,0)$ *in (1.1) for the current moments* $t^-, t^+$ *of the process.*

The cutoff transitional probabilities allow *omitting complete implementation of initial formalism (Sec.1.1).*
*The Bayes probabilities' impulses provide the entropy measure for each impulse uncertainty (Sec.1.3.2), which the EF integrates along the process. The real cutoff converts this entropy to elementary information a Bit. The multiple real No-Yes actions convert the EF to equivalent the IPF measure changing observing random process to information process.*
*The IPF integrates the impulse's cutting of information in information process.*
Information is a physical entity, which distinguishes from entropy which is the observer's virtual-imaginable.
Interactive process within each impulse sequentially connects the imaginable with physical reality along the multi-dimensional cuts. In the observable multi-dimensional Markov process, each probing a priori probability turns to the following a posteriori probability, cutting off uncertainty and converting it to certainty. The real step-wise controls, acting on the process all dimensions, sequentially stops and starts the process, evaluating the multiple functional information collected by the IPF. Impulses delta-function $\delta u_t$ or its discreet Kronicker' $\delta u_{\tau_k}$ implement transitional transformations (1.1), initiating the Feller kernels along the process and extracting total kernel information for $n$-dimensional process with $n$ cuts off. Their maximal sum measures the interstates information connections held by the process along the trajectories during its real time, which are hidden by the random process correlating states.

The EF functional information measure on trajectories is not covered by traditional Shannon entropy measure.
The dissolved element of the process' functional correlation matrix at the cutoff moments provides independence of the process cutting off fractions, leading to the orthogonal correlation matrix for these cutoff fractions.
Cutting probability of random ensemble is symmetrical.
The EF connects the observing a priori and a posteriori probabilities with the observing increment of correlations.
Let us have ratio of a posteriori $P_t(\omega)$ and a priori probabilities $P_s(\omega)$ for elementary events $\omega_o = (s, \tilde{x})$ preceeding current observation of events $\omega_o = (s, \tilde{x})$ and $\omega = (t, \tilde{x})$ with their ratio $P_t(\omega)/P_s(\omega_o) = p_{s,t}(\omega)$ satisfying relation

$$S_{s,t} = -\int_s^\tau \ln(p_{s,t}(\omega))P_s(\omega_o)d\omega = -\int_s^\tau \ln(P_t(\omega)/P_s(\omega_o))P_s(\omega_o)d\omega = 1/2\int_s^\tau u^2 \dot{r}_t / r_s dt, s < t < \tau \; , \tag{5.1}$$

or

$$-\int_s^\tau \ln(P_t(\omega)/P_s(\omega_o))P_s(\omega_o)d\omega = 1/2\int_s^\tau u^2 \dot{r}_t / r_s . \tag{5.2}$$

The considered cutting action allows its equivalent representation by delta-functions $d\omega = \delta(\omega - \omega_o)d\omega_o$ and $u^2 = \delta(t - \tau)$. Applying the first delta-function to left integral (5.2) and second to the right integral instantaneously, leads on the right to ratio of the increment of posterior correlation to a prior correlation $\dot{r}_t / r_s \cong \delta r_t / r_s \delta t$ (where $\delta t$ is time interval of the increment of the correlation at the cut-off action), which directly evaluates the relations for priory and posteriori probabilities for current random events $\omega_o, \omega$ along trajectory $\tilde{x}(s,t)$:

$$\dot{r}_t / r_s = -2P_s(\omega_o)\ln(P_t(\omega)/P_s(\omega_o)) \tag{5.3}$$

And vice versa, these probabilities can identify the ratio of these correlations.

*Example.* At $P_t(\omega)/P_s(\omega_o)=1/4$ it leads to $\dot{r}_t/r_s = -2\times 3.6889 P_s(\omega_o)$.

The connection of the probability density with additive functional's functions drift and diffusion (Eq.1.1.4) allows using these functions for solving Ito Eqs. [26]. The resulting connections through the Ito Eqs.:

$$p(\omega) \to \varphi_s^t(\omega) \to [a(t,x), b(t,\tilde{x})] \to \mathrm{x}_u \tag{5.4}$$

determines Markovian state $\mathrm{x}_u$ which can control other Markov states along this process instead of $p(\omega)$.

The Markov process becomes self-controlling thru the identified self-observing a priori-a posteriori probabilities.

In the considered Markov diffusion process, each local time intervals $\delta t$ spent in vicinity of each random $\tilde{x}$, decrease, and the distance from the origin of the random path inclines. Such Markov process is Levy Walk [21, Ref.18,p.370].

*Let us compare* the EF with definitions of Boltzmann entropy:

$$S_B = k_B \log W \tag{5.5}$$

where $k_B$ is the Boltzmann constant, and W is the number of accessible microstates of a system having a fixed energy, volume, and number of particles.

And let's compare the EF with Boltzmann H-function called H-entropy:

$$H(t) = \int f(v,t) \log[f(v,t)] \mathrm{d}v \tag{5.6}$$

which is defined for a distribution of velocities in volume $f(v,t)$.

The Boltzmann entropy (5. 5) acquires form of Gibbs entropy

$$S_G = k_B \sum_i p_i \log p_i \tag{5.7}$$

when $p_i$ are the probabilities of finding the system with fixed energy and volume, or fixed energy and number of particles in equilibrium.

Both (5.5, 5.7) are time independent, while (5.6) depends on time and volume of the distributed microstates in an equilibrium.

Entropy (5.7) is directly connected with Shannon entropy for any given probability distribution $p_i$ of states in local equilibrium [68]:

$$H = -K \sum_i p_i \log p_i. \tag{5.8}$$

It's seen that all (5.5, 5.7, 5.8) do not integrate random process entropy, while Boltzmann H-entropy measure integrates only deterministic speed of the process molecules in equilibrium, and none on them directly measures time of the measurement.

*The EF integral entropy measure, which average both speed and diffusion of observing Markov process, differs not only from (5.6) but from all (5.5,5.7, 5.8) by providing also the integral time of the measured process' non-equilibrium entropy. The IPF directly measures information along a path of the interacting impulses, currently cutting information.*

**1.6. Imposing the minimax law. Invariant impulse logic. The measure of probabilistic and information causations.**

The impulse delta-function of the step-down cut generates maximal information, while the step-up action delivers minimal information from impulse cut to next impulse step-down within the Markov diffusion process (Sec.1.3.2).

Within each Markovian impulse, the cutting action, delivered by the field, maximizes the cutting entropy of observing impulse, while the reaction, currying minimum of the cutting maximum, minimizes the following Markov impulse entropy. That runs max principle for each observing impulse, and mini principle along the Markov interacting impulses.

Specifically, when a preceding No action cuts a maximum of entropy (and a minimal probability), then following Yes action gains the maximal entropy reduction-its minimum (with a maximal probability) during the impulse cutoff.

The impulse' maximal cutting No action minimizes absolute entropy that conveys Yes action (rising its probability), which leads to a maxmin of relational entropy between the impulse actions transferring the probabilities.

Extracting maximum of minimal impulse information and transferring minimal entropy between impulses express maxmin-minimax principle of converting process entropy to information.

*As soon as the initial impulse 0-1 actions involve, the minimax principle is imposed.*

The variation problem, formulating for this principle and its solution [32], brings invariant entropy increment of each discrete impulse preserving its probability measure, and synchronizes an ad joint local time measure for n-process' dimensions in an absolute time scale. In physical terms, the sequence of opposite interactive actions models reversible micro-fluctuations, produced within observable irreversible macroprocess (like push-pull actions of piston moving gas in cylinder). More simple example, when a rubber ball hits ground, the energy of this interaction partially dissipates that increases interaction's total entropy, while the ball's following reverse movement holds less entropy (as a part of the dissipated), leading to max-min entropy of the bouncing ball. Adding periodically small energy, compensating for the interactive dissipation, supports the continuing bouncing.

The maximin-minimax principle does not contradict the Second Law.

The observing Intervals between the impulses are imaginary-potential for getting information, since no real double controls are applying within these intervals. The minimized increments of the cutting entropy functional between the cutoffs intervals allow prediction each following cutoff with maximal conditional probability.

Under this principle, the observing sequence of the functional a priori-a posteriori probabilities grows providing Bayesian entropy that *measures* the probabilistic causality, which is transforming to *physical* causality when the growing posteriori probabilities of the process approaching its maximum.

The observer logic depends on the sending sequence of probing impulses requested in observation.

The multiple symbols ↓↑ hold inner probabilistic logic following certain logic which integrates the EF-IPF measures.

Each mathematical triad in the field which measures independent sets of the events, start an independent observation.

Each of them defines the initial conditions for the observing process dimension in the multi-dimensional observation.

The EF-IPF of such observing process integrates logic of the multi-dimensional observations.

Sum of cutting information contributions, extracted from the EF, approaches its theoretical measure (1.1.10) which evaluates the upper limit of the sum (4.4.6) of the observing logic.

## 1.7. The arrow of time in the interactive observation

Within the probability field, the emerging initial random time has a discrete probability measure satisfying the Kolmogorov law, which interacts through these probabilities. The random time runs in many directions simultaneously.

The observing Bayes a priori-a posteriori probabilities determine probabilistic arrow of the time course in the observing process with continuous correlations and a probabilistic causality.

In quantum mechanics' microprocess, time is reversible until interaction–measurement (observation) affects quantum wave function. A jumping cut of continuous correlations of merging impulses starts an information microprocess (Sec.2).

The real (physical) arrow of time arises in a macroprocesses which average multiple microprocesses over their local time intervals. Natural arrow of time ascends along the multiple interactions and persists by the process correlations, which carry a correlation' causality. Both virtual and information observers hold own time arrow: the virtual-symmetric, temporal, the information- asymmetric physical which memorizes encoding information in an observer. Even if the time direction holds, a non-locality of quantum microprocess provides reversible time-space holes, when the irreversible time-space process, passing the holes, acquires the field energy. Since particular observation accesses only a fraction of the entire random field, both its time interval and time arrow distinguish along such observations. A common information measure of the process' interactive impulse' actions ↓↑ standardizes a scale of the impulse time intervals measure.

The real time memorizes the process information processing memory and encoding the information process.

The process' persistent probabilistic logic, connecting the time impulses, brings logical causality for the observer time.

## 2. Emerging the Space-Time Observer, Constrains and a Microprocess

### 2.1. The evaluation of an impulse in the interactive observing process

#### 2.1.1. Discrete control action on the entropy functional

Let us find a class of step-down $u_-^t = u_-(\tau_k^{-o})$ and step-up $u_+^t = u_+(\tau_k^{+o})$ functions acting on discrete interval $o(\tau_k) = \tau_k^{+o} - \tau_k^{-o}$, which will preserve the Markov diffusion process' additive and multiplicative functions within the cutting process of each impulse.

Lemma 1.1.

1. Opposite discrete functions $u_-^t$ and $u_+^t$ in form

$$u_-(\tau_k^{-o}) = \downarrow_{\tau_k^{-o}} \bar{u}_-, u_+(\tau_k^{+o}) = \uparrow_{\tau_k^{+o}} \bar{u}_+ \tag{1.1}$$

satisfy conditions of additivity

$$[u_+^t - u_-^t] = U_a \text{ (a) or } [u_+^t + u_-^t] = U_a \text{ (b)} \tag{1.1A}$$

and multiplicativity

$$[u_+^t - u_-^t] \times [u_+^t + u_-^t] = U_m \tag{1.1B}$$

at

$$U_a = U_m = U_{am} = c^2 > 0, \tag{1.1C}$$

where instance-jump $\downarrow_{\tau_k^{-o}}$ has time interval $\bar{u}_-$ and instance jump $\uparrow_{\tau_k^{+o}}$ has high $\bar{u}_+$ for relation (1.1A)(a) at real values

$$\bar{u}_- = 0.5, \bar{u}_+ = 1, \bar{u}_+ = 2\bar{u}_-, \tag{1.2a}$$

and for relation (1.1A) (b) the considered intervals at real values hold

$$\bar{u}_-^o = \bar{u}_+^o = 2. \tag{1.2b}$$

2. Complex functions

$$u_t(u_\pm^{t1}, u_\pm^{t2}), u_\pm^{t1} = [u_+ = (j-1), u_- = (j+1)], j = \sqrt{-1} \tag{1.2c}$$

satisfy conditions (1.1aA), (1.1B) in forms

$$u_+ - u_- = (j-1) - (j+1) = -2, u_+ \times u_- = (j-1) \times (j+1) = (j^2 - 1) = -2,$$

which however do not preserve positive (1.1C). Therefore it holds $c^2 < 0$ and imaginary opposite complex functions

$$u_t(-u_\pm^{t1}) = u_t(u_\pm^{t2}), u_\pm^{t2} = [u_+ = (j+1), u_- = (j-1)], \tag{1.2d}$$

satisfying (1.1bA)-(1.C).

At equal absolute values of actions $|u_+^t| = |u_-^t|$, imaginary functions

$$u_+^t = j\sqrt{2}, u_-^t = -j\sqrt{2} \tag{1.2d1}$$

satisfy only multiplicative part $U_m = u_+^t \times u_-^t = -2$ when the impulse additive meausure holds $U_a = 0$.

Proofs are straight forward.

Assuming both opposite functions apply on borders of impulse interval $o(\tau_k) = (\tau_k^{+o}, \tau_k^{-o})$ in forms

$$u_-^{t1} = u_-(\tau_k^{-o}), u_+^{t1} = u_+(\tau_k^{-o}) \text{ and } u_-^{t2} = u_-(\tau_k^{+o}), u_+^{t2} = u_+(\tau_k^{+o}), \tag{1.2e}$$

at $u_-^{t1} u_+^{t1} = c^2(\tau_k^{-o})$, $u_-^{t2} u_+^{t2} = c^2(\tau_k^{+o})$, $t = \tau_k^{+o}$, (1.2f)

then it follows that only by end of this time interval at $t = \tau_k^{+o}$ both Markov properties (1.1A,B) satisfy, while at beginning $t = \tau_k^{-o}$, the starting process satisfies only (1.1A). ∎

Corollary 1.1.

1. Conditions 1.1A-1.1C imply that $c^2(\tau_k^{-o}), c^2(\tau_k^{+o})$ are discrete functions (1.1a), (1.2f) switching on interval $\Delta_\tau = \tau_k^{+o} - \tau_k^{-o}$.

Requiring $\Delta_\tau = \delta_o$ leads to discrete function $\delta^o u_t$ which for $\delta_o = (\tau_k^{+o} - \tau_k^{-o})$ holds

$\delta^o u_{t=\tau_k} = [u_-(\tau_k^{-o}) - u_+(\tau_k^{+o})]/(\tau_k^{+o} - \tau_k^{-o})$ and using (1.1) and (1.2b) for $\bar{u}_+^o = \bar{u}_+ = 2$ brings

$u_-(\tau_k^{-o}) = -1_{\tau_k^{-o}} \bar{u}_-, u_+(\tau_k^{+o}) = +1_{\tau_k^{+o}} \bar{u}_+$, at $\bar{u}_- = 0.5$, $\bar{u}_+ = 2$, (1.3)

when positivity of $c^2 > 0$ implies equality

$\delta^o u_{t=\tau_k} = [u_+(\tau_k^{+o}) - u_-(\tau_k^{-o})]/(\tau_k^{+o} - \tau_k^{-o}) > 0$. (1.3a)

2. Discrete functions on $\Delta = (\tau_k^{-o} - s_k^{+o})$:

$u_+(s_k^{+o}) = +1_{s_k^{+o}} \bar{u}_+, u_-(\tau_k^o) = -1_{\tau k} \bar{u}_-$ (1.3b)

are multiplicative: $(u_-(\tau_k^{-o}) - u_+(s_k^{+o})) \times (u_-(\tau_k^{-o}) - u_+(s_k^{+o})) = [u_-(\tau_k^{-o}) - u_+(s_k^{+o})]^2$.

2a. Discrete functions (1.2e) in form

$\bar{u}_+ = j\bar{u}, \bar{u}_- = -j\bar{u}, \bar{u} \neq 0$ (1.3c)

satisfy only condition (1.1A) which for functions (1.3b) holds

$[u_-(\tau_k^{-o}) - u_+(s_k^{+o})]^2 = -(j\bar{u})^2[-1_{\tau_k^{-o}} - 1_{s_k^{+o}}]^2 > 0$. ∎ (1.3d)

Let us find discrete analog of the integral increments under *discrete* function (1.3a) in form of a delta-function on $\delta_o$:

$\delta[u_-(\tau_k^{-o}), u_+(\tau_k^{+o})], u_-(\tau_k^{-o}) = -1_{\tau_k^{-o}} \bar{u}_-, u_+(\tau_k^{+o}) = +1_{\tau_k^{+o}} \bar{u}_+$. (1.3e)

Proposition 1.2.

1. Applying discrete delta-function (1.3e) to the EF integral in form (1.3.1.5) leads to

$$\Delta S[\tilde{x}_t / \varsigma_t]\big|_{t=\tau_k^{-o}}^{t=\tau_k^{+o}} = \begin{cases} 0, t < \tau_k^{-o} \\ 1/4 u_-(\tau_k^{-o}) o(\tau_k^{-o}) / \tau_k^{-o}, t = \tau_k^{-o}, 1/4 \downarrow 1_{\tau_k^{-o}} \bar{u}_{ko} \\ 1/2 (u_-(\tau_k^{-o}) - u_+(\tau_k^{+o})) o(\tau_k) / (\tau_k^{+o} - \tau_k^{-o}), t = \tau_k, \tau_k^{-o} < \tau_k < \tau_k^{+o}, 1/2 (\downarrow 1_{\tau_k^{-o}} - \uparrow 1_{\tau_k^{+o}}) \bar{u}_{km} \\ 1/4 u_+(\tau_k^{+o}) o(\tau_k^{+o}) / \tau_k^{+o}, t = \tau_k^{+o}, 1/4 \uparrow 1_{\tau_k^{+o}} \bar{u}_{k1} \end{cases}$$ (1.4)

where

$\bar{u}_{ko} = \bar{u}_- \times o(\tau_k^{-o}) / \tau_k^{-o}, \bar{u}_{km} = (\bar{u}_+ - \bar{u}_-) \times o(\tau_k) / (\tau_k^{+o} - \tau_k^{-o}), \bar{u}_{k1} = \bar{u}_+ \times o(\tau_k^{+o}) / \tau_k^{+o},$

$\bar{u}_{km} = 1/2(\bar{u}_+ - \bar{u}_-) = 0.75, o(\tau_k) = \tau_k^{+o} - \tau_k^{-o}, o(\tau_k^{-o}) / \tau_k^{-o} = 0.5, o(\tau_k^{+o}) / \tau_k^{+o} = 0.1875$ , (1.5)

and $|\bar{u}_- \times \bar{u}_+| = |1/2 \times 2| = |\bar{u}_k| = |1|_k$ is multiplicative measure of impulse $(\downarrow 1_{\tau_k^{-o}} - \uparrow 1_{\tau_k^{+o}}) \bar{u}_k$.

Measuring middle interval in (1.4), (1.5) by single impulse information unit $\bar{u}_k = |1|_k$, determines finite size of the impulse unit parameters $\bar{u}_{ko}, \bar{u}_k, \bar{u}_{k1}$ in (1.5), which estimate value $\bar{u}_{km}$ on the unit border:

$$\bar{u}_{ko} = 0.25 = 1/3\bar{u}_{km}, \bar{u}_{k1} = 2 \times 0.1875 = 0.375 = 0.5\bar{u}_{km} \; . \bullet \tag{1.6}$$

*Proofs* follow from Proposition 1.3 below.

Let us introduce an entropy unit impulse $\bar{u}_s = |1|_s$ with moments $(s_k^{-o}, s_k^o, s_k^{+o})$ prior to impulse $\bar{u}_k = |1|_k$, which measures interval of impulse entropy $\bar{u}_{sm}$.

Then we will find increment of $\Delta S[\tilde{x}_t / \varsigma_t]|_{s_k^+}^{\tau_k^{-o}}$ on border of impulse $\bar{u}_k$ at prior $\Delta_{\tau s+} = \delta_{sk\pm} = (s_k^{+o} - \delta_k^{\tau-})$ and posterior $\Delta_{\tau s-} = \delta_{sk\mp} = (\delta_k^{\tau-} - \delta_k^{\tau+})$ moments under the impulse functions with unit $\bar{u}_s = |1|_s$:

$$\delta^o u_{\tau=(s_k^{+o}-\delta_k^{\tau-})} = (u_+(s_k^{+o}) - u_-(\delta_k^{\tau-}))(s_k^{+o} - \delta_k^{\tau-})^{-1} = \uparrow 1_{s_k^{+o}} \bar{u} - \downarrow 1_{\delta_k^{\tau-}} \bar{u} = [\uparrow 1_{s_k^{+o}} - \downarrow 1_{\delta_k^{\tau-}}]\bar{u}, \tag{1.7}$$

$$\delta^o u_{\tau=(\delta_k^{\tau-}-\delta_k^{\tau+})} = (u_-(\delta_k^{\tau-}) - u_+(\delta_k^{\tau+}))(\delta_k^{\tau-} - \delta_k^{\tau+})^{-1} = \downarrow 1_{\delta_k^{\tau-}} \bar{u} - \uparrow 1_{\delta_k^{\tau+}} \bar{u} = [\downarrow 1_{\delta_k^{\tau-}} - \uparrow 1_{\delta_k^{\tau+}}]\bar{u}, \tag{1.8}$$

$$\delta^o u_{\tau=(\delta_k^{\tau+}-\tau_k^{-o})} = (u_+(\delta_k^{\tau+}) - u_-(\tau_k^{-o}))(\delta_k^{\tau+} - \tau_k^{-o})^{-1} = \uparrow 1_{\delta_k^{\tau+}} \bar{u} - \downarrow 1_{\tau_k^{-o}} \bar{u} = [\uparrow 1_{\delta_k^{\tau+}} - \downarrow 1_{\tau_k^{-o}}]\bar{u}. \tag{1.9}$$

Here $\bar{u}$ evaluates each of these impulse interval, which according to the optimal principle is an invariant.

Since the EF is additive functional, applying functions (1.7)-(1.9) leads to additive discrete sum of its' increments:

$$\Delta S[\tilde{x}_t / \varsigma_t]|_{s_k^+}^{\tau_k^{-o}} = \Delta S[\tilde{x}_t / \varsigma_t]|_{s_k^+}^{\delta_k^{\tau-}} + \Delta S[\tilde{x}_t / \varsigma_t]|_{\delta_k^{\tau-}}^{\delta_k^{\tau+}} + \Delta S[\tilde{x}_t / \varsigma_t]|_{\delta_k^{\tau+}}^{\tau_k^{-o}} \tag{1.10}$$

along time interval

$$\Delta_{\tau sk\pm} = s_k^{+o} - \delta_k^{\tau-} + \delta_k^{\tau-} - \delta_k^{\tau+} + \delta_k^{\tau+} - \tau_k^{-o} = s_k^{+o} - \tau_k^{-o} = \Delta_{\tau s}. \tag{1.10a}$$

Proposition 1.3.

A. The increments of entropy functional (1.10) collected on intervals (1.10a), satisfying (1.4) uder functions (1.7)-(1.9), bring the following entropy contributions:

$$\Delta S[\tilde{x}_t / \varsigma_t]|_{s_k^+}^{\delta_k^{\tau-}} = 1/2(u_+(s_k^{+o}) - u_-(\delta_k^{\tau-}))o(s_k^{+o} - \delta_k^{\tau-}))(s_k^{+o} - \delta_k^{\tau-})^{-1} = 1/2[\uparrow 1_{s_k^{+o}} - \downarrow 1_{\delta_k^{\tau-}}]\bar{u}_{ks} \tag{1.11}$$

at $\bar{u}_{ks} = \bar{u}(o(s_k^{+o} - \delta_k^{\tau-})(s_k^{+o} - \delta_k^{\tau-})^{-1}$; \hfill (1.11a)

$$\Delta S[\tilde{x}_t / \varsigma_t]|_{\delta_k^{\tau-}}^{\delta_k^{\tau+}} = 1/2(u_-(\delta_k^{\tau-}) - u_+(\delta_k^{\tau+}))o(\delta_k^{\tau-} - \delta_k^{\tau+}))(\delta_k^{\tau-} - \delta_k^{\tau+})^{-1} = 1/2[\downarrow 1_{\delta_k^{\tau-}} - \uparrow 1_{\delta_k^{\tau+}}]\bar{u}_{k\delta s}, \tag{1.12}$$

at $\bar{u}_{k\delta s} = \bar{u} \times (o(\delta_k^{\tau-} - \delta_k^{\tau+}))(\delta_k^{\tau-} - \delta_k^{\tau+})^{-1}$ \hfill (1.12a)

and

$$\Delta S[\tilde{x}_t / \varsigma_t]|_{\delta_k^{\tau+}}^{\tau_k^{-o}} = 1/2(u_+(\delta_k^{\tau+}) - u_-(\tau_k^{-o}))o(\delta_k^{\tau+} - \tau_k^{-o})(\delta_k^{\tau+} - \tau_k^{-o})^{-1} = 1`/2[\uparrow 1_{\delta_k^{\tau+}} - \downarrow 1_{\tau_k^{-o}}]\bar{u}_{k\delta}, \tag{1.13}$$

at $[\uparrow 1_{\delta_k^{\tau+}} - \downarrow 1_{\tau_k^{-o}}]\bar{u}_{k\delta} = [\uparrow 1_{\delta_k^{\tau+}} + \uparrow 1_{\tau_k^{-o}}]\bar{u}_{k\delta}$. \hfill (1.13a)

Here each impulse interval acquires specific entropy measure:

$$\bar{u}_{k\delta} = \bar{u} \times (o(\delta_k^{\tau+} - \tau_k^{-o}))(\delta_k^{\tau+} - \tau_k^{-o})^{-1} = \bar{u} \times (o(\delta_k^{\tau+})(\delta_k^{\tau+} - \tau_k^{-o})^{-1}) + \bar{u} \times (o(\tau_k^{-o})(\tau_k^{-o})^{-1}(\delta_k^{\tau+} - \tau_k^{-o})^{-1}\tau_k^{-o} \tag{1.14}$$

on the impulse invariant interval $\bar{u}$.

Relation (1.14) leads to the impulse interval

$$\bar{u}_{k\delta} = \bar{u}_{k\delta o} + \bar{u}_{k\delta 1} \tag{1.14a}$$

with its parts

$$\bar{u}_{k\delta o} = \bar{u} \times (o(\delta_k^{\tau+}))(\delta_k^{\tau+} - \tau_k^{-o})^{-1}), \; \bar{u}_{k\delta 1} = \bar{u}_{ko1} \times \bar{u}_{ko2}, \tag{1.14b}$$

$$\bar{u}_{ko1} = \bar{u} \times (o(\tau_k^{-o}))(\tau_k^{-o})^{-1}, \ \bar{u}_{ko2} = \bar{u}^{-1} \times \tau_k^{-o}(\delta_k^{\tau+} - \tau_k^{-o})^{-1}. \qquad (1.14c)$$

B. Intervals $\bar{u}_{ko1}$ and $\bar{u}_{ko2}$ are multiplicative parts of impulse step-up interval $\bar{u}_{k\delta1}$, which satisfies relations

$$\bar{u}_{k\delta o} = \bar{u}_{k\delta 1} = 1/2\bar{u}_{k\delta}, \ \bar{u}_{k\delta 1} = \bar{u}_{ko1}, \qquad (1.15)$$

where invariant impulse $|\bar{u}_{k\delta}| = |1|_s$, acting on time interval $\delta_k^{\tau+} = 2\tau_k^{-o}$, measures

$$\bar{u}_{k\delta} = \bar{u}_{ks} \text{ at } |\bar{u}_{k\delta 1}| = 1/2\bar{u}_{sm}, \qquad (1.15a)$$

and the relative time intervals of $\bar{u}_{ko}$ and $\bar{u}_{k1}$ accordingly are

$$o(\tau_k^{-o})(\tau_k^{-o})^{-1} = 0.5, \ o(\tau_k^{+o})/\tau_k^{+o} = 0.1875. \qquad (1.15b)$$

Step-controls of impulse $\bar{u}_{k\delta}$ apply on two equal time intervals:

$$(\delta_k^{\tau+} - \tau_k^{-o}) = \delta_k^{\tau+}/2 \ (4.16a) \qquad \text{and} \qquad \tau_k^{-o} = \delta_k^{\tau+}/2. \qquad (1.16b)$$

On first (1.16a), its step-up part $[\uparrow 1_{\delta_k^{\tau+}}]$ captures entropy increment

$$\Delta S[\tilde{x}_t / \varsigma_t]|_{\delta_k^{\tau+}}^{\tau_k^{-o}} = 1`/2[\uparrow 1_{\delta_k^{\tau+}}]\bar{u}_- = 1`/8[\uparrow 1_{\delta_k^{\tau+}}], \qquad (1.16)$$

on second (1.16b), its step-down multiplicative part in (1.14b) at $\bar{u}_{ko2} = \bar{u}^{-1}$ transfers entropy (1.16) to starting impulse action $[\downarrow 1_{\tau_k^{-o}}]$ which cuts is within impulse (1.4) at $\bar{u}_{ko1} = 1/2\bar{u}_{ko}$; where $\bar{u}_{k\delta 1}$ multiplies

$$\bar{u}[\uparrow 1_{\delta_k^{\tau+}}]\delta_k^{\tau+}/2 \times \bar{u}^{-1}[\downarrow 1_{\tau_k^{-o}}]\tau_k^{-o}. \qquad (1.16c)$$

Both equal time intervals in (1.16b) are orthogonal to opposite inverse entropy increments on the reated impulse border.

C. The applied *extremal* solution (Proposition I.3.1.), decreasing time intervals (I.3.1.b). brings minimal (1.10a) and
(a)-persistence continuation a sequence of the process impulses;
(b)-the balance condition for the entropy contributions;
(c)-each impulse invariant unit $\bar{u}_k = |1|_k$, supplied by entropy unit $\bar{u}_s = |1|_s$, *triples* information that increases information density in each following information unit. •

*Proofs.* The additive sum of entropy increments under invariant impulses (1.7-1.9) satisfies balance condition:

$$\Delta S[\tilde{x}_t/\varsigma_t]|_{s_k^+}^{\tau_k^{-o}} = \Delta S[\tilde{x}_t/\varsigma_t]|_{s_k^+}^{\delta_k^{\tau-}} + \Delta S[\tilde{x}_t/\varsigma_t]|_{\delta_k^{\tau-}}^{\delta_k^{\tau+}} + \Delta S[\tilde{x}_t/\varsigma_t]|_{\delta_k^{\tau+}}^{\tau_k^{-o}} =$$
$$1/2[\uparrow 1_{s_k^{+o}} - \downarrow 1_{\delta_k^{\tau-}}]\bar{u}_{ks} + 1/2[\uparrow 1_{\delta_k^{\tau-}} - \uparrow 1_{\delta_k^{\tau+}}]\bar{u}_{k\delta s} + 1/2\uparrow 1_{\delta_k^{\tau+}}\bar{u}_{k\delta o} - 1/2\downarrow 1_{|\tau|_k^{-o}}\bar{u}_{k\delta 1} = 0 \qquad (1.17)$$

where action $1/2 \downarrow 1_{|\tau|_k^{-o}} \bar{u}_{k\delta 1} \Rightarrow 1/4 \downarrow 1_{|\tau|_k^{-o}} \bar{u}_{ko}$ transfers entropy increment $\Delta S[\tilde{x}_t/\varsigma_t](\tau_k^{-o}) = 1/4 \downarrow 1_{|\tau|_k^{-o}} \bar{u}_{ko}$ on discrete locality $|\tau|_k^{-o}$ by step-down action $\downarrow 1_{|\tau|_k^{-o}} \bar{u}_{k\delta 1}$.

Fulfillment the relations

$$[\uparrow 1_{s_k^{+o}}\bar{u}_{ks} - \downarrow 1_{\delta_k^{\tau-}}\bar{u}_{ks} + \uparrow 1_{\delta_k^{\tau-}}\bar{u}_{k\delta s} - \uparrow 1_{\delta_k^{\tau+}}\bar{u}_{k\delta s} + \uparrow 1_{\delta_k^{\tau+}}\bar{u}_{k\delta o} - \downarrow 1_{|\tau|_k^{-o}}\bar{u}_{k\delta 1}] = 0$$

$$[\uparrow 1_{s_k^{+o}}\bar{u}_{ks} + \uparrow 1_{\delta_k^{\tau-}}[\bar{u}_{k\delta s} - \bar{u}_{ks}] + \uparrow 1_{\delta_k^{\tau+}}[\bar{u}_{k\delta o} - \bar{u}_{k\delta s}] = \downarrow 1_{|\tau|_k^{-o}}\bar{u}_{k\delta 1}, \downarrow 1_{|\tau|_k^{-o}}\bar{u}_{k\delta 1} = -1/2\downarrow 1_{|\tau|_k^{-o}}\bar{u}_{ko}$$

leads to sum of the impulse intervals:

$$\bar{u}_{ks} - \bar{u}_{ks} + \bar{u}_{k\delta s} + \bar{u}_{k\delta s} - \bar{u}_{k\delta s} + \bar{u}_{k\delta o} - \bar{u}_{k\delta 1} = 0, \text{ and } \bar{u}_{k\delta 1} = -1/2\bar{u}_{ko},$$

or to $\bar{u}_{k\delta o} = \bar{u}_{k\delta 1}$. (1.17a)

Impulse $[\uparrow 1_{\delta_k^{\tau+}} \bar{u}_{k\delta o} - \downarrow 1_{|\tau|_k^{-o}} \bar{u}_{k\delta 1}] = [\uparrow 1_{\delta_k^{\tau+}} + \uparrow 1_{|\tau|_k^{-o}}] \bar{u}_{k\delta}$ contains intervals $\bar{u}_{k\delta} = \bar{u}_{k\delta o} + \bar{u}_{k\delta 1}$,

where from (1.9), (1.13a) follows $\bar{u}_{k\delta} = \bar{u}$, and (1.17a) leads to

$$\bar{u}_{k\delta o} = \bar{u}_{k\delta 1} = 1/2\bar{u}.$$ (1.17b)

Interval $\bar{u}_{k\delta} = \bar{u} \times [(o(\delta_k^{\tau+}))(\delta_k^{\tau+} - \tau_k^{-o})^{-1}) + (o(\tau_k^{-o}))(\tau_k^{-o})^{-1}(\delta_k^{\tau+} - \tau_k^{-o})^{-1}\tau_k^{-o}]$ (1.17c)

consists of $\bar{u}_{k\delta}$ components:

$$\bar{u}_{k\delta o} = \bar{u} \times (o(\delta_k^{\tau+}))(\delta_k^{\tau+} - \tau_k^{-o})^{-1}) \text{ and } \bar{u}_{k\delta 1} = \bar{u}_{ko1} \times \bar{u}_{ko2} / \bar{u},$$ (1.17d)

where $\bar{u}_{ko1} = \bar{u} \times (o(\tau_k^{-o}))(\tau_k^{-o})^{-1}$, $\bar{u}_{ko2} = \bar{u}^{-1} \times \tau_k^{-o}(\delta_k^{\tau+} - \tau_k^{-o})^{-1}$.

Intervals $\bar{u}_{ko1}$ and $[\bar{u}_{ko2} / \bar{u}]$ are multiplicative parts of impulse interval $\bar{u}_{k\delta 1}$ covered by starting interval $|\tau|_k^{-o}$.
From (1.17b) and relations (1.17d) it follows

$$\bar{u}_{k\delta o} = \bar{u} \times (o(\delta_k^{\tau+}))(\delta_k^{\tau+} - \tau_k^{-o})^{-1}) = 1/2\bar{u},$$

$$(o(\delta_k^{\tau+}))(\delta_k^{\tau+} - \tau_k^{-o})^{-1}) = 1/2$$ (1.18)

and $\bar{u}_{ko1} = \bar{u} \times (o(\tau_k^{-o}))(\tau_k^{-o})^{-1} = 1/2\bar{u}$. (1.18a)

That leads to

$$(o(\tau_k^{-o}))(\tau_k^{-o})^{-1} = 1/2,$$ (1.18b)

and from (1.18) to relation

$$(\delta_k^{\tau+} - \tau_k^{-o})^{-1}) = (\tau_k^{-o})^{-1}, \delta_k^{\tau+} - \tau_k^{-o} = \tau_k^{-o}, \tau_k^{-o}(\delta_k^{\tau+} - \tau_k^{-o})^{-1} = 1,$$ (1.18c)

then to

$$\tau_k^{-o} = 1/2\delta_k^{\tau+}.$$ (1.18d)

From (1.18c) it follows

$$\bar{u}_{ko2} = \bar{u}^{-1}.$$ (1.18f)

Applying the sequence of Eqs (1.7-1.9), at fixed invariant $\bar{u}$, leads to

$$\bar{u} = u_+(s_k^{+o}) - u_-(\tau_k^{-o}),$$ (1.19)

$$\bar{u} = u_+(\delta_k^{\tau+}) - u_-(\tau_k^{-o}) \text{ at } u_+(\delta_k^{\tau+}) - u_-(\tau_k^{-o}) = \bar{u}_{kb},$$ (1.19a)

which brings invariant $|\bar{u}_s| = |1|_s$ to both impulses (1.19) and (1.19a).
Relation

$$u_+(s_k^{+o}) + u_-(\tau_k^{-o}) = 2[u_-(\delta_k^{\tau-}) + (u_+(\delta_k^{\tau+})] = 0$$

following from the sequence of Eqs (1.7-1.9) leads to

$$u_-(\delta_k^{\tau-}) = -u_+(\delta_k^{\tau+}),$$ (1.19b)

or to reversing and mutual neutralizing these actions on related moments $\delta_k^{\tau-} \cong \delta_k^{\tau+}$.

Impulse interval $\bar{u}_{k\delta}$, with $\bar{u}_{k\delta o}$ and $\bar{u}_{k\delta 1}$, starts interval of applying step-down action $o(\tau_k^{-o})(\tau_k^{-o})^{-1} = 0.5$ in (1.4) at $\bar{u}_{k\delta 1} = \bar{u}_{k\delta o} = 1/2\bar{u}_{k\delta}$.

Invariant impulse $|\bar{u}_s| = |1|_s$, consisting of two step-actions $[\uparrow 1_{\delta_k^{\tau+}} \downarrow 1_{|\tau|_k^{-o}}]\bar{u}_{k\delta}$, which measures intervals

$$\bar{u}_{kb} = \bar{u}_{sm} = \bar{u}_s \text{ at } \bar{u}_{k\delta 1} = 1/2\bar{u}_{sm}. \tag{1.19c}$$

At conditions (1.18c, d), limiting time-jump in (1.13a), step-actions of impulse $\bar{u}_{k\delta}$ applies on two equal time intervals following from (1.19c). On the first interval

$$(\delta_k^{\tau+} - \tau_k^{-o}) = \delta_k^{\tau+}/2$$

step-up part of $\bar{u}_{k\delta}$-action $[\uparrow 1_{\delta_k^{\tau+}}]$ captures entropy increment

$$\Delta S[\tilde{x}_t / \varsigma_t]\big|_{\delta_k^{\tau+}}^{\tau_k^{-o}} = 1/2[\uparrow 1_{\delta_k^{\tau+}}]\bar{u}_- = 1/8[\uparrow 1_{\delta_k^{\tau+}}]. \tag{1.20}$$

On the second interval $\tau_k^{-o} = \delta_k^{\tau+}/2$, the captured entropy (1.20) through the step-down multiplicative part (1.17c,d) delivers to cutting action $\bar{u}_{ko} = \bar{u}_- \times o(\tau_k^{-o})(\tau_k^{-o})^{-1}$ the equal contribution

$$\Delta S[\tilde{x}_t / \varsigma_t]\big|_{\delta_k^{\tau+}}^{\tau_k^{-o}} = 1/4[\downarrow 1_{\tau_k^{-o}}]\bar{u}_- = 1/8[\downarrow 1_{\tau_k^{-o}}]. \tag{1.20a}$$

The control action $[\downarrow 1_{\tau_k^{-o}}]$ at $\bar{u}_- = 0.5$ cuts external entropy of correlation in impulse (1.4) at $\bar{u}_{ko1} = 1/2\bar{u}_{ko}$.

Comment.

Action $[\uparrow 1_{\delta_k^{\tau+}}]$ cuts the captured entropy from impulse $\bar{u}_s = |1|_s$, while multiplicative step-down part (1.17b) transforms the captured entropy to the cutting action in (1.4) at $\bar{u}_{ko2} = \bar{u}^{-1}$. •

At the end of $k$ impulse, control action $\bar{u}_+$ transforms entropy (1.20) on interval $\bar{u}_{kio} = \bar{u}_- \times (o(\tau_k^{+o})/\tau_k^{+o})$ to information

$$\Delta I[\tilde{x}_t / \varsigma_t]\big|_{\delta_{k+}^{\tau+}}^{\tau_k^{+o}} = 1/4[\uparrow 1_{\tau_k^{-o}}]\bar{u}_{kio}\bar{u}_+ = 1/4 \times (-2\bar{u}_{kio})[\uparrow 1_{\tau_k^{-o}}] \tag{1.21}$$

and supplies it to $k+1$ impulse.

(If between these impulses, the entropy increments on the process trajectory are absent (cut)).

That leads to balance equation for information contributions of $k$-impulse:

$$\Delta I[\tilde{x}_t / \varsigma_t]\big|_{\delta_k^{\tau+}}^{\tau_k^{-o}} + \Delta I[\tilde{x}_t / \varsigma_t]\big|_{\tau_k^{-o}}^{\tau_k} + \Delta I[\tilde{x}_t / \varsigma_t]\big|_{\tau_k}^{\tau_k^{+o}} = \Delta I[\tilde{x}_t / \varsigma_t]\big|_{\delta_{k+}^{\tau+}}^{\tau_k^{+o}}, \tag{1.21a}$$

where interval $\bar{u}_{kio}$ holds information contribution $\Delta I[\tilde{x}_t / \varsigma_t]\big|_{\tau_k}^{\tau_k^{+o}} = 1/4\bar{u}_{km}$ satisfied (1.4) at $\bar{u}_+ = -2$, which measures $\bar{u}_{km} = 0.75$ (1.5). That brings relations

$$0.125 + 0.75 + \bar{u}_{kio} = -2\bar{u}_{kio}, 0.125 + 0.75 + 3\bar{u}_{kio} = 0, \bar{u}_{k1} = 3\bar{u}_{kio} = 0.375 = \bar{u}_- \times o(\tau_k^{+o})/\tau_k^{+o}) \tag{1.22}$$

and $o(\tau_k^{+o})/\tau_k^{+o} = 0.1875$, \hfill (1.22a)

$$\bar{u}_{ko} + \bar{u}_{km} + \bar{u}_{k1} = 1.25 = 5/3\bar{u}_{km} \tag{1.22b}$$

from which and (1.21a) it follows

$$\Delta I[\tilde{x}_t / \varsigma_t]\big|_{\tau_k}^{\tau_k^{+o}} = 3\Delta I[\tilde{x}_t / \varsigma_t]\big|_{\delta_k^{\tau+}}^{\tau_k^{-o}}. \tag{1.23}$$

Ratio $\bar{u}_{k1}/2\bar{u}_{kio} = 3/2$ at $2\bar{u}_{kio} = 0.25$ evaluates part of $k$ impulse' information transferred to $k+1$ impulse.

Relations (1.17b,d), (1.18b,d,f), (1.19c), and (1.22a) *prove* the Proposition parts A-B (including (1.16b)). •

Since $\bar{u}_{-} = 0.5$ is cutting interval of impulse $\bar{u}_k$, it allows evaluate the additive sum of the discrete cutoff entropy contributions (1.4) during entire impulse $(\downarrow 1_{\tau_k^{-o}} - \uparrow 1_{\tau_k^{+o}}) = \delta_k$ using $\bar{u}_{-} = \bar{u}_k$:

$$\Delta S[\tilde{x}_t/\varsigma_t]\big|_{\tau_k^{-o}}^{\tau_k^{+o}} = 1/4\bar{u}_k/2 + 1/2\bar{u}_k + 1/4 \times 3/2\bar{u}_k = \bar{u}_k , \qquad (1.24)$$

That determines the impulse cutoff information measure

$$\Delta S[\tilde{x}_t/\varsigma_t]_{\delta_k} = \Delta I[\tilde{x}_t/\varsigma_t]_{\delta_k} = (\downarrow 1_{\tau_k^{-o}} - \uparrow 1_{\tau_k^{+o}})\bar{u}_k = 1|\bar{u}_k, \bar{u}_k = 1|_k \; Nat \qquad (1.24a)$$

equals to $\cong 1.44$ Bit, which the cutting entropy functional of that random process generates.

That single unit impulse $\bar{u}_k = 1|_k$ measures the relative information intervals

$$\bar{u}_{ko} = 1/3\bar{u}_{km},\; \bar{u}_{km} = 1,\; \bar{u}_{kio} = 1/3\bar{u}_{km} = \bar{u}_{ko},\; \text{and } \tau_k^{+o}/\tau_k^{-o} = 3 . \qquad (1.24b)$$

From relations

$\bar{u}_{ko1} = 1/2\bar{u}_{sm}$ and $\bar{u}_{ko1} = 1/2\bar{u}_{ko} = 1/6\bar{u}_{km}$ it follows

$$\bar{u}_{km} = 3\bar{u}_{sm} , \qquad (1.25)$$

which shows that impulse unit $\bar{u}_k = 1|_k$ triples information supplied by entropy unit $\bar{u}_s = 1|_s$ or interval $\bar{u}_k$ compresses three intervals $\bar{u}_s$.

At satisfaction of the extremal principle, each impulse holds invariant interval size $|\bar{u}_k| = 1|_k$ proportional to middle impulse interval $o(\tau)$ with information $\bar{u}_{km}$ which measures $o(\tau)$, and vice versa, time $o(\tau)$ measures the information.

Condition of decreasing $t - s_k^{+o} = o(t) \to 0$ with growing $t \to T$ and squeezing sequence $s_k^{+o} \to \tau_{m-1}^{+o}, k = 1,2....m$ leads to persistence continuation of the impulse sequence with transforming previous impulse entropy to information of the following impulse: $\bar{u}_s = 1|_s \to \bar{u}_k = 1|_k$.

The sequence of growing and compressed information increases at

$$\bar{u}_{k+1} = |3\bar{u}_k| = 1|_{k+1} . \qquad (1.25a)$$

The persistence continuation of the impulse sequence links intervals between sequential impulses $(\bar{u}_{ks}, \bar{u}_{k\delta s}, \bar{u}_{k\delta o})$ whose imaginary (virtual) function $[\uparrow 1_{s_k^{+o}} - \downarrow 1_{\delta_k^{\tau-}} + \uparrow 1_{\delta_k^{\tau+}}]u$ prognosis entropies increments (1.11), (1.12), (1.10).

Information contributions at each cutting interval $\delta_{k-1}, \delta_k, k, k+1,...,m$: $\Delta I[\tilde{x}_t/\varsigma_t]_{\delta_{k-1}}, \Delta I[\tilde{x}_t/\varsigma_t]_{\delta_k},....$

determine time distance interval $\tau_k^{-o} - \tau_{k-1}^{+o} = o_s(\tau_k)$, when each entropy increment

$\Delta S[\tilde{x}_t/\varsigma_t]\big|_{\tau_{k-1}^{+o}}^{t \to \tau_k^{-o}} = 1/2uo(\tau_k) = \bar{u}_s \times o_s(\tau_k)$ supplies each $\Delta I[\tilde{x}_t/\varsigma_t]_{\delta_k}$ satisfying

$\bar{u} \times o(\tau_k) = \Delta I[\tilde{x}_t/\varsigma_t]_{\delta_k}$ at $\bar{u} \times o(\tau_k) = \bar{u}_k(\tau_k^{+o} - \tau_k^{-o})$.

Hence, impulse interval

$$\bar{u}_k = \Delta I[\tilde{x}_t/\varsigma_t]_{\delta_k}/(\tau_k^{+o} - \tau_k^{-o}) \qquad (1.26)$$

measures density of information at each $\delta_k = \tau_k^{+o} - \tau_k^{-o}$, which is sequentially increases in each following Bit.

Relations (1.25a,b), (1.26) confirm part C of Proposition 1.3. •

Such Bit includes three parts:

-the first delivers multiplicative action (1.16c) by capturing entropy of random process;

-the second delivers the impulse step-down cut of the process entropy;

-the third is information, which delivers the impulse step-up action and then transfers to nearest impulse.

That keeps information connection between the impulses and provids persistence continuation of the impulse sequence during the process time $T$.

Corollaries 1.2.

A. The additive sum of discrete functions (1.4) during the impulse intervals determines the impulse information measure equals to Bit, generated from the cutting entropy functional of random process.

The step-down function generates $1/8 + 0.75 = 0.875 Nat$ from which it spends $1/8$ $Nat$ for cutting correlation while getting $0.75$ $Nat$ from the cut. Step-up function holds $1/8$ $Nat$ while $0.675$ $Nat$ it gets from cutting $0.75$ $Nat$, from which $0.5 Nat$ it transfers to next impulse leaving $0.125$ $Nat$ within $k$ impulse.

The impulse has $1/8 + 0.75 + 1/8 = 1 Nat$ of total $1.25 Nat$ from which $1/8$ $Nat$ is the captured entropy increment from a previous impulse. The impulse actually generates $0.75 Nat \cong 1 Bit$, while the step-up control, using $1/8 Nat$, transfers $2/8 Nat$ information to next $k$ impulse, capturing $1/8 Nat$ from the entropy impulse between $k$ and $k+1$ information impulses (on interval $o_s(\tau_k)$).

B. From total maximum $0.875$ Nat, the impulse cuts minimum of that maximum $0.75$ $Nat$ implementing minimax principle, which validates variation condition (I.1.7) and results (I.2.17a,b).

By transferring overall $0.375 Nat$ to next $k+1$ impulse, that $k$ impulse supplies it with its maximum of $1/3 \times 0.75 Nat$ from the cutting information, thereafter implementing principle maximum of minimal cut.

C. Thus, each cutting Bit is *active information unit* delivering information from previous impulse and supplying information to following impulse, which transfers information between impulses.

Such Bit includes: the cutting step-down control's information delivered through capturing external entropy of random process; the cutoff information, which the above control cuts from random process; the information delivered by the impulse step-up control, which being transferred to the nearest impulse, keeps information connection between the impulses providing persistence continuation of the impulse sequence.

D. The amount of information that each second Bit of the cutoff sequence condenses is growing in three times, which sequentially increases the Bit information density. At invariant increments of impulse (1.4), every $\bar{u}_k$ compresses three previous intervals $\bar{u}_{k-1}$ thereafter sequentially increase both density of interval $\bar{u}_k$ and density of these increments for each $k+1$ impulse. •

## 2.2. Information path functional in $n$-dimensional Markov process under $n$-cutoff discrete impulses

The IPF unites information contributions extracting along $n$ dimensional Markov process:

$$I[\tilde{x}_t / \varsigma_t]\Big|_{s_k^-}^{\tau_n^{+o} \to T} = \lim_{k=n \to \infty} \sum_{k=1}^{k=n} \Delta I[\tilde{x}_t / \varsigma_t]_{\delta_k}, \qquad (2.1)$$

where each dimensional information contribution $\Delta I[\tilde{x}_t / \varsigma_t]_{\delta_k} = |1| \bar{u}_k Nat$, satisfying ratios of intervals

$$\bar{u}_{k+1} = |3\bar{u}_k| = |1|_{k+1}, \bar{u}_k = |3\bar{u}_{k-1}|, \qquad (2.2)$$

increases in each third interval in $\bar{u}_{(k+1)}/\bar{u}_{(k-1)} = 9$ times, concentrating in impulse $|1|_{k+1}$, at

$$\bar{u}_k = \Delta I[\tilde{x}_t/\varsigma_t]_{\delta_k}/(\tau_k^{+o} - \tau_k^{-o}) \tag{2.3}$$

which measures density of information at each $\delta_k = \tau_k^{+o} - \tau_k^{-o}$.

Since sequence of $\bar{u}_k, k=1,...,\infty$ is limited by information contributions of the sequence in (2.1):

$$\lim_{k\to\infty}|1|\bar{u}_k \leq |1|_{k\to\infty}, \tag{2.3a}$$

which converges to the finite integral impulse?

Each Bit $|1|_k$ distinguishes from other $|1|_{k+1}$ by condensing the impulse space-time geometry depending on $\bar{u}_k$ density. Each one-dimensional cutoff enables converting entropy increment $\Delta S[\tilde{x}_t/\varsigma_t]_{\tau_k}$ to kernel information contribution $\Delta I[\tilde{x}_t/\varsigma_t]_{\tau_k}$ with optimal density (2.3) satisfying (2.2). The following Propositions detail the IPF specifics.

Proposition 2.1.

A. Optimal distance between nearest $k-1,k$ information impulses, measured by the difference between each previous ending finite intervals $\bar{u}_{k-1}$ and following starting finite interval $\bar{u}_k$ relatively to following interval $\bar{u}_k$:
$\Delta^*_{uk} = (\bar{u}_{k-1} - \bar{u}_k)/\bar{u}_k$, decreases twice for each fixed $k-1,k$:

$$\Delta^*_{uk} = 1/2^k. \tag{2.4}$$

Indeed,

$$\Delta^*_{uk} = (\bar{u}_{k-1} - \bar{u}_k)/\bar{u}_k = (3/2\bar{u}_{k-1m} - 1/3\bar{u}_{km})/1/3\bar{u}_{km} = 1/2.$$

Here $3/2\bar{u}_{k-1m}$ measures information of $k-1$ impulse's ending interval, $1/3\bar{u}_{km}$ measures information of $k$ impulse' starting interval. Time intervals $o_s(\tau_k)$ are along the EF measure, the $\delta_k$ are impulse intervals on along the IPF measure.

B. With growing $k \to n$, the between impulses distance (2.4) decreases by $\Delta^*_{un} = 1/2^n$, which at very high process dimension $n \to \infty$, approaches limit:

$$\lim_{n\to\infty}[\Delta^*_{un} = 1/2^n] \to 0. \tag{2.4a}$$

The impulses' finite entropy increment, located between the information impulses:

$$\Delta S[\tilde{x}_t/\varsigma_t]\big|_{\tau_{k-1}^{+o}}^{t\to\tau_k^{-o}} = 1/2uo(\tau_k) = \bar{u}_s \times o_s(\tau_k), \tag{2.4b}$$

at finite impulses time $o_s(\tau_k)$, determines density measure for the impulse of invariant size of $\bar{u}_s = |1|_s$:

$$\Delta S[\tilde{x}_t/\varsigma_t]\big|_{\tau_{k-1}^{+o}}^{t\to\tau_k^{-o}}/o_s(\tau_k) = \bar{u}_s. \tag{2.5}$$

With growing $k \to n$, the decrease of interval $o_s(\tau_k) \to 0$ is limited by minimal physical time interval.

Eq. (I.3.1.9) shows that a source of entropy increment (2.4b) between impulses is *time course*

$$\Delta_k = (\tau_k^{-o} - \tau_{k-1}^{+o}) \to o_s(\tau_k), \tag{2.5a}$$

which moves the nearest impulses closer. Moreover, each moment $\delta_k^{\tau+}$ of this time course $\Delta_k$ pushes for automatic conversion its entropy density to information density $\bar{u}_k = \Delta I[\tilde{x}_t / \varsigma_t]_{\delta_k} / (\tau_k^{+o} - \tau_k^{-o})$ in information impulse $\bar{u}_k = |1|_k$, where the relative time intervals between impulses (2.4) measures also information density (1.26).

Distance between nearest information impulses (2.5a) evaluates interval of forming entropy increments

$$\Delta_{ks} = 2\tau_{k-1}^{-o} \to o_s \ . \tag{2.5b}$$

Finite instances (2.5), (2.5a,b) limit both information density and equivalence of the entropy and information functionals. Time course intervals (2.5a,b) also runs to convert entropy increment (2.5) in kernel information contribution $\Delta I[\tilde{x}_t / \varsigma_t]_{\delta_k}$ for each cutoff dimension and drives the sequential *integration* for all contributions.

C. With decreasing $\Delta_t = t - s_k^{+o} = o(t)$ at $t \to T$, both $\Delta_k$ and $\delta_k$ are reduced in limit to zero:

$$\lim_{k \to \infty} \Delta_k = 1/2 \lim_{k \to \infty} \delta_k \to 0, \tag{2.6}$$

which follows from

$$\Delta_t = t - s_k^{+o} = o(t) \to \Delta_k = o(\tau_k) \tag{2.6a}$$

at $t \to \tau_k^{-o}, \tau_{k-1}^{+o} \to s_k^{+o}$, and reduces $o(t)$ at $t \to T$.

C. Total sum of the descending time distances at satisfaction (2.4-2.6):

$$\lim_{n \to \infty} \sum_{k=1}^{k=n} \Delta_k = 1/2 \lim_{n \to \infty} \sum_{k=1}^{k=n} \delta_k = T - s \tag{2.7}$$

is finite, converging to total interval of integrating entropy functional (I.1.10).

D. Sum of information contributions $\Delta I[\tilde{x}_t / \varsigma_t]_{\delta_k}$ on whole $(T - s)$ is converging to both path functional integral and the entropy increments of the initial entropy functional:

$$\lim_{k=n \to \infty} \sum_{k=1}^{k \to n} \Delta I[\tilde{x}_t / \varsigma_t]_{\delta_k} \to I[\tilde{x}_t / \varsigma_t]_s^T = S[\tilde{x}_t / \varsigma_t]_s^T, \tag{2.8}$$

limiting the converging integrals at the finite time interval (2.7).

The integral time course contributions run integration of the impulse contributions in (2.1). Information density $\bar{u}_k$ of each dimensional information contribution $\Delta I[\tilde{x}_t / \varsigma_t]_{\delta_k}$ grows according to (2.3), approaching infinity at limit (2.3a).•

Comments 2.1.
1. Sequence of the EF small finite fractions of integrant (I.3.1.5):
$$\delta s_k[\tilde{x}_t / \varsigma_t] = 1/2 u_k o(t_k), k = 1,....,\infty \tag{2.8a}$$
at limited $u_{k \to \infty} = c^2 > 0$ approaches to
$$\lim_{k \to \infty} \delta s_k[\tilde{x}_t / \varsigma_t] = 1/2 u_k o(t_k) = 0, \tag{2.8b}$$

where each integrant (2.8a) is an entropy density, which impulse control $u_k$ converts to information density.
Hence, the information density at infinite dimensions is finite.
2. Sum of the invariant information contributions on discrete intervals increases, while (2.8) integrates all previous contributions. This allows integrates any number of the process' connected information Bits, providing total process information including both random inter-states' and inter-Bits connections.
The IPF information concentrates the integrated Bit •

Increments of correlation functions for the impulse extreme (optimal) process: within its interval $\Delta_k = \tau_k^{-o} - s_k^{+o}$ and on cutoff time borders $\tau_k^{-o}, \tau_k^{+o}$ determine

Proposition 2.2.

A. Correlation function on discreet interval $\Delta_t = t - s_k^{+o}$ for the extreme process holds

$$r_k^-(t) = 1/2 r_k (s_k^{+o})[t^2/(s_k^{+o})^2 + 1]|_{s_k^{+o}}^{t \to \tau_k^{-o}}, \tag{2.9}$$

ending with correlation on the cutoff left border $\tau_k^{-o}$:

$$r_k^-(\tau_k^{-o}) = 1/2 r_k (s_k^{+o})[(\tau_k^{-o}/s_k^{+o})^2 + 1]. \tag{2.9a}$$

After the cutoff, correlation function on following time interval $(\tau_{k+1}^{-o} - \tau_k^{+o})$ holds

$$r_k^+(t) = 1/2 r_k (\tau_k^{+o})[t^2/(\tau_k^{+o})^2 + 1]|_{\tau_k^{+o}}^{t \to \tau_{k+1}^{-o}}. \tag{2.10}$$

B. Correlation on right border $\tau_k^{+o}$ in the finite cutoff at $\tau_k^{+o}/\tau_k^{-o} = 3$ holds:

$$r_k^+(\tau_k^{+o}) = 1/2 r_k (\tau_k^{-o})[(\tau_k^{+o}/\tau_k^{-o})^2 + 1]|_{\tau_k^{-o}}^{\tau_k^{+o}} = 5 r_k (\tau_k^{-o}). \tag{2.11}$$

C. Difference of these correlations at $\delta_k^r = \tau_k^{+o} - \tau_k^{-o} = 1/2 o(\tau_k)$ is

$$r_{ko}^+(\tau_k^{+o}) - r_{ko}^-(\tau_k^{-o}) = \Delta r_{ko}(\delta_k), \ \Delta r_{ko}(\delta_k) = 5 r_k (\tau_k^{-o}) - r_k (\tau_k^{-o}) = 4 r_k (\tau_k^{-o}). \tag{2.12}$$

Its relative value during that finite cutoff holds

$$\Delta r_{ko}(\delta_k)/r_k (\tau_k^{-o}) = 4. \tag{2.13}$$

Correlation within cutoff moment $\tau_k = 1/2 \delta_k^r = 1/4 o(\tau_k)$ evaluates

$$r_k^+(\tau_k) \to 0 \text{ at } o(\tau_k) \to 0. \tag{2.13a}$$

<u>Proof A, B, C.</u> Relation $t = s_k^{+o} b_k (t)/b_k (s_k^{+o})$, at $b_k (t) = 1/2 \dot{r}_k (t)$, determines functions

$\dot{r}_k (t) = 2 b_k (s_k^{+o}) t / s_k^{+o}$ at $b_k (s_k^{+o}) s_k^{+o} = 1/2 r_k (s_k^{+o})$ and solution

$$r_k (t) = \int_{s \to s_k^{+o}}^{t \to \tau_k^{-o}} 2 b_k (s_k^{+o}) t / s_k^{+o} = b_k (s_k^{+o}) t^2 / s_k^{+o} + C_1, \ C_1 = 1/2 r_k (s_k^{+o}). \tag{2.14}$$

From (2.14) follows correlation function (2.9) on this interval and its end (2.9a) for the extreme process.

After the cutoff, correlation function on the next time interval $(\tau_{k+1}^{-o} - \tau_k^{+o})$ holds (2.10).

The correlation, preceding the current cut on the impulse left border $\tau_k^{-o}$:

$$r_{ko}^-(\tau_k^{-o}) = 1/2 r_k (s_k^{+o})[3^2 + 1] = 5 r_k (s_k^{+o}), \tag{2.15}$$

grows in five time of the optimal correlation for previous cutoff at $s_k^{+o}$.

Correlation on the impulse right border $\tau_k^{+o}$ in finite cutoff (2.11) allows finding both difference of these correlations (2.12) on $\delta_k = \tau_k^{+o} - \tau_k^{-o} = 1/2 o(\tau_k)$ and its relative value (2.13) during the finite cutoff. •

Let us find the entropy increments under control $u_-(\tau_k^{-o}), u_+(\tau_k^{-o} - \delta_k^{\tau+})$ near a left border of the cut $t = \tau_k^{-o} - \delta_k^{\tau+}$.

Applying (2.10), (2.11) at $t = \tau_k^{-o} - \delta_k^{\tau+}$ leads to

$$\Delta S[\tilde{x}_t / \varsigma_t]|_{\tau_k^{-o} - \delta_k^{\tau+}}^{t \to \tau_k^{-o}} = -1/2 (u_-(\tau_k^{-o}) - u_+(\tau_k^{-o} - \delta_k^{\tau+}))(\tau_k^{-o} - \delta_k^{\tau+})(\delta_k^{\tau+})^{-1}(\tau_k^{-o} - \delta_k^{\tau+}) =$$

$$-1/2 u_-(\tau_k^{-o})(\tau_k^{-o} - \delta_k^{\tau+}) - u_+(\tau_k^{-o} - \delta_k^{\tau+})(\tau_k^{-o} - \delta_k^{\tau+})(\tau_k^{-o} - \delta_k^{\tau+}))(\delta_k^{\tau+})^{-1}. \tag{2.16}$$

If both the entropy measure of these controls:

$$\Delta S_u = 1/2[u_-(\tau_k^{-o}) - u_+(\tau_k^{-o} - \delta_k^{\tau+})](\tau_k^{-o} - \delta_k^{\tau+}) \tag{2.17}$$

and interval $(\tau_k^{-o} - \delta_k^{\tau+})$ are finite, then entropy increment near the border is infinite:

$$\Delta S[\tilde{x}_t / \varsigma_t]\big|_{\tau_k^{-o} - \delta_k^{\tau+}}^{t \to \tau_k^{-o}} = \Delta S_u (\tau_k^{-o} - \delta_k^{\tau+})(\delta_k^{\tau+})^{-1} \to \infty, \text{ at } \delta_k^{\tau+} \to 0. \tag{2.18}$$

Entropy of control (2.17):

$$\Delta S_u = 1/2[u_-(\tau_k^{-o}) - u_+(\tau_k^{-o} - \delta_k^{\tau+})](\tau_k^{-o} - \delta_k^{\tau+}) = 1/2(2j[\uparrow 1_{\delta_k^{\tau+}} + \downarrow 1_{\tau_k^{-o}}])(\tau_k^{-o} - \delta_k^{\tau+}) = j[\uparrow 1_{\delta_k^{\tau+}} + \downarrow 1_{\tau_k^{-o}}](\tau_k^{-o} - \delta_k^{\tau+})$$

at $(\tau_k^{-o} - \delta_k^{\tau+}) = \delta_k^{\tau+}$, \hfill (2.19)

compensates for the infinity in (2.18), when an imaginary Bit of potential control $j[\uparrow 1_{\delta_k^{\tau+}} + \downarrow 1_{\tau_k^{-o}}]$ applied on interval $(\tau_k^{-o} - \delta_k^{\tau+}) = \delta_k^{\tau+}$ compensates for relative interval $(\delta_k^{\tau+})^{-1}(\tau_k^{-o} - \delta_k^{\tau+})$.

Real controls $u_-(\tau_k^{-o})$ and $u_+(\tau_k^{-o} - \delta_k^{\tau+})$ generates real Bit $(-1_{\tau_k^{-o}} + 1|_{\delta_k^{\tau+}/2})$ which compensates for this infinite increment.

The opposite actions of functions $u_+(\delta_k^{\tau+}/4)$ and $u_-(t = \delta_k^{\tau+}/2) \to u_-(\tau_k^{-o})$ model an interaction on $\delta_k^{\tau+}/2$ with applied control $u_-(\tau_k^{-o})$, which provides external influx entropy that this control captures.

If interactive action $u_+(\delta_k^{\tau+}/4)$ proceeds $u_-(\tau_k^{-o})$, then this control is a reaction on $u_+(\delta_k^{\tau+}/4)$, while the control information covers the influx of entropy within interval

$$\delta_k^{\tau+}/2 = \tau_k^{-o}. \tag{2.20}$$

Opposite symmetric actions $u_+(\delta_k^{\tau+}/4) = j(+1_{\delta_k^{\tau+}/4})$ and $u_-(t = \delta_k^{\tau+}/2) = j(-1_{\delta_k^{\tau+}/2})$, at

$$(t - \delta_k^{\tau+}/4)^{-1}(\delta_k^{\tau+}/4)^2, t = \delta_k^{\tau+}/2, (t - \delta_k^{\tau+}/4)^{-1}(\delta_k^{\tau+}/4)^2 = \delta_k^{\tau+}/4, \tag{2.21}$$

bring total imaginary entropy (a potential) influx:

$$\Delta S[\tilde{x}_t / \varsigma_t]\big|_{\delta_k^{\tau+}/4}^{\delta_k^{\tau+}/2} = -1/2[j(-1_{\delta_k^{\tau+}/2}) - j(+1_{\delta_k^{\tau+}/4})](\delta_k^{\tau+}/4) = 1/4 j[+1_{\delta_k^{\tau+}/4}) - 1_{\delta_k^{\tau+}/2}]\bar{u}_k \delta_k^{\tau+} = 1/4 j[1_{\delta_k^{\tau+}/4}^{\delta_k^{\tau+}/2}]\bar{u}_k \delta_k^{\tau+} \tag{2.22}$$

with two opposite imaginary entropies fractions:

$$S_+[\tilde{x}_t / \varsigma_t]\big|_{\delta_k^{\tau+}/2}^{t \to \tau_k^{-o}} = 1/8 j[1_{\delta_k^{\tau+}/4}^{\delta_k^{\tau+}/2}]\bar{u}_k \delta_k^{\tau+}, \tag{2.23a}$$

$$S_-[\tilde{x}_t / \varsigma_t]\big|_{s_k^+}^{t \to \tau_k^{-o}} = -1/8 j[1_{\delta_k^{\tau+}/2}^{\tau_k^{-o}}]\bar{u}_k \delta_k^{\tau+}. \tag{2.23b}$$

Action $u_-(t = \delta_k^{\tau+}/2) = j(+1_{\delta_k^{\tau+}/2})$ coincides with start of real control $u_-(\tau_k^{-o})$, while entropy (2.16) with

$S_+[\tilde{x}_t / \varsigma_t]\big|_{\delta_k^{\tau+}/2}^{t \to \tau_k^{-o}} = 1/8$ Nat evaluates difference between interactive action $u_+(\delta_k^{\tau+}/4)$ and potential reaction $u_-(t = \delta_k^{\tau+}/2)$.

Multiplicative relation $S_+ \times S_- = 1/2 S_-^+$ (following (1.14b)) evaluates the equivalent interactive entropy of these interactive actions. For the interactive contributions (2.23a) and (2.23b) this relation leads to

$$S_-^+ = -1/32. \tag{2.24}$$

Information of control, starting at $\delta_k^{\tau+}/2 = \tau_k^{-o}$ with its impulse wide $\bar{u}_k \times \delta_k^{\tau+}$, implements this interactive action spending part of its information

$$\Delta I_-^+ = 0.25 \times 1/32 = 0.0078 \cong 0.008 Nat \quad (2.25)$$

on compensating for the interactive entropy (2.24), while capturing (2.22) in a move to the cut.

Thus, information covering (2.23b) includes $\Delta I_-^+$ with total contribution

$$\Delta I_-^o = \Delta I_- + \Delta I_-^+ = 0.125 + 0.008 = 0.133 Nat \ . \quad (2.26)$$

The conjugated components $S_+, S_-$ start not simultaneously but with equal values (2.23a,b), acquiring by moment $\delta_k^{\tau+}/2 = \tau_k^{-o}$ dissimilarity between the entropy and information parts according to:

$$S_-^o = 0.117 Nat \text{ and } \Delta I_-^o = 0.133 Nat \quad (2.27)$$

which satisfies minimal difference between direct action and its reaction at time shift

$$\delta_k^{\tau+}/4 \ . \quad (2.28)$$

Real control $u_-(\tau_k^{-o})$, applied instead of imaginary action $u_-(\delta_k^{\tau+}/2)$, converts total entropy (2.16) on interval (2.28) to the equal control information and compensates for (2.22). That includes (2.23b) and (2.24).

Entropy gap between the anti-symmetric actions (2.23a, b) is imaginable, as well as time interval $\Delta_t = (t - s_k^{+o})$, compared with $t = \tau_k^{-o}$, while control $u_-(\tau_k^{-o})$ of real impulse applies and covers the gap.

At satisfaction (2.24), the delivered information compensates for entropy

$$\Delta S[\tilde{x}_t / \varsigma_t]|_{t \to \delta_k^{\tau+}/4}^{\tau_k^{-o}} \to -1/4 Nats \ . \quad (2.29)$$

Results (2.16)-(2.29) *extend Prop.1.3 specifying information process of capturing external entropy influx in interaction.*

## 2.3. The emerging microprocess

At growing Bayes a posteriori probability along the impulse observations, neighbor impulses may merge, generating interactive jump on each impulse border.

The merge meets action with reaction, superimposing cause and effect and their probabilities.

Mathematically the jump increases Markov drift (speed) up to infinity (Sec.1.1.3.2).

A starting jumping action ↑ interacting with opposite ↓ action of the bordered impulses initiates the impulse inner process $\tilde{x}_{otk} = \tilde{x}(t \in o(\tau_k)))$ called a microprocess.

(Because the merge squeezes the inter-action interval to a micro-minimum).

### 2.3.1. The conjugated entropy increments in the microprocess

The microprocess is developing under step-function $u_\pm^{t1}$, $u_\pm^{t2}$ within the bordered impulse with the step-function $u_t(u_-^t, u_+^t) = c^2(t \in o(\tau_k))$ on a fixed impulse interval $o(\tau_k)$ within the discrete impulse (1. 4).

The impulse step-down $u_-^t = u_-(\tau_k^{-o})$ and step-up $u_+^t = u_+(\tau_k^{+o})$ functions, acting on discrete interval $o(\tau_k) = \tau_k^{+o} - \tau_k^{-o}$ satisfying (1.1A-1.1C) and (1.2a-1.2d), generates the EF (2.1) increments:

$$\Delta S_- = \Delta S_-[u_-^t], \Delta S_+ = \Delta S_+[u_+^t], \quad (3.1)$$

which preserve the additive and multiplicative properties within the Markov process.
(But these merging actions may not simultaneously possess both these Markov properties).

Here, step functions $u_\pm^{t1}$ (1.1c) is analog of $\bar{u}_{k\delta 1}$ in (1.16c) at locality $\delta_k^{\tau+}/2$ of beginning impulse moment $\tau_k^{-o}$.

Opposite functions $u_\pm^{t1}(t^*)$ of jumps ↑↓, starting at beginning of the process with relative time

$$t^* = [\mp \pi/2 \times \delta t^{\pm*} / o(\tau_k)], \delta t^{\pm*} \in (\delta t_{ok}^\pm \to 1/2 o(\tau_k)), \quad (3.2)$$

hold directions of opposite impulses

$$u_{\pm}^{t1} = [u_+ = \uparrow_{t_o^{*-}} (j-1), u_- = \downarrow_{t_o^{*+}} (j+1)] \quad (3.3)$$

on interval $\delta_o[t_o^{*-}, t_o^{*+}] = \delta t^* < o(\tau)$ at a locality of the impulse initial time $\tau_k^{-o}$.

Controls (3.3), holding $u = c^2 < 0$ (1.2f), brings imaginable $u$ and minimal time interval

$$o = (\delta_k^{\tau+}/2)^2 = (\tau_k^{-o})^2. \quad (3.3a)$$

The microprocess increments at interval $o$ do not possess Markov properties (1.C).
The jumps (3.3) initiate relative increments of entropy:

$$\frac{\delta S}{S}/\delta t^* = u_{\pm}^{t1}, [u_+ = \uparrow_{t_o^{*-}} (j-1), u_- = \downarrow_{t_o^{*+}} (j+1)], \quad (3.4)$$

which in a limit leads to differential Eqs

$$\dot{S}_+(t^*) = (j-1)S_+(t^*), \dot{S}_-(t^*) = (j+1)S_-(t^*). \quad (3.5)$$

The applied (3.3) with symbol $j$ of orthogonality to the microprocess entropy increments rotates them.

Solutions of (3.5) describe the microprocess with opposite conjugated entropies functions on relative time $t^*$:

$$S_+(t^*) = [exp(-t^*)(\cos(t^*) - j\sin(t^*))]|_{t_o^{*-}}^{1/2o(\tau_k)}, S_-(t^*) = [exp(t^*)(\cos(t^*) + j\sin(t^*))]|_{t_o^{*+}}^{1/2o(\tau_k)} \quad (3.6)$$

with initial conditions $S_+(t_o^{*-}), S_-(t_o^{*+})$ at moment

$$t_o^{*+} = t_o^{*-} = [\mp \pi / 2\delta t_{ok}^{\pm}]. \quad (3.6a)$$

The relative wide of step-function $u_{\pm}^{t1}: \delta t_o^{\pm}/o(\tau_k) = 0.2 + 0.005 = 0.205$ and the impulse initial relative interval of this function $\tau_k^{-o}/o(\tau_k) = 0.25$ determine starting relative moment $\delta t_{ok}^{\pm} = \delta t_o^{\pm}/\tau_k^{-o} = \pm 0.82$.

From that numerical solutions of (3.6) by this moment $\delta t_{ok}^{\pm} = \pm 0.82$ follow:

$$S_+(t_o^+) = [exp(-\pi/2 \times 0.82)(\cos(\pi/2 \times 0.82)) - j\sin(\pi/2 \times -0.82))] \approx 0.2758 \times 1,$$
$$S_-(t_o^-) = [exp(\pi/2 \times -0.82)(\cos(-\pi/2 \times -0.82) + j\sin(-\pi/2 \times -0.82))] \approx 0.2758 \times 1 \quad (3.7)$$

The numerical solutions by moment $\delta t^{*\mp} = 1/2o(\tau_k)$ at at time

$$t^{*-} = -\pi/2 \times 1/2o(\tau_k)/o(\tau_k) = -\pi/4, t^{*+} = \pi/2 \times 1/2o(\tau_k)/o(\tau_k) = \pi/4 \quad (3.8)$$

are
$$S_+(t^{*-}) = S_+(t_o^{*-}) \times exp(-\pi/4)[\cos(\pi/4) - j\sin(\pi/4)],$$
$$S_-(t^{*+}) = S_-(t_o^{*+}) \times exp(-\pi/4)[\cos(-\pi/4) + j\sin(-\pi/4)] = S_-(t_o^{*+}) \times exp(-\pi/4)[\cos(-\pi/4) - j\sin(-\pi/4)] \quad (3.9)$$

These vector-functions at opposite moments (3.6a) hold opposite signs of their angles $\mp \pi / 4$ with values:

$$S_+(t^{*-}) \cong 0.2758 \times 0.455 \cong +0.125, S_-(t^{*+}) \cong 0.2758 \times 0.455 \cong -0.125. \quad (3.10)$$

Function $u_{\pm}^{t2}$ (1.2d), starting these opposite increments, turns them on angle $\varphi_-^2 - \varphi_+^2 = \pi/2$ that equalizes the increments and starts entangling both equal increments with their angles within interval $t = \tau_k \mp 0$:

$$S_-^2(t = \tau_k + 0) = \delta S_-^1(t = \tau_k - 0) \times \downarrow_{\tau_k+0} \pi/2 = S_-^1(t = \tau_k - 0) \times exp(\pi/2 \times t_{\tau_k+0}^{*+})[\cos(\pi/2 \times t_{\tau_k+0}^{*+}) + j\sin(\pi/2 \times t_{\tau_k+0}^{*+})], (3.11)$$
$$S_+^2(t = \tau_k + 0) = \delta S_+^1(t = \tau_k - 0) \times \uparrow_{\tau_k+0} \pi/2 = S_-^1(t = \tau_k - 0) \times exp(-\pi/2 \times t_{\tau_k+0}^{*-})[\cos(-\pi/2 \times t_{\tau_k+0}^{*-}) + j\sin(-\pi/2 \times t_{\tau_k+0}^{*-})]$$

at moments

$t^{*\pm}_{\tau_k+0} = [\mp \pi \times 2\delta t^{\pm}_{1k}], \delta t^{\pm}_{1k} = \delta t^{\pm}_1 /1/2\tau_k \cong 0.4375, \delta t^{\pm}_1 = \pm(0.5 - \delta t^{k1}_{\pm}), \delta t^{k1}_{\pm} = \tau^{-o}_k/\tau_k + \delta t^{ko}_{\pm}/\tau_k = 0.25 + 0.03125 = 0.2895$ (3.12)

where $\delta t^{ko}_{\pm}/\tau_k \cong 32^{-1}$ evaluates dissimilarities (following (2.24)) between functions

$u'^2_{\pm} = [u_+ = (j+1), u_- = (j-1)]$ switching from moment $t = \tau_k - 0$ to moment $t = \tau_k$.

The resulting values at $t = \tau_k + 0$ are

$S^2_-(t = \tau_k + 0) = 0.125\exp(\pi/2 \times 0.4375) \times 1 \cong 0.25, S^2_+(t = \tau_k + 0) = 0.125\exp(\pi/2 \times 0.4375) \times 1 \cong 0.25$ (3.13)

which, being in the same direction, are summing at that locality:

$S^o_{\mp} = 2S^2_{\mp}[(\delta t^{ko}_{\pm}/\tau_k)] \cong \mp 0.5$. (3.14)

The entanglement, starting with entropy (3.13), continues at (3.14) up to cutting all entangled entropy increments.

Thus, the entanglement starts at angle $(\pi/2) \times 0.4375 < \pi/4$ takes relative time interval of the impulse $\delta t^{ko}_{\pm}/\tau_k \cong 0.03125$ to ends on angle $\pi/2$.

Since only at angle $\pi/2$ the space interval within impulse begins, it means that *the entanglement starts before the space is formed and ends with beginning the space.*

Here $\tau_k = 1/2o(\tau), o(\tau) = 1Nat$ and $\delta t^{ko}_{\pm} = 0.03125 \times 1/2o(\tau) = 0.015625o(\tau) = \varepsilon_{ok}$. (3.14a)

*Moreover, the entanglement may even create the space during that time interval which is reversible.*

Comments. A potential path during creation both entanglement and space could be a *wormhole*-a short cut in space-time predicted by general relativity. But real *space curvature do not exists during this time*. It may emerge only after entanglement by a moment of forming a Bit at the end of the impulse. Hence, space curvature may form at the *end of microprocess-analog of quantum process-* when the Bit, as an elementary unit of macroprocess, emerges.

*Since the entanglement has no space measure, the entangled states can be everywhere in a space.* •

The $t = \tau_k \mp 0$ locality evaluates the $0_k$-vicinity of action of inverse opposite functions (3.9), whose signs imply the signs of increments in (3.14) and in the following formulas.

The subsequent step-up function changes increment (3.14) according to Eqs

$S_{\mp}(\tau^{+o}_k) = S^o_{\mp}(\delta t^{ko}_{\pm}/\tau_k) \times \exp(t^{*+}_{\tau^{+o}_k}), t^{*+}_{\tau^{+o}_k} = [\pi/2\delta t^{*o}_k], \delta t^{*o}_k \in (\delta t^{*o}_{1k} \to \tau^{+o}_k/\tau_k)$, (3.15) at

$\delta t^{*o}_{1k} = \delta t^{\pm}_{1k}/1/2\tau, \delta t^{\pm}_{1k} = \pm(0.5 - \delta t^{k1}_{\pm}), \delta t^{k1}_{\pm} = \delta t^{ko}_{\pm}/\tau_k + \tau^{+o}_k/\tau_k = 0.25 + 0.03125 = 0.2895, \delta t^{\pm}_{1k} = \delta t^{\pm}_1/1/2\tau_k \cong 0.4375$

with resulting value

$S_{\mp}(\tau^{+o}_k) = \mp 0.5\exp(\pi/2 \times 0.4375) = \mp 0.5 \times (\cong 2) \cong \mp 1$, (3.16)

which measures total entropy of the impulse $\bar{u}_k = |1|_k = 1Nat$. (3.17)

Trajectories (3.10-3.16) describe anti-symmetric conjugated dynamics of the microprocess within the impulse, which up to the cutting moment is reversible, generating the entangled entropy increments (3.16).

Comments. From relation (3.4) and Jacobi-Hamiltonian variation equation $\partial S/\partial t = -\tilde{H}$ it follows that the microprocess Hamiltonian gets form'

$\tilde{H}(t^*) = -u'^2_{\pm}S(t^*)$. (3.17a)

That Eq. admits the conjugated Hamiltonian with both real and imaginary parts:

$\tilde{H}(t^*) = -[(j+1)S + (j-1)S] = -[\dot{S}_+(t^*)/S_+(t^*) + \dot{S}_-(t^*)/S_-(t^*)]S(t^*)$ \(3.17b)

At the entanglement, $S_+(t^*+) = S_-(t^*+), S = S_+(t^*+) + S_-(t^*+) = 2S_+(t^*+) = 2S_-(t^*+)$ and Hamiltonian acquires view
$$\tilde{H}(t^*+) = -[(\dot{S}_+(t^*)/2 + \dot{S}_-(t^*)/2] = -\dot{S}(t^*+) \tag{3.17c}$$•

Cutting this joint entropy at moment $\tau_k^+ \cong 0_k + \tau_k^{o+}$ coverts it to equal information contribution
$$S_\mp^o[\tau_k^+] = \Delta I[\tau_k^+] \cong 1.44 \text{ bit} \tag{3.18}$$
that each $\overline{u}_k$ impulse produces.

Interacting impulse outside of the impulse microprocess delivers entropy on the above $0_k$-vicinity of the cutting moment:
$$S_c^*(\tau_k^+) = \exp 0_k = 1. \tag{3.19}$$
Each current impulse requests an interaction for generating information bit from the microprocess reversible entropy, since the impulse contains the requested step-up action $[\uparrow_{\tau_k^{+o}} \overline{u}_+^o]$ (in (3.6)).

Thus, the jumping actions provide minimal discrete displacement (3.3a,3.2), which rotates the entropy opposite increments. The interactive jump generates a pair of random interactive action on the bordered impulses, which are equal probable, reversible within the probabilities of multiple random interactive actions.

The curving shift initiates microprocess within the bordered impulse running the superposition and entanglement of conjugates entropy fractions during time interval starting with the jump. The entanglement starts before the space of the shift is formed and ends with beginning the space shift, being small part of impulse reversible time interval.

The real cut involves an interaction which imposes irreversibility on information process with multiple cutting bits.

## 2.3.2. The rotating conjugated dynamics of the microprocess

*Starting step functions* $u_\pm^{t1}$ initiates increments of the entropies on interval $o(\tau_k - 0)$ by moment $t = \tau_k - 0$:
$$\delta S_+[u_+^{t1}] = \delta S_+^1(t = \tau_k - 0)) = \delta S_+^1(t = \tau_k^{-o}) \uparrow_{\tau_k^{+o}} (j-1), \delta S_-[u_-^{t1}] = \delta S_-^1(t = \tau_k - 0) = \delta S_-^1(t = \tau_k^{-o}) \downarrow_{\tau_k^{-o}} (j+1). \tag{3.20}$$

Step functions $u_\pm^{t2}$ (1.2d) starting at $t = \tau_k - 0$ contribute the entropy increments on interval $o(\tau_k)$ by moment $t = \tau_k$:
$$\delta S_+[u_+^{t2}] = \partial S_+^2(t = \tau_k) = \partial S_+^2(t = \tau_k - 0)) \uparrow_{\tau_k} (j+1), \delta S_-[u_-^{t2}] = \partial S_-^2(t = \tau_k) = \partial S_-^2(t = \tau_k - 0)) \downarrow_{\tau_k} (-j+1). \tag{3.21}$$

Complex function $u_+^{t1}$ turns on the multiplication of functions $\delta S_+^1(t = \tau_k^{-o})$ on angle $\varphi_+^1 = -\pi/4$, and function $u_-^{t1}$ turns on the multiplication function $\partial S_-^1(t = \tau_k^{-o})$ on angle $\varphi_-^1 = \pi/4$ by moment $t = \tau_k - 0$:
$$\delta S_+^1(t = \tau_k - 0)) = \delta S_+^1(t = \tau_k^{-o}) \times \uparrow_{\tau_k^{+o}} -\pi/4, \delta S_-^1(t = \tau_k - 0)) = \delta S_-^1(t = \tau_k^{-o}) \times \downarrow_{\tau_k^{+o}} \pi/4. \tag{3.22}$$

Analogously, step functions $u_\pm^{t2}$, starting at $t = \tau_k - 0$, turn entropy increments (3.22) on angles $\varphi_-^2 = \pi/4$ by moment $t = \tau_k$ and on angle $\varphi_+^2 = -\pi/4$ the following entropy increments by moment $t = \tau_k$:
$$\delta S_-^2(t = \tau_k) = \delta S_-^2(t = \tau_k - 0) \times \downarrow_{\tau_k} \pi/4, \delta S_+^2(t = \tau_k) = \delta S_+^2(t = \tau_k - 0) \times \uparrow_{\tau_k} -\pi/4. \tag{3.23}$$

The difference of angles between the functions in (3.22): $\varphi_+^1 - \varphi_-^1 = -\pi/2$ is overcoming on time interval $o(\tau_k - 0) = \tau_k^{-o} + 1/2 o(\tau_k)$.

After that, control $u_\pm^{t2}$, starting with opposite increments (3.23), turns them on angle $\varphi_-^2 - \varphi_+^2 = \pi/2$ equalizing (3.23).

That launches *entanglement* of entropies increments and their angles *within* interval $o(\tau_k)$ (on a middle of the impulse):
$$\delta S_+^2(t = \tau_k) = \delta S_-^2(t = \tau_k) = \delta S_\mp^2. \tag{3.24}$$

Control $\overline{u}_- = 0.5$, turning the initial time-located vector-function

$$u^t_- = u_-(\tau_k^{-o}) : \xleftarrow{\tau_k^{-o}, \bar{u}_-=0.5,} \delta\varphi_1 = 0 \quad (3.25)$$

on angle

$\delta\varphi_1 = \varphi^1_+ - \varphi^1_- = \pi/2$, transforms it to space vector $u_+(\tau_k - 0) = \uparrow_{\tau_k - o} \bar{u}_+ = 1$ during a jump from moment $t = \tau_k^{-o}$ to moment $t = \tau_k - 0$ on interval $o(\tau_k - 0)$ in (3.22).

Then vector-function $\downarrow_{\tau_k} \bar{u}^o_- = 2$, starting on time $t = \tau_k - 0$ by space interval $\bar{u}^o_- = 2$, jumps to vector-function $\uparrow_{\tau_{k+0}} \bar{u}^o_+ = 2$ forming on time interval $o(\tau_k + 0) = 1/2 o(\tau_k) + \tau_k^+$ the additive space-time impulse

$$u_{\mp} = [\downarrow_{\tau_k+0} \bar{u}^o_-] + [\uparrow_{\tau_k^{+o}} \bar{u}^o_+]. \quad (3.26)$$

The first part of (3.26) equalizes (3.24) within *space-time* interval $\bar{u}_- \times 1/2 o(\tau_k)$, then joins, summing them on $\bar{u}_- \times o(\tau_k + 0)$, which finalizes the entanglement. The last part of impulse (3.26) cuts-kills the entangled increments on interval $\bar{u}_+ \times \tau_k^+$ at ending moment $\tau_k^+$. Section 2.3.4 details the time –space relation and their measures.

Relations (3.1-3.26) lead to following specifics of the microprocess.

3.1a. Step-functions $u^{t1}_\pm$ initiate microprocess $\tilde{x}_{otk1} = \tilde{x}(t \in o(\tau_k - 0))$ on beginning of the impulse discrete interval $o(\tau_k - 0)$ with only additive increments (3.2).

Opposite step functions $u^{t2}_\pm$ continue the microprocess within interval $o(\tau_k + 0)$ at $\tilde{x}_{otk2} = \tilde{x}(t \in o(\tau_k + 0))$ with both additive and multiplicative increments (3.3) preserving the process Markov properties.

3.1b. Space-time impulse (3.16) within interval $o(\tau_k + 0)$ processes entanglement of increments (3.25) of microprocess $\tilde{x}_{otk2} = \tilde{x}(t \in o(\tau_k + 0))$ summing these increments on $o(\tau_k)$ locality of $t = \tau_k$:

$$S^o_{\mp} = 2\delta S^2_{\mp}[(o(\tau_k)]. \quad (3.27)$$

Then it kills entropies (3.27) at ending moment $\tau_k^{o+} \to \tau_k^+$:

$$S^o_{\mp}[\tau_k^+] = 0. \quad (3.27a)$$

The microprocess, producing entropy increment (3.27) within the impulse interval, is reversible before killing which converts the increments in equal information contribution

$$S^o_{\mp}[\tau_k^+] \Rightarrow \Delta I[\tau_k^+]. \quad (3.27b)$$

The information emerging at the ending impulse time interval accomplices injection of an energy with step-up control $[\uparrow_{\tau_k^{+o}} \bar{u}^o_+]$, which starts at transitional impulse. The energy injection can be a result of the impulses middle inter-action with environment. From the impulse ending moment starts an irreversible information process.

3.1c. Transferring the initial time-located vector to equivalent space-vector $\uparrow_{\tau_k - o} \bar{u}_+$ transforms a transition impulse, concentrating within jumping time $\tau_k^{-o}$ on interval of $\bar{u}_- = 0.5$ up to space interval of $\bar{u}_+ = 1$.

The opposite space vector $\downarrow_{\tau_k} \bar{u}^o_- = 2$, acting on relative time interval $1/2 o(\tau_k)/(\tau_k^{+o} - \tau_k^{-o}) = 0.5$, forms space-time function $\downarrow_{\tau_k} \bar{u}^1_- : \bar{u}^1_- = 2 \times 0.5 = 1$, which, as inverse equivalent of opposite function $\uparrow_{\tau_k - o} \bar{u}_+$, neutralizes it to zero. While both time duration of $\bar{u}_- = 0.5$ and $\bar{u}_+ = 1$ concentrate them in transition interval $\tau_k - (\tau_k - 0) = 0_k$.

Within this impulse, only step-down functions $[\downarrow_{\tau_k^{-o}} \bar{u}_-]$ on time interval $\bar{u}_- = 0.5$ and step-up function $[\uparrow_{\tau_k^{+o}} \bar{u}^1_+]$ on space-time interval $\bar{u}^1_+ = \bar{u}_+ \times \tau_k^+ = 2_{\tau_k^+}$ are left. That determines size of the discrete $1-0$ impulse by multiplicative measure $U_m = |0.5 \times 2| = |1|_k = \bar{u}_k$ generating an information bit.

Therefore, functions $u_+(\tau_k - 0) = \uparrow_{\tau_k - o} \bar{u}_+$ and $\downarrow_{\tau_k} \bar{u}_-^o$ are transitional during formation of that impulse and creation time-space microprocess $\tilde{x}_{otk} = \tilde{x}(t \in 1/2o(\tau_k), h_k \in 2_{\tau_k^+})$ with final entropy increment (3.27), and a virtual logic; functions that starting the microprocess transits from $\tau_k$ connecting it to actual information (3.27b).

### 2.3.3. Probabilities functions of the microprocess

Amplitudes of the process probability functions at $S_\mp^*(\tau_k^{+o}) = |S_+^*| = |S_-^*| = 1$ are equal and independent:

$$p_{+a} = 0.3679, p_{-a} = 0.3679. \tag{3.28}$$

That leads to

$$p_{+a}p_{-a} = p_{\pm a}^2 = 0.1353, S_{\mp a}^* = -\ln p_{a\pm}^2 = 2,$$

or at

$$S_{\mp a}^* = 2, \text{ to } p_{a\pm} = \exp(-2) = 0.1353, \tag{3.28a}$$

where $S_{\mp a}^* = S_\mp^*(\tau_k^{+o}+) + S_c^*(\tau_k^+ +)$

includes the interactive components at $\tau_k^{+o} +$ following $k$ impulse.

Functions $u_+ = (j-1), u_- = (j+1)$, satisfying (1.IA), fulfill the additive property at the impulse starting interval $o[t_o^\mp]$, running the anti-symmetric entropy fractions.

Opposite functions $u_+ = (1+j), u_- = (1-j)$, satisfying (1.IB) by the end of impulse at $\uparrow_{\tau_{k+}^{+o}} \bar{u}_\pm$, mount entanglement of these entropy fractions within the impulse' $|1/2 \times 2| = |\bar{u}_k| = |1|_k$ space interval $\bar{u}_\pm = \pm 2$.

The entangling fractions hold the equal impulse probabilities (3.28), which indicates appearance of both entangled anti-symmetric fractions simultaneously with starting space interval.

Interacting probability amplitudes $p_{+a}, p_{-a}$ of $p_{\pm a}$ satisfy multiplicative relation $p_{\pm a} = \sqrt{p_{+a}p_{-a}}$.

However sum of the non-interacting probabilities: $p_+ + p_- = \exp(-S_+^*) + \exp(-S_-^*) = p_\pm \neq p_{a\pm}$ does not comply with it.

The summary probability $p_{\pm am} = 0.7358$ of the non-interacting entropies components is unequal to probability $p_{\pm a}$ of interacting entropies. The interacting probabilities in transitional impulse $[\uparrow 1_{\tau|_k^-} \downarrow 1_{\tau|_k^+}]\bar{u}_k$ on $\tau_k$-locality violates their additive property, but preserves additivity of the entropy increments.

The impulse microprocess on the ending interval preserves both additivity and multiplicativity only for the entropies.

These basic results are the impulse' entropy and probability' equivalents for the quantum mechanics (QM) probability amplitudes relations. However, the impulse cutting probabilities $p_+, p_-$ are probability of random events in the hidden correlations, while probability amplitudes $p_{+a}, p_{-a}$ are attributes of the microprocess starting within the cutting impulse.

That distinguishes the considered microprocess from the related QM equations, considered physical particles.

The entropy of multiple impulses integrate microprocess along the observing random distributions.

With minimal impulse entropy ½ Nat starting a virtual observer, each following impulse' initial entropy $S_\pm(t_o) = 0.25 Nat$ self-generates entropy $S_{\mp a}^* = 0.5 Nat$. Thus, the virtual observer' time–space microprocess starts with probability $p_{a\pm} = \exp(-0.5) = 0.6015$. Probability $p_{a\pm} = 0.1353$ is relational to the impulse initial conditions, which evaluates appearance of time–space actual impulse (satisfying (3.26)) that *decreases* its initial entropy on $S_{\mp a}^* = 2$ Nat.

The impulse's invariant measure, satisfying the minimax, preserves $p_{a\pm}$ along the time-space microprocess for multiple time-space impulses. Reaching probability of appearance the time-space impulse needs $m_p = 0.6015/0.1353 \cong 4.4457 \approx 5$ multiplications of invariant $p_{a\pm} = 0.1353$, which predicts a priori probability of a previous impulse's reactive action.

Space interval, beginning the displacement shift, starts within interval of entanglement (3.15a) having probability

$$P_\Delta^*(\delta t_\pm^{k1}) = 0.821214, P_\Delta^*(\delta t_\pm^{k1}) = \exp(-|S_\mp^*(\delta t_\pm^{k1})| \, \delta t_\pm^{k1} = 0.2895, S_\mp^*(\delta t_\pm^{k1}) = \mp 0.125 \exp(\pi/2 \times \delta t_\pm^{k1}) = \mp 0.1969415, \quad (3.28b)$$

continues during the shift, and extends to the space part of the impulse multiplicative measure after the diplcement ends.

Hence, each reversible microprocess within the impulse generates invariant increment of entropy, which enables sequentially minimize the starting uncertainty of the observation.

Assigning the entropy minimal uncertainty measure $h_\alpha^o = 1/137$ - physical structural parameter of energy, which includes the Plank constant's equivalent of energy, leads to relation:

$$S_{\mp a}^* = 2h_\alpha^o, p_{\pm a} = \exp(-2h_\alpha^o) = 0.98555075021 \to 1, \quad (3.29)$$

which evaluates probability of real impulse' physical strength of coupling independently chosen entropy fractions.

The initially orthogonal non-interacting entropy fractions $S_{+a}^* = h_\alpha^o, S_{-a}^* = h_\alpha^o$ at mutual interactive actions, satisfy multiplicative relation

$$S_{\mp a}^* = (h_\alpha^o)^2 [\cos^2(\bar{u}t) + \sin^2(\bar{u}t)]|_{t_o^\mp}^{t=1/2\tau} = (h_\alpha^o)^2 = inv \quad (3.29a)$$

which at $S_{\mp a}^* = (h_\alpha^o)^2 \to 0$ approaches $p_{\pm a}^* = \exp[-(h_\alpha^o)^2] \to 1$.

The impulse interaction adjoins the initial orthogonal geometrical sum of entropy fractions in linear sum $2h_\alpha^o$.

Starting physical coupling with double structural $h_\alpha^o$ creates an initial information triple with probability (3.29).

*The microprocess initiates the merge that starts with the jumping actions' multiplication on the bordered impulse time according to (1.16b) succeeding displacement (3.3a) during the merge. Both follow from the EF extreme.*

*The multiplication violates Markov property (1.1B) leading to complex control (1.2c), which starts the microprocess within the displacement and rotates the starting conjugated entropy increments.*

*Examples.* Let us find which of the entropy functional expression meets requirements (1.1A,B) within discrete intervals $\Delta_t = (t-s) \to o(t)$, particularly on $\Delta_k = (\tau_k^{-o} - s_k^{+o}) \to o(\tau_k^{-o})$ under opposite functions $u_+, u_-$:

$$u_+(s_k^{+o}) = +1_{s_k^{+o}} \bar{u} = \bar{u}(s_k^{+o}), u_- = -u_+(s_k^{+o}) = -1_{s_k^{+o}} \bar{u}. \quad (3.30)$$

Following relations (1.11), we get entropy increments

$$S_+[\tilde{x}_t/\varsigma_t]|_{s_k^+}^{t\to\tau_k^{-o}} = -1/2[u_+(s_k^{+o})](\tau_k^{-o} - s_k^{+o})^{-1}(s_k^{+o})^2 = -1/2[u_+(s_k^{+o})(s_k^{+o})^2/s_k^{+o}(3-1)] = -1/4[u_+(s_k^{+o})s_k^{+o}] \quad (3.30a)$$

$$S_-[\tilde{x}_t/\varsigma_t]|_{s_k^+}^{t\to\tau_k^{-o}} = 1/2[u_-(s_k^{+o})](\tau_k^{-o} - s_k^{+o})^{-1}(s_k^{+o})^2 = 1/2[u_-(s_k^{+o})(s_k^{+o})^2/s_k^{+o}(3-1)] = 1/4[u_-(s_k^{+o})s_k^{+o}],$$

which satisfy

$$S_+[\tilde{x}_t/\varsigma_t]|_{s_k^+}^{t\to\tau_k^{-o}} = -S_-[\tilde{x}_t/\varsigma_t]|_{s_k^+}^{t\to\tau_k^{-o}}, \quad (3.31a)$$

$$S_+[\tilde{x}_t/\varsigma_t]|_{s_k^+}^{t\to\tau_k^{-o}} - S_-[\tilde{x}_t/\varsigma_t]|_{s_k^+}^{t\to\tau_k^{-o}} = -1/2[\bar{u}(s_k^{+o})s_k^{+o}] = \Delta S[\tilde{x}_t/\varsigma_t]|_{s_k^+}^{t\to\tau_k^{-o}}. \quad (3.31b)$$

Relations

$$4S_+[\tilde{x}_t/\varsigma_t]|_{s_k^+}^{t\to\tau_k^{-o}}/s_k^{+o} = -\bar{u}(s_k^{+o})(s_k^{+o}) = -2\times 1_{s_k^{+o}},$$

satisfy conditions

$$4S_+[\tilde{x}_t/\varsigma_t]|_{s_k^+}^{t\to\tau_k^{-o}}/s_k^{+o} \times 4S_-[\tilde{x}_t/\varsigma_t]|_{s_k^+}^{t\to\tau_k^{-o}}/s_k^{+o} = -\bar{u}(s_k^{+o}) \times \bar{u}(s_k^{+o}) = -(2\times 1_{s_k^{+o}})\times(2\times 1_{s_k^{+o}}) = -4\times 1_{s_k^{+o}}, \quad (3.32a)$$

$$4S_+[\tilde{x}_t/\varsigma_t]|_{s_k^+}^{t\to\tau_k^{-o}}/s_k^{+o} - 4S_-[\tilde{x}_t/\varsigma_t]|_{s_k^+}^{t\to\tau_k^{-o}}/s_k^{+o} = -\bar{u}(s_k^{+o}) - \bar{u}(s_k^{+o}) = -(2\times 1_{s_k^{+o}}) - (2\times 1_{s_k^{+o}}) = -4\times 1_{s_k^{+o}}. \quad (3.32b)$$

These entropy expressions at any *current* moment $t$ within $\Delta_t = (t - s_k^{+o})$ do not comply with (1.1A,B).

The same results hold true for the entropy functional increments under functions
$$u_+ = +1_{s_k^{+o}} \bar{u}, u_- = -1_{\tau_k^{-o}} \bar{u} .$$  (3.33)

Indeed. For this functions on $\Delta_t = (t - s_k^{+o})$ we have
$$\Delta S[\tilde{x}_t / \varsigma_t]|_{s_k^+}^t = -1/2(u_-(t) - u_+(s_k^{+o}))(t - s_k^{+o})^{-1}(s_k^{+o})^2$$  (3.34)

which for $t \to \tau_k^{-o}$ holds
$$\Delta S[\tilde{x}_t / \varsigma_t]|_{s_k^+}^{t \to \tau_k^{-o}} = -1/2(u_-(\tau_k^{-o}) - u_+(s_k^{+o}))(\tau_k^{-o} - s_k^{+o})^{-1}(s_k^{+o})^2 ,$$

and satisfies relations
$$S_+[\tilde{x}_t / \varsigma_t]|_{s_k^+}^{t \to \tau_k^{-o}} - S_-[\tilde{x}_t / \varsigma_t]|_{s_k^+}^{t \to \tau_k^{-o}} = \Delta S[\tilde{x}_t / \varsigma_t]|_{s_k^+}^{t \to \tau_k^{-o}} ,$$  (3.34a)

$$S_+[\tilde{x}_t / \varsigma_t]|_{s_k^+}^{t \to \tau_k^{-o}} = -S_-[\tilde{x}_t / \varsigma_t]|_{s_k^+}^{t \to \tau_k^{-o}}$$  (3.34b)

which determine
$$S_+[\tilde{x}_t / \varsigma_t]|_{s_k^+}^{t \to \tau_k^{-o}} = -1/4(u_-(\tau_k^{-o}) - u_+(s_k^{+o}))(\tau_k^{-o} - s_k^{+o})^{-1}(s_k^{+o})^2$$  (3.35a)

$$S_-[\tilde{x}_t / \varsigma_t]|_{s_k^+}^{t \to \tau_k^{-o}} = 1/4(u_-(\tau_k^{-o}) - u_+(s_k^{+o}))(\tau_k^{-o} - s_k^{+o})^{-1}(s_k^{+o})^2 .$$  (3.35b)

We get the entropy expressions through opposite directional discrete functions in (3.35a,b):
$$S_+[\tilde{x}_t / \varsigma_t]|_{s_k^+}^{t \to \tau_k^{-o}} 4(\tau_k^{-o} - s_k^{+o})^{-1}(s_k^{+o})^2 = -(u_-(\tau_k^{-o}) - u_+(s_k^{+o})) ,$$

$$S_-[\tilde{x}_t / \varsigma_t]|_{s_k^+}^{t \to \tau_k^{-o}} 4(\tau_k^{-o} - s_k^{+o})(s_k^{+o})^{-2} = u_-(\tau_k^{-o}) - u_+(s_k^{+o}) ,$$

which satisfy additivity at
$$-2(u_-(\tau_k^{-o}) - u_+(s_k^{+o})) = -2(-1_{\tau_k^{-o}} \bar{u} - 1_{s_k^{+o}} \bar{u})] = 2\bar{u}[1_{\tau_k^{-o}} + 1_{s_k^{+o}}] = 4[1_{\tau_k^{-o}} + 1_{s_k^{+o}}] .$$  (3.36)

While for each
$$S_+[\tilde{x}_t / \varsigma_t]|_{s_k^+}^{t \to \tau_k^{-o}} 4(\tau_k^{-o} - s_k^{+o})(s_k^{+o})^{-2} = -\bar{u}[-1_{\tau_k^{-o}} - 1_{s_k^{+o}}] ,$$  (3.36a)

$$S_-[\tilde{x}_t / \varsigma_t]|_{s_k^+}^{t \to \tau_k^{-o}} 4(\tau_k^{-o} - s_k^{+o})(s_k^{+o})^{-2} = \bar{u}[-1_{\tau_k^{-o}} - 1_{s_k^{+o}}]$$  (3.36b)

satisfaction of both 1.1A, B:
$$-(u_-(\tau_k^{-o}) - u_+(s_k^{+o})) \times (u_-(\tau_k^{-o}) - u_+(s_k^{+o})) = -[u_-(\tau_k^{-o}) - u_+(s_k^{+o})]^2 ,$$

requires
$$\bar{u} = -2j ,$$  (3.37)

when $-[u_-(\tau_k^{-o}) - u_+(s_k^{+o})]^2 = (-2j)^2[-1_{\tau_k^{-o}} - 1_{s_k^{+o}}]^2 .$  (3.37a)

Simultaneous satisfaction of both 1.1.A, B leads to
$$\Delta S[\tilde{x}_t / \varsigma_t]|_{s_k^+}^{t \to \tau_k^{-o}} 2(\tau_k^{-o} - s_k^{+o})(s_k^{+o})^{-2} = -2\bar{u}[-1_{\tau_k^{-o}} - 1_{s_k^{+o}}] = 4j[1_{\tau_k^{-o}} + 1_{s_k^{+o}}] ,$$

$$-(-1_{\tau_k^{-o}} \bar{u} - 1_{s_k^{+o}} \bar{u}) \times (-1_{\tau_k^{-o}} \bar{u} - 1_{s_k^{+o}} \bar{u}) = (-2j)^2(-1_{\tau_k^{-o}} + 1_{s_k^{+o}})^2 .$$  (3.37b)

At $o(t) \to 0$, these admit an instant existence of both $(-1_{\tau_k^{-o}} \bar{u}, +1_{s_k^{+o}} \bar{u})$.

Thus, under function (3.35), the entropy expressions are imaginary:
$$S_+[\tilde{x}_t / \varsigma_t]|_{s_k^+}^{t \to \tau_k^{-o}} 4(\tau_k^{-o} - s_k^{+o})(s_k^{+o})^{-2} = -2j[-1_{\tau_k^{-o}} + 1_{s_k^{+o}}] = -2j[1_{s_k^{+o}}^{\tau_k^{-o}}] ,$$  (3.38a)

$$S_-[\tilde{x}_t/\varsigma_t]|_{s_k^+}^{t\to\tau_k^{-o}} 4(\tau_k^{-o}-s_k^{+o})(s_k^{+o})^{-2}=2j[-1_{\tau_k^{-o}}+1_{s_k^{+o}}]=2j[1_{s_k^{+o}}^{\tau_k^{-o}}], \quad (3.38b)$$

at their multiplicative and additive relations:

$$S_-[\tilde{x}_t/\varsigma_t]|_{s_k^+}^{t\to\tau_k^{-o}} 4(\tau_k^{-o}-s_k^{+o})(s_k^{+o})^{-2} \times S_+[\tilde{x}_t/\varsigma_t]|_{s_k^+}^{t\to\tau_k^{-o}} 4(\tau_k^{-o}-s_k^{+o})(s_k^{+o})^{-2}=4, \quad (3.39a)$$

$$S_+[\tilde{x}_t/\varsigma_t]|_{s_k^+}^{t\to\tau_k^{-o}} 4(\tau_k^{-o}-s_k^{+o})(s_k^{+o})^{-2} - S_-[\tilde{x}_t/\varsigma_t]|_{s_k^+}^{t\to\tau_k^{-o}} 4(\tau_k^{-o}-s_k^{+o})(s_k^{+o})^{-2}=-j2[1_{\tau_k^{-o}}+1_{s_k^{+o}}]=-j2[1_{s_k^{+o}}^{\tau_k^{-o}}],$$

$$S_+[\tilde{x}_t/\varsigma_t]|_{s_k^+}^{t\to\tau_k^{-o}} - S_-[\tilde{x}_t/\varsigma_t]|_{s_k^+}^{t\to\tau_k^{-o}}=1/2j[1_{s_k^{+o}}^{\tau_k^{-o}}]. \quad (3.39b)$$

Relations (3.36),(3.36a,b) satisfy additivity only at points $\tau_k^{-o}, s_k^{+o}$.

Between these points, within $\Delta_t=(t-s_k^{+o})\to o(t)$, entropy expressions (3.38a, b), (3.39b) are imaginary.
Time direction may go back within this interval until an interaction occurs.
Within this interval, entropy $S_t=S_+-S_-$ satisfies relations

$$(S_+-S_-)^2=S_+^2+S_-^2-2S_+\times S_-, S_+=-1/2jS_t^+, S_-=1/2jS_t^-, S_+^2=-1/4S_t^{\pm 2}, S_-^2=1/4S_t^{\pm 2}, \quad (3.40)$$

leading to

$$-2S_+\times S_-=-2(-1/4)jjS_t^{\pm 2}=-1/2S_t^{\pm 2}, \text{ while } S_+^2+S_-^2=-1/2S_t^{\pm 2} \text{ and } (S_+-S_-)^2=(-jS_t^+)^2=-(S_t^+)^2.$$

At fulfillment of 1.1A, B, the entropy satisfies relations

$$S_{t=\tau}^2=-1/2jS_{t\to\tau}, S_{t=\tau}^2=S_+\times S_-=-1/4S_{t\to\tau}^{\pm 2}, S_{t\to\tau}=\pm 2jS_{t=\tau}, \quad (3.40a)$$

from which also follows

$$S_{t\to\tau}=2j. \quad (3.40b)$$

These examples concur with *(3.5), (3.6) and illustrate it. Results show that a window of interaction with an environment opens only on the impulse border twice: at begging, between moments $\delta_k^{\tau+}/4$ and $\tau_k^{-o}$, when the entropy flow with energy accesses impulse, and at the end of a gap when an entangle entropy converts to equivalent information.*

### 2.3.4. The relation between the curved time and equivalent space length within an impulse

Let us have a two-dimensional rectangle impulse with wide $p$ measured in time $[\tau]$ unit and high $h$ measured in space length $[l]$ unit, with the rectangle measure

$$M_i=p\times h. \quad (3.41)$$

The problem: Having a measure of wide part of the impulse $M_p$ to *find* high $h$ at equal measures of both parts:

$$M_p=M_h \text{ and } M_p+M_h=M_i. \quad (3.42)$$

From (3.42) it follows

$$M_h=1/2M_i=1/2p\times h. \quad (3.43)$$

Assuming that the impulse has only wide part $1/2p$, it measure equals $M_p=(1/2p)^2$.

Then from $M_p=(1/2p)^2=M_h=1/2p\times h$ it follows

$$h/p=1/2. \quad (3.44)$$

Let us find a length unit $[l]$ of the curved time unit $[\tau]$ rotating on angle $\pi/2$ using relations

$$2\pi h[l]/4=1/2p[\tau] \quad (3.45a), \quad [\tau]/[l]=\pi h/p. \quad (3.45)$$

Substitution (3.44) leads to ratio of the measured units:

$$[\tau]/[l]=\pi/2. \quad (3.46)$$

Relation (3.46) sustains orthogonality of these units in time-space coordinate system, but since initial relations (3.42) are linear, ratio (3.46) represents a linear connection of time-space units (3.45).

The impulse-jumps curve the time unit in (3.8).

According to Proposition 1.3, the impulse' invariant entropy implies the multiplication, starting the rotation.

The microprocess, built in rotation movement, curving the impulse time, adjoins the initial orthogonal axis of time and space coordinates (Fig.1a).

The curving impulse illustrates Fig.1b.

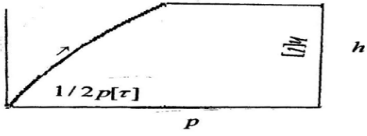  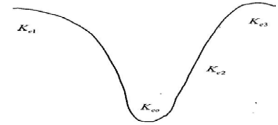

Fig.1(a)  Fig.1 (b)

**Fig.1(a). Illustration of origin the impulse space coordinate measure $h[l]$ at curving time coordinate measure $1/2p[\tau]$ in transitional movement.**

**Fig.1(b). Curving impulse with curvature $K_{e1}$ of the impulse step-down part, curvature $K_{eo}$ of the cutting part, curvature $K_{e2}$ of impulse transferred part, and curvature $K_{e3}$ of the final part cutting all impulse entropy.**

The impulses, preserving the multiplicative and additive measures, have common ratio of $h/p = 1/2$, whose curving wide part $p = 1/2$ brings universal ratio (3.46), which concurs with Lemma 1.1, (1.2a).

At above assumption, measure $M_h$ does not exist until the impulse-jump curves its only time wide $1/2p$ at transition of the impulse. This transition is measured only in time. The following impulse transition is measured both in time $1/2p$ and space coordinate $h$. According to (3.43), measure $M_h$ emerges only on a half of that impulse' total measure $M_i$.

The transitional impulse could start on border of the virtual impulses $\downarrow\uparrow$, where the transition, curving time $\delta t_p = 1/2p$ under impulse-jump during $\delta t_p \to 0$, leads to

$$M_p \to 0 \text{ at } M_h \Rightarrow M_i = p \times h .  \qquad (3.47)$$

If a virtual impulse $\downarrow\uparrow$ has equal opposite functions $u_-(t), u_+(t+\Delta)$, at $\bar{u}_+ = \bar{u}_-$, the additive condition for measure (1.2a): $U_a(\Delta) = 0$ is violated, and the impulse holds only multiplicative measure $U_m(\Delta) \neq 0$ in relation (1.2C): $U_m(\Delta) = U_{am}$ which is finite only at $\bar{u}_+ = \bar{u}_- \neq 0$.

If any of $\bar{u}_+ = 0$, or $\bar{u}_- = 0$, both multiplicative $U_m(\Delta) = 0$ and additive $U_a(\Delta) = 0$ disappear. At $\bar{u}_- \neq 0$, measure $U_a(\Delta)$ is a finite and positive, specifically, at $\bar{u}_- = 1$ it leads to $U_a(\Delta) = 1$ preserving measure $U_{amk} = |U_a|_k$.

Existence of the transitional impulse has shown in (Secs. 2.3.2, 2.3.3).

An impulse-jump at $o[t_o^{\mp}] \to \delta t_p \to 0$ curves a "needle pleat" space at transition to the finite form of the impulse.

The Bayes probabilities measure may overcome this transitive gap.

Since entropy (I. 3.2.1) is proportional to the correlation time interval, whose impulse curvature $K_s = h[l]^{-1}$ is positive, this curved entropy is positive. The curving needle cut changes the curvature sign converting this entropy to information.

## 2.4. Curvature of the impulse

An external step-down control carries entropy which evaluates:

$$\delta_{ue}^i = 1/4(u_{io} - u_i), \qquad (3.48)$$

where $u_{io} = \ln 2 \cong 0.7 Nat$ is total cutoff entropy of the impulse and $u_i \cong 0.5 Nat$ is its cutting part.

The same entropy-information carries the impulse step-up control, while both cutting controls carry $\delta_{ueo}^i \cong 0.1 Nat$.

That evaluates information wide of each single impulse control's cut which the impulse carries:
$$\delta_{ue}^i \cong 0.05 Nat. \qquad (3.48a)$$

To create information, the starting step-down part and the step-up part transfer entropy to the final cutting part generating information, carrying these entropy measures accordingly:
$$\delta_{ue1}^i \cong 0.025 Nat, \delta_{ue2}^i \cong 0.02895, \delta_{ue3}^i \cong 0.01847 Nat.$$

These relations allow estimate Euclid's curvature $K_{e1}$ of the impulse step-down part, related to currying entropy $0.25 Nat$ and its increment $\delta K_{e1}$:
$$K_{e1} = (r_{e1})^{-1}, r_{e1} = \sqrt{1 + (0.025/0.25)^2} = \mp 1.0049875, K_{e1} \cong -0.995037, \delta K_{e1} \cong -0.004963. \qquad (3.49)$$

The cutting part's curvature estimates relations
$$K_{eo} = (r_{eo})^{-1}, r_{eo} = \mp\sqrt{1 + (0.1/0.5)^2} = 1.0198, K_{eo} \cong -0.98058, \delta K_{eo} \cong -0.01942. \qquad (3.49a)$$

The transferred part's curvature estimates relations
$$K_{e2} = (r_{e2})^{-1}, r_{e2} = \sqrt{1 + (0.02895/0.25)^2} \cong 1.0066825, K_{e2} \cong +0.993362, \delta K_{e2} \cong 0.006638 \qquad (3.49b)$$

which is opposite to the step-down part.

The final part cutting all impulse entropy estimates curvatures
$$K_{e3} = (r_{e2})^{-1}, r_{e3} = \sqrt{1 + (0.01847/\ln 2)^2} \cong \pm 1.014931928, K_{e3} \cong +0.99261662, \delta K_{e3} \cong -0.00738338.$$

whose sign is same as the step-down part.

Thus, the entropy impulse is curved with three different curvature values (Fig.1b).

These values estimate each impulse' curvatures holding the invariant entropies.

The entropies emerge in minimax cutoff of the impulse carrying entropy $S_{ki} = 0.5$ and a priori probability $p_{a\pm} = \exp(-0.5) = 0.6015$ after multiple numbers $m_p$ of probing impulses observe this probability.

Since the rectangle impulse, cutting a time correlation, has measure $M = |1|_M$, the curving impulse, cutting the curving correlation, determines measure
$$r_{iM} = M \times K_{ei}. \qquad (3.50)$$

The rectangle impulse not cutting time-correlations possess Euclid's curvature $K_{iM} = 1$.

Accordingly, the impulse with both time and space measure $|M_{io}| = \pi$, which could appear in transitional impulse curvature of cutting part $K_{eo}$, determines correlation measures
$$r_{icM} = M_{io} \times K_{eo}. \qquad (3.50a)$$

At appearance of the impulse with emerging space coordinate, the increment of the curved impulse correlations measure ratio of measures for the curved correlation to one with only time correlation:
$$r_{icM}/r_{iM} = \pi/|1| K_{ei}/K_{eio} \qquad (3.50b)$$

Counting (3.50b) leads to

$r_{icM} / r_{iM} = (\pi / |1|) K_{eo} / K_{e3} \cong 3.08$.

Relative increment of correlation:

$\Delta r_{iM} / r_{iM} = (r_{iM} + r_{icM}) / r_{iM} = 1 + r_{icM} / r_{iM} = 4$

concurs with (2.12), which in limit:

$\lim_{\Delta r(\Delta t), \Delta t \to 0} [\Delta r_{iM} / r_{iM}) = \dot{r}_{icM} / r_{iM}$

brings the equivalent contribution to integral functional IPF (I.4.4.7).

Measure $|M_{io}| = |[\tau] \times [l]| = \pi$ satisfies

$[\tau] = \pi / \sqrt{2}, [l] = \sqrt{2}$ at $[\tau] / [l] = \pi / 2$ . (3.50c)

Shortening the cutting time intervals triples density (1.26) of each invariant curving correlation for the minimax impulse (1.25), preserving its measure (3.50).

Since any virtual cutting impulse preserves its virtual measure (3.50b), the related virtual time correlation is able to create the space during the entanglement that triple density measures.

For the invariant impulse that compresses the impulse curvature, the probability of both cutting time interval and emerging space coordinate increases.

After accumulating energy these information curvatures evaluate the impulse information gravity.

## 2.5. How the observation's cutting jump rotates the microprocess time and creates space interval

Each observation, processing the interactive impulses, cuts the correlation of random distributions.

The virtual impulse' curved cutting correlations evaluates entropy measure of the curvature, which with growing probability eventually bring information-physical curvature to real impulse.

The curved jump of the cutting correlation rotates the impulse time interval starting the impulse microprocess.

The jump initiates multiplicative impulse action $\uparrow_{\delta_k^+} \downarrow_{\tau_k^{-o}}$ on the edge of starting instance $\tau_k^{-o}$, applied to the opposite imaginary conjugated entropies, rotates the entropies with enormous angular speed [45] up the entanglement.

The edge of interval $\tau_k^{-o}$ determines both the jump width-displacement and the curvature forming in the rotation.

The curved time interval $\delta t_\pm^{ko} / \tau_k \cong 0.03125$ relative to the impulse time, formed during *the entanglement, turns on beginning a space before the entanglement ends at angle $\pi / 2$* . That illustrates Fig.1a.

Thus, the time and then space intervals emerge in the interacting impulse as a phase interval, whose probabilistic functions of frequencies enclose a fractional probability of the field available for the observation.

The negative curvature of the curved impulse (Fig1b) attracts an observing positive curvature of an interacting impulse. The attraction in the interacting virtual impulses measures the entropy increment of the interacting curvatures as an analogy of a virtual gravitation.

A real impulse' negative curvature attracts energy from the random field necessary to create information, which causes gravitation attraction. Hence, *the attracting gravitation starts with creation space at the entanglement.* We detail it below.

The interactive impulse microprocess rotates in a transitive movement holding transitive action $\uparrow$.

This action, starting on angle of rotation $|\pi/4|$, initiates entanglement of the conjugated entropies.

The rotation movement, rotating action $\uparrow$ on additional angle, approaching $|\pi/4|$, conveys action $\downarrow$ that settles a transitional impulse, which finalizes the entanglement at angle $|\pi/2|$.

The transitional impulse holds temporal actions $\uparrow\downarrow$ opposite to the primary impulse $\downarrow\uparrow$ which intends to generate the conjugated entanglement, involved, for example in left and rights rotations ($\mp$).

The transitional impulse, interacting with the opposite correlated entanglements $\mp$, reverses it on $\pm$.

Since the hidden entropy' impulse is virtual, transition action within this impulse is also virtual, and its interaction with the forming correlating entanglement is reversible.
Such interaction logically erases each previous directional rotating the entangle entropy units of entropy volume. This erasure emits minimal energy $e_l$ of quanta

$$\varepsilon_o = \hbar \omega_o, e_l = \varepsilon_o [\exp(\varepsilon_o / k_B \theta) - 1]^{-1}$$

which lowers the energy quality compared with the injected energy ($\hbar$ is Plank's constant, $\omega_o$ is frequency, $k_B$ is Boltzmann constant, and $\theta$ is absolute temperature).
The transitional impulse absorbs this emission inside of the virtual impulse, which logically memorizes the entangle units making their mirror copy.
The curvature of interacting impulses creates asymmetry and allows encoding qubits in memorized bit.
Such operations perform function of logical Maxwell Demon [4] specified below.
The entangle logic is memorized temporary until the rotating step-up action $\uparrow$, ending the transitional impulse, moves to transfer the entangle entropy volume to the ending step-up action that kills and finally memorizes the joint entangle qubits in the impulse ending state as the information Bit.
The killing is the irreversible erasure encoding the Bit, satisfying the Landauer principle [7] which requires inject of energy from an external irreversible process to compensate for the real cost of Maxwell Demon(DM).

Logical energy $e_l$, in a classical-macroprocess' limit, at $\hbar \to 0$, is transformed to real energy of the elementary Bit: $e_r = k_B \theta$ acting as the microprocess' inner thermodynamics.
The interacting movement along the impulse boundary ends with cutting the impulse correlation, which carries the potential erasure, becoming a real with delivering an external energy.
Since the entropy' impulse is virtual, transition action within this impulse $\uparrow\downarrow$ is also virtual and its interaction with the forming correlating entanglement is reversible, as well as the space and attractive entropy gravitation.
Comments.
The time-interactions are emerging actions of the initial probability field of interacting events beginning the random microprocesses. From this field, emerges first time-correlation and then space coordinates on a middle of the impulse. It implies possible delivering the field energy on the impulse' middle with emerging the space.
Within the probability field, the emerging initial time has a discrete probability measure, satisfying Kolmogorov law. ●
Thus, the time-interactions hold a discrete sequence of impulses carrying entropy from which emerges a space in the sequence: interactions-correlations–time-space. The sequence of the impulses replicates frequencies of observation creating a wave function. The information form of Schrodinger equation was established in [71] and published in [87].

## 2.6. The interacting curvatures of step-up and step-down actions, and memorizing a bit

Each impulse (Fig.1a) step-down action has negative curvature (3.49,3.49a) corresponding attraction, step-up reaction has positive curvature (3.49b) corresponding repulsion, the middle part of the impulse having negative curvature transfers the attraction between these parts.
In the probing virtual observations, the rising Baeys probabilities increase reality of interactions bringing energy.
When an external process interacts with the entropy impulse, it injects energy capturing the entropy of impulse' ending step-up action (Sec.2.3.4). The inter-action with other (an internal) process generates its impulse' step-down reaction, modeling 0-1 bit (Figs.2A, 2B).
When the interactive process provides Landauer's energy [15,47,47a] with maximal probability (certainty) 1, the interactive impulse' step-down action ending state memorizes the Bit.
Such certain interaction injects the energy overcoming the transitive gap including the barrier toward creation the Bit.

The step-up action of a natural process brings curvature $+K_{e3}$ enclosing potential entropy $e_o = 0.01847 Nat$, which carries entropy $\ln 2$ of the impulse total entropy $1 Nat$ and may transit it to the interacting (an internal process).

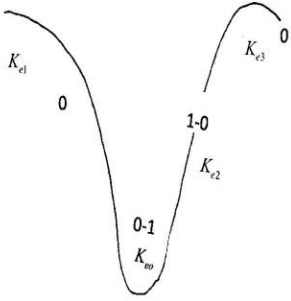
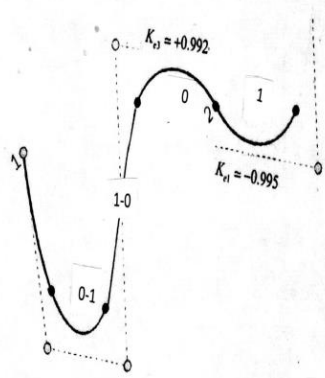

**Figure 2A**   **Figure 2B**

A virtual impulse (Fig.2A) starts step-down action with probability 0 of its potential cutting part; the impulse middle part has a transitional impulse with transitive logical 0-1; the step-up action changes it to 1-0 holding by the end interacting part 0, which, after the inter-active step-down cut, transforms the impulse entropy to information bit.

In Fig. 2B, the impulse Fig. 2A, starting from instance 1 with probability 0, transits at instance 2 during interaction to the interacting impulse with negative curvature $-K_{e1}$ of this impulse step-down action, which is opposite to curvature $+K_{e3}$ of ending the step-up action ($-K_{e1}$ is analogous to that at beginning the impulse Fig.2A). The opposite curved interaction provides a time–space difference (a barrier) between 0 and 1 actions, necessary for creating the Bit.

The interacting step-down part of the internal process impulse' invariant entropy $1 Nat$ has potential entropy $1 - \ln 2 = e_1$. Actually, this step-down opposite interacting action brings entropy $-0.25 Nat$ with anti-symmetric impact $-0.025 Nat$ which carries the impulse wide $\sim -0.05 Nat$ (Sec. 2.3.5) with total $\sim -0.3 Nat$ that is an equivalent to $-e_1$.

Thus, during the impulse interaction, the initial energy-entropy $W_o = k_B \theta_o e_o$ changes to $W_1 = -k_B \theta_1 e_1$, since the interacting parts of the impulses have opposite-positive and negative curvatures accordingly; the first one repulses, the second attracts the energies.

The internal process needs minimal entropy $e_{10} = \ln 2$ for erasing the Bit which corresponds Landauer's energy

$$W_B = k_B \theta \ln 2.$$

If the internal interactive process accepts this Bit by memorizing (through erasure), the above Landauer energy should compensate the difference of these energies-entropy: $W_o - W_1 = W_B$ in balance form

$$k_B \theta_o e_o + k_B \theta_1 e_1 = k_B \theta \ln 2. \tag{3.51}$$

Assuming the interactive process supplies the energy $W_B$ at moment $t_1$ of appearance of the interacting Bit, we get $k_B \theta_1(t_1) = k_B \theta(t_1)$. That brings (3.51) to form

$$k_B \theta_o 0.01847 + k_B \theta (1 - \ln 2) = k_B \theta \ln 2, \theta_o / \theta = (2\ln 2 - 1)/0.01847 = 20.91469199. \tag{3.52}$$

The opposite curved interaction decreases the ratio of above temperatures on increment $\ln 2 / 0.0187 - (2\ln 2 - 1)/0.01847 = 16.61357983$, with ratio $(2\ln 2 - 1)/\ln 2 \cong 0.5573$. (3.52a)

Natural impulse with maximal entropy density $e_{do} = 1/0.01847 = 54.14185$ interacting with internal curved impulse transfers minimal entropy density $e_{d1} = \ln 2 / 0.01847 = 37.52827182$.

Ratio of these densities $k_d = e_{do} / e_{d1} = 1.44269041$ equals to

$$k_d = 1/\ln 2 \tag{3.53}$$

which identifies a single impulse 1 measured by $k_d$ bits-or $1Nat$.

Hence, that ratio, enables producing the information impulse $1Nat = 1.44bit$.

Here the interacting curvature, enclosing entropy density (3.53), lowers the initial energy and the related temperatures in the above ratio. From that follow

*Conditions creating a bit in interacting curved impulse*

1. The opposite curving impulses in the interactive transition require keeping entropy ratio 1/ln2.
2. The interacting process should possess the Landauer energy by the moment ending the interaction.
3. The interacting impulse hold invariant measure M=[1] of entropy 1 Nat whose the topological metric preserves the impulse curvatures. • The last follows from the impulse' max-min mini-max law under its stepdown-stepup actions, which generate invariant [1]Nat's time-space measure' topological metric π(1/2circle) preserving opposite curvatures. •

Results [48] prove that physical process, which holds the invariant entropy measure for each phase space volume (for example, minimal phase volume $v_{eo} \cong 1.242$ per a process dimension ([45], Sec.3.3)) characterized by the above topological invariant, satisfies Second Thermodynamic Law.

By the moment $t_1$ of appearance of the interacting Bit, ratio (3.52a) selects part of the information impulse $i_{11} \cong (1.44 - \ln 2) \times 0.5573 \cong 0.2452bit$ which the curve interaction deducts from the internal impulse' Bit.

The antisymmetric interaction involves middle part of the internal impulse with the asymmetry of curvature $K_{e2} \cong +0.993362$ which encloses entropy $0.02895Nat$.

Difference $0.02895 - 0.025 = 0.00395Nat$ adds asymmetry to the starting transitional entropy, while $0.02895 - 0.01845 = 0.0105$ estimates the difference between the final asymmetry of the main impulse and ended asymmetry of transitional impulse.

Taking into account the asymmetry information $i_{13} \cong 0.0105 \times 1.44 = 0.015bit$, we get information

$$i_f \cong 0.2452 - 0.015 \cong 0.23bit \tag{3.53a}$$

evaluating the total asymmetrical increment of the curved interaction. This is free information created in addition to the Bit, which measures the attracting action of the asymmetrical interaction. It amount simply evaluates~1/3/Bit.

Since the movement within the internal impulse ends at the impulse step-up stopping states, the thermodynamic process delivering this energy should stop in that state. Hence, the erased impulse cutoff entropy memorizes the equivalent information $1.44bit$ in the impulse ending state. It includes $1.44 - 1.23 = 0.21$ where $0.21 \times 1.44 \cong 0.3Nat$ is ransferred to next interacting impulse as the equivalent to is $-e_1$.

In the ending observer's probing logic, such curving interaction moving along negative curvature of its last a priori step-up action to overcome the gap by moving along positive curvature of the posteriori the step-down my acquires information (3.53a) that compensates for the movements logical cost.

Thus, the attractive logics of an invariant impulse, converting its entropy to information within the impulse, performs function of *logical Demon Maxwell* (DM) in the microprocess. (More details are in [46.47]).

*Topological transitivity at the curving interactions*

The impulse of the observer internal process holds its 1 Nat transitive entropy until its ending curved part interacts, creating information bit during the interaction.

Theoretically, when a cutting maximum of entropy reaches a minimum at the end of the impulse, the interaction can occur, converting the entropy to information by getting energy from the interactive process.

The invariant' topological transitivity has a duplication point (transitive base) where one dense form changes to its conjugated form during orthogonal transition of hitting time. During the transition, the invariant holds its measure (Fig.2) preserving its total energy, while the densities of these energies are changing.

The topological transition separates (on the transitive base) both a primary dense topological form and its conjugate dense form, while this transition turns the conjugated form to orthogonal.

At the turning moment, a jump of the time curvature switches to a space curvature with potential rising a space waves, Sec.2.3.2. This is what a real DM does using for that an energy difference of the forms temperature [47].

Forming transitional impulse with entangled qubits leads to possibility memorizing them as a quantum bit.

That requires first to provide the asymmetry of the entangled qubits, which starts the anti-symmetric impact by the main impulse step-down action ↓ interacting with opposite action ↑ of starting transitional impulse.

That primary anti-symmetric impact $-0.025 \times 2 = 0.05 Nat$ starts curving both main and transitional impulses with curvature $K_{e1} = -0.995037$, enclosing $0.025 Nat$, while the starting step-up action of the transitional impulse generates curvature $K_{e2} = +0.993362$ enclosing $e_o = 0.01847 Nat$. Difference $(0.025-0.01847) Nat$ estimates the entropy measuring total asymmetry of main impulse $0.0653 Nat = S_{as}$. The entangled qubits in the transitional impulse evaluates entropy volume 0.0636 Nat [45], which the correlated entanglement can memorize in the equivalent information of two qubits. That is the information "dimon cost" for the entangled correlation.

Starting asymmetric impact brings minimal asymmetry $0.05 Nat$ beginning the transition asymmetry.

The middle part of the main impulse generates curvature $K_{e2} \cong +0.993362$ which encloses entropy $0.02895 Nat$.

Difference $0.02895 - 0.025 = 0.00395 Nat$ adds asymmetry to the starting transitional entropy, while $0.02895 - 0.01845 = 0.0105$ estimates the difference between the final asymmetry of the main impulse and ended asymmetry of transitional impulse.

With starting entropy of the curved transitional impulse $0.05 Nat$, the ending entropy of the transitional impulse asymmetry would be $0.0653-0.0105-0.00395=0.05085 Nat$.

Memorizing this asymmetry needs compensation with a source of equivalent energy. It could be supplied by opposite actions of the transitional step-down ↓ and main step-up interacting action ↑ ending transitional impulse.

That action will create the needed curvature at the end of the main impulse, adding $0.0653 - 0.05085=0.01445, 0.01445-0.0105=0.00395 Nat$ to entropy of transitional impulse curvature sum $0.05085$.

Another part 0.0105 will bring the difference of entropy' curvature $0.02895 - 0.01845 = 0.0105$ with total 0.0653.

Thus, $0.05085 Nat = s_{as}$ is entropy of asymmetry of entropy volume $s_{ev} = -0.0636$ of transitional impulse (see (3.3.5)), whereas $0.0653 Nat = S_{as}$ is entropy of asymmetry of the main impulse. This asymmetry generates the same entangled entropy volume that step-action of the main impulse transfer for interacting with external impulse.

Thus, $s_{as}$ is the information "demon cost" for the entangled correlation, which the curvature of the transitional impulse encloses. It measures information memorizing two qubits in the impulse measure 1 Nat.

When the posteriori probability is closed to reality, the impulse positive curvature of step-up action, interacting with the merging impulse' negative curvatures of step-down action, transits a real interactive energy, which the opposite asymmetrical curvatures actions enfolds.

During the curved interaction, asymmetrical curvature of virtual step-up action compensates the asymmetrical curvature of step-down' real impulse, and that real asymmetry memorizes the erasure of the supplied external Landauer's energy.

The ending action of the internal impulse creates classical bit with probability

$$P_k = \exp-(0.0636^2) = 0.99596321. \qquad (3.54)$$

Since the entanglement in the transitional impulse creates entropy volume $0.0636$, the potential memorizing pair of qubits has the same probability.

Therefore, both memorizing classical bit and pair of quits occur in probabilistic process with high probability but less than 1, so it happens and completes not always.

The question is how to memorize entropy, enclosed in the correlated entanglement, which naturally holds this entropy and therefore has the same probability?

If transitional impulse, created during interaction, has such high probability, then its curvature holds the needed asymmetry. It should be preserved for multiple encoding with the identified difference of locations of the entangled qubits.

Information, as the memorized qubits, can be produced through the interaction generating the qubits within a material - devices (a conductor-transmitter) that preserve curvature of the transitional impulse in a Black Box, by analogy with [49].

At such invariant interaction, the multiple connected conductors memorize the qubits' code.

The needed memory of the transitional curved impulse encloses entropy $0.05085 Nat$.

*The time intervals of the curved interaction*

If the natural space action curves the internal interactive part, the joint interactive time-space curved action measures its interactive impact.

If the interaction at moment $t_o$ creates internal curvature $K_{e1} \simeq -0.995037$ enclosing $-0.025 Nat$ by moment $t_{o1} = 0.01845 Nat$, then interacting time-space interval measures the difference of these intervals

$$|t_{o1}| - t_o \triangleq 0.0250 - 0.01847 = 0.00653 Nat. \quad (3.55)$$

For that case, the internal curved inter-action attracts the energy of natural interactive action.

If No part of the interacting impulse emerges at $t_o$ and Yes part arises by $t_1$, then the invariant interacting impulse will spend $1 - \ln 2 + \ln 2 = 1$ Nat on creation bit ($\ln 2 Nat$).

If inter-action of the natural process on the internal process delivers energy $W_B$ by moment $t_1$, this energy will erase the bit and memorize it according to the balance relations.

The interacting impulse spends ~1 Nat on creating and memorizing bit $\ln 2$ holding free information $(1 - \ln 2) \cong 0.3 Nat$.

The curved topology of interacting impulses decreases the needed energy ratio according to the above balance relation.

Thus, time interval $t_o - t_1$ creates the bit and performs the DM function.

Coordination of an observer external time-space scale with its internal time-space scale happens when an external step-down jump action interacts with observer inner thermodynamics' time-space interval, which, in the curved interaction, measures the difference of the time (3.55).

Ratio $[\tau]/[l] = \pi/2$ leads to $\Delta l_{10} = 2\Delta t_{10}/\pi, \Delta l_{10} \cong 0.00415 Nat$ (in Nats equivalent measure).

*Thus, curvature of rotating impulse encloses its time and space.*

The interacting jump injects energy capturing the entropy of impulse' ending step-up action. This inter-action models 0-1 bit. The opposite curved interaction provides a time–space difference (an asymmetrical barrier) between 0 and 1 actions, necessary for creating the Bit. The interactive impulse' step-down ending state memorizes the Bit when the observer interactive process provides Landauer's energy with maximal probability.

The coordination in the observer's macrolevel is in Sec.6.5, where the interactive memorizing runs information network.

*How to find an invariant energy measure, which each bit encloses starting the DM?*

Since its minimal energy is $W_B = k_B \theta \ln 2$, it's possible to find such temperature $\theta_1^o$ that is equal to inverse value of $k_B$. If the interacting process carries this temperature, then its minimal energy holds $W_1^o = \ln 2$ at $\theta_1^o = 1/k_B$, which becomes equal to the bits' time-space Nat measure' entropy invariant.

Let us evaluate $\theta_1^o$ at $k_B = 8617 \times 10^{-5} eV/K$ and Kelvin temperature measure $K = 20/293 = 0.0682259386^{oC/K}$

equivalent to $20^{oC}$. Then $\theta_1^o = 588.19 \times 10^5 /eV$.

If we assume that this a primary natural energy brings $eV$ amount equivalent to quanta of light $e_q = 1240 eVnm$, $1nm = 10^{-9}m$, then we come to related temperature

$$\theta_1^o = 588.19 \times 10^5 \times 1.240 \times 10^3 / e_q \times 10^{-9} m \cong 72.9356^{oC/m} / e_q.$$

Or each quant should bring temperature' density $\theta_1^o = 72.9356^{oC/m}$, which is reasonably real.

At this $\theta_o^o$, the interacting impulse will bring energy $W_B^o = \ln 2$ to create its bit.

Following the balance relation, the intrernal process at this $\theta_o^o$ should have temperature $\theta_o^o = 20.91469199 \theta_1^o = 1525,42^{oC/m}$ brought by a quant.

This energy holds an invariant impulse measure $|1|_M = 1 Nat$ with metric π, or each such impulse has entropy density $1 Nat / \pi$. The interacting impulse' bit has minimal density energy equivalent to $\ln 2 / \pi = 0.22$ at temperature $\theta_1^o$.

In cognitive dynamics [44], it allows spending at each above interaction minimal cognitive quantity equal to Landauer's energy ln2 for erasure the observing bits and memorizes each bit by the equal neuron information bits.

*With such energy, the information attraction-gravitation imitates free information $0.23 bit$ enables attracting actions.*

Applying the Jarzynski equality (JE) of irreversible thermodynamic transition [80] to conversion energy in information, and using results of its experimental verification [87], lead to the JE form

$$e^{\Delta F / k_B \theta} < e^{W/k_B \theta} >= \gamma, 0 \leq \gamma \leq 2,$$

where $\Delta F$ is increment of free energy needed to produce energy $W$, $\gamma$ is parameter of the verification, which defines sum of probabilities that trajectory in the experiments are observed. At $\gamma = 1$, the JE satisfies exactly:

$$e^{\Delta F / k_B \theta} - < e^{W/k_B \theta} >= 1.$$

Thermodynamic process, satisfying the JE for all its states sequence, is evolving irreversible.

Quantity of information $I_\delta$ equivalent to average energy $W$ which $\Delta F$ compensates during a fixed transition time $\delta$ satisfies Eq.

$$\Delta F = k_B \theta I_\delta.$$

An average thermodynamic energy $< W >= W$, which produces the multiple impulse dissipation (measured by Markov diffusion), integrates the EF equivalent entropy.

Such dissipative energy has high entropy value compared with the considered natural source energy.

Erasing high entropy' energy by the natural source energy brings a non-random information $I_\delta$ equal to the entropy erased during fixed $\delta$.

Taking logarithm from both side of JE leads to relation

$$\Delta F / k_b \theta - \ln < \exp W / k_b \theta >= \ln \gamma,$$

where $< \exp W / k_b \theta >$ is average exponential energy collected during natural interactions of multiple impulses.

Applying this formula to the averaging exponential energy collected during random impulse interactions, leads to

$$< \exp W / k_b \theta >= \exp \Delta S_{\delta t},$$

where $\Delta S_{\delta t}$ is the EF entropy emerging on cutting time intervals $\delta_t$.

By that time, a certain logic with information $I_\delta$ appears (as it's assumed).

Influx of energy $\Delta F = \Delta F_{\delta t}$ at $\delta_t$ enables converting entropy $\Delta S_{\delta t}$ to equivalent information $I_{\delta t}$.

After substituting, we get equation that connects it to JE in form:

$$\Delta F_{\delta t} / (k_b \theta \times I_{\delta t}) - 1 = \ln \gamma / I_{\delta t}.$$

The equivalence of JE in both formulas for the information transition requires $I_{\delta t} = 1$, where $I_{\delta t} = [1]$, is a unit of information per impulse to compensate for the DM energy at the time interval of the transition.

Indeed. At $I_{\delta t} = \ln <\exp W / k_b \theta>$, the previous Eq. forms equality

$\Delta F_{\delta t} / (k_b \theta) - I_{\delta t} = \ln \gamma$ which leads to the JE.

Forthermore, relations $I_{\delta t} = [1], \gamma = 1$ lead to

$\Delta F_{\delta t} / (k_b \theta) = I_{\delta t} = [1], F_{\delta t} = (k_b \theta)[1]$

which also agrees with [46].

Thus, to satisfy the DM, information producing by each impulse time interval should be invariant holding constant unit (Bit, Nat) in $I_{\delta t}$ above.

It confirms that the impulse minimax extremal principle (EP) satisfies the JE for impulse information transition, or vice versa, each impulse time interval enables encoding invariant unit of information.

Or, the EP follows from the JE in the physical process whose interactive time interval is an equivalent of the impulse information cutting from the correlation carrying the energy.

The cutting correlation's time intervals hold the information equivalent of this energy, and any real time interval of interaction brings the entropy equivalent of energy $\Delta F_{\delta t}$ which compensates for the DM while producing information during the interaction.

In interactive random process, whose sequences of cuts satisfy the EP, each impulse encodes the cutting correlation. And all information of the process cutting correlations encodes the information process, which fulfills the minimax law independently on size of any impulse.

Additionally, sum of probabilities of the inverse trajectories of the merging interacting impulses in the microprocess are part of the natural interacting process, which exactly satisfy the JE initial conditions [80,81] according to [45]. The evolving microprocess starts with such probabilities and relational entropies of inverse states $S^*_{\mp a} = 2$. The impulse inner correlation in the microprocess connects 0-1 entropy units (potential bit), while the microprocess entanglement binds 0-1-0-1 entropy entities (a potential qubit).

The multiple microprocesses in the natural encoding generate statistical thermodynamic process where the JE automatically measures energy of these impulses discrete units.

*The above information-physical relations had applied the JE for the first time in [91] for measuring energy within the impulse microprocess (quantum), which encodes this process' information measure. That connects the JE with the natural encoding information measure at the cutting correlation. The random interactions on the path to generation information naturally average the impulse microprocess' dissipative work in the JE thermodynamics.*

This approach distinct from other JE applications by averaging the work in the JE during the evolving natural encoding, while others need multiple experiments and specific procedure of averaging their results.

The information process' last cutting impulse encodes the process total information integrated in its IPF.

Such an impulse natural encoding merges memory with the time of memorizing information and compensates the cutting cost by running time intervals of encoding. The information process, preserving the invariant cutting information, holds the invariant irreversible thermodynamics in its information dynamics.

*Multiple interactions generate a code of the interacting process at the following conditions:*

1. Each impulse holds an invariant probability–entropy measure which the natural Bit code also satisfies.

2. The impulse interactive process, which delivers such code, must be a part of a real physical process that keeps this invariant entropy-energy measure π. That process memorizes the bit and creates information process of multiple encoded bits, which build the process information dynamic structure through the attracting free information.

For example, water, cooling natural drops of hot oils in the found ratio of temperatures, enables spending a part of the water energy' chemical components to encode other chemical structures. Or the water kinetic energy will carry the accepting multiple drops' bits as an arising information dynamic flow.

3. Building the multiple Bits code requires increasing the impulse information density in three times with each following impulse acting on the interacting process (Sec 2.1.1). Such physical process generating the code should supply it with the needed energy. To create a code of the bits, each interactive impulse, produced a Bit, should follow three impulses measure $\pi$, i.e. frequency of interactive impulse should be f=1/3 $\pi$=~0.1061. •

The interval 3 $\pi$ gives opportunity to joint three bits' impulses in a triplet as elementary macro unit and combats the noise, redundancies from both internal and external processes. Ratio of the impulse density Nat/Bit=1.44 to its average curvature $K_{e2}$ is equivalent to a relative information mass: $M^m = 1.44 / K_{e2}$.

For the asymmetrical impulse, its curvature $K_{e2} = 0.993362$, $M^m = 1.452335645$.

The opposite curved interaction lovers potential energy, compared to other interactions for generating a bit, and generates the information mass.

The multiple curving interactions create topological bits code, which sequentially forms moving spiral structure [46]. Therefore, the curving interaction dynamically encodes bits in a *natural process*, developing information structure Fig.3 (Sec.3.2) of the interacting information process. Growing impulse' curvature during rotation increases density of the bit.

*The rotating thermodynamic process with minimal Landauer energy performs natural memorizing of each natural bit.*

In natural processes, interacting through impulses with gradient of temperatures, emerges information and physical process producing code analogous to DNA.

The found mathematical, information, and thermodynamic condtions of generation logical information bits, their memorizing and encoding in helix triplet code (Ch.8).

## 2.7. Explaining some known paradoxes and problems

1.The lottery card magical paradox [52, p.1-12].

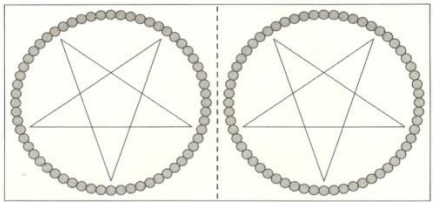
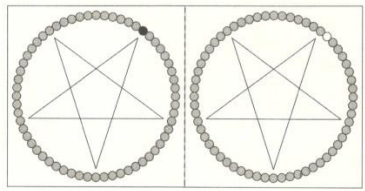

FIGURE 1-1 Lottery card.   FIGURE 1-2 Winning lottery card.

Let two people, as partners, select one of the lottery card ([51, Fig.1-1], and tear it in half between them.
Each take its half of the card and scratch off 1 of the 60 silver spots on the clock face to reveal color, either black or white. It's assumed that half of all spots are black and half white, Fig.1-2.
If both partners scratch off the *same* spot on each clock face, both spots would be black or both white.
But if the partners were to scratch off spots exactly $\pi/2$ degree apart from each other, the partners would always get opposite colors: white and black, or black and white.
These partners could be in different geographic locations, but result will be the same. How can that be?
Physicist [51] explained it through polarization of a photon, while the partners eyes' neurons measure the photon spin by scratching off the lottery card in spots exactly $\pi/2$ degree.

**2.** The formal name of that magic is EPR paradox after paper [17], originating from "Schrödinger Cat Paradox".

Quantum physics justify this spooky actions link by introducing Schrödinger wave operator.

But where it comes from? From the quantum conjugated probabilities? How they arise? Both are true or false?

Modern quantum information-based views of physics [49,51,52, others] also have no answer what information *is* and how it originates. We confirm, the field of probability is source of information and physics.

Even from physical quantum field [18], with vacuum quantum fluctuations, a natural probability of this fluctuation is source of physical particles.

The information observer, created during the objective Kolmogorov-Bayes probabilities observations, obtains a posterior' entropy impulse with rotation of starting entropy.

When the increment of angle of the rotation reaches $\pi/4$, the conjugated entropies become entangled creating a transitional impulse. Continued rotation, reaching angle $\pi/2$ degree, brings equal conjugated entropies with half of entropy bit each - a potential qubit by the end of transitional impulse.

When measurement of that qubits takes place, it converts to information quibits with probability $P_m \cong 0.9855075$ on the edge of reality.

Therefore, the described information approach explains the paradox without using quantum particle theory and the formally established Schrödinger wave operator.

Information observer self-originates itself in observation from uncertainty.

**3**. Problem with the Aspect-Bell's tests [53].

Total multi-dimensional probability for each anti-symmetric entangled local space entropies, rotating on angle $\varphi_\mp^2 = \mp \pi/4$ (Sec.2.3.2), integrates its local conditional Kolmogorov's–Bayes probabilities in final process probability; that is consistent with the Aspect-Bell's tests.

The correlated values of the entropies, starting on these angles, emerge as a process probabilistic logic created through the observer probes-observations, which enables prediction next observation locality.

The Bell's experimental test, based on a *partial* representation of the process probabilities in the field, violates Bell's inequality [54], while the multi-dimensional process' interactions cover all probabilities, assigning the probability spaces to each experimental context and the probabilistic logic.

Comments 1.7.

The "No" finite probability covers an impulse of random process, which inside its locality may hold some events of observing random process, or hidden variables. Such variables satisfy the additive sigma algebra [19] for their random process probabilities, which satisfies a non-additive fraction of the EF. As it's shown, such No impulse' entropy can cover a beginning of the probabilistic microprocess with only multiplicative properties of Markov process. The "No" finite probability does not measure its entropy until a following "Yes" action reveals a potential information of the cutting variables. ●

**4.** The known paradox between truth: yes(no) and lie: no truth-yes (no) or lie: no(yes) solves each particular observer by sending probes to requested answers from the imaginary probes up to real information. That mathematically leads to imaginary impulse, where the time rotation on angle $\pi/2$ creating a space corresponds its multiplying on imaginary symbol $j$. Virtual observer, located inside each sending probe, rotates its imaginary time enclosing logic of this paradox. Within the impulse reversible microprocess, its time is imaginary; its arrow is reversing temporally, while the impulse information generates irreversible time.

Comments.

In recent publication [128], physicists create first direct images of the square of the wave function of a hydrogen molecule below.

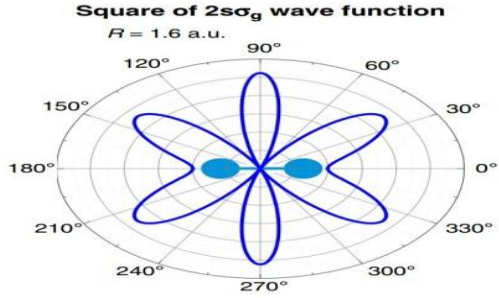

This is image of "electron-electron correlations" which approximates the wave function of a hydrogen molecule.

It's seen that the wave function correlation rotates on angle equal $\pm \pi/4$ from primary $0$ angle.

This result *confirms* our formula (3.9), (322), (3.23) and other related in Sec.2.3 regarding rotation of the equivalent entropies increments. These formulas were earlier in [34, 2015].

Image below, published in [126], shows "interaction of quantum particles with environment when they're *not* being observed". Correlations can trace such random interactions which a probabilistic wave observes.

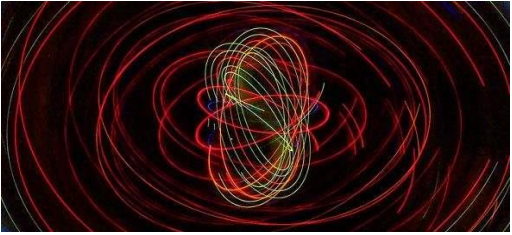

## 3. Information dynamic processes determined by extreme of EF and IPF functionals

### 3.1. The EF-IPF connection

Since the IPF functional integrates a finite information and converges with the entropy functional, which had been expressed through the additive functional, the EF covers both cutoff information contributions and entropy increments between them. The IPF at $n \to \infty$ integrates unlimited discrete sequence of the EF cutoff fractions.

The integration of the discrete fractions and solving a classical variation problem for the IPF to find continuous extreme dynamics presents a *difficult mathematical task*.

The EF presents potential information functional of the Markov process until the applied high probability impulse control, carrying the cutoff increments, transforms it to physical IPF.

The IPF maximal limit approaching EF at $o(t \to T) \to 0, n \to \infty$ avoids the direct access to Markov random process.

Extreme of this integral provides dynamic process $x(t)$, which minimizes distance $\Delta_t$ and dynamically approximates movement $\tilde{x}_t$ to $\varsigma_t$ evaluating transition to Feller kernel.

Process $x(t)$ carries information, collected by maximal IPF at $n \to \infty$, and describes the IPF information dynamic macroprocess at decreasing each following interval

$$\Delta_t = (t-s) \to o(t). \qquad (1.1)$$

Increment of the EF at the end of interval $o_m \to 0$ approaches zero satisfying:

$$\lim_{t_m = T} \Delta S_m[\tilde{x}_t(\tau_m \to t_m))] \to 0. \qquad (1.2)$$

The IPF extracts the finite amount of integral information on all cutoff intervals, approaching initial $S[\tilde{x}_t / \varsigma_t]$ before its cutting.

The sequential cuts on $(T-s)$ breaks the process correlations and the EF functional connections, which transforms initial random process to a limited sequence of independent states.

In the (1.2) limit, the IPF extracts a deterministic process, which approaches the EF extremal trajectories, while the IPF collected information approaches its source, measured by the EF.

Only the impulse Yes-No actions, releasing information from the EF, make the IPF information feasible for the observer in form of Bits, allowing communication with the EF observing process.

### 3.2. Estimation of the extreme process

Mathematical expectations of Ito's Eqs (1.16):

$$E[a] = \dot{\bar{x}}(t) = E[c\tilde{x}(t)] = cE[\tilde{x}(t)] = c\bar{x}(t) \tag{2.1}$$

approximates a regular differential Eqs

$$\dot{\bar{x}}(t) = c\bar{x}(t), \tag{2.2}$$

whose common solution averages the random movement by a dynamic *macroprocess* $\bar{x}(t)$:

$$\bar{x}(t) = \bar{x}(s)\exp ct, \bar{x}(s) = E[\tilde{x}(s)]. \tag{2.3}$$

Within discrete $o(t) = \delta_o$, the opposite controls $u_+, u_-$, satisfying relation (Sec.2.1):

$$c^2 = |u_+ u_-| = c_+ c_- = \bar{u}^2, c_+ = u_+, c_- = u_-, \ |u_+ u_-| = \bar{u}^2$$

are imaginable, presenting an opposite discrete complex:

$$u_+ = j\bar{u}, u_- = -j\bar{u}. \tag{2.4}$$

Conditions 2.1.1A, B are fulfilled at

$$\bar{u} = -2j. \tag{2.4a}$$

when

$$u_+ u_- = j\bar{u}(-j\bar{u}) = \bar{u}^2, \ -j\bar{u} - (+j\bar{u}) = -2j\bar{u}, \ \bar{u}^2 = -2j\bar{u}, \bar{u} = -2j. \tag{2.4b}$$

The controls are real when

$$u_+ = j(-2j) = 2, u_- = -j(-2j) = -2. \tag{2.5}$$

Relations (2.3), (2.4) satisfy two differential equations

$$\dot{x}_+(t) = j\bar{u}x_+(t), \dot{x}_-(t) = -j\bar{u}x_-(t) \tag{2.6}$$

describing microprocess $x_+(t)$, $x_-(t)$ under controls (2.4, 2.5) on time interval $\Delta_t = t - s, \Delta_t \to o(t)$.

Solutions of (2.6) takes forms

$$\ln x_+(t) = Cu_+ t, \ln x_-(t) = Cu_- t, x_+(t) = C\exp(j\bar{u}t), x_-(t) = C\exp(-j\bar{u}t), C = x_-(s^{+o}) = x_+(s^{+o}), \tag{2.7}$$

$$x_+(t) = x_+(s^{+o})(\cos\bar{u}t + j\sin\bar{u}t), x_-(t) = x_-(s^{+o})(\cos\bar{u}t - j\sin\bar{u}t). \tag{2.7a}$$

Correlation function for microprocess (2.7a) at $\cos^2(\bar{u}t) + \sin^2(\bar{u}t) = 1$ holds

$$r(x_+(t), x_-(t)) = r_s = x_+(s^{+o}) \times x_-(s^{+o}). \tag{2.7b}$$

During this fixed correlation, the conjugated anti-symmetric entropies (2.3.6) interact producing entropy flow (2.3.16). It is possible an interactive resonance in attractive correlating movement in forming space entanglement.

Correlation (2.7b) depends on interaction at moment $\delta^{+o}/4, \delta^{+o}/2$ on an edge of the impulse.

Applying formula [21, p.27] for the correlation between moments $\delta^{+o}/4, \delta^{+o}/2$ leads to

$$r_s = \sqrt{(\delta^{+o}/4)/(\delta^{+o}/2)}, \ r_s = \sqrt{0.5}. \tag{2.7c}$$

Within this correlation acts impulse $[1_{\delta_k^{\tau+}/4}^{\delta_k^{\tau+}/2}]\bar{u}_k, \bar{u}_k = |\pm 1|_k$ which includes both opposite controls instantly.

If real control, cutting the influx of entropy at this correlation, does not compensate it, the states' correlation is not dissolved, and the states, carrying both opposite controls, will hold during the process' correlation.

These correlated states might be a hidden microprocess entangling states within the entropy-information gap balancing the anti-symmetric actions.

Solutions (2.7a) describe microprocess on $o(t) \to o(\tau_n^{-o})$, compared to impulse macroprocess (2.3) averaging (2.7a) within these intervals.

The microprocess becomes an inner part of the dynamics process, minimizing distance (1.1), when its time intervals satisfy optimal time (Prop.2.12) running between the impulse cutoff information at

$$\tau_k^{-o}/\tau_k^{+o} = 3, \delta_k = \tau_k^{+o} - \tau_k^{-o} = 2\tau_k^{-o}, \tau_k^{-o} = 3\delta_k/2. \tag{2.8}$$

It implies that imaginary time interval between impulses triples the cutoff discrete intervals:

$$\Delta_k = 3\delta_k, \tag{2.8a}$$

while microprocess (2.7a) locates within these cutoff discrete intervals.

To find dynamic process $x(t)$ we consider solution of (2.3) under real control (2.5) starting at moment $t = t^e$:

$$x_{\pm}(t^e) = x(s^+)\exp(u_{\pm}t^e), \ t^e = s_k^{+o}b_k(t)/b_k(s_k^{+o}), \ u_{\pm} = \pm 2, \tag{2.9}$$

which approximates an extreme of the entropy functional within each $\Delta_t = t - s$.

The solutions in form

$$dx(t)/x(t) = cdt, \ln x(t) = ct, t = s_k^{+o}b_k(t)/b_k(s_k^{+o}), \ln x(t) = cs_k^{+o}b_k(t)/b_k(s_k^{+o}) \tag{2.10}$$

starting on time $t = t^e$, integrate, on minimal time distance $\Delta_t = t - s$, the process

$$x(t) = \exp[cs_k^{+o}b_k(t)/b_k(s_k^{+o})], x(s_k^{+o}) = \exp(cs_k^{+o}), c = \ln(x(s_k^{+o}))/s_k^{+o}, x(s_k^{+o}) = \overline{x}(s_k^{+o}),$$

$$x(t) = \exp[\ln(x(s_k^{+o}))b_k(t)/b_k(s_k^{+o})] = \exp[\ln(x(s_k^{+o}))t/s], \ln x(t) = \ln(x(s_k^{+o}))t/s, \tag{2.10a}$$

which at $t \to T$ approaches

$$\ln x(T) = \lim_{t \to T}[\ln(x(s_k^{+o}))T/s], x(T) \to x(s_k^{+o}))T/s. \tag{2.10b}$$

Process $x(t)$ in (2.3), (2.10b) is the extremal solution of *macroprocess* $\overline{x}(t)$, which averages solution of Ito Eq. under the optimal controls. That multiples the cutoff the EF for $n$-dimensional Markov process within intervals

$$\Delta_t = (t - s) \to o(t). \tag{2.11}$$

Process $x(t)$ carries the EF increments, while the information dynamic macroprocess collects the maximal IPF contributions on each (2.11) at $o(t) \to 0$, $n \to \infty$.

Information, collected from the diffusion process by the IPF, approaches the EF entropy functional.

Finding process $x(t)$ which the EF generates requires solution of the EF variation problem.

### 3.3. The solution of variation problem for the entropy functional

Applying the integral functional form of entropy functional (1.1.10):

$$S = \int_c^T L(t, x, \dot{x})dt = S[x_t], \tag{3.1}$$

leads to variation problem minimizing the entropy functional of the diffusion process:

$$\min_{u_t \in KC(\Delta, U)} \tilde{S}[\tilde{x}_t(u)] = S[x_t], \ Q \in KC(\Delta, R^n). \tag{3.1a}$$

Specifically, for integral (1.3.2.1), at (13.22), it leads to the variation problem

$$\text{extr } S[\tilde{x}_t/\varsigma_t] = \underset{c^2(t)}{\text{extr}} 1/2 \int_s^T c^2(t)A(t,s)dt, c^2(t) = \dot{x}(t). \tag{3.1b}$$

Proposition 3.1.

1. An *extremal solution* of variation problem (3.1a, 3.1) for entropy functional (1.1.10) brings the following equations of extremals for vector $x$ and conjugate vector $X$ accordingly:

$$\dot{x} = a^u, \ a^u = a(u,t,x) \ (t,x) \in Q, \tag{3.2}$$

$$\dot{X} = -\partial P / \partial x - \partial V / \partial x . \tag{3.3}$$

Where function

$$P = (a^u)^T \frac{\partial S}{\partial x} + b^T \frac{\partial^2 S}{\partial x^2} , \tag{3.4}$$

is a potential of functional (1.1.10) that depends on function of action $S(t,x)$ on extremals (3.2, 3.3); $V(t,x)$ is integrant of additive functional (1.1.7), which defines the probability function (1.1.3).

<u>Proof.</u> Using the Jacobi-Hamilton (JH) equations for function of action $S = S(t,x)$, defined on the extremals $x_t = x(t)$, $(t,x) \in Q$ of functional (3.1), leads to

$$-\frac{\partial S}{\partial t} = H, H = \dot{x}^T X - L, \tag{3.5}$$

where $X$ is a conjugate vector for $x$ and $H$ is a Hamiltonian for this functional.
(All derivations here and below have vector form).
From (3.1a) it follows

$$\frac{\partial S}{\partial t} = \frac{\partial \tilde{S}}{\partial t}, \frac{\partial \tilde{S}}{\partial x} = \frac{\partial S}{\partial x}, \tag{3.6}$$

where for the JH we have

$$\frac{\partial S}{\partial x} = X, -\frac{\partial S}{\partial t} = H . \tag{3.6a}$$

The Kolmogorov Eq. for functional (1.1.10) on diffusion process allows joining it with Eq. (3.6a) in the form

$$-\frac{\partial \tilde{S}}{\partial t} = (a^u)^T X + b \frac{\partial X}{\partial x} + 1/2 a^u (2b)^{-1} a^u = -\frac{\partial S}{\partial t} = H , \tag{3.7}$$

where dynamic Hamiltonian $H = V + P$ includes differential function $V = d\varphi / ds$ of additive functional (1.1.7) and potential function (3.4) which, at satisfaction (3.7), imposes on transforming diffusion process' extremals the constraint:

$$P(t,x) = (a^u)^T X + b^T \frac{\partial X}{\partial x} . \tag{3.8}$$

Applying Hamilton equations $\frac{\partial H}{\partial X} = \dot{x}$ and $\frac{\partial H}{\partial x} = -\dot{X}$ to (3.7) brings the extremals for vectors $x$ and $X$ in forms (3.2) and (3.3) accordingly. •
More details in [32 and 43].
<u>Proposition 3.2.</u>
A *minimal solution* of variation problem (3.1a,3.1) for the entropy functional determines the equations of extremals for $x$ and $X$ accordingly:

$$\dot{x} = 2bX_o , \tag{3.9}$$

satisfying condition
$$\min_{x(t)} P = P[x(\tau)] = 0 . \tag{3.10}$$

Condition (3.10) is a dynamic constraint, which is imposed on solutions (3.2), (3.3) at some set of the functional's field $Q \in KC(\Delta, R^n)$ holding the $\tau$-localities of cutting process $x(t)_{t=\tau} = x(\tau)$ :

$$Q^o \subset Q, Q^o = R^n \times \Delta^o, \Delta^o = [0,\tau], \tau = \{\tau_k\}, k = 1,...,m . \tag{3.11}$$

Hamiltonian
$$H_o = -\frac{\partial S_o}{\partial t} \tag{3.12}$$

defines function of action $S_o(t,x)$, which on extremals Eq.(3.9) satisfies condition
$$\min(-\partial \tilde{S}/\partial t) = -\partial \tilde{S}_o/\partial t. \tag{3.13}$$
Hamiltonian (3.12) and Eq. (3.9) determine a second order differential Eq. of extremals:
$$d^2 x/dt^2 = dx/dt[\dot{b}b^{-1} - 2H_o]. \tag{3.14}$$

Proof. Using (3.4) and (3.6) allows finding the equation for Lagrangian in (3.1) in form
$$L = -b\frac{\partial X}{\partial x} - 1/2\dot{x}^T(2b)^{-1}\dot{x}. \tag{3.15}$$
On extremals (3.2, 3.3), both functions drift and diffusion in (1.1.10) are nonrandom.
After substitution the extremal Eqs (3.2,3.9) and (3.7) to (3.1), the integral functional $\tilde{S}$ on the extremals holds:
$$\tilde{S}[x(t)] = \int_s^T 1/2(a^u)^T(2b)^{-1}a^u dt, \tag{3.15a}$$
which should satisfy variation conditions (3.1a), or
$$\tilde{S}[x(t)] = S_o[x(t)], \tag{3.15b}$$
where both integrals are determined on the same extremals.
From (3.15) and (3.15a,b) it follows
$$L_o = 1/2(a^u)^T(2b)^{-1}a^u, \text{ or } L_o = \dot{x}^T(2b)^{-1}\dot{x}. \tag{3.16}$$
Both Lagrangian expressions (3.15) and (3.16) coincide on the extremals where potential (3.7) satisfies condition (3.10):
$$P_o = P[x(t)] = (a^u)^T(2b)^{-1}a^u + b^T\frac{\partial X_o}{\partial x} = 0, \tag{3.17}$$
while Hamiltonian (3.12) and function of action $S_o(t,x)$ satisfies (3.13).
From (3.15b) it also follows
$$E\{\tilde{S}[x(t)]\} = \tilde{S}[x(t)] = S_o[x(t)]. \tag{3.17a}$$
Applying Lagrangian (3.16) to the Lagrange's equation
$$\frac{\partial L_o}{\partial \dot{x}} = X_o, \tag{3.17b}$$
leads to the equations for vector
$$X_o = (2b)^{-1}\dot{x} \tag{3.17c}$$
and extremals (3.9).
Both Lagrangian and Hamiltonian here are the *information forms* of JH solution for the EF.
Lagrangian (3.16) satisfies the maximum principle for functional (3.1,3.1a), from which also follows (3.17a).
Functional (3.1) reaches its minimum on extremals (3.8), while it is a maximal on extremals (3.2,3.3) of (3.6).
Hamiltonian (3.7), at satisfaction of (3.17), reaches minimum:
$$\min H = \min[V + P] = 1/2(a^u)^T(2b)^{-1}a^u = H_o \tag{3.18}$$
from which it follows (3.10) at
$$\min_{x(t)} P = P[x(\tau)] = 0. \tag{3.19}$$
Function $(-\partial \tilde{S}(t,x)/\partial t) = H$ in (3.6) on extremals (3.2,3.3) reaches a *maximum* when constraint (3.10) is not imposed.
Both the minimum and maximum are conditional with respect to the constraint imposition.
Variation conditions (3.18), imposing constraint (3.10), selects Hamiltonian

$$H_o = -\frac{\partial S_o}{\partial t} = 1/2(a^u)^T (2b)^{-1} a^u \qquad (3.20)$$

on the extremals (3.2,3.3) at discrete moments $(\tau_k)$ (3.11).

The variation principle identifies two Hamiltonians: $H$-satisfying (3.6) with function of action $S(t,x)$, and $H_o$ (3.20), whose function action $S_o(t,x)$ reaches absolute minimum at moments $(\tau_k)$ (3.11) of imposing constraint $P_o = P_o[x(\tau)]$.
Substituting (3.2) and (3.17b) in both (3.16) and (3.20), leads to Lagrangian and Hamiltonian on the extremals:

$$L_o(x, X_o) = 1/2 \dot{x}^T X_o = H_o . \qquad (3.21)$$

Using $\dot{X}_o = -\partial H_o / \partial x$ brings $\dot{X}_o = -\partial H_o / \partial x = -1/2 \dot{x}^T \partial X_o / \partial x$,
and from constraint (3.10) it follows

$$\partial X_o / \partial x = -b^{-1} \dot{x}^T X_o \text{ and } \partial H_o / \partial x = 1/2 \dot{x}^T b^{-1} \dot{x}^T X_o = 2 H_o X_o, \qquad (3.22)$$

which after substituting (3.17b) leads to extremals (3.9).
From the Eq. for conjugate vector (3.3), Eqs. (3.7), (3.8), and (3.17c) the constraint (3.10) acquires form

$$\frac{\partial X_o}{\partial x} = -2 X_o X_o^T . \qquad (3.23)$$

Differentiating (3.9) leads to a second order differential Eqs on the extremals:

$$\ddot{x} = 2b \dot{X}_o + 2 \dot{b} X_o, \qquad (3.24)$$

which after substituting (3.22) leads to (3.14). ● This solution simplifies proof of Theorem 3.1[32].
Comments.
The entropy force-gradient of entropy (3.6a) arises as a covariant coordinate-unit of displacement, or a distance in each observation when the microprocess appears. Until the space coordinate emerges, such force does not exist.
With appearance of the logic Bit in certain logic, the certain time-space emerges. In such time–space, time is a contravariant vector and the force is covariant vector in the rotating coordinate system dependent of its probabilistic choice in the time course in the observation. In the symmetric Lorentz group' transformation and Riemann geometry, the time and space are independent contravariant and covariant vectors accordingly, whose product is a scalar independent on the coordinate system. Within the microprocess of virtual observer, the time-space emerges as probabilistic, whose probabilities are symmetrical. In the certain logic of the observer, such time-space is symmetrical and certain along with Hamiltonian Dynamics above. With emerging information Bits, the created information dynamics is irreversible and asymmetrical. Observation of the emerging multiple information bits accompanies with a peace –wise Hamiltonian when its symmetry (reversibility) alternates with asymmetry (irreversibility).
Thus, each observer holds own real coordinates system emerging with observing the information macrodynamics.
Universal–single coordinate system may emerge for all multiple observers in the Universe. ●

**3.4. The initial conditions for the entropy functional and its extremals**
The *initial conditions for the EF* determines ratio of primary a priori- a posteriori probabilities beginning the probabilistic observation:

$$p(o_s^p) = \frac{P_{s,x}^a}{P_{s,x}^p}(o_s^p) . \qquad (4.1)$$

Start of the observation evaluates minimal probability and entropy [45]:

$$p(o_s^p) \cong 1.65 \times 10^{-4} \qquad (4.2)$$

$$\Delta s_{ap}(o_s^p) = -\ln p(o_s^p) = 0.5 \times 10^{-4} \qquad (4.3)$$

with minimal posterior probability $P_{poo} \approx 1 \times 10^{-4}$. (4.4)

That gives estimation of an average initial entropy of the observation:

$$S(o_s^p) = [-\ln p(o_s^p) \times P_{poo}] \cong 0.5 \times 10^{-8} Nat.$$ (4.5)

Based on physical coupling parameter $h_\alpha^o = 1/137$, physical observation theoretically starts with entropy:

$$S(o_{rs}^p) = 2/137 \cong 0.0146 Nat,$$ (4.6)

while entropy of the first real 'half-impulse' probing action starts at moment $t^{oe}$:

$$S_{ko}(t^{oe}) = 0.358834 Nat$$ (4.7)

with a priori-a posteriori probabilities $P_{ako} = 0.601, P_{pko} = 0.86$.

In the presenting approach, involving no material entities, physical process begins with the real probing action, and the physical coupling may start with minimizing this entropy at beginning the information process.

At real cut of posteriori probability $P_{po} \to 1$, ratio of their priori-a posteriory probabilities $P_{ao}/P_{po} \cong 0.8437$ determines minimal entropy shift between interacting probabilities $P_{ao} \to P_{po}$ during real cut: $\Delta s_{apo} = -\ln(0.8437) \cong 0.117 Nat$, which after averaging at $P_{po} = 1$ leads to

$$S(o_r^p) = 0.117 Nat.$$ (4.8)

That conforms (2.2.2.7).

Minimal entropy cost on covering the gap during it conversion to information is $s_{ev} \cong 0.0636 Nat$ (3.35).

Theoretical start of observation (4.6) is a potential until an Observer gets information from that entropy.

If we *define* the launch of real Information Observer by minimal converting entropy (4.8), then opening real observation *defines* entropy of first real 'half-impulse' probing actions (4.7), which could be multiple for a multi-dimensional process.

Since the potential observing physical process with the logical structural entities is possible (virtual) approaching entropy (4.6), this observation is a *virtual* until real observation generating information Observer starts and confirms it.

(The term 'virtual' here associates with physically possibility, until this physical objectivity becomes information-physical reality for the information Observer.)

Specifically, if the start of virtual observation associates with entropy binding primary a priori-a posteriori probability (4.5), then Virtual Observer identifies the entropy of appearing potential coupling structures (4.6).

Here, we have *identified* both beginning of virtual and physical observations and the virtual and real information Observers, based on the actual quantitative parameters, which are independent for each particular Observer.

However it may impose specific limitations after the Observer has formed [45, Sec.2.6].

The initial conditions for the EF *extremals* determine function $x_\pm(t^e) = x(s^+)\exp(u_\pm t^e)$ (from (2.9)), which at moment $t^e = s_k^{+o} b_k(t^e)/b_k(s_k^{+o})$, starts the virtual or real observations, depending on required minimal entropy of related observations (4.5-4.8). It brings functions

$$x_\pm(t^e) = x(s^+)\exp(\pm 2t^e) \quad (4.9a), \qquad t^e = s_k^{+o} b_k(t^e)/b_k(s_k^{+o}),$$ (4.9b)

where (4.9b), at known dispersion function $b_k(t^e, s_k^{+o})$, identifies its time dependency $t^e = t^e(s_k^{+o}, b_k)$, while (4.9a) identifies initial conditions for the EF extreme conjugated process:

$$x_\pm(t^e) = x(s^+)\exp(\pm 2t^e(s_k^{+o}, b_k)).$$ (4.9)

Applying the EF solutions (2.3.6) at opposite relative time $t_-^* = -t_+^*$ lead to the entropy functions:

$$S_\pm(t_\pm^*) = 1/2 S_+(t_+^*) \times S_-(t_-^*) = 1/2[\exp(-2t_+^*)(Cos^2(t_+^*) + Sin^2(t_+^*) - 2Sin^2(t_+^*))] =$$
$$1/2[\exp(-2t_+^*)((+1 - 2(1/2 - Cos(2t_+^*))))] = 1/2\exp(-2t_+^*)Cos(2t_+^*) \quad (4.10)$$

This interactive entropy $S_\pm(t_\pm)$ becomes minimal interactive threshold (4.8) at $t_+^* = t_*^e$, which starts the information Observer. From (4.8) it follows:

$$S(t_+^* = t_*^e) = 1/2\exp(-2t_*^e)Cos(2t_*^e) = 0.117 \quad (4.11)$$

with relative time $t_*^e = \pm\pi/2t^e$.

Solution (4.11) will bring real $t^e$ which, after substitution in (4.9b), determines initial moment $s_k^{+o} = s_k^{+o}(t^e, b_k(t^e, s_k^{+o}))$ if dispersion functions $b_k(t^e, s_k^{+o})$ are known. Substituting relative to $t_*^e$ moment $s_{k*}^{+o} = s_k^{+o}(t_*^e, b_k)$ to

$$S(s_{k*}^{+o}) = 1/2\exp(-2s_{k*}^{+o})Cos(2s_{k*}^{+o}) \quad (4.12)$$

allows finding unknown posteriori entropy $S_\pm(s_k^{+o})$ starting virtual Observer at $s_k^{+o} = s_{k*}^{+o}/(\pi/2)$.

To find the moment of time, starting virtual Observer at maximal uncertainty measure (4.5) when dispersion functions unknown, only the joint pre-requirements (4.12) and (4.5) are available.

Eqs (2.10) for initial conditions $\ln x(s_k^{+o}) = x(s_k^{+o})u_\pm s_k^{+o}$ after integration lead to

$1/2[\ln x(s_k^{+o})]^2 = u_\pm(s_k^{+o})^2$, $\ln x(s_k^{+o}) = \sqrt{2u_\pm}(s_k^{+o})$ and to

$x_\pm(s_k^{+o}) = \exp(\pm\sqrt{2u_\pm})(s_k^{+o}), x_+(s_k^{+o}) = \exp(\pm\sqrt{2\times 2})s_k^{+o}, x_-(s_k^{+o}) = \exp(\pm j\sqrt{2\times 2})s_k^{+o}, u_+ = 2, u_- = -2$.

It brings both real and complex initial conditions for starting extremal processes in virtual observer:

$$x_+^i(s_k^{+o}) = \exp(\pm 2s_k^{+o}), x_-^i(s_k^{+o}) = \exp(\pm 2js_k^{+o}) = Cos(2s_k^{+o}) \pm jSin(2s_k^{+o}), \quad (4.13)$$

where first one deescribes virtual trajectory before the impulse generates the complex microproces (2.3.6).

Finally the trajectories of extreme processes (4. 9a) by moment $t_i^e$ of each $i$-dimension takes form

$$x_\pm^i(t_i^e) = x_\pm^i(s_k^{+o})[Cos(2s_k^{+o}) \pm jSin(2s_k^{+o})]\exp(\mp 2t_i^e). \quad (4.14)$$

Let us numerically validate the results (4.11-4.14). Solution of (4.11):

$\ln 1/2 - 2t_*^e + \ln[Cos(2t_*^e)] = \ln 0.117, -0.693 + 2.1456 = 2t_*^e - \ln[Cos(2t_*^e)], 1.4526 = 2t_*^e + \ln[Cos(2t_*^e)]$

leads to $2t_*^e \approx 1.45, t^e = 1.45/\pi \approx 0.46$ - as one of possible answer.

Applying condition (4.6) to (4.12):

$$S(s_{k*}^{+o}) = 1/2\exp(-2s_{k*}^{+o})Cos(2s_{k*}^{+o}) = 2/137 \quad (4.14a)$$

leads to solution

$-0.693 + 4.22683 = 2s_{k*}^{+o} - \ln[Cos(2s_{k*}^{+o})], 3.534 = 2s_{k*}^{+o} - \ln[Cos(2s_{k*}^{+o})]$

with result $s_{k*}^{+o} \approx 1.767, s_k^{+o} \approx 1.12$.

Applying condition of beginning virtual observation (4.5) at relative time $o_{s*}^p$ to

$$S(o_{s*}^p) = 1/2\exp(2o_{s*}^p)Cos(2o_{s*}^p) = 0.5\times 10^{-8} \quad (4.14b)$$

leads to $-0.693 + 12.8 = 2o_{s*}^p - \ln[Cos(2o_{s*}^p)]$ with solutions $2o_{s*}^p \approx 12$ and $o_s^p \approx 3.85$.

Applying condition of starting information observation (4.7) with eq. (4.10) at relative time $t_*^{oe}$, leads to solution $2t_*^{oe} \approx 0.33, t^{oe} = 2t_*^{oe}/\pi \approx 0.1$.

These time moments are counting from the real Observer after the observing process overcomes the threshold (4.8).

This means that $o_s^p \approx 3.85$ evaluates time interval of virtual observation, while virtual observer starts on time interval $s_k^{+o} \approx 1.12$, and real observer starts on $t^e \approx 0.46$.

Whereas observing the first real 'half-impulse' probing action takes *part* of this time: $t^{oe} \approx 0.1$.

States $x_+^i(s_k^{+o}) = \exp(\pm 2 \times 1.12)$ hold probability $P_{ako} = 0.601$ and have multiple correlations $r_{\pm}^x(s_k^+) = x_+^i(s_k^{+o})x_-^i(s_k^{+o})$, starting virtual observer (4.12, 4.12a). Conjugates processes (4.14) interact through correlation

$$r_{\pm}^x(t_i^e) = x_+^i(t_i^e) \times x_-^i(t_i^e) = r_{\pm}^x(s_k^+)[Cos(2s_k^{+o})^2 + jSin(2s_k^{+o})^2]\exp(-2t_i^e)\exp(+2t_i^e) = 1 \qquad (4.14c)$$

reaching information observer' threshold (4.8) with relative probability $P_{ao}/P_{po} \cong 0.8437$ and two conjugated entropies

$$S_{\pm}(t^e) = 1/2\exp(\pm\pi/2 \times t^e)Cos(\pm\pi/2 \times t^e), \qquad (4.15)$$

following from (4.11). Processes (4.15) moves the entangled entropies.

The starting extemal process (4.14) evaluate two pairs of real states for the conjugated process:

$$x_{\pm}^r(t_i^e) = 9.39 \times 0.999 \times 0.3985 = 3.738,\ x_{\mp}^r(t_i^e) = 0.1064 \times 0.999 \times 2.509 = 0.2666$$
$$x_{\mp}^{r1}(t_i^e) = 0.1064 \times 0.999 \times 0.3985 = 0.042,\ x_{\pm}^{r1}(t_i^e) = 9.39 \times 0.999 \times 2.509 = 23.536 \qquad (4.15a)$$

Imaginary initial conditions evaluate four options:

$$x_+^{im1}(t^e) = \exp(2 \times 1.12)[\pm jSin(2 \times 1.12)] \times \exp(-0.92) = j3.064 \times 0.039 \times 0.3985 \cong \pm j0.0475 \qquad (4.15b)$$

$$x_-^{im2}(t^e) = \exp(-2 \times 1.12)[\mp jSin(2 \times 1.12)] \times \exp(0.92) = \mp j0.323 \times 0.039 \times 2.509 \cong \mp j0.0316. \qquad (4.15c)$$

**4.5. Applying equation of extremals $\dot{x} = a^u$ to a dynamic model's traditional form**:
$$\dot{x} = Ax + u, u = Av, \dot{x} = A(x+v), \qquad (5.1)$$
where $v$ is a control $u$ reduced to state vector $x$. Solving the initial variation problem (VP) for this model allows finding optimal control $v$ and identifying matrix $A$ under this control's action.

Proposition 5.1.

The reduced control is formed by a feedback function of macrostates $x(\tau) = \{x(\tau_k)\}, k = 1,...,m$ in the form:
$$v(\tau) = -2x(\tau). \qquad (5.2)$$
Or using (5.1), it applies to $u = u(x(\tau)) = u(\tau)$ as function of speed of the macroprocess in (5.1):
$$u(\tau) = -2Ax(\tau) = -2\dot{x}(\tau), \qquad (5.3)$$
at the localities of moments $\tau = (\tau_k)$ (3.3.11). Then matrix $A$ determines equations
$$A(\tau) = -b(\tau)r_v^{-1}(\tau), r_v = E[(x+v)(x+v)^T], b = 1/2\dot{r}, r = E[\tilde{x}\tilde{x}^T] \qquad (5.4)$$
and $A$ identifies the correlation function with its derivative, or directly dispersion matrix $b$ from (1.2.1):
$$|A(\tau)| = b(\tau)(2\int_{\tau-o}^{\tau} b(t)dt)^{-1}, \tau - o = (\tau_k - o), k = 1...,m. \qquad \bullet \qquad (5.5)$$

Proof. Using the variation Eq. for the conjugate vector (3.3) allows writing constraint (3.10) in the form
$$\frac{\partial X}{\partial x}(\tau) = -2XX^T(\tau), \qquad (5.6)$$
for model (5.1). It leads to Eqs for the conjugate vector in this model:
$$X = (2b)^{-1}A(x+v), X^T = (x+v)^TA^T(2b)^{-1}, \frac{\partial X}{\partial x} = (2b)^{-1}A, b \neq 0. \qquad (5.7)$$
After substituting (5.7) to (5.6) it acquires form
$$(2b)^{-1}A = -2E[(2b)^{-1}A(x+v)(x+v)^TA^T(2b)^{-1}], \qquad (5.8)$$
from which, at a nonrandom $A$ and $E[b] = b$, the identification equations (5.4) follow strait.

Completion of (5.6),(5.7) performs the control's action. Using (5.4), leads to(5.8) under the control in form
$A(\tau)E[(x(\tau)+v(\tau))(x(\tau)+v(\tau))^T] = -E[\dot{x}(\tau)x(\tau)^T]$, at $\dot{r} = 2E[\dot{x}(\tau)x(\tau)^T]$.

This relation after substituting (5.1) leads to
$A(\tau)E[(x(\tau)+v(\tau))(x(\tau)+v(\tau))^T] = -A(\tau)E[(x(\tau)+v(\tau))x(\tau)^T]$ and then to
$E[(x(\tau)+v(\tau))(x(\tau)+v(\tau))^T + (x(\tau)+v(\tau))x(\tau)^T] = 0$,
which is fulfilled at applying control (5.2).

Since $x(\tau)$ is a discrete set of states, satisfying (3.11), (3.13), the control has a discrete form.

Each stepwise control (5.2), with its inverse value of doubling controlled state $x(\tau)$, applied to both Eqs (5.7), implements (5.6); while (5.6), following from variation conditions (3.1a), fulfills this condition.

This control, applied to additive functional (1.1.7), imposes constraint (3.8,3.10) which limits transformation of random process' segments to the process extremals.

By applying the control step-down and step-up actions to satisfy conditions (3.7) and (3.10), the control sequentially starts and terminates the constraint, while extracting the cutoff hidden information on the $x(\tau)$-localities. Performing the transformation, this control initiates the identification of matrix $A(\tau)$ (5.4, 5.5) during its time interval $\tau-o,\tau,\tau+o$ (Sec.1.3.2), solving simultaneously the identification problem [70]. •

Obtaining this control here *specifies* some results of Theorems 4.1 [32].

Corollary.5.1.

The control, which turns on the constraint (3.8,3.10), creates the Hamilton dynamic model with complex conjugated eigenvalues of matrix $A$. After the constraint's termination, the control transforms this matrix to its *real* form (on the diffusion process' boundary point [71]), which identifies diffusion matrix in (5.4). Thus, within each extremal segment, the dynamics is reversible; irreversibility rises at each constraint termination between the segments. The EF-IPF Lagrangian integrates both the impulses and constraint information on its time space-intervals. •

Proposition 5.2.

Let us consider controllable dynamics under feedback control (5.3), described by operator $A^v(t,\tau)$ with eigenfunctions $\lambda_i^v(t_i,\tau_k)_{i,k=1}^{n,m}$, whose matrix equation:

$\dot{x}(t) = A^v x(t)$,   (5.9)

includes the feedback control (5.3).

The drift vector for both models (5.1) and (5.9) has same form:
$a^u(\tau,x(\tau,t)) = A(\tau,t)(x(\tau,t)+v(\tau)); A(\tau)(x(\tau)+v(\tau)) = A^v(\tau)x(\tau)$ .   (5.9a)

Then the followings hold true:

(1)-Matrix $A(t,\tau)$ under control $v(\tau_k^o) = -2x(\tau_k^o)$, applied during time interval $t_k = \tau_k^1 - \tau_k^o$ of the identification $A(t_k)$, depends on initial matrix $A(\tau_k^o)$ at the moment $\tau_k^o$ according to Eq

$A(t_k) = A(\tau_k^o)\exp(A(\tau_k^o)t_k)[2-\exp(A(\tau_k^o)t_k)]^{-1}$ .   (5.9b)

(2)- The identification Eq.(5.4) at $\tau_k^1 = \tau$ holds
$A^v(\tau) = -A(\tau) = 1/2 b(\tau)r_v^{-1}(\tau), b(\tau) = 1/2\dot{r}_v(\tau)$,   (5.9c)

whose covariation function $r_v(\tau_k^o)$, starting at moment $\tau_k^o$ by the end of this time interval $\tau_k^1$, acquires form
$r_v(\tau_k^1) = [2-\exp(A(\tau_k^o)t_k)]r(\tau_k^o)[2-\exp(A^T(\tau_k^o)t_k)]$.   (5.9d)

(3a)-At moment $\tau_k^o + o$ following $\tau_k^o$, at applying control $v(\tau_k^o) = -2x(\tau_k^o)$, the controllable matrix gets form
$$A^v(\tau_k^1)_{t_k \to 0} = A^v(\tau_k^o + o) = -A(\tau_k^o), \tag{5.9e}$$
changing the initial matrix sign.

(3b)- When this control, applied at the moment $\tau_k^1$, ends the dynamic process on extremals in following moment $\tau_k^1 + o$, at $x(\tau_k^1 + o) \to 0$, function $a^u = A^v x(t)$ in (5.9a) turns to
$$a^u(x(\tau_k^1 + o)) \to 0; \tag{5.9f}$$
which brings (5.9a) to its dynamic form $a^u = A(\tau_k^1 + o)v(\tau_k^1 + o) \to 0$ that requires turning the control off.

(3c)-At fulfilment of (5.9f), the Ito' stochastic Eq. [26] includes only its diffusion component, which identifies dynamic matrix $A(\tau_k^1 + o)$. This matrix, transforming in following moment $\tau_{k+1}^1 : A(\tau_k^1 + o) \to A(\tau_{k+1}^1)$ through the identified correlation matrix $r(\tau_{k+1}^1)$, holds
$$A(\tau_{k+1}^1) = 1/2\dot{r}(\tau_{k+1}^1)r^{-1}(\tau_{k+1}^1) \text{ at } r(\tau_{k+1}^1) = E[\tilde{x}(t)\tilde{x}(t+\tau_{k+1}^1)^T] = r^v(\tau_{k+1}^1)_{v(\tau_k^1+o) \to 0}.$$

(3d)-The dispersion matrix on the extremals, identifies by covariation matrix (5.4), acquires form
$$\partial r_v(\tau_k^1)/\partial t_k = -A(\tau_k^o)\exp(A(\tau_k^o)t_k)r(\tau_k^o)[2 - \exp(A^T(\tau_k^o)t_k)] + \\ [2 - \exp(A(\tau_k^o)t_k)]r(\tau_k^o)[-A^T(\tau_k^o)\exp(A^T(\tau_k^o)t_k)] \tag{5.10}$$
which for symmetric matrix $A(\tau_k^o)$ leads to relations
$$\partial r_v(\tau_k^1)/\partial t_k |_{t_k=\tau_k^1} = -2A(\tau_k^o)\exp(A(\tau_k^o)\tau_k^1)r(\tau_k^o), b(\tau_k^1) = -A(\tau_k^o)\exp(A(\tau_k^o)t_k)r(\tau_k^o),$$
$$b(\tau_k^1 = \tau_k^o) = -A(\tau_k^o)\exp(A(\tau_k^o)0_k)r(\tau_k^o) = -A(\tau_k^o)r(\tau_k^o), b(\tau_k^1)/b(\tau_k^o) = \exp(A(\tau_k^o)\tau_k^1)A(\tau_k^o)^{-1}. \tag{5.10a}$$

Last ratio in (5.10a) for a single dimension, at $A(\tau_k^o) = \alpha_1(\tau_{k1}^o)$, leads to
$$b(\tau_{k1}^1)/b(\tau_{k1}^o) = \exp\alpha_1(\tau_{k1}^o)/\alpha_1(\tau_{k1}^o) \tag{5.10b}$$
which after applying relation (1.3.2.12) in form $b(\tau_{k1}^1)/b(\tau_{k1}^o) = \tau_k^1/\tau_{k1}^o$ leads to
$$\tau_k^1 = \tau_{k1}^o \exp\alpha_1(\tau_{k1}^o)/\alpha_1(\tau_{k1}^o), \tag{5.10c}$$
connecting interval $t_k = \tau_k^1 - \tau_k^o$ with eigenvalue at $\tau_{k1}^1 = \tau_{k1}^o[\alpha_1(\tau_{k1}^o)]$ that measures information speed on the interval (5.10c).

(3e)- Equation for conjugated vector (5. 7) on each extremal segments follows from relation:
$$X_o(t_k) = 2b(t_k)^{-1}\dot{x}(t_k) = 2A(\tau_k^o)\exp(A(\tau_k^o)t_k)x(\tau_k^o)A^T(\tau_k^o)\exp(A^T(\tau_k^o)t_k). \tag{5.11}$$

(3f)- Entropy increment $\Delta S_{io}$ on optimal trajectory at
$$E[\frac{\partial \tilde{S}}{\partial t}(\tau)] = 1/4Tr[A(\tau)] = H(\tau), A(\tau) = -1/2\sum_{i=1}^n \dot{r}_i(\tau)r_i^{-1}(\tau), (r_i) = r, \tag{5.11a}$$
measured on the cutting localities, determines the time interval ratio $\tau_i/\tau_{i-1}$ of the nearest segments:
$$\Delta S_{io} = I_{x_t}^p = -1/8\int_s^T Tr[\dot{r}r^{-1}]dt = -1/8Tr[\ln(r(T)/\ln r(s)], (s = \tau_o, \tau_1, ..., \tau_n = T) \cdot \bullet \tag{5.12}$$

*Proof* (1). Control $v(\tau_k^o) = -2x(\tau_k^o)$, imposing the constraint at $\tau_k^o$ on both (5. 6) and (5.9) and terminating it at $\tau_k^1$ on time interval $t_k = \tau_k^1 - \tau_k^o$, brings solutions of (5.1) by the end of this interval:
$$x(\tau_k^1) = x(\tau_k^o)[2 - \exp(A(\tau_k^o)t_k)]. \tag{5.13}$$

Substituting this solution to the right side of $\dot{x}(\tau_k^1) = A^v(\tau_k^1)x(\tau_k^1)$ and to its derivative on the left side leads to

$-x(\tau_k^o)(A(\tau_k^o)t_k)\exp(A(\tau_k^o)t_k) = A^v(\tau_k^o)x(\tau_k^o)[2 - \exp(A(\tau_k^o)t_k)])]$,

or to connection of both matrixes $A^v(\tau_k^1)$ and $A(\tau_k^1)$ (at the interval end) with matrix $A(\tau_k^o)$ (at the interval beginning):

$$A^v(t_k) = -A(\tau_k^o)\exp(A(\tau_k^o)t_k)[2 - \exp(A(\tau_k^o)t_k)], \quad (5.14)$$

and to $A^v(\tau_k^1) = -A(\tau_k^1)$ by moment $\tau_k^1$, from which follows (5.9e).

Other *Proofs* of the parts are straight forward. •

The identified functions drift and diffusion of the EF through the cutoff feedback automatically transforms the EF to IPF revealing the integrated information hidden in the observing process cutting correlations.

The initial conditional probability (1.1.2) determines the probability measure along the extremal trajectory:

$$p[x(t)] = p[x(s)]\exp(-S[x(t)]) \quad (5.14a)$$

where starting probability

follows from (4.12) and numerical values (4.14a),(4.15,4.15a,b).

### 3.5.1. Finding the invariant relations

Using (5.3) in form $u(\tau) = -2Ax(\tau) = -2\dot{x}(\tau)$, and $c^2 = |u_+ u_-| = c_+ c_- = \bar{u}^2, c_+ = u_+, c_- = u_-$ from (2.2.4, 2.2.10) leads to

$$c^2 = \dot{x}(\tau) = -2Ax(\tau), \ln x_k(\tau)/\ln x_k(s) = \alpha_{ko}(\tau - s) = u_\pm(\tau - s) \quad (5.15)$$

where $\alpha_{io}$ is engenvalue of matrix $A(\tau_k^o) = \|\alpha_{ko}\|$ starting information speed on interval $t_{ko} = (\tau - s)$ of imposing constraint (3.10.318) on each inavriant interval.

Applying constraint (3.10) to (5.15) brings

$$\alpha_{ko}(\tau - s) = u_\pm(\tau - s) = inv = \mathbf{a}_o, \quad (5.15a)$$

where invariant $\mathbf{a}_o = \mathbf{a}_o(\gamma_k)$ depends on ratio of imaginary to real eigenvalues of matrix (5.5): $\gamma_k = \beta_{ko}/\alpha_{ko}$.

From (5.15a), and (2.9) in particular, it follows numerical value for real eigenvalues and matrix

$$\alpha_{io} = \pm 2(\tau - s), A(\tau_k^o) = \|\pm 2(\tau - s)\|. \quad (5.15b)$$

Since $u_\pm = 2$ is a real act of the interacting Markov process, it determines the real eigenvalue $\alpha_{ko} = 2$ at $k = n = 1$.

For optimal model with information invariant $\mathbf{a}_o(\gamma_k \to 0.5) = \ln 2$, it leads to $\mathbf{a}_o/\alpha_{ko} = (\tau - s) = \ln 2/2 = 0.346$, which evaluates $\delta_k = \tau_k^{+o} - \tau_k^{-o} \cong 0.35$ [45].

Invariant $\mathbf{a}_o = \ln 2 \cong 0.7$ measures information generating at each impulse cut (Secs. 3.3, 3.4).

Correlation matrix (5.9d), measured by the `optimal model's invariant, takes form

$$r_v(\tau_k^1) = [2 - \exp\mathbf{a}_o)]r(\tau_k^o)[2 - \exp(\mathbf{a}_o)] = r(\tau_k^o) \times [1.5^2], \quad (5.16)$$

where vector $x(\tau_k^1) = x(\tau_k^o) \times [1.5]$ measures its extreme relation (5.13). Conjugate vector

$$X_o(\tau_k^1) = 2A(\tau_k^o)\exp(\mathbf{a}_o)x(\tau_k^o)A^T(\tau_k^o)\exp(\mathbf{a}_o) \quad (5.16a)$$

for a single dimension holds

$$X_{o1}(\tau_k^1) = 2\alpha_1(\tau_{k1}^o)^2 \exp(2\mathbf{a}_o)x(\tau_{k1}^o), \quad (5.16b)$$

or at $\alpha_1(\tau_{k1}^o) = 2\mathbf{a}_o/t_k, t_k \to 0, t_k \to 0, X_{o1}(\tau_k^1) = (2\mathbf{a}_o/t_k)^2 \exp(2\mathbf{a}_o)x(\tau_{k1}^o) \to \infty$.

It means at decreasing time intervals $t_k$ between the impulse' generated information invariants $\mathbf{a}_o$, information force (5.16b) grows infinitely. The force grows in square function of time intervals, which involve a potential overrunning the impulse with a minimal $t_k$.

It leads to possibility of pulling together the real action and its result for the minimal time impulse,(Sec.2.7).

The EF-IPF *estimate* invariant's measure $\mathbf{a}_o(\gamma_k)$, counting both the segment's and inter-segment's increments:

$$\tilde{S}^i_{\tau m} = \sum_{k=1}^{m}(\mathbf{a}_o(\gamma_k) + \mathbf{a}_o^2(\gamma_k)), \tilde{S}_\tau = \sum_{i=1}^{n}\tilde{S}^i_{\tau m}, \qquad (5.17)$$

where $m$ is number of the segments, $n$ is the model dimension (assuming each segment has a single $\tau_k$-locality). However, to *predict* each $\tau_k$-locality, where information should transfer the feedback, only invariant measure $\mathbf{a}_o(\gamma_k)$ needs. Sum of process's invariants

$$\tilde{S}^{io}_{\tau m} = \sum_{k=1}^{m}\mathbf{a}_o(\gamma_k), \quad \tilde{S}^o_\tau = \sum_{i=1}^{n}\tilde{S}^{io}_{\tau m} \qquad (5.18)$$

estimates EF entropy with maximal process' probability (5.14a) expressed through $\mathbf{a}_o = \mathbf{a}_o(\gamma_k)$.

This entropy allows encoding the observing process using Shannon's formula for an average optimal code-word length:

$$l_c \geq \tilde{S}^o_\tau / \ln D, \qquad (5.19)$$

where $D$ is the number of letters of the code's alphabet, which encodes $\tilde{S}^o_\tau$ (5.18).

An elementary code-word to encode the optimal process' segment is

$$l_{cs} \geq \mathbf{a}_o(\gamma_k) / \log_2 D_o, \qquad (5.20)$$

where $D_o$ is code alphabet, which implements the microstates invariant connections on the extremal segment.

At $\mathbf{a}_o(\gamma_k \to 0.5) \cong 0.7$, $D_o = 2$, it follows $l_{cs} \geq 1$, or (5.20) encodes a bit per encoding the alphabet letter.

With the values of $x^r_\pm(t^e), x^{im1}_-(t^e), x^{im2}_-(t^e)$ (4.15, 4.15a,b,c) start Hamiltonian process (Sec.3) and correlation (5.16), which identify the initial segment's state $x(\tau^o_{k1o}) = x(\tau^o_k)$ (in (5.13)) and the eigenvalues of matrix (5.14) in process forming information units. The connection to each following segment on the extremal trajectory determines segment state $x(\tau^o_{k1})$ (5.13). Each $x(\tau^o_{k1o}), x(\tau^o_{k1})$ enfolds the impulse microprocess and starts the macroprocess segment.

Moment $t^e$ identifies starting dispersion and correlation on the optimal trajectories that determine operator of information speed $A(t, \tau_k)$, Hamiltonian, and both EF-IPF on the trajectory segments.

Optimal control (5.3) starts with each segment initial states.

The details of information micro-macrodynamics (IMD), based on the *invariant* $\mathbf{a}_o(\gamma_k)$ *description*, are in [43,44], where scale parameter $\gamma^\alpha_{k,k+1} = \alpha_k / \alpha_{k+1}$ of the dynamics depends on frequency spectrum of observations detecting through the identified invariant parameter $\gamma_k$.

The observer self-scaled observation initiates its time-space IMD which builds distributed information network [72].

*The above equations finalize both math description of the micro-macro processes and validate them numerically.*

## 4. Arising the observer collective information

### 4.1. Rising a cooperative attraction

Cooperative rotation and ordering start with entangling entropy increments in the entropy volumes.

Then, the sequential impulse' entropy increments with their volumes involve in collective rotating movement.

Since each following impulse may start only after the previous impulse cutting time $\delta^t_{ei}$ (3.3.3b) will triple, time interval between the impulse cutoff actions:

$$\Delta_t = 3\delta^t_{ei} \cong 1.2 \times 10^{-15} \text{ sec} \qquad (1.1)$$

imposes limitation on adjoining a pair of the impulse entropies in collective movement in imaginary time course $\Delta_t$.
This limitation is applicable to virtual Observer with its virtual impulse, which can initiate a virtual collective movement of the adjoining entangled volumes. That requires an entropic force

$$X_e = \Delta s_o / \Delta l_o, \tag{1.2}$$

where $\Delta s_o \cong 0.25 Nat$ is a minimal entropy increment between the impulse, and $\Delta l_o \cong \Delta_{lo}$ (3.2.14) is a distance between nearest impulses with that entropy. At these relations, the entropic force measures

$$X_e \cong 0.25/14.4 \times 10^{-5} \cong 0.1736 \times 10^4 [Nat/m]. \tag{1.3}$$

Speed of rotating moment $\delta M_e / \delta t$ defined by force (1.3) and velocity (3.2.12a) in relation

$$\delta M_e / \delta t = X_e w_o \tag{1.4}$$

characterizes intensively of the rotation which evaluates

$$\delta M_e / \delta t = 0.344 \times 10^6 [Nat/m\sec]. \tag{1.4a}$$

Since both rotating moment's speed and entropy force proceed between the impulses, relation (1.4a) evaluates intensively of attracting rotation intended on capturing next impulse's entropy increment in joint distributed rotating movement.
It measures cooperative connection of the impulses entropy increments prior forming next information units.
Thus virtual collective movement (between the entangle increments entropy increments) may emerge before a following impulse kills the entropy volumes.
Perhaps, that movement engages a cooperative transition of the entangled entropy volumes (with starting speed (3.3.6)) to the cutting gap. It makes possible a collective entanglement that not requires spending energy.
Study [80] "demonstrates that the spreading of entanglement is much faster than the energy diffusion in this nonintegrable system". Such a cooperative virtual distribution rotates the involving entangled groups-an ensemble for a joint preparing them to the following killing-cut producing information units.
Therefore, the distributed rotation primary involves the entropies of the Bayesian linked probabilities of virtual interactive impulses currying different frequencies. The entangled entropies' volumes, in a cooperative rotation, engage in the transition movement up to cutting them on information units. The rotation continues connecting the forming information units.

**4.2. Forming a triple consolidation of information units during cooperative rotation**

Imaginary entropy in each virtual impulse predefines the real information of a certain impulse starting movement with information speed $\alpha_{io}$, time interval $t_{io} = (\tau - s)$ and invatriant $\mathbf{a}_{io} = \alpha_{io} t_{io}$.

Speed $\alpha_{io}$ of generation $\mathbf{a}_{io}$ starts a single real information unit (after killing entropy volume).
Speed of cutting correlation is an initial source of real information speed, while imaginary information speed arises in the microprocess at $s$-locality on the minimax trajectory of observing process.
Ending value of this speed $c_{ev}$ transits the entropy speed $\beta_{io}$ to $\mathbf{a}_{io}$.

$$\beta_{io} = c_{ev}. \tag{2.1}$$

Imaginary $\beta_i$ and real $\alpha_i$ speeds are components of the dynamic process' complex eigenvalues:

$$\text{Re}\lambda_{io} = \alpha_{io}, \text{Im}\lambda_i = \beta_{io} \; \lambda_{io} = \alpha_{io} \pm j\beta_{io}, \tag{2.2}$$

which determines the ratio of their starting imaginary and real components $\beta_{io}, \alpha_{io}$: $\gamma_{io} = \beta_{io}/\alpha_{io}$.
When the unit gets complete information $\mathbf{a}_{io}$ with real speed $\alpha_{io}(t_{io})$, the imaginary speed will be turned to zero by the end of time interval $t_{io}$. That requirement leads to Eq. connecting and $\gamma_{io}$ [43]:

$$2\sin(\gamma_i \mathbf{a}_{io}) + \gamma_i \cos(\gamma_i \mathbf{a}_{io}) - \gamma_i \exp(\mathbf{a}_{io}) = 0. \tag{2.3}$$

From that, at $\mathbf{a}_{io} \cong \ln 2$ follows ratio
$$\gamma_{io} = \beta_{io} / \alpha_{io} \to (0.4142 - 0.5). \tag{2.4}$$
which imposes limitation on $\beta_{io}$ by minimal speed of transferring entropy volume (3.3.6):
$$\beta_{io} = c_{ev} = 0.596 \times 10^{15} \, Nat/\sec. \tag{2.5}$$
Then, from (2.4)-(2.5) follow information speed of starting single real information (after killing the entropy volume):
$$\alpha_{io} = c_{iv} \cong 2.4143 \times 0.596 \times 10^{15} \, Nat/\sec \cong 1.44 \times 10^{15} \, Nat/\sec, \tag{2.6}$$
which determines minimal time interval of completion information unit $\mathbf{a}_{io}$:
$$t_{io}^o \cong 0.48 \times 10^{-15} \sec, \tag{2.6a}$$
satisfying the minimax. Estimation (2.6a) is close to $\delta_{ei}^t$ (3.3.3b).

Invariant of imaginary information $\mathbf{b}_o^{'} = \beta_{io} t_{io}$ satisfies condition of turning the real component to zero by the end of time interval $t_{io}$ at completion this imaginary information.
From that condition follows Eq.
$$2\cos(\gamma \mathbf{b}_o^{'}) - \gamma \sin \gamma(\mathbf{b}_o^{'}) - \exp(\mathbf{b}_o^{'}) = 0 \tag{2.7}$$
connecting $\mathbf{b}_o^{'}$ with ratio $\gamma = \gamma_{io}$.

Solution (2.7): $\beta_{io} t_k = \pi/6$ at (2.6) evaluates $t_{ko} : t_{ko} = \pi/6/0.596 \times 10^{15} = 0.8785 \times 10^{-15} \sec$,
which determines minimal interval of turning inaginary speed (2.6) to zero; $t_{ko}$ is close to the constraint interval between impulses (1.1). It's seen that imaginary (virtual) time interval is wider than real (2.6a).
Ratio of the invariant
$$\beta_{io} t_{ko} / \alpha_{io} t_{io} = \gamma_{io} t_{ko} / t_{io} = \gamma_{io}^{ko}, \gamma_{io}^{ko} \cong 0.915, \tag{2.7a}$$
evaluates *attractiveness* a real speed by the imaginary speed. Or how, at the same fixed time $\delta_{tio}$, imaginary entropy $\delta_{Eio} = \beta_{io} \delta_{tio}$ (2.7b) enables *attracting real information* at forming a single unit:
$$\delta_{Iio} = \alpha_{io} \delta_{tio}. \tag{2.8}$$
According to (2.6), and (2.7b), at a minimal coefficient of attraction:
$$\delta_{Eio} = 0.4142 \delta_{Iio}, \tag{2.8a}$$
a unit of entropy $\delta_{Eio} = 1$ may attract 0.4142 $\delta_{Iio}$ units of information.

Entropy volume $s_{ve} = 0.0636 \, Nat$, moving with minimal speed $c_{ev}$ (3.3.6), may attract potential-information (entropy) $i_v \cong 0.02634 \, Nat$ of a nearest impulse on minimal time interval $\Delta_t$ with attracting information speed
$$c_{ivo} = i_v / \Delta_t, \, c_{ivo} \cong 0.0548 \times 10^{15} \, Nat/\sec. \tag{2.9}$$
That might increase minimal transition speed $c_{ev}$ up to
$$c_{ev} + c_{ivo} \cong 0.65 \times 10^{15} \, Nat/\sec. \tag{2.9a}$$
With transition speed (2.9a), the rotating entropy volume moves to the cutting gap, engaging other entangled entropy volumes in a joint (collective) rotation with the speed of rotating moment (1.4).
Impulse, carrying $\cong 0.25 \, Nat$, cuts random process with entropy volume which evaluates entropy (3.3.3c):
$$\delta_e \cong 0.4452 \, Nat. \tag{2.9b}$$
The control cutting the volume should compensate for the interacting entropy increments $0.117 \, Nat$ (3.2.16a), which requires information of the applied control $0.25 + 0.117 = 0.367 \, Nat$.
This control, cutting the phase volume (3.3.1a), could bring information

$$s_{evo} = 0.367 \times 1.272 = 0.466824 \, Nat. \tag{2.10}$$

Additional potential information conveys the entropy of transferring volume $s_{ve} = 0.0636 \, Nat$, which may decrease the amount of attracting potential information (entropy) $s_v = i_v \cong 0.02634 \, Nat$ that follows from (2.8a).

The difference of the entropy volumes is
$$\delta s_{ve} = s_{ve} - s_v = 0.0374. \tag{2.10a}$$
Sum $\delta s_{ve} + s_{evo} \cong 0.5 \, Nat$ \hfill (2.10b)

coincides with (3.3.5a), which determines the amount of information delivered by the impulse cutting the random process. This information compensates for entropy of the virtual probing that delivers the entangled entropy volume. Potential cut of this volume brings total entropy (2.10b), which the real impulse converts to equivalent information thru memorizing.

Assuming virtual impulse spends part $\delta s_{ve}$ of the entropy volume on transition to cutting gap, the real impulse overcomes entropy threshold $s_{evo}$ by the real cut, producing information $0.5 \, Nat$, which includes information compensation for the virtual entropy volume
$$\delta s_{ve} = \delta s_{iv}. \tag{2.10c}$$

In multi-dimensional virtual process, correlations grow similarly in each dimension under manifold of impulse observation. The correlations, accumulated sequentially in time, increase with growing number of currently observed process' dimensions, entropy volumes, and collective attractions.

Each impulse cuts the increasing entropy volume, leading to rising density of the cutting entropy even at the invariant impulse size. The current density, which measures the ratio of the impulse volume to the cutting impulse wide, increases the impulse speed. Killing the distinct volumes densities converts them in the Bits distinguished by information density. Between these different Bits, an information gradient of attracting force rises, minimizing the difference, which prompts a collective memory. That connects elementary information Bits in units of information process.

**4.3. Conditions of forming optimal triplet with stable cooperative information structure. Emerging information macroprocess**

Information of each previous impulse starts attracting next cutting information in the rotating movement during the impulse imaginary time interval (1.1) with entropy force, moment's speed (1.4, 1.4a), and potential attracting information
$$i_v \cong 0.02634 \, Nat \tag{3.1}$$
Information ratio in (2.8a) allows evaluate the attracting information brought by each cutting impulse:
$$i_{vo} = 0.5 \times 0.4142 = 0.207 \, Nat. \tag{3.1a}$$
Sum of the values in (3.1) and (3.1a) adds information for the attraction:
$$i_{vf} = i_v + i_{vo} = 0.23334 \, Nat. \tag{3.1b}$$

As a part of information delivered with the impulse, $i_{vf}$ evaluates its *free* information enables attracting next cutting information until this information will spend on adjoining the following cutting information. The $i_{vf} = i_f$ in Sec.2.6.

Each step-down cut requests information (of ~1/3 of the impulse):
$$1/3 (\ln 2 + 0.05) Nat \cong 0.24766 \, Nat \approx 0.25 \, Nat. \tag{3.2}$$

That, concurring with (3.1b), measures the cooperative attraction, which could deliver the free information (3.1).

Elementary information unit Bit is an impulse, which encloses the equivalent of entropy $\ln 2$ Nat, whose cut converts this entropy to information. The nearest impulses' step-down and step-up actions, while cutting entropy of the final probe, memorizes it delivering an information equivalent cost of energy (Sec.2.6). The impulse step-up action limits the cutting information to equivalent Nat that satisfies the minimum of maximal cutting entropy volume.

The Bit encloses the probing impulses logic, conserving through information cost of energy as a logical equivalent of Maxwell Demon's energy spent on this conversion, and generates the free information of attraction.

Thus, the information Bit is a self-participator in both converting entropy to equivalent information, which memorizes logic of its entropy prehistory, and in extending a posterior logic during persistence attraction, involving the multiple impulses' Bits sequence in the information process. The primary self-participating inter-action holds the difference between the entropy of last virtual step-up and real step-down cut actions within the ending impulse actions. That could emerge in natural or artificial processes. The real interactive impulse carries both real microprocess, attracting next Bit, and the information cost of getting the Bit. That distinguishes the real impulses from the virtual impulses of random activity.

The information equivalent of the impulse wide $\delta_{ue}^i \cong 0.05 Nat$ limits its size and the extension minimal time interval $\delta_{te} \approx 1.6 \times 10^{-14}$ sec. Potential information speed of attraction within the impulse:

$$c_{ia} = 1/3(\ln 2 + 0.05)/\delta_{te} \cong 0.1548 \times 10^{14} Nat/\sec, \quad (3.3)$$

is less than both maximal speed $\alpha_{io}$ (2.6), starting a single real information from imaginary entropy, and the speed between the nearest impulses on time interval $\Delta_t$:

$$c_{ika} \cong 0.0516 \times 10^{14} Nat/\sec. \quad (3.3a)$$

Maximal speed (3.3a) conveys a flow of the formed information Bits in the information process, which carries the enclosed entropy, memory, energy, and logic, while enclosing free information (3.1) between the impulses.

A single Bit, moving with speed (3.3a), spends its free information of $\sim 1/3 Nat$ to attract next Bit on time interval $t_{ika}$:

$$t_{ika} = 1/3 \ln 2 / c_{ika} \cong 1/3 \ln 2 / 0.0516 \times 10^{-14} = 4.477 \times 10^{-14} \text{ sec}. \quad (3.3b)$$

Minimal distance $\Delta_{lo} \approx 14.4 \times 10^{-15} m$ to a next Bit limits a maximal dynamic spatial speed of information attraction:

$$c_{lo} \approx 14.4 \times 10^{-15} m / 4.477 \times 10^{-14} \sec = 3.216 \times 10^{-1} m/\sec. \quad (3.3c)$$

In this process, the impulses pair adjoins in a doublet which encloses bound free information spent on the attraction.

The cutoff attracting bits, start collecting each three of them in a primary basic triplet unit at equal information speeds, which resonates and coheres joining in triplet units of information process.

The triple resonace concurs with Efimov's Scenario [81-83], early proposed in Borromean Universal three-body relation, including Borromean knot [84] and ring. Ancient Borromean Rings represent symbols for strength in unity.

Information triplet, satisfying the minimax, is forming during the cooperative rotation of information units, applied to each eigenvector of information dynamics, as well as to the unit groups: doublets and triplets.

### 4.4. Space-time trajectory solving the minimax variation problem

Since the entropy functional is defined on Markov diffusion, which includes both micro and macroprocessses, the EF extremals of the VP, describe the Hamiltonian dynamic, time-space movement, rotating the opposite directional-complimentary conjugated trajectories $+\uparrow SP_o$ and $-\downarrow SP_o$. The trajectories form spirals located on conic surfaces Figs.3, 3a.

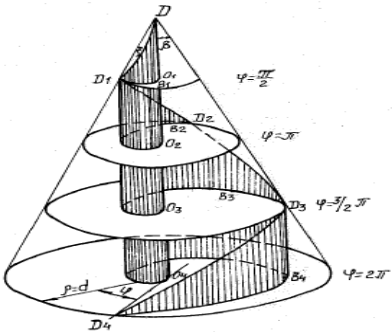

**Fig. 3. Forming a space-time spiral trajectory with radius $\rho = b\sin(\varphi \sin \beta)$ on the conic surface at the points D, D1, D2, D3, D4 with the spatial discrete interval DD1=$\mu$, which corresponds to the angle $\varphi = \pi k/2$, $k = 1, 2, \ldots$ of the radius vector's $\rho(\varphi, \mu)$ projection of on the cone's base (O1, O2, O3, O4) with the vertex angle $\beta = \psi^o$.**

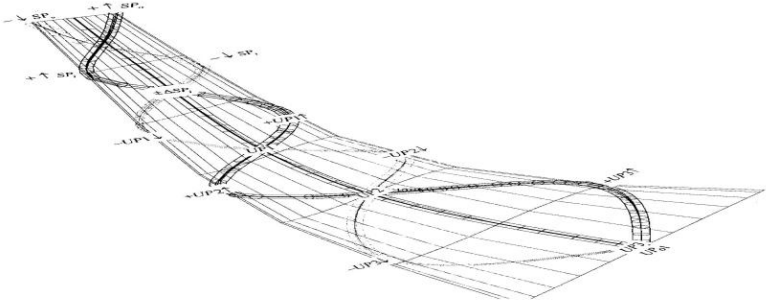

**Fig.3a.** Time-space opposite directional-complimentary conjugated trajectories $+\uparrow SP_o$ and $-\downarrow SP_o$ of Hamiltonian process (Sec. 3), forming the spirals located on the surfaces(Fig.3). Trajectory on the join bridges $\pm \Delta SP_i$ binds the contributions of process unit $\pm UP_i$ through the impulse joint No-Yes actions, which model a line of switching inter-actions(the middle line between the spirals).

In an observing process, the trajectories entangle its opposite *segments carrying potential qubits with growing probabilities.* When the probability approach the gap, the trajectory holds a high probable pair of qubits entropy, which the step-down action on the pair first kills its entropy, creating information qubits, and then joins them in bit.

Such jumping step-down action, ending the microprocess, is analogous of the jumping step-up actions starting the microprocess. This means, the microprocess proceeds between the merging No and Yes probabilistic actions of observation. It implies that each of such action covers a sub-markov opposite jumping actions.

Or, the Yes-No probabilistic actions are actually not separated until the virtual actions of growing probabilities, displays border of each impulse with No-Yes actions.

The virtual qubits measure the impulse virtual observation ending in the entangled entropies.

Until the impulses are separated, the microprocess exists within a probability "eroding a fuzzy border" of the impulse, and the microprocess disappears automatically when the between borders probability approaches zero.

Or, the impulse ending border rises with probability approaching 1, when classical physical bit emerges.

Fig.3a illustrates how a pair of qubits from the opposite segments (at end of the conic tube) binds a bit under the step-down actions within the gap. If the entangled qubits hold their probabilities, their joining in the bit not occurs.

The illustration also indicates and explains that the conjugated entropies entangle $\pm$ or $\mp$ complementary parts of the trajectory units $\pm UP$ or $\mp UP$, which enable assembling by mutual attractions along the observing trajectory.

Such opposite triple qubits can assemble a triplet bit, spending on binding $(3 \times \ln 2 / 2 - 1) Nat \cong 0.0397 Nat$ to assemble 1 Nat. That Bit would have free information $(1 - \ln 2 - 0.0397) \cong 0.267 Nat$ which will be enough for spending it on attraction and binding another two qubits for structuring new triplet.

When the real triple qubits assemble a triplet, it also joins each virtual observation with the microprocess along the multiple trajectories in the information process.

The trajectories assemble opposite *segments of information process* $\pm SP_i, i=1,...,n$ dimensions compiling them up to maximal $n$. Each $\pm SP_i$ segment averages the microprocesses entangling the attracting qubits.

The impulse step-up and step-down actions selects sections $-SP_i, \pm SP_i$ ending on the trajectories bridge $\pm \Delta SP_i$ which binds the spiral of each segment $i$.

Each opposite directional segment's bridge enables attracting half of each process unit $\pm UP_i$.

The impulse, joining No-Yes action, connects opposite units $-UP_i$ and $+UP_i$ in unit $UP_i$-Bit through the bridge $\pm\Delta SP_i$. Dynamics on the bridge borders describes trajectory of switching actions located on middle between the opposite spirals (Fig.3a). Each pair of opposite sections $+SP_i$ and $-SP_i$ forms local circles ↑o↓ with sequentially reverse directions of their movement during their assembling, while total directions along conjugated trajectories $+\uparrow SP_o$ and $-\downarrow SP_o$ preserve. The cyclic process temporary exists.
The persistent attraction assembles Bits in information process, whose information integrates and measures information path functional (IPF) on the macroprocess trajectories.
Transfer from the VP extremal's maximum to minimum limits the dynamic constraint (Sec.3.3).

Each $UP_i$ integrates into $UP_{i+1}$ along the space-time trajectory, and when $-SP_i$ transfers to $+SP_i$, the local EF maximum transforms to the minimum on the bridge. The doublets integrate in triplets.

The EF-IPF linear and nonlinear equations describe information flows initiated via the gradients (Sec.3.3) as information forces, which become physical analogies of the thermodynamics flows and forces accordingly in Irreversible Thermodynamics [85, 43]. The qubits doublets, integrated in triplets, form elementary structural units of the emerging information process. The observation process builds the conjugated dynamics of microprocess, which disappear after the information qubits emerge, composing triple units of the macroprocess segments.

In multidimensional observations, the recursive step-down-step-up actions between segments feed the following segment with microlevel entropy, currently converting in information, and connect it with each previous one. The conjugated trajectories describe the EF extremals, while the emerging bits on the bridge integrates the IPF enclosing integral information in its final bit. Each bit, memorized in the conjugated interactive bridge, divides the trajectory on reversible process section, excluding the bit' bridge, and irreversible bridge between the reversible segments. Thus, the observer irreversible dynamic trajectory includes the reversible segments ending with each bridge, where each irreversible bit emerges from the current observation. It brings irreversibility to the composing conjugated Hamiltonian dynamics.

*Information mechanism of assembling information units in triplets*

While the ending microprocess binds each pair $\pm UP_i$ in information unit $UP_i$ on the bridge, each three *information units* $UP_i$ assemble new formed triplet's units $UP_{oi}$ through their ending minimal information speeds having opposite directions.

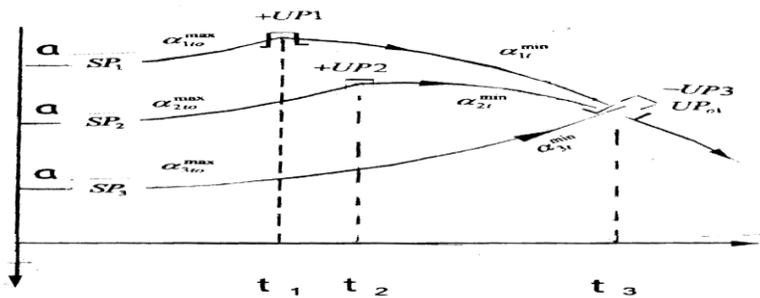

**Fig.4. Illustration of assembling a triple qubit units** $+UP1, +UP2, -UP3$ **and adjoining them to the composed triplet unit** $UP_{o1}$ **at changing information speeds on the space-time trajectory from** $\alpha_{1to}^{\max}$, $\alpha_{2to}^{\max}$, $-\alpha_{3to}^{\max}$ **to** $\alpha_{1t}^{\min}$, $\alpha_{2t}^{\min}$, $-\alpha_{3t}^{\min}$ **accordingly; a is dynamic information invariant of an impulse.**

Fig.4 illustrates simplified dynamics of assembling tree units $UP_i$ $i = 1, 2, 3$ and adjoining them in triplet $UP_{o1}$ along the segments of space–time trajectory $\pm SP_i$ at changing the opposite information speeds on the trajectory between the segments from maximal $|\mp\alpha_{ito}|^{max}$ to $|\pm\alpha_{it}|^{min}, i=1,2,3$, while the bridge of unit $UP3$ connects these units with $UP_{o1}$ (Fig.4 shows a single symmetrical part of the conjugated dynamics).

Suppose segment $-SP_1\downarrow$ starts moving $-UP1\uparrow$ with maximal speed $|-\alpha_{1t}|$, and segment $+SP_1\uparrow$ starts moving $+UP1\uparrow$ with speed $|+\alpha_{1t}|$. The rotating movement of both units involves a local circle rotating these segments speeds in right direction $\uparrow O \downarrow$. The next segment $-SP_2\downarrow$ starts moving $-UP2\downarrow$ with maximal speed $|-\alpha_{2t}|<|-\alpha_{1t}|$ and $+SP_2\uparrow$ starts moving $+UP2\uparrow$ with speed $|+\alpha_{2t}|<|+\alpha_{1t}|$. Each of these speed forms second local circle $\downarrow O \uparrow$ rotating in left direction with the absolute speed less than those in the previous circle.

Third segment $-SP_3\downarrow$ starts moving $-UP3\downarrow$ with maximal speed $|-\alpha_{3t}|<|-\alpha_{2t}|$, and $+SP_3\uparrow$ starts moving $+UP3\uparrow$ with maximal speed $|+\alpha_{3t}|<|+\alpha_{2t}|$.

Third circle rotates in right direction $\uparrow O \downarrow$ with absolute values of the speeds satisfying relation
$$|\alpha_{3t}|<|-\alpha_{2t}|<|\alpha_{1t}|. \qquad (4.1)$$

The opposite directional speeds within each circle attract each $-UP_i$ to $+UP_i$, minimizing the ending speeds down to $|\alpha_{it}|=\alpha_{it}^{min}$.

When these pairs approach, the starting attracting force (1.2) can bind them in related units $UP_i$ by sequence
$$\alpha_{1t}^{min}\to\alpha_{2t}^{min}\to\alpha_{3t}^{min}. \qquad (4.2)$$

Current units $UP1, UP2$, have been made at the growing time-space intervals $t_{o1}<t_{o2}<t_{o3}$, at $t_{o3}\geq t_{o1}+t_{o2}$, automatically integrate the primarily memorize unit $UP3$, and then condenses the units forming the triple knot $UP_{o1}$.

The IPF is sequentially summing information of $UP_i, i=1,2,3$ and memorizing only current $UP_{o1}$, while the previous $UP_i$ are erased as their information integrates $UP_{o1}$.

When the condensed information of $UP_{o1}$ integrates sequence $UP1\to UP2\to UP3$ in a primary triplet, it automatically implements the IPF with minimization of total time of building the triplet.

When speeds $+\alpha_{13t}$ and $-\alpha_{13t}$ of each opposite segments move close to speeds $+\alpha_{3t}, -\alpha_{3t}$ by the end of interval $t_3$, according to (4.1, 4.2), it makes possible connecting the conjugated information segments
$$a_{+23}=+\alpha_{23t3}t_3 \text{ with } a_{-23}=-\alpha_{23t3}t_3. \qquad (4.3)$$
This concludes the process of joining each of two complementary units in a third time-space loop, which synchronizes their equal speeds-frequencies. The triple knot generates free information, which forms the concluding loop of the complimentary triplet's processes, allowing self-formation of the joint triplet in the following consecutive attraction.

Entropy of virtual impulse through the bridges between the segments is converting to information, which, closing the loop in the cycle, transforms this information to information of triptet.

Such a triple self-supporting cyclic process indeed requires an initial flow of entropy converted to information.

At forming $UP3$, new triplet Bit may appear if the $UP1$ ending speed, minimized by the opposite speed of $UP2$, equalizes with third segment minimal speed $\alpha_{3t}^{min}$; and the $UP2$ ending speed, being also minimized in the opposite movement, equalizes with third segment minimal speed $\alpha_{3t}^{min}$ by the moment of forming $UP3$.

Adjoining the two with the third allows forming $UP3$ *during* formation of $UP1$ and $UP2$, which are joining with a minimal information speed equals to ending speeds of the two moving segments.

These speeds minimize the attracting information $\mathbf{a}=1/3bit\approx 0.23 Nat$ of each $UP1, UP2, UP3$ whose sum $3\mathbf{a}\cong\mathbf{a}_o$ can join all three in new triplet Bit with information $\mathbf{a}_o$.

(More precise calculation brings free information $\mathbf{a}^* = (1 - \ln 2 + 0.0397)/1.44 \cong 0.24 bit$ which includes both attracting and binding information $\mathbf{a}_b = 0.0397/1.44 \cong 0.02757 bit$ in forming the triplet). That confirms (3.2).

Actually, attracting information $\mathbf{a}$ of unit $UP1$ decreases speed of its starting movement $\alpha_{13t} = \alpha_{1t} - \Delta\alpha_{1t} \to \alpha_{1t}^{\min}$ on such increment $\Delta\alpha_{1t}$ which can equalize the unit speed with $\alpha_{3t}^{\min}$ for $UP3$.

Attracting information $\mathbf{a}$ of $UP2$ decreases speed of its starting movement

$$-\alpha_{23t} = -\alpha_{2t} - \Delta\alpha_{2t} \to -\alpha_{2t}^{\min} \qquad (4.4)$$

on such increment $\Delta\alpha_{2t}$ which allows also equalize this unit speed with $\alpha_{3t}^{\min}$. Here the speed signs hold the directions of rotation in each local circle. The attracting movement connects these speeds in the triple movement arranging in relation

$$+\alpha_{1t}^{\min} \Rightarrow -\alpha_{2t}^{\min} \Leftarrow +\alpha_{3t}^{\min} \qquad (4.5)$$

when ending speed $-\alpha_{2t}^{\min}$ emanated from $UP2$ joins equal minimal speeds $+\alpha_{1t}^{\min}, +\alpha_{3t}^{\min}$ forming the end of triple knot which is assembling the triplet unit $UP_{o1}$.

The knot binding information memorizes the accumulated information at moving equal speed

$$\alpha_{1t}^{\min} = |\alpha_{2t}^{\min}| = \alpha_{3t}^{\min} = \alpha_{uo1} . \qquad (4.5a)$$

An example of assembling a triplet space structure shows Fig.5.

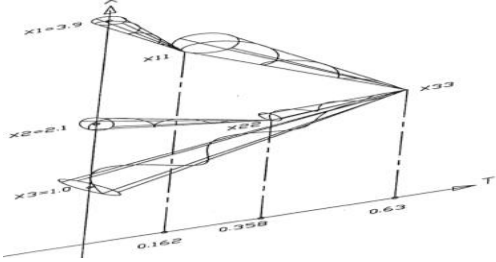

**Fig. 5. Forming a triplet's space-time structure.**

Let us identify the time of formation of this space-time structure.

Assume forming $UP1$ Bit on segment $SP_1$ needs its moving time interval $t_1$ and time interval $\Delta t_{13}$ to attract $UP3$.

Forming second $UP2$ Bit on segment $SP_2$ needs its moving time interval $t_2$ and time interval $\Delta t_{23}$ to attract $UP3$.

Forming $UP3$ Bit takes time interval $t_3$ of moving third segment $SP_3$.

Since all three segments move decreases information speeds, it is possible to reach equality

$$t_{13} = t_3 = t_{23} + t_{12} \qquad (4.6)$$

where $t_{12}$ is time interval between $UP1$ and $UP2$.

Satisfaction of (4. 6) allows adjoining all three segments during time interval $t_3$ of forming $UP3$.

Since intervals $t_1, t_2, t_3$ are connected by the same free information invariant $\mathbf{a} = 1/3 bit \approx 0.23 Nat$, it brings joint relation

$$\alpha_{1to}^{\max} t_1 = \alpha_{2to}^{\max} t_2 = \alpha_{3to}^{\max} t_3 = \mathbf{a}_o \cong 3\mathbf{a} . \qquad (4.7)$$

The dynamics of simulated speeds determine information delivered at each of these intervals [101]:

$$\mathbf{a}^{13} = \alpha_{13t} t_{13} \cong 0.232, \ \mathbf{a}^{23} = \alpha_{23t} t_{23} \cong 0.1797 \text{ and } \mathbf{a}^{33} = \alpha_{3t} t_3 \cong 0.268. \qquad (4.7a)$$

At $t_{13} = t_3$ and $\alpha_{3to} t_3 \cong \mathbf{a}_o, \alpha_{13t} t_{13} \cong 1/3 \mathbf{a}_o \cong \mathbf{a}$, we get $\alpha_{3to}/\alpha_{13t} \cong 3$. (4.7b)

To satisfy (4. 7) at $\mathbf{a}^{13} = \mathbf{a}^{33} = \mathbf{a}$, the information spent on attraction and assembling triplet unit $UP_{o1}$, should also be $\mathbf{a}$.

Therefore, if information spent on attraction $UP3$ is $\alpha_{23t} t_{23}$, the difference

$$\Delta \mathbf{a}^{23} = \mathbf{a} - \alpha_{23t} t_{23} \cong 0.23 - 0.1797 \cong 0.05 \tag{4.7c}$$

spends on assembling $UP_{o1}$ using the delivered information $2\mathbf{a} = \mathbf{a}^{13} + \mathbf{a}^{33}$.

Following that, relations

$$\Delta \mathbf{a}^{23} = |\alpha_{23t}| \delta t_{23}|, |\alpha_{23t}| \cong 1/3 \alpha_{3ot}, \alpha_{3ot} = \mathbf{a}_o / t_3, \tag{4.8a}$$

determine the time interval on assembling $UP_{o1}$:

$$\delta t_{23} \cong 3 t_3 \Delta \mathbf{a}^{23} / \mathbf{a}_o, \tag{4.8b}$$

which evaluates

$$\delta t_{23} \cong 0.214 t_3. \tag{4.8c}$$

Assembling information $\mathbf{a}_o \cong 3\mathbf{a}$ in the knot with speed $-\alpha_{23t}$ determines $-UP_{o1}$ sign.

The assembling loop, connecting speeds (4.5a), builds the triplet knot $-UP_{o1}$, which binds primary information of $UP1, UP2$, $UP3$. The knot free information will spend on the information attraction of a second forming triplet $+UP_{o1}$.

The attracting free information from triplet $-UP_{o1}$ or $+UP_{o1}$ forms secondary information unit $UP_{o1}$. The time of building $UP_{o1}$ minimizes the time of sequential connections of units $UP_i$, $i = 1, 2, 3$ in the information process.

The $n$-dimensional process trajectory locates multiple conjugated pairs $+\uparrow SP_{oi}$ and $-\downarrow SP_{oi}$, which could assemble each triple $UP_i$, $i = 1, 2, 3$ in cooperative $\mp UP_{oi}$ Bit, where the sign of each unit depends on the sign each second segment $SP_{2i}$ which binds each of these units in the triple segment knot.

This attracting movement assembles their three Bits in new formed $\mp UP_{i0}$ Bit forming new circular loops on higher (second) level which connects the equal speeds of each triple (Fig.6).

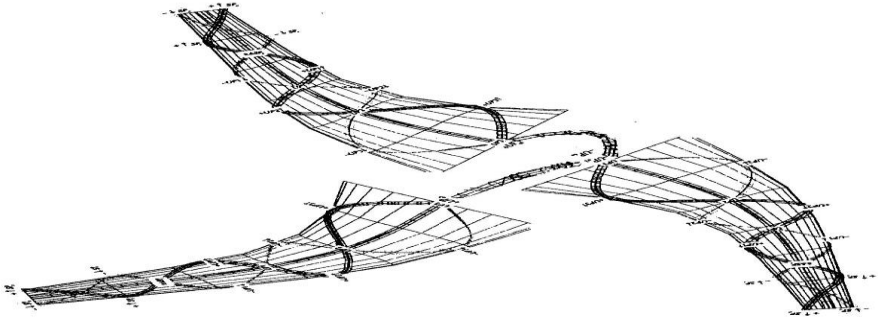

**Fig.6. Assembling three formed units $\pm UP_{i0}$ at a higher triplet's level connecting the units' equal speeds (Fig.4).**
**The attracting circles, analogous to that on Fig.5, are not shown here.**

Particularly, the attractive motion of rotating triples units $-1_o(+UP_{o1}, -UP_{o2}, +UP_{o3})$ can cooperate next one to form triplets unit $-UP_{4o}$ depending on $-SP_{2i}$. The sign of $+SP_{2i}$ at cooperating opposite triple $+2_o(-UP_{o5}, +UP_{o6}, -UP_{o7})$ participates in forming triplets unit $+UP_{5o}$, other three rotating triple $-3_o(+UP_{o8}, -UP_{o9}, +UP_{o10})$ cooperates in unit $-UP_{6o}$. That closes each of three rotating circles. Cooperative motion of units $+4_{1o}(-UP_{4o}, +UP_{5o}, -UP_{6o})$ assembles them in new formed triplet $+UP_{I0}$ Bit composing third levels of triplets. If above units $-1_o, +2_o, -3_o$ have equal attracting speeds by the moment of cooperation, they may join simultaneously in composite unit $+4_{10}$ during building this triple.

Information of units $UP_{o1}, UP_{o2}$, currently formed at the growing time-space intervals $t_{io1} < t_{o2} < t_{io3}$, at $t_{io3} \geq t_{io1} + t_{io2}$, automatically integrate and memorize $UP_{oit}$ which along with $UP_{o3}$ forming the triple knot $UP_{4ot} = UP_{oit}$.

The sequential built triplet knots, memorizing only current $UP_{oit}$, while the previous units information have erased, automatically implement the IPF minimax, minimizing total time of building each composite information unit.

The minimax leads to sequential decreasing ending information speed of each $UP_{oi}$, and therefore to decreasing starting information speed of a next cooperative unit. The minimax requires the ordered connection of binding speeds at forming triplets, which memorizes the bound information units, sequentially structures the observer's information dynamics, building and connecting each $UP_{oi}$ in new cooperating triple units.

Building the units of dynamic information structure from multiple $\pm SP_{ij}$ segments of $n$-dimensional process trajectory requires assembling each $UP_{ij}$ from two segments, taking from conjugated segments on the trajectory of one dimension, with other opposite directional segment of the conjugated pair which belongs to other dimension.

Building a cooperative forming triplet $UP_{oij}$ involves three such complimentary parts from three different process' dimensions illustrated on **Figs.3a,6**. In the minimax attracting movement of $\pm UP_{ij}$, three of the time–space segments spend only the time-space interval of a third segment, joining three segments simultaneously, while both the first and second segments attract third segment during their segments moving intervals accordingly. The opposite moving segments on the trajectory cooperate the symmetrical-complimentary $\pm UP_{oij}$ triplets, enclosing half of each complementary Bits, which through self-joining enable creating complete Bit whose attracting information can assemble other $UP_{oij}$ triplets.

The minimax movement decreases ending information speed of each complementary units of $\pm UP_{oij}$. The units enclosed in the knot increase information density of each following Bit. Each such triple unit contains four Bits including the forth Bit, which encloses the triplet information in a triple knot and provides free information for subsequent assembling.

The attracting space-time movement selects, orders, and assembles the rotating segments' cooperating speeds along the time-space dynamic trajectories. Indeed. Each primary $\pm UP_{o1}$ from different conjugated pairs emanates three time–space segments $\pm SP_{o1}, \pm SP_{o2}, \mp SP_{o3}$ selected on the minimax trajectory. The symmetry of opposite moving segments on the trajectory actually generates half of each complementary $\pm UP_{o1}$ unit's $\pm BitI4, I = 0,1,....m$ which could self-joins, creating complete Bit $|BitI4|$ of $UP_{o1}$, whose attracting information can assemble other triplets units $\pm UP_{Io1}$ and so on.

The triplets' knots create *new class* of information Bits that distinct from the first class information Bits of the assembling units, which were generated via virtual probes with entropies of cutting random process.

Each forming Bit of the following class grows density of the enclosed information, geometrical density, and the curvature. Forming the closing loop in processing a complimentary triplet allows self-formation of such joint triplet.

The triple knot generates free information for a next consecutive attraction, increasing number of cooperating triplets.

The joint triplet with free information produces cooperative assembling which brings new information in each triple knot.

The basic mechanism builds each dynamic triple cooperation through equalization speeds of information moving impulses, carrying coherent speeds, whose frequencies resonate rotating in a cycle that assembles the triplet.

### 4.5. The conditions for building space-time information network, observer geometry, and restriction on its parameters

Multiple moving triplets, sequentially equalizing their speeds-frequencies in resonance, assemble nested layers of bound triplet units-nodes. The attracting nodes' logic self-organizes information networks (IN) in logical information structure of the nested hierarchy of the nodes. The self-building continues during the time-space information dynamics, which self-assemble space-time information nested structure of the IN (Fig.7), enclosing the growing number of triplets.

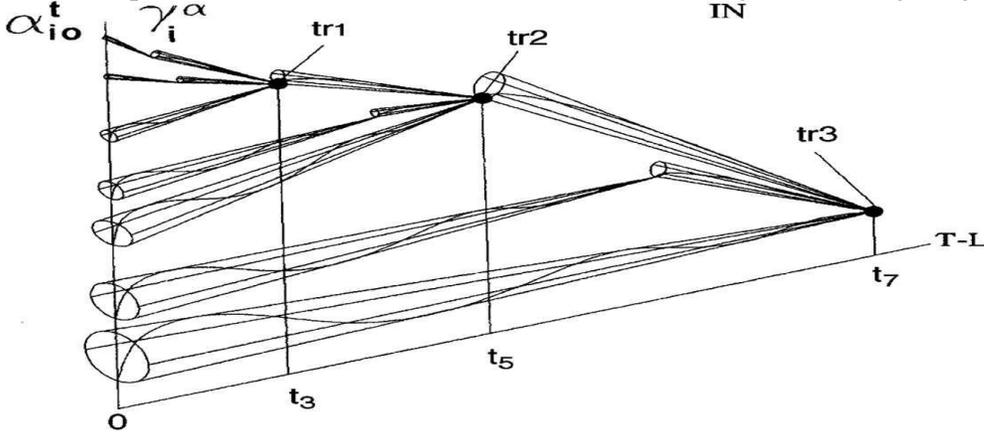

Fig.7. The IN information geometrical structure of hierarchy of the spiral space-time dynamics of triplet nodes (tr1, tr2, tr3,...); $\{\alpha_{io}\}$ is a ranged string of the initial eigenvalues, cooperating on (t1, t2, t3) locations of T-L time–space, $\{\gamma_i^\alpha\}$ is parameter measuring ratio of the IN nodes space-time locations.

The attracting process of assembling triplet's knots forms a rotating loop shown on Fig.7 at forming each following cooperative triplet tr1,tr2, . The scales of curves on Fig.6 distinct from the interacting knots' curves on Fig. 7, since these units, at reaching equal speeds, resonates, which increases size of the curves on Fig. 6.

Each triplet accumulates three Bit's information logic enfolded in the knot's information Bit.

Such self-forming structure automatically implements the IPF integration of the process information in a last IN node.

The *triplet information dynamics* start with information speed of its first segment's eigenvalue

$$\alpha_{io} = c_{iv} \cong 2.4143 \times 0.596 \times 10^{15} \, Nat/\sec \cong 1.44 \times 10^{15} \, Nat/\sec , \qquad (5.1)$$

where $\alpha_{io} = \alpha_{1o}$ is potential information speed predicted by the moving entangled entropy volume.

Each triplet self-formation joins two trajectory segments with positive eigenvalues by reversing their unstable eigenvalues and attracting a third segment with negative eigenvalues whose rotating trajectory moves it up to the two opposite rotating eigenvectors and cooperates all three information segments in a triplet's knot (**Figs 3,4,5,7).**

Prior to joining, the segment entropy-information, satisfying the minimax, should end with its minimum, which evaluates relations (2.9b) and (2.10).

The minimax optimal triplet requires minimal time interval (1.1) spent on equalization of each eigenvalue with the following segment eigenvalue.

At forming triplet, the cooperation of two segments may bind $2i_{vf} = 2(i_v + i_{vo}) = 0.4668 Nat$ of both segments attracting information, which equals to that for overcoming entropy threshold $\delta_e \cong 0.4452 Nat$ (3.2.9b) for the third segment.

Third segment encloses information of the two triplet segments, including information that binds them.

Its ending free information attracts next information segment of following triplet, which builds the attracting information units in the subsequent information dynamics.

Dynamic invariant $\mathbf{a}(\gamma)$ that connects the ending eigenvalues of triplet's segments determines ratios of starting information speeds $\gamma_1^\alpha = \alpha_{io}/\alpha_{i+1o}$ and $\gamma_2^\alpha = \alpha_{i+1o}/\alpha_{i+2o}$ needed to satisfy (2.3), (2.4) according to formulas [43]:

$$\gamma_1^\alpha = \frac{\exp(\mathbf{a}(\gamma)\gamma_2^\alpha) - 0.5\exp(\mathbf{a}(\gamma))}{\exp(\mathbf{a}(\gamma)\gamma_2^\alpha/\gamma_1^\alpha) - 0.5\exp(\mathbf{a}(\gamma))}, \gamma_2^\alpha = 1 + \frac{\gamma_1^\alpha - 1}{\gamma_1^\alpha - 2\mathbf{a}(\gamma)(\gamma_1^\alpha - 1)} \quad (5.2)$$

where multiplication $\gamma_1^\alpha \times \gamma_2^\alpha = \gamma_{13}^\alpha$ holds the eigenvalue ratio $\gamma_{13}^\alpha = \alpha_{io}/\alpha_{i+3o}$.

Invariants $\mathbf{a}_i(\gamma) = \alpha_i^t t_i$ and $\mathbf{a}_{io}(\gamma)$, at $\lambda_i^t = \lambda_{io}^t \exp(\lambda_{io}^t t_i)[2 - \exp(\lambda_{io}^t t_i)]^{-1}$, connect Eq.

$$\mathbf{a} = \mathbf{a}_o \exp(-\mathbf{a}_o)(1+\gamma^2)^{1/2}[4 - 4\exp(-\mathbf{a}_o)\cos\gamma\mathbf{a}_o + \exp(-2\mathbf{a}_o)]^{-1/2}]. \quad (5.3)$$

In dynamics of real eigenvalue: $\alpha_i^t = \alpha_{io}^t \exp(\alpha_{io}^t t_i)[2 - \exp(\alpha_{io}^t t_i)]^{-1}$, the invariants $\mathbf{a}_i(\gamma_i), \mathbf{a}_{io}(\gamma_i)$ connect Eq:

$$\mathbf{a}_i(\gamma_i) = \mathbf{a}_{io}(\gamma_i)\exp\mathbf{a}_{io}(\gamma_i)(2 - \exp\mathbf{a}_{io}(\gamma_i))^{-1}. \quad (5.4)$$

Optimal ratio $\gamma_{io} = 0.4142$ corresponds to the minimax with $\mathbf{a}_{io}(\gamma_{io} = 0.4142) \cong 0.73$ and $\mathbf{a}_i(\gamma_{io}) \cong 0.23$.

Each triplet structure identifies first minimax invariants $\gamma_i, \mathbf{a}_{io}(\gamma_i)$ and then both $\gamma_1^\alpha, \gamma_2^\alpha$.

At known starting eigenvalue $\alpha_{io} = c_{iv} \cong 2.4143 \times 0.596 \times 10^{15} Nat/\sec \cong 1.44 \times 10^{15} Nat/\sec$ and $\alpha_{io} = \alpha_{1o}$, next such speeds in the triplet are $\alpha_{1o}/\gamma_1^\alpha = \alpha_{2o}, \gamma_1^\alpha(\gamma_{io}) = 2.236$, at. $\alpha_{2o} \cong 0.644 \times 10^{15} Nat/\sec$.

Actual starting information speed of first eigenvalue $c_{ika} = \alpha_{1o}, c_{ika} \cong 0.0516 \times 10^{14} Nat/\sec$ leads to second

$\alpha_{2o} \cong 0.02345 \times 10^{14} Nat/\sec = 0.2345 \times 10^{13} Nat/\sec$.

Known $\alpha_{2o}, \gamma_{32}^\alpha$ settles third starting eigenvalue's speed:

$$\alpha_{3o} \cong \alpha_{2o}/\gamma_2^\alpha = 0.02345/1.6 \times 10^{14} Nat/\sec = 0.1465 \times 10^{13} Nat/\sec. \quad (5.4a)$$

Invariants $\gamma_1^\alpha, \gamma_2^\alpha$ limits each triplet's time intervals and related intervals of rotating space movement of IMD at

$$\alpha_{io}t_{io} = \alpha_{i+1o}t_{i+1o} = \alpha_{i+2o}t_{i+2o} = \mathbf{a}_{io}(\gamma_i), \gamma_1^\alpha = \alpha_{io}/\alpha_{i+1o} = t_{i+1o}/t_{io} \text{ and } \gamma_2^\alpha = \alpha_{i+1o}/\alpha_{i+2o} = t_{i+2o}/t_{i+1o}. \quad (5.5)$$

Particular observations leads to specific ratios of the triplet's initial eigenvalues $\alpha_{1o}/\alpha_{2o} = \gamma_1^\alpha, \alpha_{2o}/\alpha_{3o} = \gamma_2^\alpha$, satisfying the invariant relations (5.2-5.4).

Each iinteracting impulse with information measure $\mathbf{a}_{io} = \ln 2 Nat$ has multiplicative information measure

$$U_m = (\mathbf{a}_{io})^2, \quad (5.6)$$

which binds the following double connection in a triplet, providing invariant information measure of each interaction.
Total information at

$$(\mathbf{a}_{io}(\gamma_{io}))^2 + \mathbf{a}_i(\gamma_{io}) \cong 0.7 \cong \mathbf{a}_{io}(\gamma_{io}). \quad (5.7)$$

approaches the delivered information from each previous impulse.
Forming a stable triplet, which enables the attraction, limits the maximal ratio

$$4.8 \geq \gamma_1^\alpha \geq 3.45. \quad (5.8)$$

That ratio determines a boundary of the triplet scale factor $\gamma_1^\alpha$. Approaching $\gamma_1^\alpha = \gamma_2^\alpha \to 1$ leads to repeating the triplet's eigenvalues that limits related theoretical admissible $|\gamma_i| \in (0.0-1.0)$.

Approaching information locality $\mathbf{a}_o(\gamma_i = 1 - o)$ of $\gamma_i \cong 1$ indicates both a jump of an event with that information, moving to the event information $\mathbf{a}_o(\gamma_i)$, and rising the time ratio of the following to preceding intervals.

For example, when $\gamma_i$ approaches 1, $\mathbf{a}_o(\gamma_i)$ is changing from $\mathbf{a}_o(\gamma_i = 1-o) = 0.56867$ to $\mathbf{a}_o(\gamma_i) = 0$, and the above time's ratio of reaches limit $\tau_{i+1}/\tau_i = 1.8254$.

This changes sign of the eigenvalues' ratio: $\alpha_{io}/\alpha_{it} \cong -1.9956964$ (following from (5.5) at $\gamma_i = 1$) and leads to $\mathbf{a}_o(\gamma_i = 1) = 0$ which brings information contributions for regular control $\mathbf{a}(\gamma_i = 1) = 0$ and impulse contraint (Sec3.3).

It leads to cutting off the model's dynamics from the initial random process. with a possibility of getting more uncertainty and rising chaotic diffusion dynamics. The appearance of an event, carrying $\gamma_i \to 1$, leads to *decoupling* the events chain and rising chaotic diffusion dynamics. Whereas the moment of this event's occurrence *predicts* measuring a current event's information $\mathbf{a}_o(\gamma_i) \neq 0$ and using it to compute $\gamma_i$ applying (5.2-5.4). Practically admissible maximal $\gamma_{ia} \to 0.8$ leads to a *minimal stable* triplet with $\gamma_1^\alpha = \gamma_2^\alpha \to 1.65$ which limits the acceptable $\gamma_i \to (0-0.8)$.

That limitations for $\gamma_{io} \to 0$ determines $\mathbf{a}_{io}(\gamma_{io}) = 1.1$ bit and $\mathbf{a}_i(\gamma_{io}) = 0.34$ bits.

The triplet dynamics with $\mathbf{a}_i(\gamma \to 0) \cong 0.23$ hold $\gamma_1^\alpha \cong 2.460, \gamma_2^\alpha \cong 1.817$ and $\gamma_{13}^\alpha \cong 4.6$.

Optimal $\gamma_{io} = 0.4142$ brings $\gamma_1^\alpha \cong 2.21, \gamma_2^\alpha \cong 1.76, \gamma_{13}^\alpha \cong 3.89$, and $\gamma_i = 0.8$ brings $\gamma_1^\alpha \cong 1.96, \gamma_2^\alpha \cong 1.68, \gamma_{13}^\alpha \cong 3.3$.

Sequence of the model eigenvalues ($\ldots \alpha_{i-1,o}^t, \alpha_{io}^t, \alpha_{i+1,o}^t$), satisfying the triplet's formation, is *limited* by boundaries for $\gamma \in (0 \to 1)$, which at $\gamma \to 0$ forms a *geometrical progression* with

$$(\alpha_{i-o}^t)^2 = (\alpha_{io}^t)^2 + \alpha_{i-1,o}^t \alpha_{io}^t, \tag{5.9}$$

representing the geometric "gold section" at $\alpha_{io}^t \cong 0.618\, \alpha_{i-1,o}^t$ and ratio

$$G = \frac{\alpha_{i+1,o}^t}{\alpha_{io}^t} \cong 0.618. \tag{5.10}$$

At $\gamma \to 1$ the sequence $\alpha_{io}^t, \alpha_{i+1,o}^t, \alpha_{i+2,o}^t, \ldots$ forms the Fibonacci series, where the ratio $\alpha_{i+1,o}^t / \alpha_{i+2,o}^t = \gamma_2^\alpha$ determines the "divine proportion" $PHI \cong 1.618$, satisfying

$$PHI \cong G + 1; \tag{5.11}$$

and the eigenvalues' sequence loses its ability to cooperate.

*Indeed*. At $\gamma \to 0$, solutions of (5.2-5.4) determine

$\mathbf{a}_o(\gamma_i \to 1) = 0.231$, $\gamma_1^\alpha \cong 2.46$, $\gamma_2^\alpha \cong 4.47$, $\gamma_2^\alpha / \gamma_1^\alpha = \gamma_{23}^\alpha \cong 1.82$.

Ratio $(\gamma_1^\alpha)^{-1} = \alpha_{io}^t / \alpha_{i-1,o}^t \cong (2.46)^{-1} \cong 0.618$ for the above eigenvalues forms a "golden section" (5.10) at

$G = (\gamma_1^\alpha)^{-1} = \alpha_{io}^t / \alpha_{i-1,o}^t \cong (2.46)^{-1} \cong 0.618$ *and* the "divine proportion" $PHI \cong 1.618$.

Above relations hold true for each primary pair of the triplets' eigenvalues sequence, while the third eigenvalue has ratio

$$\alpha_{i+1,o}^t / \alpha_{io}^t = (\gamma_{23}^\alpha)^{-1} \cong 0.549. \tag{5.11a}$$

Solution of the equation for invariants(5.2-5.4) at $\mathbf{a}(\gamma_i \to 1) = 0$) brings $\gamma_1^\alpha = \gamma_2^\alpha = 1$ with $\alpha_{i-1,o}^t = \alpha_{io}^t = \alpha_{i+1,o}^t = \alpha_{i+2,o}^t = ,...$

The eigenvalues' sequence loses ability to cooperate, disintegrating to equial and independent eigenvalues.
The network, built through the attracting resonance, has limited stability and therefore each IN encloses a finite structure.
That's why the observing process might self-build only multiple limited IN.
*The restrictions on the space-time rotating trajectory grow with increasing dimensions* $1,...,i,...,n$.

Each rotating movement presents $n$-three-dimensional parametrical equations of helix curve located on a conic surface (**Figs.3, 8**).

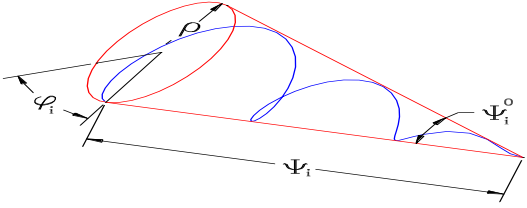

**Fig. 8. The cone's parameters (the indications are in the text).**

A projection of the space trajectory' radius-vector $\bar{r}(\rho,\varphi,\psi^o)$ on $i$-cone's base is spiral trajectory with radius

$$\rho_i = b_i \sin(\varphi_i \sin\psi_i^o). \tag{5.12}$$

At angle $\varphi = \pi k/2, k = 1,2,...$, the trajectory transfers from one cone to another cone trajectory on the cone space points $l_i$ which satisfy the extreme condition for the equation (5.12) at the angles

$$\varphi_i(l_i)\sin\psi_i^o = \pi/2, \tag{5.13}$$

where $\psi_i^o$ is angle on the cone vertex, and the base' radious is

$$\rho_i = b_i \sin(\pi/2) = b_i. \tag{5.14}$$

The angle at the cone vertex takes values

$$\sin\psi_i^o = (2k)^{-1}, k = 1,2,..,m,m+1,... \tag{5.15}$$

at $k=1$, $\psi_i^o = \pi/6$.

The angle on the cone space points $l_i$ takes values

$$\varphi_i(l_i(\tau_i)) = k\pi. \tag{5.15a}$$

The minimax imposes optimal condition on these angles:

$$\varphi_i = \pm 6\pi, \psi_i^o = \pm 0.08343. \tag{5.15b}$$

Projection of moving vector $l(\bar{r}) = l(\rho,\varphi,\psi^o)$ on the cone base satisfies Eq

$$dl = [(\frac{d\rho}{d\varphi})^2 \sin^{-2}\psi^o + \rho^2]^{1/2} d\varphi. \tag{5.16}$$

Spiral space angle $\psi$, depends on angle $\psi_i^o$ (5.15) according to Eq.

$$tg\psi = \frac{(1-\sin\psi^o \cos\psi^o + \sin^2\psi^o)}{(1\pm\sin\psi^o \cos\psi^o + \sin^2\psi^o)} \tag{5.17}$$

which for $\psi_i^o = \pi/6$ brings $\psi = 0.70311$. This angle, for the right directional spirals, at small angle $\psi^o$, satisfies

$$\psi = \pi/4 - \psi^o. \tag{5.17a}$$

For the spirals with the opposite directions this angle is $\psi^1 = \pi/4$.

A relative increment of information volumes $\Delta V_{m,m+1}$ (Fig.8) between the volumes of two sequential triplets' $m$ and $(m+1)$:
$V_m, V_{m+1}$: $\Delta V^*_{m,m+1} = (V_{m+1}/V_m - 1)$ depends on these triplets scale factor $\gamma^\alpha_{m,m+1}$:

$$\Delta V^*_{m,m+1} = (\gamma^\alpha_{m,m+1})^3 - 1 . \quad (5.18)$$

Where the triplet initial volume determines

$$V_c = 2\pi c^3 / 3(k\pi)^2 tg\psi^o \quad (5.19)$$

which depends on angle $\psi_i^o$, initial space speed $c_{io}$

$$c_{lo} \approx 14.4 \times 10^{-15} m / 4.477 \times 10^{-14} \sec = 3.216 \times 10^{-1} m/\sec, \quad (5.20)$$

and parameter $k$ (5.15) of sequential consolidation of the $m$-volumes in $V_{m+1}$, starting with (5.19) at $k = 1$.

Velocity of rotation attraction $\omega_i[l_i(t_{ika})] = 0.1646 \times 10^{-14} radian/\sec$ determines space angle $\psi' = \pi/4$ at moment $t_{ika}$ (3.3b), where $\omega_i$ relates to formulas (3.2.12a), (3.3c).

The information dynamics at moment $t_{ika}$ brings Eqs (5.12-5.19) and above rotating angles, which determine space interval $l_i(\tau_i)$ for the rotating volumes, and the eigenvalue ratios (speed) depending on triplet parameter $\gamma_1^\alpha, \gamma_2^\alpha$ in (5.2).

The triplet joint three eigenvalues form a first speed on its cone vertex, which delivers information to next triple units that join in next triplet in the proceeding rotating movement, generating an observer time course and space intervals.

Transfer from one cone's trajectory to another one locates on the cone's base, where each location satisfies extreme condition for entropy –information. The sequential transfer requires rotating each spiral on the space angle up to adjoin a next optimal trajectory and relocate it in cooperation (Fig.6). The space-time trajectories, rotating on the cones and cooperating in the triplet, shapes its geometrical structure (Figs.5, 7) evolving during each triplet formation with growing parameter $k$. Both information dynamics and its space structure evolve concurrently, producing each other (Fig.9).

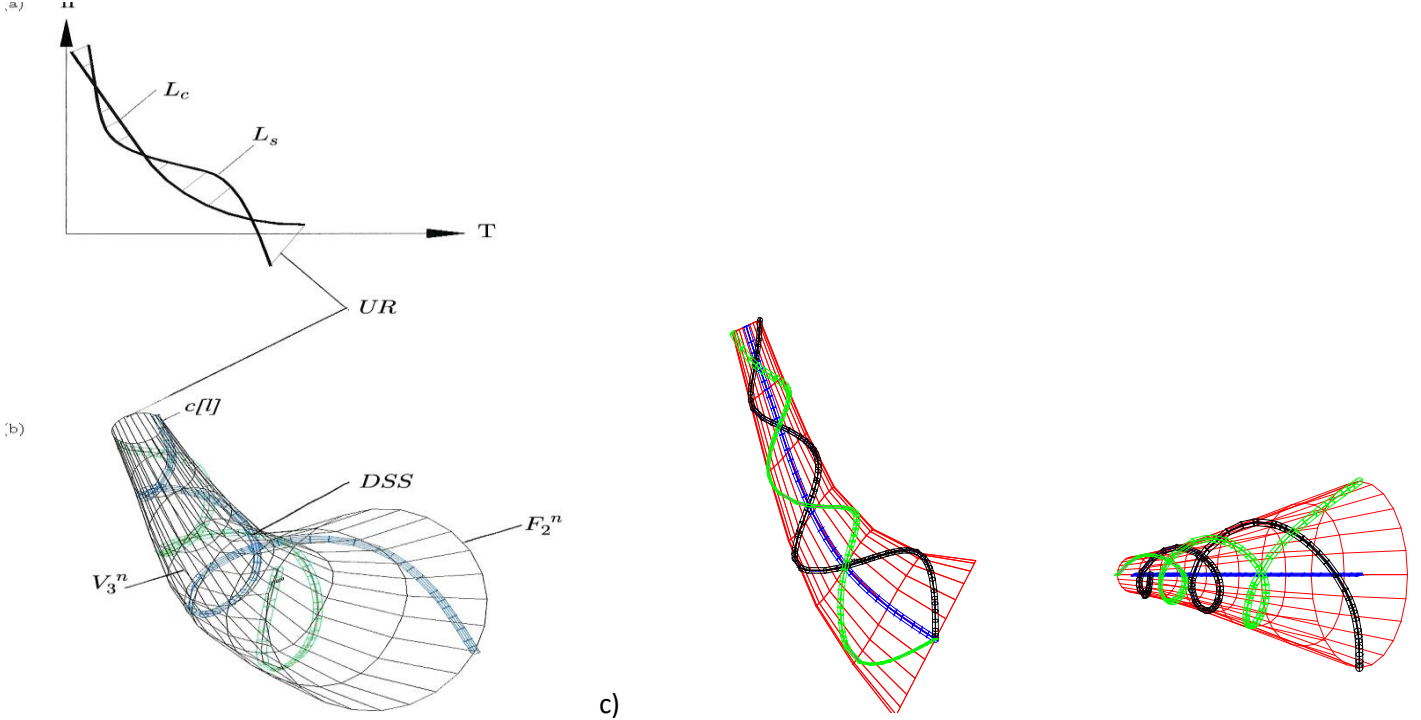

**Fig. 9.** Simulation of the double spiral cone's structure (*DSS*) with the cells (c[l]), arising along the switching control line *Lc* (a); with surface $F_2^n$ of uncertainty zone (*UR*) (b), surrounding the *Lc*-hyperbola in form of the *Ls*-line, which in space geometry enfolds a volume $V_3^n$ (b,c).

Each IN triplet accumulates three Bit's enfolded in its knot, which forms the IN node. The nested nodes enclose information logic enfolded in an ending IN knot. This ending triplet in every network contains the maximum amount of free information.
The INs can be self-connected through the attraction of their ended triplet's logic.
The multiple moving IN, sequentially equalizing the speeds-frequencies of the attracting information logic in resonance, assemble total INs logic.
Each forming IN emerges with a logic of assembling triplets, which encodes a triplet code-cell.
The code of multiple IN holds geometrical double spiral structure (DSS), Fig.9 enfolding each triple informational cell.
Each IN ending knot encloses the cell which condenses its local DSS code.
Since each code holds energy of cognitive thermodynamics (Sec.2.6), it physically organizes the multiple IN with their local codes in coding information structure of information Observer.
The Egs of rotating time-space trajectory on the cones and the space volume determine observer geometry, generated by the information dynamics (ID). The IN scale parameter $\{\gamma_i^\alpha\}$ identifies the rotating velocity and knots of cooperating volumes, transferred to next triplet.
The multiple IN geometry structures the Observer's information geometry by a manifold of the cellular DSS (Fig.10).

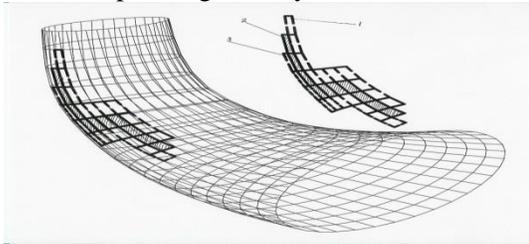

**Fig. 10. Structure of the cellular geometry, formed by the cells of the DSS triplet's code, with a portion of the surface cells (1-2-3), illustrating the observer space formation.**

**5.1. Stages and levels of the emerging observer self-organization of information and the evolution dynamics**
Assuming an observable random field of probable events conceals a randomly distributed energy, we show how the field's interactive impulses start self-connection, developing in interactive virtual observer, generate information bit of information observer, which self-organizes observation in evolving information dynamics, creating cooperative networks, and finally. an intelligence of self-evolving observer with self-supporting feedback acquisition of new information.
I. Observable process.

Multiple interactive actions ($\downarrow\uparrow$) are random events-variables forming a random process of the impulses in a surrounding random probability field. In the random field, the occurrence of specific events starts each sequence of the probabilistic observation. Manifold of the specific events provides manifold of the observing sequences.
These discrete events can be virtually observed through the discrete yes-no probabilities of the field.
In a probability field of interacting events, an infinite sequence of independent events satisfying Kolmogorov 0-1 law, affect a Markov diffusion process' probabilities, distributed in this field.
The Markov transitional probabilities change the process a priori-a posteriori Bayes probabilities, the probability density of random No-Yes impulses 0-1 or 1-0. That links the Kolmogorov's 0-1 probabilities, the Markov process' Bayes probabilities, and the Markov No-Yes impulses in common Markov diffusion process. These abstract objective probabilities quantify the probabilistic link of objective measures. The random interactive actions may randomly shift each impulse 0-1 to a following 0-1 or to 1-0, which connects them in a correlation through the Markov process drift and diffusion.
The probabilistic impulse' Yes-No actions represent an act of a virtual observation where each observation measures a probability of potential events. The arising correlation reduces the conditional entropy measures connecting the probabilistic observations in a virtual observing process with No-Yes actions. That defines first level of this stage.
This correlation connects the Bayesian a priori-a posteriori probabilities in a temporal memory that does not store virtual connection, but renews, where any other virtual events (actions) are observed.

Memorizing this action indicates start of observation with following No-Yes impulse at level two.

The starting observation limits minimal entropy of virtual impulse, which depends on minimal increment of process correlation overcoming a maximal finite uncertainty at level three.

II. The impulse' max-min self-action.

Each impulse' opposite No-Yes interactive actions (0-1) carries a virtual impulse which potentially cuts off the random process correlation (at first level) whose conditional (Bayes) entropy decreases with growing the cutting correlations.

If a preceding No action cuts a maximum of the cutting entropy (and a minimal probability), then following Yes action gains the maximal entropy reduction-its minimum (with its maximal probability) during the impulse cutoff (at level two).

The impulse' maximal cutting No action minimizes absolute entropy that conveys Yes action (raising its probability).

Thus, the cut action, delivered by the field, maximizes the cutting entropy, while reaction, spending an entropy on it creation, minimizes the cutting entropy. That provide max-min principle for conditional (relational) entropy between the impulse No-Yes actions. The following Yes-No actions transfer the probabilities and minimum of the impulse cutoff entropy to a next impulse, initiating minimax principle between the impulses along multiple observing impulses. .

The maximin-minimax rules the impulse observations (at level three).

The virtual impulse, transferring the Bayes probabilistic observation, virtually probes an observable random process, processing the maxmin observation. The probing impulse, consisting of step-down No and step-up Yes actions, preserves probability measure of these maxmin actions along the observation process.

This sequence of interacting impulses, transforming opposite No-Yes actions, increases each following Bayesian posterior probability and decreases the relative entropy reducing the entropy along the observation.

The observation under random probing impulses with opposite Yes-No probability events reveals hidden correlation [71], which connects the process' Bayesian probabilities increasing each posterior correlation (at level three). The maximin - minimax self-defines the variation principle which *formalizes* description of the observing evolution path (level four).

III. Virtual Observer

If the observing process is self–supporting through automatic renewal these virtual probing actions, it calls a Virtual Observer which acts until these actions resume up to the emergence of real Information Observer (if it appears).

Such virtual observer belongs to a self–observing process, whose Yes action virtually starts next impulse No action, etc. Both process and observer are temporal, ending with stopping the virtual observation.

Starting the virtual self-observation limits a *threshold* of the impulse's connection on first level on this stage.

The virtual observations may not link up to the real ones, but are preceding them.

The sequentially reduced relational entropy conveys probabilistic causality along the process with temporal memory collecting correlations, which interactive impulse innately cuts. The cutting entropy defines second level of this stage.

IV. Emerging the observer time

The correlation holds entrance of a *time interval* of the impulse–observation (at level one).

The time interval, connecting the probability-entropy in correlation, measures an uncertainty distance between nearest current observations (at level two).

The measuring, beginning from the starting observation, identifies an interval from the start, which is also virtual, disappearing with each new connection that identifies a next interval temporally memorized in that correlation connection.

The difference of the probabilities actions temporary holds memory of the correlation, as a virtual measure of an *adjacent distance* between the impulses' No-Yes actions.

It indicates a probabilistic accuracy of measuring correlation in a *time interval's unit* (at level three).

The impulses of the observable random process hold virtually observing random time intervals.

V. Emergence of the impulse space interval and space-time geometry in an Observer structure.

With growing correlations, the intensity of entropy per the interval (as entropy density) increases on each following interval, indicating a shift between virtual actions, measured in a time interval's unit invariant measure $|1|_M$. (Level one).

The dense shift merges interactive actions ↑↓ of bordered impulses, which generate an interactive jump of a high entropy density with curving action ↓.

The jump' curving action ↓ *curves* an emerging *½ time unit* of the border impulse' time interval, initiating a *displacement* of the curved action.

The displaced curved jump action starts rotation of the *½ time unit*. (Level two).

This originates a curved space interval within the impulse, quantified by the impulse $\bar{u}_k$ discrete probability measure $p[\bar{u}_k]$ (1 or 0).

Preserving invariant measure $|1|_M$ of the impulse with ½ time rotating time unit requires its such jumping rotation which originates *two space units* at the probabilistic transformation $M[\bar{u}_k] = |1/2 \times 2| \xrightarrow{p[\bar{u}_k]} |1|_{M_k}$.

That transformation changes the primary impulse' time measure $|1|_M$ to equal measure $|1|_{M_k}$ for the impulse with both time and space units, as a counterpart to the primary impulse curved time interval measure $|1|_M$, Fig.1a). The emerging time-space coordinate system rotates in a following time-space correlated curved movement which conserves the transferring impulse invariant measure. (Level three).

The virtual observer, being displaced from the initial virtual process, sends the discrete time-space probabilistic impulses as virtual probes to *self-test* the preservation of Kolmogorov probability measure in the observing process with probes' frequencies defined by the probabilities (Level four).

The Observer *self-supporting* probes increase frequencies with a growing probability. Such probes test checks this probabilities' symmetry condition indicating the probability correctness. (Level four).

The memory of the correlating time-space intervals temporary holds a difference of the impulses' space-time, identifying both accuracy of their closeness and the virtual Observer' location.

The location space-time measures a time-space shape of the virtual Observer geometry (Level six).

The evolving shape gradually confines the running rotating movement, which *self-supports* by developing both the shape and Observer geometry. The Observer *self-develops* its space-time virtual geometrical structure during virtual observation, which gains its real form with sequential transforming the integrated entropy to equivalent information. (Level seven).

VI. The microprocess.

The microprocess emerges inside Markov diffusion process, which, therefore, should possess the Markovian additive and multiplicative properties. Satisfaction the following conditions indicates the levels of the emerging microprocess below.

1. Growing Bayes a posteriori probabilities along observations intensify the entropy force drawing together the neighbor impulses action ↓ and reaction ↑. This changes a wide of time intervals and curves it.

The curving squeezes the time interval up to merging the neighbor interactive actions ↓↑ on the bordered impulses, which are equal probable, reversible within the probabilities of multiple random interactive actions.

Jumping action ↓ starts entropy increment on the time beginning a microprocess. The curving jump ↑ initiates extreme gradient entropy on the curving time interval, identifying a virtual radius of rotation which displaces the curving action.

The microprocesses opposite entropies emerge with starting time on displacement.

The displacement' space emerges during the jump's curving time. The interactive jump identifies the impulse curvature, entropy measure, and time-space measure equal to π, which minimax principle holds invariant. The jumping impulse ↓↑ develops that impulse time-space volume.

When that displacement rises between the actions with real probabilities 0 or 1, the displacement has no real probabilities.

The process within the displacement has studied as a sub-markov process Markov path, loops [88], connected with Schrodinger bridge [90] which is unique Markov process in the class of reciprocal processes introduced by Bernstein [89].

2. When the sub-markov process gets negative entropy measure $S^*_{\mp a} = -2$ through jumping action ↓, with relative probability $p_{a\pm} = \exp(-2) = 0.1353$, it initiates the microprocess with the minimal time on the verge of the displacement.

3. The developing anti-symmetric entropy increments connect the observing entropies in correlation which are increasing with growing the classical prior probabilities of multiple random actions ↓ and the posterior probabilities of the random action ↑. Both of them are virtual, whose manifold decreases with growing the probability measure.

The sub-markov process with its opposite random actions disappears from the observation.

At the maximal probabilities, only a pair of the Markov additive entropies increments with axiomatic symmetric probabilities (which contains symmetrical-exchangeable states) advances in the correlated superposition of both actions ↓↑. These random actions, belonging the Markov process, measure a multiplicative probability.

At satisfaction of the symmetry condition, the actions axiomatic probabilities transform the microprocess 'quantum' probability with pairs of conjugated entropies to their correlated movements.

4. The conjugated entropies increments rotating on angle $\pi/4$ raise the space interval with virtual transitive action ↑ within the microprocess, which initiates the correlated entanglement.

Maximal correlation adjoins the conjugated symmetric entropies, uniting them in a running pair entanglement.

The conjugated entropies, rotating the space interval on angle $-\pi/4$, transforms the transitive action ↑ to action ↓ that settles a *transitional impulse* ↑↓ which finalizes the entanglement at total angle $\pi/2 = \pi/4 - (-\pi/4)$.

The transitional impulse, holding actions ↑↓ opposite to the primary jumping impulse ↓↑, intends to generate an inner conjugated entanglement, involved, for example in left and rights rotations ($\mp$).

The transitional impulse, interacting with the opposite correlated entropies $\mp$, reverses it on $\pm$.

Since the correlated entropies are virtual, transition action within this impulse ↑↓ is also virtual and its interaction with the forming correlating entanglement is reversible.

Within the impulse time interval, the *entanglement starts before the space is formed and ends with beginning the space.* It occurs *during reversible relative time interval* $0.015625\pi$ *being* part of the impulse invariant measure $\pi$ with time interval $\tau = 1 nat$. *Since the entanglement has no space measure, the entangled states can be everywhere in a space.*

*The space emerges with probability* $P_\Delta^*(\delta t_\pm^{k1}) = 0.8212145$.

5. The following interaction logically erases (cuts) each previous directional rotating the entangle entropy increment of the entropy volume. This erasure emits minimal energy $e_l$ of quanta [91].

The transitional impulse absorbs this emission inside the virtual impulse, which logically memorizes the entangle units making their mirror copy. Such operation performs function of logical Maxwell Demon.

7. The entangle logic is memorized, when the rotating step-up action ↑ of the transitional impulse moves to transfer the entangle entropy volume to ending step-up action ↑ of the jump, which follows real step-down action ↓.

The last interactive action kills and finally memorizes the joint entangle units in the action ending state as the information Bit. The killing is the irreversible erasure encoding the Bit, which requires energy.

The energy, capturing between these interacting actions ↑↓, carries the last of these actions.

That action, cutting the entropy volume, initiates an irreversible process, satisfying the Landauer principle and compensating for the cost of Maxwell Demon. The memorized bit freezes the energy spent on the erasure for its creation as the bit equivalent ln2.

8. The microprocess is different from that in quantum mechanics (QM), because it arises inside the evolving probing impulse under No–Yes virtual and final real actions. The microprocess' superposing rotating anti-symmetric entropy increments, which the cutting EF defines, have additive time–space complex amplitudes correlated in time–space entanglement that do not carry and bind energy, just connects the entropy in joint correlation.

These complex amplitudes models elementary interaction with no physics, while real cut brings the physical bit. Whereas the QM probabilistic particles carry analogous conjugated probability amplitudes correlated in time-space entanglement.

Theoretically, Kolmogorov's probability measure at quantum mechanics' entanglement, when both additivity and symmetry of probability for mutual exchangeable events vanish, challenges predictability of the QM probability. It happens in this Observer probabilistic approach with appearance of the microprocess predictable probability by cutting the entangle entropy.

Memorizing the process information by cutting the entangle entropy ends the microprocess in the evolving observation.

9. The logical operations with information units achieve the approach a goal: integrating the discrete information hidden in the cutting correlations in the creating structure of Information Observer.

The relational entropy conveys probabilistic causality with temporal memory of correlations, while the cutoff memorizes certain information causality during the objective probability observations.

The self-observing virtual observer self-generates elementary Bit self-participating in building self-holding geometry and logic of its prehistory, and predicting evolving dynamics without any physical law.

VII. Specific of memorizing and encoding elementary information through interacting curved impulse

The interaction of the impulse Fig.2A and Fig.2B holds the opposite action ↑↓, curving a displacement between them, which provides a time–space asymmetry (a barrier) between 0 and 1 action, necessary for creating the Bit.

The step-down state of real action ↓ (carrying the energy) supplies Landauer's minimal energy equivalent $\ln 2$ with maximal probability.

This action initiates irreversible process killing entropy and erasing it, which memorizes a classical bit.

That process, starting with creation of the entangled entropy volume, freezes $-\ln 2$ memorizing two opposite qubits.

Conclusively, the impulse Fig.2A step-down cut ↓, extracting each Bit hidden position, erases it at a cost of the cutting real time interval, which encloses the energy of an interactive process.

The impulse Fig.2B step-up action ↑, stopping state at the end of the impulse's time interval, memorizes (encodes) the information Bit of the impulse logic.

Each impulse encoding merges its memory with the time of encoding that minimizes this time.

Forming transitional impulse with entangled qubits leads to possibility memorizing them as quantum bit.

The needed memory of the transitional curved impulse encloses entropy $0.05085 Nat$ (Sec.2.2.5).

The curving topological geometry can enclose minimal energy ln2.

VIII. The gap between entropy and information.

As maximal a priori probability approaches $P_a \to 1$, both the entropy volume and rotating moment grow. Still, between the maximal a priori probability of virtual process and a posteriori probability of real process $P_p = 1$ is a small microprocess' gap, associated with time-space probabilistic transitive movement, separating entropy and its information (at $P_a < 1$ throughout $P_p \to 1$).

It implies a distinction of statistical possibilities, with the entropies of virtuality, from the information-certainty of reality.

The gap holds a hidden real locality which the impulse cuts within the hidden correlation.

The rotating potential momentum, growing with the increased entropy volume, intensifies the time–space volume transition over the gap. That momentum acquires physical property near the gap end when last posterior probability $P_p$ overcomes last prior virtual probability, and the momentum curves a physical cut of the transferred entropy volume.

It is impossible to reach a reality in quantum world without overcoming the gap between entropy-uncertainty and information-certainty, which is located on edge of reality.

Within the gap, the entangled microprocess' conjugated entropies $S^*_{\mp a} = 2h^o_\alpha$, limited by minimal uncertainty measure $h^o_\alpha = 1/137$ - structuralhe  parameter of energy, confine the entangling qubits.

Injection of the energy has probability
$p_{\pm a} = \exp(-2h^o_\alpha) = 0.9855507502$.

The energy starts the erasure of entropy and creation memory, while the actual killing with probability $P_k = 0.99596321$ ends the erasure and starts memory. A gap to reality evaluates probability $1 - P_k \cong 0.004$.

Memorizing the classical bit and qubits have high probability but less than 1, which does not allow reaching absolute reality.

## IX. Information process

This process emerges from the observing process of virtual observer (Level 1) evolving from the microprocess of conjugated entropies within a merging interacting impulse (Level 2).

Information arises from multiple random interactive impulses when some of them erase-cut other providing Landauer's energy (on first level of this stage). Asymmetrical inter-action, erasing the impulse, becomes bit of information, finalizing next level of the stage (Level 3).

The impulse cutoff correlation sequentially converts the cutting entropy to information that memorizes the probes logic in Bit, participating in next probe-conversions, which generate interactive information process at Level 4.

The origin of information, thus, associates with the impulse ability of both cut and persist observing process, generating information under the cut, whose memory holds the impulse' cutting time interval.

Since the curved topology of the interacting impulses decreases total needed energy, this energy, at the found ratio of the impulses' external and internal temperatures (Sec.2.6), can deliver the minimal Landauer's energy equivalent to ln2.

The increment of information covering the asymmetric interaction evaluates free information $i_f = 0.23 bit$, which enables connecting a multiple bits through *information attraction* (Level 5).

Multiple cuts of the more probable posterior correlations in the interactive multi-dimensional observation are a source of persistent $i_f$ that composes multiple Bits with memories of the collecting impulse' cutting time intervals which freeze the observing events dynamics in information processes. (Level 6).

Integration of the cutting Bits time intervals along the observing time course converts it to the *Information Observer* inner time course. That time is opposite to the virtually observable process time course in which the process entropy increases (Level 7).

On information process trajectory, each previous impulse' information distinguish from that in the following impulse.

The difference measure a random interval between these impulses which is predictable by the EF on this interval. That random difference can model "mutation" in evolving information process, which the EF-IPF measure estimates (Level 8).

The difference holds the imaginary entangled entropy of a microprocess which proceeds along the EF.

Each impulse observation creating a bit requires frequency off probing impulses $F_{im} = 10^{-4} \times 1.13276$, while frequency $F_{imo} = 10^{-4} \times 1.13636$ anticipates memorizing the bit. That evaluates at level 9.

## X. The emerging macroprocess composing the basic triplet units

The rotating movement (Fig.3) connects the microprocess imaginary entropy with information Bits *in a rising macroprocess*, where the free information of multiple in impulses binds the Bits in *collective* information time-space macro trajectories (Figs.3,5) (Level1).

The observing information moves the macroprocess through the rotation which depends on forming the entropy gradient (as a potential Coriolis force).

Minimum three rotating Bits join in triplet unit (UP) which measures macroprocess information $\mathbf{a}_{io}(\gamma_i)$ (at level 2).

The free information' parameter $\gamma_i = \beta_{toi} / \alpha_{tio}$ connects the imaginary UP entropy part $\beta_{toi} = c_{ev}$ with the forming UP real part $\alpha_{tio}$ holding the ending information speed of the volume that transits through the gap via the rotation speed. The entropy volume rotating with speed $c_{ev}$ measures frequency $\beta_{toi} \to f_{io}$.

Since that, each killed entropy has a frequency which carries the creating bit.

It implies that such frequency $f_{io}$ will periodical appear as a bleaching signal from the microprocess.

The microprocess within impulse is reversible, holding transitional measure $\pi$ of the curved impulse, until the impulse cutting action transforms to information bit composing the irreversible macroprocess with frequency $f_{io} = \pi/3$.

Each bit is memorized in a local cyclic rotation (Sec. 5.4).

At forming UP, the information rotating speeds of two cooperating bits, selected automatically during the attracting minimax movement, should coincide with the rotating speed of third bit (at level 3), which joins the two cutoff Bits and third Bit.

This attracting process joins the free information of triple bits, rotating in the local cycles, in a loop which harmonize the equal speeds-information frequencies in a coherent (resonance) movement.

The rotating loop is analogous to the Efimoff scenario [81] of resonance frequencies which gives a rise of three units systems, Fig.6. The loop includes Borromean knot [84] and ring.

The free information of triple bits $3\mathbf{a}(\gamma_i)$ supply the minimal Landauer energy (ln2, Sec.2.26) necessary to memorize the triple bits in the triplet knot. Switching the movement of third bit to opposite direction not only attracts the two cooperating bit to the third. It also provides both a time–space asymmetry (a barrier) and the interactive action whose energy of ending state initiates killing and erasure, memorizing the total tree bits free information in the triplet knot.

The memorized triplet' bit encodes it in the knot.

The information binding third bit in the triplet knot, provides stability of the formed UP. (Level 4).
Forming UP depends on the bit's *fitness* for the triple cooperation in the UP. The fitness implies ability of altering direction the bit's moving speed. The conjugated Hamiltonian dynamics of the opposite process segments brings such ability for the segments attracting with free information.

The 'mutation' between the bits brings a variety of the bits with different moving speeds, which changes its fitness for particular UP. Variety of the finesses for three bits could produce different $UP_i$, which the minimax selects (at level 4). Or not produce any UP if they do not fit its triple self-connection, ending triple cooperation at level 4.

The information triplet can build a pair of qubits from the observing process opposite segments under the step-down actions within the gap attracting other qubits from the complementary segments. Such opposite triple qubits can assemble a triplet bit (Fig.4). That allows forming UP from the qubits at level 4.

Memorizing each triple bit preempts the dynamic loop that assembles each triplet tr1, tr2,... on Fig.7 (Level 5).

The macroprocess free information integrates the Bits in an information path functional (IPF), which encloses the bits, joining their UP time–space geometry in the process' information structure. (Level 6).

XI. Emerging Information networks

1. Through assembling $UP_i$ in the triplet cooperative units $UP_{oi}$ during the attractive macro movement which cooperates time-space hierarchical network (IN) (Fig.7) at level 1.

Particularly, information speeds of primary triplets' $UP1, UP2$ connect them to $UP3$ building the triplet's knot $UP_{o1}$ by the information spent on their attractive movement.

The knot' attracting free information forms the rotating loop, which attracts next forming $UP_{o2}$ and then $UP_{o3}$, possibly, from different observing process dimensions. That happens when the triple fits the cooperative minimax conditions analogous to forming the starting UP (Level 2).

Free information of cooperating knots $UP_{oi}$ builds hierarchical IN information structure of nested knots-nodes. (Level 3).

Each $UP_{oi}$ has *unique position* in the IN hierarchy, which defines exact location of the $UP_{oi}$ information logical structures.

The position depends on each unit' information measure $\mathbf{a}_{iuo}(\gamma_{iu})$ which identifies parameter $\gamma_{iu}$.

The IN node positions on its hierarchical level classifies *quality* of assembled information. The ending IN node integrates the assembled information enfolding all IN's levels. (Level 4)

New information for each IN delivers the node requesting information through interactive impulse, providing feedback impact on the cutoff memorized entropy of observing data events. (Level 5).

The appearing new information on that level currently renovate the IN, building its temporary hierarchy of changing qualities of information. The IN high level enfolds information logic that requests new information extending the IN hierarchy and logic. (Level 6).

The IN ending triplet integrates these cooperative qualities, holding information, frequency, and space-time location, which evaluate quality of all IN. (Level 7).

In the IN hierarchy where the node quality grows with the node hierarchical level, the ending triplet-node has higher quality, compared to other nodes.

The IN nested structure harmonizes its nodes quality which the ending node enfolds. That cooperates the observer' running IN.

Variety of distinctive $UP_k$ can cooperate multiple different networks $IN_{oj}$; each of them harmonizes the ending nodes' specific cooperative qualities. The ending nodes enfold all IN's particular quality information. The node' time-space positions identify the value of each nested node information quality $\mathbf{a}_{iuo}(\gamma_{iu})$. (Level 8).

2. Through assembling $UP_{oi}$ in the cooperative structures with higher levels $1_{oi}$ of the information networks.

The attracting minimax movement assembles each three of ending of the network $IN_{oj}$ nodes $UP_{oi}$ in new formed $1_{oi}$ unit forming new loops on the higher structural organization level, which connects the cooperating speeds of each triple (Fig.6) in coherent movement.

Particularly, the attractive motion of rotating triples units $(+UP_{o1}, -UP_{o2}, +UP_{o3})$, emerging from opposite (conjugated) ending nodes of the running networks, can cooperate the units in composite triplet $-1_o(+UP_{o1}, -UP_{o2}, +UP_{o3})$. Then, cooperate the same way the opposite composite triplet $+2_o(-UP_{o5}, +UP_{o6}, -UP_{o7})$. After the analogous formation of a third composite triplet, adjoining the first two to the third rotating triplet. Triplet unit $-1_o$ enfolds qualities of the above three networks though a loop of resonance frequencies, which depend on the nodes location and information values.

Thus, this units' resonance frequency joins the qualities of the IN ending nodes in new quality which adjoins qualities at next level network that enfolds the new quality.

Since the resonance frequency is only a part of spectrum of these network nodes frequencies, it's less than the highest of them. The highest of these frequencies can adjoin new higher information level, compared to any of the above three whose frequencies hold a lesser information and related quality.

The resonance frequency encloses higher information quality from all three that it composes.

As a result, information qualities of each triplet units $-1_o, +2_o, -3_o$ grow, when they join in rotating cooperative circle forming new triplet unit $1_{o3}$. This unit encloses the composite units from the above three which fit the cooperative circle.

Multiple triplets $1_{oj}, j = 3,5,7,...$ sequentially cooperate a new network $IN_1$ which harmonizes its nodes higher qualities and enfolds the highest of these qualities in its ending node (Level 9).

3. Each harmonized $IN_1$ forms a *domain* of an observer with its specific qualities and high density of information enclosing all IN node densities. (Level 10).

The sequentially built triplet knots memorize only the IN current composite unit $1_{oj}, j = 3,5,7,...$, while the previous units hold only logical information. The sequentially memorized information automatically implements the IPF concurrent maximum minimizing total time of building each composite information unit. The minimax leads to sequential decreasing ending information speed of each node, and therefore to decreasing starting information speed on next cooperative unit of growing structural organization.

It restricts spectrum of information frequencies for each self-built IN, decreasing the spectrum with growing the IN level of structural organization. Each self-built IN has limited number of cooperating triplets and the IN nodes. These restrict the ability of each IN ending node to next cooperation. The violation of the restriction leads to the IN instability with rising chaotic movement.

Building each higher level cooperative unit toughens the requirements to the fitness for the variety of primary bits, units $UP, UP_i, -UP_{o1}$, etc., which sequentially decreases their variety (at the end of level 10).

Any forming INs should satisfy invariant relations for ratios of starting information speeds $\gamma_1^\alpha = \alpha_{io}/\alpha_{i+1o}$ and $\gamma_2^\alpha = \alpha_{i+1o}/\alpha_{i+2o}$ connected by dynamic invariant $\mathbf{a}(\gamma)$ that binds the ending eigenvalues of triplet's segments (Sec.4.)

**5.2. The self-control of the multiple IN cooperative domains integrating quality of the domain information**

The main information mechanism, governing both each IN cooperation, the domain, and multiple IN domains cooperation, is the space–time spiral rotation.

The basic parameters of this mechanism are invariant $\gamma_i^\alpha$ for the each vertex angle $\beta = \psi^o$ of conic rotation (Fig.3), which is changing at growing the quality domain level.

That depends on each resonance frequency $f_i = f_i(\gamma_i^\alpha)$ which determines quality of the domain information.

A primary radius of rotation determines the emerging displacement distance $d_a = r_{e1}$ from which follows initial angle $\beta_o$ of the rotation trajectory of cone **Figs.3.8,** using relation [43]:

$$r_{e1} = \rho = b\sin(\varphi \sin \beta_o) \text{ at } \varphi = \pi k/2, k=1, b=1/4. \quad (2.1)$$

According to [46], the vertex angle can be changed discretely following formula
$\sin \psi_i^o = (2k)^{-1}, k=1, 2,..$

where at $k=1$, $\psi_i^o = \pi/6$, and the following $k=2$ determines

$$\psi_{k=2}^o = \pi/2, \psi_{k=3}^o = 2\pi/3. \quad (2.1a)$$

Here $k$ is the number of current domain level, where total numbers of the $k$ domains levels determine the numbers of changing resonance frequencies. For wide diversity of observers, the specific frequencies for each observer are different.

Since the maxmin leads to growing the particulate observer' quality of the domain information, it implies grows of the domain quantity.

Each switch to the following domain level increases the angle of cone rotation, extending the growing IN levels in the domain.

Transferring the angle of the rotating mechanism to the domain with decreasing frequency moves the quality of domain information on a higher information level. Whereas rotation of each existing domain continues its functioning for the current observer.

The nested domains, which composes the extending IN hierarchical structure, requests new information bits with raised information densities. Each request is sending down along the IN hierarchy to the observing bit's spectrum of information frequencies. The highest level of the IN domain sends the request for maximal density of needed information to get high frequency impulses.

The increase of information density grows quality of information to be enfolded in the higher IN domain. Each information domain requests an impulse with different information density and geometry for the requested IN.

The information equivalent of impulse wide $\delta_{ue}^i \cong 0.05 Nat$ limits its size, the minimal time interval $\delta_{te} \approx 1.6 \times 10^{-14}$ sec, and the speed between the nearest impulses on time interval $\Delta_t$:

$c_{ika} \cong 0.0516 \times 10^{14} Nat/\sec$.

That requires shortening time interval of observer inner communication for transporting the impulse information within its information structure. This also limits potential speed of information attraction that restrains obtaining the new information. Thus, requiring each Observer to create the *own time of inner communication,* depending on the requested information density, and a *time scale,* dependent on *density* of accumulated (bound) information in each IN.

Each self-organizing information triplet is a macrounit of specific self-forming information time-space cooperative distributed network enabling self-scaling, self-renovation, and adaptive self-organization.
Such unit starts IN upper level, cooperates triplets nodes along the IN hierarchy which ends with triplet node on IN lower level. Each IN has invariant parameter of its triplet cooperation $\gamma_i^\alpha \cong 3.45$.

The first triplet' cooperation begins with the microprocess real speed $\alpha_{io}$ which determines starting cooperative *information* speed after killing the entropy volume.

<u>Comments.</u> Within a reversible microproces at $(+t = -t)$, the ratio of a current imaginary entropy increments to real one follows from the equations for the opposite entropies, Sec.2.3.2):

$$S_-(t)/S_+(t) = \beta_{it}/\alpha_{it} = \gamma_i = [1+\mathrm{jtg}(-t\ )]/[1-\mathrm{jtg}(+t\ )] = 1. \qquad (2.2)$$

It means, when the entangled conjugated entropies are equal, the frequency of each 'bleaching signal' from the microprocess, depending on ratio $\gamma_i$, stays invariant (according to (2.2)). •

Information speed, starting primary bits after killing minimal entropy volume:

$$\alpha_{io} = c_{iv} \cong 2.4143 \times 0.596 \times 10^{15} \, Nat/\sec \cong 1.44 \times 10^{15} \, Nat/\sec, \qquad (2.3)$$

determines invariant $\gamma_{io} = \beta_{io}/\alpha_{io}$ as the ratio of imaginary part of the speed to its real part.

In forming a first IN triplet, speed (2.3) starts the information attraction and harmonization with the second and third bits, creating the first triplet' invariant structure. (The bleaching frequency may signal approaching the microprocess end).

The parameter $\gamma_{iko}^\alpha(\gamma_{io}) \cong 3.45 - 3.8$ connecting triplets in multiple cooperation identifies frequency,

$$f_{ik} = 1/3(\gamma_{iko}^\alpha) \cong 1.15 \qquad (2.3a)$$

of appearance a new triplet, which is closed to optimal interactive frequency $\pi/3 \cong 1.05$ (Sec.2.6).

4. Any current quality of a domain information may request the needed information increasing its IN information quality.
The requested information from any lower level of the IN domains with less information density requires longer communication time to get the observing bit with less density.

The question is: Are there any requirements to a particular observer for choosing a specific needed domain information, or creates it and then builds and develops, while satisfying the above objective limitations on getting a maximal observer quality domain with highest information density?

## 5.3. Acquisition of the IN current information via an observer' feedback with observation

The IN node' interaction with the observable information spectrum delivers the needed information frequency when the interactive impact of requested information $\mathbf{a}(\gamma_i)$ on the delivered $\mathbf{a}_\tau(t-s)$ forms an impulse function. This function step-control, carrying attracting information $\mathbf{a}(\gamma_i)$, initiates the observer dynamic process to acquire the delivering information $\mathbf{a}_\tau(t)$.

Assume the IN current $i$ node with speed $\alpha_{i\tau}$ requests an external information frequency, which will bring speed $\alpha_{k\tau}$ attracting the IN's $k$ node and spends on the request information of attraction $\mathbf{a}(\gamma_i)$.

Incoming information $\mathbf{a}_\tau(t-s)$ is delivered during time interval $\Delta_t = t - s$ by sequence of probing impulses-control functions, which on each time interval $t_k$, $k = 0,1,2,....,m$ brings invariant information $\mathbf{a}_o(\gamma_i)$ as a part of $\mathbf{a}_\tau(t-s)$, where the node parameter $\gamma_i$ is a priori unknown.

The deficit of requested information is equivalent to a related equal uncertainty which will compensate the sending probing impulses satisfying the requested node parameter $\gamma_i$. It will define the node information $\mathbf{a}(\gamma_i)$ attracting the equal requesting information that the node then attaches.

Let us find it.

An interactive impact of requested information $\mathbf{a}(\gamma_i)$ on delivered $\mathbf{a}_\tau(t-s)$ evaluates Riemann-Stieltjes integral $I_s = \int_{-\infty}^{\infty} f(t-s) dg(s)$, applied to an information function $f(t-s) \to \mathbf{a}_\tau(t-s)$ in the form:

$$I_s = \int_{-\infty}^{\infty} \mathbf{a}_\tau(t-s) \mathbf{a}(\gamma_i) \delta(s) ds = \mathbf{a}_\tau(t) \mathbf{a}(\gamma_i). \tag{3.1}$$

Solution (3.1) is found [44] for step-function $dg(s) \to \mathbf{a}(\gamma) du(s)$, which holds derivation $du(s)$ forming impulse function
$$du(s) = \delta(s) ds. \tag{3.2}$$
The requested node' information impacts on potential observing information measures of impulse (3.2) in form:
$$\mathbf{a}(\gamma_i) \mathbf{a}_\tau(t_{k+1}) = \mathbf{a}_o^2(\gamma_i) \tag{3.3}$$
which *binds* information $I_{ik} = \mathbf{a}(\gamma_i) \mathbf{a}_\tau(t_{k+1})$ according to (3.1) at moment $t_{k+1}$.

The binding impulse memorizes cutting entropy of the observing process spectrum, prompting to encode information of observing impulse in the requesting node' knot.

Applying (3.2) allows finding delivering information $\mathbf{a}_\tau(t_{k+1})$ requested by $\mathbf{a}(\gamma_i)$ with the IN known $\gamma_i$.

For example, at $\gamma_i = 0.3$, the procedure follows:

1. Parameter $\gamma_i$ that determines the requesting i-node information $\mathbf{a}_i(\gamma_i) = 0.239661$, while other node information are $\mathbf{a}_o(\gamma_i) = 0.743688, \mathbf{a}_o^2(\gamma_i) = 0.553$, and location of the i-node defines the IN node parameter $\gamma_{ik}^\alpha(\gamma_i)$.

The $\gamma_{ik}^\alpha$ identifies frequency of the k-probing impulse $f_{ik} = (\gamma_{ik}^\alpha)^{-1}$ which will interact with observing random process to observe it.

2. Suppose the requested node' triplet ends with a speed $\alpha_{io}$ which produces three multiplicative impulse control actions $u_i = 2^3 = \alpha_{io}$ (Sec.4.5) that request this information speed.

3. Assume the attracting information requires speed $\alpha_{ko}$ to bring the delivering information to start forming new triplet node, whose location along the IN determines the triplet cooperative parameter $\gamma_{ik}^\alpha$.

Then, required information speed $\alpha_{ko} = \alpha_{io} / \gamma_{ik}^\alpha$, at $\gamma_{ik}^\alpha = 3.33$, and $\alpha_{io} = 8$ lead to $\alpha_{ko} = 2.3966$.

It allows finding time interval
$$\delta_{ek}^t = \mathbf{a}_i(\gamma_i) / \alpha_{ko}, \quad \delta_{ek}^t \cong 0.1 \sec \tag{3.4}$$
of potential acceptance of the needed information $\mathbf{a}_\tau(t_{k+1})$ to be delivered during the time of communication.

4. The interactive information for any invariant impulse follows from eq.
$$\mathbf{a}_\tau(t_{k+1}) = (\mathbf{a}_{io}(\gamma_i))^2 / \mathbf{a}_i(\gamma_i) = \mathbf{a}_\tau(t_{k+1}) \cong 2.3. \tag{3.5}$$

5. Relation
$$\mathbf{a}_\tau(t_{k+1}) \cong 2.3 / \mathbf{a}_o(\gamma_i) \cong 3 \tag{3.6}$$
determines number of sending information impulses $k = 3$, which at $\delta_{ek}^t \cong 0.1 \sec$ defines time communication $3\delta_{ek}^t = \Delta_t \cong 0.3 \sec$. The sending impulse of the requested information is the IN updating feedback.

6. Three receiving impulses should build triplet, which the requested node will accept and join to its IN.
Therefore, the joint node (with sought $\gamma_k$) should satisfy balance Eq.
$$(\mathbf{a}_{io}(\gamma_k))^2 + \mathbf{a}_i(\gamma_k) \cong \mathbf{a}_{ko}(\gamma_k), \tag{3.7}$$
where interactive relation $(\mathbf{a}_{ko}(\gamma_k))^2 = (\mathbf{a}_{io}(\gamma_i))^2$ holds at $\gamma_k \neq \gamma_i$.

Since at forming triplet's node, the adjoining three information units interact, the interaction provides some uncertainty from which follows the unit distinctiveness.

7. For this example at $\mathbf{a}_{io}^2(\gamma_i) = 0.553$ we get $\gamma_k \cong 0.05$ at $\mathbf{a}_k(\gamma_k) \cong 0.2564$, $\gamma_{ik}^\alpha \cong 4.81$.

Last parameter $\gamma_{ik}^\alpha$ corrects the initial frequency of observing impulses $f_{ik} \cong 0.2$ .

Thus, the balance Eq. identifies $\gamma_k$ which defines new $\mathbf{a}_k(\gamma_k)$ obtained from the feedback probing impulses.

Since ratio $\delta_{ek}^t = \mathbf{a}_k(\gamma_k)/\alpha_{ko}$ for each invariant triplet unit' holds constant, intervals $\delta_{ek}^t \cong 0.1\sec, \Delta_t \cong 0.3\sec$ are preserved, as well as the number of requested impulse k=3.

Comparing with the request information, renewing information increases: $\mathbf{a}_k(\gamma_k) \cong 0.2564 > \mathbf{a}(\gamma_i) = 0.239661$ at increasing frequency of the probing observation. That increase grows quality of the IN renewing information. The invariant triplet builds a temporary IN, which, the requested IN node accepts after verification by (3.7). Growing $\gamma_{ik}^\alpha(\gamma_k)$ increases requesting information for next IN node.

Thus, known $\alpha_{io}$ determines $\alpha_{ko}$ and time interval $t_{k+1} = \tau_{k+1}^1 - (\tau_{k+1}^o + o_{ko}) = \delta_{ek}^t$ starting at the moment $\tau_{k+1}^o + o_{ko}$ and ending at the moment $\tau_{k+1}^1$ of acceptance needed information.

The time interval $\Delta_t$ ends communication with probing the $k$-control impulses, which deliver the needed information $\mathbf{a}_k(\gamma_k)$ building triplet unit.

The frequencies for obtaining higher quality information will increases at the same time of communication, bringing the needed information for building temporary IN with and same interval $\delta_{ek}^t$ of acceptance new information

The feedback requested information emanates from the network until it satisfies the delivering impulse requirements, building both the triple cooperations and the feedback interactive binding.

The delivering new information generally increases the observer information quality by both growing current number of network nodes and/or developing another IN with renown $\gamma_{ik}^\alpha \to f_{ik}$. That $f_{ik}$ should be able synchronizing additional information spectrum needed for the growing quality of information.

The extended observing information spectrum enables increasing the INs domain quality information and adding the INs with new domain through the feedback communication with the observing information.

What is actually initiates a particular observer to obtain the specific needed information, create and develop explicit domain having an advantage with the others?

### 5.4. Selection information and structuring a selective observer

1. Forming an information dynamic cooperative requires rising cooperative information force between the potential cooperating triplets:

$$X_{ik}^{Im} = -\frac{\delta I_{ik}^m}{\delta l_{ik}^\alpha} = \mathbf{a}_{oi}(\gamma)(\gamma_{ik}^m - 1) \tag{4.1}$$

where current IN' triplet $m_i$, currying information $\delta I_{ik}^m = \mathbf{a}_i + \mathbf{a}_{oi}^2 \cong \mathbf{a}_{oi}$, attracts $m_k$-triplet, depending on information $\mathbf{a}_{oi}(\gamma)$ with IN invariant parameter $\gamma_{ik}^m$, and on relative distance

$$(t_i^m - t_k^m)/t_i^m = (l_i^m - l_k^m)/l_i^m = \partial l_{ik}^{m*}. \tag{4.2}$$

That cooperative force measures the cooperative attraction between these triplets in Nats (bits).

2. The cooperative force, relative to information of first IN triplet, measures

$$X_{1k}^{Im1} = (\gamma_{1k}^m - 1). \tag{4.3}$$

The relative cooperative information force between the IN first and second triplets:

$$X_{12}^\alpha \geq [\gamma_{12}^\alpha - 1], \tag{4.4}$$

at limited values $\gamma_{12}^\alpha \to (4.48 - 3.45)$. That restricts the related cooperative forces by inequality
$$X_{12}^\alpha \geq (3.48 - 2.45). \tag{4.5}$$
The quantity of information, needed to provide this information force, is
$$I_{12}(X_{12}^\alpha) = X_{12}^\alpha \mathbf{a}_o(\gamma_{12}^\alpha), \tag{4.6}$$
where $\mathbf{a}_o(\gamma_{12}^\alpha)$ is invariant evaluating quantity of information concentrated in a selected triplet by $\mathbf{a}_o(\gamma_{12}^\alpha) \cong 1 bit$ at $\gamma_{12}^\alpha = \gamma_{1o}^\alpha$, From that it follows
$$I_{12}(X_{12}^\alpha) \geq (3.48 - 2.45) bits. \tag{4.7}$$
3. The invariant's quantities $\mathbf{a}_o(\gamma_{io} \to 0)$, $\mathbf{a}(\gamma_{io} \to 0)$ provide *maximal* cooperative force $X_{12}^{am} \cong 3.48$.
Minimal quantity of information, needed to form a very first triplet, estimates dynamic invariants
$$I_{o1} \cong 0.75 Nats \cong 1 bits. \tag{4.8}$$
Therefore, total information, needed to start adjoining next triplet to the IN, estimates
$$I_{o12} = (I_{o1} + I_{12}(X_{12}^\alpha)) \geq (4.48 - 3.45) bit. \tag{4.9}$$
which supports the node cooperation and can initiate the IN feedback in Sec.5.3.
That information equals or exceeds the information of the IN current node requesting the sequential cooperation with a next triplet. This information should deliver an observer to select the requesting unit.
Minimal triplets' node force $X_{1m}^\alpha = 2.45$ depends on the ratio of starting information speeds of the nearest nodes, which determines the force scale factor $\gamma_{m1}^\alpha = 3.45$ satisfying the minimax.
The observer, satisfying both minimal information $I_{o1}$ and $I_{12}(X_{12}^\alpha)$, delivers total information $(4.48 - 3.45) bit$. We call it a *minimal* selective observer.
It includes the feedback control carrying the needed information for a triplet's node.
These limitations are the observer boundaries of *admissible* information spectrum, which depends on related scale factors, and also applied to multi-dimensional selective observer.
At satisfaction of cooperative condition (4.9), each observed information speed, delivering the required density of information spectrum to an existing IN, enables creating next IN's level of triplet's hierarchy.
This leads to two conditions: *necessary*-for creating a triplet with the required information density, and *sufficient* –to provide a cooperative force, needed to adjoin this triplet with an observer's IN.
Both these conditions should satisfy the observer's ability to *select* information of growing density.
The minimal selective observer, satisfying only the necessary condition, we call an objective observer.
At satisfaction the sufficient condition, each next information units joins a sequence of triplet's information structures forming the IN, which progressively increases information bound in each following triplet, and the IN ending triplet's node conserves all IN information.
The node *location* in the IN spatial-temporal hierarchy determines *quality* of the information bound in IN node, which depends on the node enclosed information density. When acquisition of information brings new $\gamma_{ik}^\alpha \cong 4.81$, the related cooperative force $X_{1k}^\alpha = 3.48$ enables transferring to a next cooperative level, extending the IN.
All these conditions satisfy a selective *subjective* observer, limited delivered information 4.48 bit.

### 5.5. The information mechanism of selection and structuring a multiple selective observer
Extending the IN requires quantity and quality of information, which could deliver an external observer, satisfying the requested information emanating from the current IN ending node.
Let us evaluate the interactive information impact of an external observer, carrying own information, on the IN requested information with $m$ triplet to form potential new $m+1$ triplet of the current observer.

Such request carries the control (free) information $\mathbf{a}_{m+1} \sim 0.25 Nat$ with information speed $\alpha_{m+1}^t$ determined by the requested IN node, which encloses information density

$$\gamma_{m+1}^\alpha = (\gamma_{m=1}^\alpha)(\gamma_{13}^\alpha)^m, \alpha_{1o}/\alpha_m^t = \gamma_{m+1}^\alpha, \qquad (5.1)$$

where $\alpha_{1o}$ is information speed on a dynamic processs' segment of the IN initial triplet.

Ratio $\alpha_{1o}$ to speed of third segment: $\alpha_{1o}/\alpha_{3o} \cong 3.45 = \gamma_{13}^\alpha$ determines $\gamma_{m=1}^\alpha = \gamma_{13}^\alpha$ which identifies scale factor equal for cooperated $m$ triplets. Parameter $\gamma_{m+1}^\alpha = (\gamma_{13}^\alpha)^{m+1}$ identifies scale factor of $(m+1)th$ triplet for a requesting IN's node (5.1).

The requested density $\gamma_{m+1}^\alpha$ requires relative information frequency

$$f_{1m+1} = (\gamma_{m+1}^\alpha)^{-1}. \qquad (5.2)$$

Any decreasing frequency has a tendency of growing $\gamma_{m+1}^\alpha$ therefore increasing information quality in evolving IN.
That quality delivers information forces, which automatically overcome the selective observer limitations and extend the IN.
Minimal IN with single triplet node and potential speed of attraction (2.3) requests its information density with speed

$$c_{m+1}^\alpha = (3.45)^2 \times 0.1444 \times 10^{14} \approx 11.9 \times 0.1444 \times 10^{14} \approx 1.7187 \times 10^{14} Nat/\sec. \qquad (5.3)$$

where $\gamma_{m+1}^\alpha = \gamma_{1+1}^\alpha = 3.45^2$.

To adjoin the requested information in the IN, the requesting control should carry information $I_{o12}$ (4.9) with speed (5.3) which requires time interval for transporting this information:

$$t_m \cong 3.45 Nat/1.7187 \times 10^{14} Nat/\sec \cong 2 \times 10^{-14} \sec. \qquad (5.4)$$

This time interval, which carries the requested information $I_{ik}$ (4.9), approximates the impulse wide's time $\delta_{te} \approx 1.6 \times 10^{-14}$ sec with ratio $t_m/\delta_{te} \cong 1.2546$.

Time interval $t_m$, according to (5.3), increases in ratio $\sim 12$ both initial speed of attraction $0.1444 \times 10^{14} Nat/\sec$ and the impulse density.

The increased impulse information density (according to Sec.1.3) decreases the impulse time wide $\delta_{te}$ to

$$\delta_{tm} = \delta_{te}/12 \cong 0.1333 \times 10^{-14} \sec, \qquad (5.5)$$

which is less than the communication time (5.4) in ~15 times.

Impulse with time wide $\delta_{tm}$ (5.5) transfers the requested information to the observing external process where it interacts through 15 of such probing impulses.
This information action should deliver the step-down cut of external impulse that requires quantity information $0.25 Nat$ - the same as the information which carries the requested control.
Communication time (5.4) should deliver to the node the needed frequency-density (5.2) from the observing external process. This time also determines the density-frequency limiting the interval of the sending probing impulses above.
That requires to increase the initial impulse's attracting information density

$$i_{od} = 0.25 Nat/\delta_{te} = 0.25 Nat/1.6 \times 10^{-14} \sec = 0.15 \times 10^{14} Nat/\sec \qquad (5.6)$$

in ~12 times up to

$$i_{md} \cong 1.8 \times 10^{14} Nat/\sec. \qquad (5.7)$$

The observer time of inner communication between the IN initial eigenvalue $\alpha_1$ of its first triplet's and $m+1$ eigenvalue $\alpha_{m+1}$, where the requested information is transferred, depends on ratio

$$(\gamma_{13}^\alpha)^{m+1} = \gamma_{m+1}^\alpha = \alpha_1/\alpha_{m+1} = M_\tau.$$

Preserving the information invariant for each triplet including the $m+1$ eigenvalue:

$$\mathbf{a}_{om} = \alpha_1 \tau_1 = \alpha_{m+1} \tau_{m+1} \qquad (5.8)$$

leads to relation $M_\tau = \tau_{m+1}/\tau_{1o}$ and to

$$\Lambda_{M\tau} = M_\tau - 1 = (\tau_{m+1} - \tau_{1o})/\tau_{m+1} = (\gamma_{12}^\alpha)^{m+1} - 1 = inv \qquad (5.9)$$

which determines the invariant relative interval of inner communication (5.9) defining the observer time scale.

For $\gamma_{m+1}^\alpha = \gamma_{1+1}^\alpha = 3.45^2$ the time scale follows

$$\Lambda_{M\tau} \cong 11. \qquad (5.10)$$

<u>Comments</u>. For elementary objective observer with $M_\tau = \tau_{m+1}/\tau_{1o} = 3.45^2 \cong 12$, for example, each 12 hours of external communication squeezes to one hour of internal communication, which the observer needs to process inter-communication. Thus, to have the non-interruptive inter communication, this observer should open the inter communication each 12 hours. Hence, to coordinate inner and external time course, the observer has required periodically switching time of inner communication holding rhythm 12/1. That implies memorizing a code of $M_\tau = 12$ on the knot of this IN's second triplet, which the cooperated IN automatically includes. Then each one hour of innercommucation will request an external information. During this one hour, the requesting information (4.9) needs $\delta_{te} \approx 1.6 \times 10^{-14}$ sec for the observing a single IN. Therefore, such an observer should potentially have $N_\tau = 3600/1.6 \times 10^{-14} = 2.25 \times 10^{17}$ single communicating networks. The last one, which had memorized $M_\tau$ code, will automatically switch to external communication requesting external information. Consequently, the rhythm, coordinating the observer inters and external time course is part of observer information regularities, even for elementary subjective observer. (The $N_\tau$ could model number of communicating-interacting pair of particles and/or cells). •

The growing information density increases both quality of information enfolding in the current IN and the time of delivering this information. Thus, each Observer *owns the time of inner communication,* depending on the requested information, and *time scale,* depending on *density* of accumulated (bound) information.

Suppose, formation of the IN new node requires $k$ cutting multiplicative information actions. Time interval of such cuts $\delta_{eik}^t$ will depend of the IN time scale, increasing proportionally: $\delta_{eik}^t = (\Lambda_{M\tau})^k \times \delta_{eio}^t$, which, for $k = 10$ increases initial $\delta_{eio}^t \cong 0.2 \times 10^{-15}$ sec in $11^{10}$ times up to $\delta_{eik}^t \cong 2.6 \times 10^{10} \times 2 \times 10^{-15}$ sec $= 5.2 \times 10^{-5} 2.6 \times 10^9$.

This is time of the observer external communication, which in $2.6 \times 10^9$ times more than the related time of inner communication (5.4), which was counted for the IN single node. With growing the node's number, the time of inner communication increases by multiplying the node number on the invariant time scale (5.10).

Thus, free information, carrying attracting information force $X_{12}^{\alpha m} \cong 3.48$ with quantity of the force information $I_{o12}$ (4.9) delivers related quality to the forming IN node by cutting external information with density (5.1).

The cutting information enables forming new IN triplet with $\mathbf{a}_o(\gamma_{12}^\alpha) \cong 1 bit \cong \ln 2 Nat$, which should be attached to the current IN. The impulse, carrying this triplet, interacts with existing IN node through the impact, which provides information $0.25 Nat$ - the same as that at the interaction with an observer external process.

The relative information effect of the impact estimates ratio $\ln 2/0.25 \approx 3$.

The attracting information $0.231 Nat$, which carries the triplet Bit, brings the total $(3\ln 2 + 0.231) Nat \cong 3.573 bit$ increase that compensates for the current IN triplet node' requested $I_{12}(X_{12}^\alpha) \geq (3.48 - 2.45) bits$.

This is minimal *threshold* for building elementary triplet, which enables attracting and delivering the requested information to IN node. The selective observer can choose this node with the needed frequency of the probing impulses.

The triplets are elementary selective objective observers, which acquire the requested IN level's quality of information. That quality ultimately evaluates the IN level of the distinctive cooperative function building the sequential IN's nodes.

The identified information threshold *separates* subjective and objective observers.

Subjective observer enfolds the concurrent information in a temporary build IN's high level logic that requests new information for the running observer's IN and then attaches it to the running IN, currently building the nodes hierarchy.

A selective observer, working as observer-participator, builds its first triplet' structural unit with 'free information whose attraction arranges the IN hierarchy of growing quality information that automatically selects maximal quality observer.

Such selective process, emerging within the macrodynamic process, distinct from Darwinian natural selective evolution, where "only the organisms best adapted to their environment tend to survive and transmit their genetic characters in increasing numbers to succeeding generations while those less adapted tend to be eliminated."

## 5.6. The IN limited time-scale, speed of cooperation, and dimension

Minimal admissible time interval of impulse acting on observing process is limited by $\delta^o_{t\min} \cong \mathbf{a}_{io} \hat{h} \approx 0.391143 \times 10^{-15}$ sec.

Minimal wide of internal impulse $\delta_{te} \approx 1.6 \times 10^{-14}$ sec limits ratio $\delta_{te}/\delta^o_{t\min} \cong 41$ which evaluates the limited IN time scale (5.10) and scale ratio $\gamma^\alpha_{m+1} = (\gamma^\alpha_{12})^{m+1} = 3.45^{m+1}$.

That, for $m+1=3$ triplets brings the IN time scale

$$\Lambda_{M\tau} = 3.45^{m+1} - 1 = 40.06, \text{ or } 3.45^{m+1} = 41.06, \tag{6.1}$$

which limits $m+1=3$ by $m=2$.

It forms *minimal subjective observer* whose IN binds two triplets enfolding $n=5$ process dimensions.

Defining the IN *geometrical* bound scale factor by $F_n = \sqrt{(\gamma^\alpha_{m=1})^n}$, we find its value for process dimension $n=5$ and the IN scale ratio $\gamma^\alpha_{m=1} = 3.45$ binding these dimensions in the IN. Thus, $F_{n=5} = \sqrt{(\gamma^\alpha_{m1})^5} \cong 22.1$ identifies geometrical bound scale factor for the minimal subjective observer.

In Physics, three particles' bound stable resonance has been observed [82, 83] with predicted scale factor $\cong 22.7$.

If an observer enables condensing the external information in a decreased wide of its impulse, then the number of the IN enclosed triplet grows.

This number limits a relative cooperative speed of the IN's last triplet $m$ node, which determines the ratio of the triplet maximal cooperative speed $c_{am}$ to the triple node information speed $c_{oa}$ running cooperation: $c_{am}/c_{oa} = C_{ocn}$.

The $C_{ocm}$ evaluates invariant relation [45]:

$$C_{ocm} \cong 1/2 \mathbf{a}_{io}(\gamma) \mathbf{a}_i^{-1}(\gamma)(\gamma^\alpha_{i=m} - 1)(\gamma^\alpha_{i=m})^m. \tag{6.2}$$

Since each pair cooperation requires $1/2\ln 2 Nat$ of information, applied during time of cooperation $\delta_{te}$, the maximal speed of cooperation:

$$c_{ami} = 0.35 Nat/\delta_{te} = 0.21875 \times 10^{14} Nat/\sec, \tag{6.2a}$$

is closed to the maximal killing speed (2.3):

$$c_{iv} \cong 0.1444 \times 10^{14} Nat/\sec \cong 20 \times 10^{13} bit/\sec. \tag{6.3}$$

The ending node cooperative speed $c_{oci} = 1 bit/\sec \cong 0.7 Nat/\sec$ predicts ratio

$c_{ami}/c_{oci} = C_{cmi} = 0.2062857 \times 10^{14}$.

Applying to (6.2 numerical values of the optimal minimax invariants leads to:
$C_{ocm} \cong 1/2\ln 2/0.33\ln 2(2.45)(3.45)\quad 1.515\quad 2.45(3.45.77(3.45^m). \tag{6.4}$

Equalizing $C_{cmi} = C_{ocm}$ brings maximal number of IN node $m_o \cong 23.6$ $n_o \approx 48$ capable cooperate with the process dimensions $n_o \approx 48$.

A new IN, starting from such triplet, can produce another triplet, cooperating other INs, and so on.

That allows integrate all observed information in a sequentially composed INs, Figs.6, 7.

Speed $c_{ocn} \approx 10 bit/s$ at $C_{oc} \cong 10^5$ requests information frequency $B_f \cong 10^6 bit/s \approx 10^{-3} Gbit/s$.

A human being' single IN approximates maximal number of nodes' levels $m_M \cong 7$, which encloses $4 > m_{M1} > 3$ levels of minimal selective subjective observers.

The minimal self-selective observer with a single triplet adds one more triplet building the minimal IN with two triplet's levels $m_{Min} = 2$.

The observer is able building multiple information networks, when each three INs 'ending nodes (of maximal $m_M$) can form a triplet structure with enfolds all three local INs, increasing the encoded information in $3m_M$ and then multiply it on other $m_M$: $3m_M m_M$ by building new IN starting with this triplet.

That process allows progressively increase both quantity and quality of total encoded information in $(3m_M m_M) \times (3m_M m_M) \times ....... = (3m_M m_M)^{m_M} = N_m$ times of the initial IN's node information $\mathbf{a}_{1o} = \ln 2$:

$$I_m = \ln 2 (3m_M m_M)^{m_M} \tag{6.5}$$

with maximal density

$$I_m^d = \ln 2 (\gamma_{12}^\alpha)^{N_m} . \tag{6.6}$$

and time scale

$$\Lambda_{mM\tau} = (\gamma_{12}^\alpha)^{N_m} - 1 . \tag{6.7}$$

This huge quantity and quality of information is limited by maximal $m_M$.

The IPF maximal information available from an external random information process is limited at infinite dimension of the process (Sec 1.3).

### 5.7. Information conditions for self-structuring a multiple selective observer

Satisfaction of sufficient condition (4.9) in multiple IN's interactions determines a multiple selective observer, whose attracting cooperative information force grows from $X_{12}^\alpha \geq [(\gamma_{12}^\alpha) - 1]$ to

$$X_{12}^{\alpha N_m} \to (\gamma_{12}^\alpha [\gamma_m])^{N_m} \text{ at } (\gamma_{12}^\alpha [\gamma_m]) \to (\gamma_{12}^\alpha [\gamma_m])^{N_m}, \tag{7.1}$$

accumulating at $\gamma_m \to 0$ maximal information

$$I_{12}(X_{12}^{\alpha N_m}) = X_{12}^{\alpha N_m} \mathbf{a}_o [(\gamma_{12}^\alpha)^{N_m}] . \tag{7.2}$$

The communicating multiple observers send *quality messengers* (qmess), enfolding the sender IN's cooperative force (7.1) which requires access to other IN observers for obtaining more information quality [77] and logic.

This allows the observer-sender generates a collective IN's logic of the accessible multiple observers.

<u>Comments.</u> Observer's information units interact on an information channel, producing randomness, entropy that generates a nonrandom information of the channels errors, which may corrupt the sender information. •

Each observer's IN memorizes its ending node information, while total multi-levels hierarchical IN memorizes information of the whole hierarchy. The observer, requesting maximal quality information by its information forces, generates probing impulses, which select the needed density-frequency's real information among imaginary information of virtual probes.

That brings to the observer-sender all current IN logical information. Whereas the IN information dynamics enable renovate the existing IN through the feedback process of exchanging the requested information with environment, and rebuild the IN by encoding and re-memorizing the recent information.

Since the whole multiple IN information is *limited* as well as a total time of the IN existence, the possibility of the IN self-replication arises (Sec.6.5).

*The observer ability of self- organizing each next level of the evolution dynamics we call the observer creativity which limits each observer's integrated information.*

### 6. The observer emerging self-encoding, cooperative complexity, information mass, and information space curvature
### 6.1. Encoding information units in the IN code-logic, and the observer's computation using this code

Observer code serves for common external and internal communications, allowing encoding different interactions in universal information measure, and conducts cooperative operations both within and outside. That unites different observers.

### 6.1.1. The code, program, and information quality of code-logic

Each triplet unit generates three symbols from three segments of information dynamics and one impulse-code from the control, composing a minimal *logical code* that encodes the observing information process. The segment's attracting dynamics join them in a triple unit, which binds knot, releasing free information, that, working as the control logic, transfers this triple code to next triplet node.

That forms next level of the IN code in the IN hierarchy.

Thus, the observing logical units enable encoding each triplet knot binding three bits during the triple self-cooperation.

The knot free information, attracting the next observing bit, forms a duplet which self-cooperates a next bit in the following triplet unit. The knot of that triple attracts another incoming bit, and so on. That sequentially creates the nested structure of information network (IN) whose nodes-knots can concurrently self-encode the observing information units.

The IN triplet's space-time dynamics holds minimal logic information structures of each three triplets. Each information unit has its unique position in the time-spaced information dynamics, which defines the exact location of each triple code in the IN.

Even though the code impulses are similar for each triplet, their time-space locations allow the *discrimination* of each code and the formed logics, and distinction both codes and its unit's geometry.

The IN code includes digital time intervals, enclosed in the unit code's geometry, which depends on digits of the IPF collected information. The IPF progressively shortens time intervals increasing code density.

The shortened process intervals in both the IPF and IN condense the observing information in the rotating space-time triplets' knots, whose nodes cooperate and integrates the IN information logical structure.

The observing process, chosen by the observer's (0-1) probes, and the following integration of the information logic determines the IN information code units that encode the IN logic.

The code logic emerges from both probabilistic and information causality which brings the self –participating bits.

The IPF collects the information units, while the IN performs logical computing operations using the doublet-triplet code of the observer's created program, which is a specific for each observer, and therefore is self-encrypting by the observer code.

Such operations, performing by the entangled memorized information units, model quantum computation [49, 64,73].

The operations with information physical macro-units, emerging from microlevel (quantum) information units, run the units cooperation in the IN, which model a classical computation of the observer logic.

An observer, that unites logic of quantum micro- and macro- information processes, enables composing quantum and/or classical computation on different IN levels.

The computing program holds a distributed space-time processing generated by the observer information dynamics.

Information logic, emanating from different IN nodes, encloses distinctive quality measure, which allows encoding the specific observing process in an observer' IN *genetic code*.

The triplets are elementary logical units enclosing the IN genetic code that composes helix geometrical structure (Fig.3a,6) analogous to DNA [91-95].

The genetic code can reproduce the encoded system by decoding the IN final node and the specific position of each node within IN structure.

Finally, the observing interacting process *naturally encodes the impulse interaction with environment.*

### 6.1.2. The observer information density

Each observing *specific* information quality depends on information density $N_b^{sc}$, defined by the number of information units (bits) that each of this unit encodes (compresses) from a logical source-code.

Since each triplet's bit encodes 3 bits, it information density is $N_b^1 = 3$. The following triplet also encodes 3 bits, but each of its bit encodes 3 bits of the previous triplet's bits. Thus, information density of such two triplets is $N_b^2 = 9$ and so on for more. Hence, for the $m$-th triplet density is $N_b^m = 3^m$ bits which encodes $3^m$ bits from all previous triplet's codes.

The IN's final node with $m = n/2$ has density $N_b^m = 3^{n/2}$ determined by the process' dimension $n$.

The information density, related to the IN's level of its hierarchy, measures also the *value* of information obtained from this level. For such a code, its information density also measures its valueability.

*For example*, an extensive architecture of an ARM chip provides the enhanced code density: it stores a subset of 32-bit instructions compressing 16-bit instructions and decompressing them back to 32 bits upon execution.

In each particular IN, the triplet elementary logical unit-cell self-organize and compose the observer Logical Structure, satisfying the IN limitations (Secs.5.6,5.7).

## 6.2. The emerging IN cooperative complexity

The IN nested structure holds cooperative complexity (CC), which decreases initial complexity of not cooperative information units.

The CC measures the *origin of* complexity in the interactive dynamic *process,* cooperating elementary duplet-triplet, whose free information anticipates new information, requests it, and automatically builds the hierarchical IN [101].

The CC emerges from both the probabilistic and information logics of observing natural processes.

Existing complexity focuses on measuring complexity of already formed complex system and processes [96-100].

The CC is an *attribute* of the process's *cooperative dynamics and its logics.*

Bound information of cooperating units is *potential* source of decreasing the CC, which measures the concealed *information,* accompanied by creation of new observing phenomena.

We study both complexity of information macrodynamic process (MC) and CC, arising in interactive dynamics of changing information flows from $I_i$ to $I_k$, accompanied changing their shared volume from $V_i$ to $V_k$: $\Delta I_{ik} = I_i - I_k, \Delta V_{ik} = V_i - V_k$.

The MC defines an increment of concentration of information in the information flow before and after interaction, measured by the flow's increment per the changed information volume: $MC_{ik} = mes[\Delta I_{ik} / \Delta V_{ik}]$, while the flow's increment measures the increment of entropy speeds $mes\Delta I_{ik} = \partial \Delta S_{ik} / \partial t$, in form

$$MC_{ik} = (\partial \Delta S_{ik} / \partial t) / \Delta V_{ik} . \tag{2.1}$$

This complexity determines an instant entropy's concentration in this volume: $\frac{\partial \Delta S_{ik}}{\Delta V_{ik} \partial t}$ (the entropy production), which evaluates the specific information contribution, transferred during the interactive *dynamics* of the information flows.

Complexity (2.1) is measured *after* the interaction occured, assuming that both increments of speeds and volumes are known.

To evaluate complexity, arising *during* interactive dynamics, *information measure of a differential interactive complexity* $MC_{ik}^\delta$ *is introduced, defined by* increment of information flow $-\frac{\partial \Delta S_{ik}}{\partial t}$ per small volume increment $\delta V_{ik}^\delta$ (within the shared volume $\Delta V_{ik}$):

$$MC_{ik}^\delta = \frac{\partial H_{ik}}{\partial t} / \frac{\partial \Delta V_{ik}}{\partial t}, \tag{2.2}$$

where (7.2) defines ratio of the speeds, measured by the increments of information Hamiltonian and volume accordingly.

The $MC_{ik}^\delta$ automatically includes both $MC_{ik}$ and its increment $\delta MC_{ik}$.

The $MC_{ik}$ measures the differential increment of information of interactive elements $i,k$ whose current *information difference* $\Delta S_{ik}$ and shared volume $\Delta V_{ik}$ -*before joining* will be reduced to increment $\delta S_{ik}$ and volume $\delta V_{ik}^{\delta}$ accordingly after their cooperation in a macrodynamic process.

Thus, within the IN nested structure emerges the IN cooperative complexity, which arises from the both $MC$ complexity and differential $MC^{\delta}$ complexity of the information process, integrating the observing bits through the information path functional (IPF).

Applying the IMD allows measuring both $MC_{ik}$ and $MC_{ik}^{\delta}$ through the VP *information invariants and direct evaluating these complexities in the bits of information code.* Both complexities measure during each triplet's dynamics through their eigenvalues which connect them to the related geometrical structure's volume.

The complexity (2.2) for a triplet is defined at the moment of three segments eigenvalues' equalization. That measure is

$$M_{i,i+1,i+2}^{\delta} = 3\dot{\alpha}_{i+2,t} / \dot{V}_{i,i+1,i+2}, \tag{2.3}$$

at $\dot{\alpha}_{i+2,t}|_{t=t_{i+2,t}} = [\alpha_{i+2,o}^2 t_{i+2,t}^2 \exp(\alpha_{i+2,o} t_{i+2,t})(2 - \exp\alpha_{i+2,o} t_{i+2,t})^{-1} - \alpha_{i+2,o}^2 t_{i+2,t}^2 \exp 2(\alpha_{i+2,o} t_{i+2,t})]/t_{i+2,t}^2$ holding invarint form

$M_{i,i+1,i+2}^{\delta} = \mathbf{a}_o^2 [exp\mathbf{a}_o (2 - exp\mathbf{a}_o)^{-1} - exp2\mathbf{a}_o]/t_{i+2,t} = (\mathbf{a}_o\mathbf{a} - \mathbf{a}_o^2 exp\mathbf{a}_o)/t_{i+2,t}, \mathbf{a} = exp\mathbf{a}_o(exp\mathbf{a}_o(2 - exp\mathbf{a}_o)^{-1}$.(2.3a)

And the relative volumes hold increments

$\dot{V}_{i,i+1,i+2} = \delta V_{i,i+1,i+2}/\delta t, \delta V_{i,i+1,i+2} = V_c \delta t^3, \ \delta V_{i,i+1,i+2}/\delta t = V_c \delta t^2 = V_c t_{i+2,t}^2 \delta t^2 / t_{i+2,t}^2 = V_c t_{i+2,t}^2 \varepsilon(\gamma)^2$ .(2.3b)

Here $\varepsilon^2(\gamma)$ is space area' invariant at fixed $\gamma$, and $V_c = 2\pi c^3 / 3(k\pi)^2 tg\psi^o$ is invarint volume at angle $\psi^o = \pi/6$ on the vertex of each cone (Fig.3), $c$ is a speed of rotation of each cone's spiral which produces angle $\pi/2$.

We get differential complexity $M_m^{\delta} = M_{i,i+1,i+2}^{\delta}$ for any joint *m*-th triplet-node in invariant form:

$$M_{i,i+1,i+2}^{\delta} = 3(\mathbf{a}_o\mathbf{a} - \mathbf{a}_o^2 \exp 2\mathbf{a}_o)/V_c t_{i+2,t}^4 \varepsilon(\gamma)^2, \tag{2.4}$$

Applying $M_m^{\delta}$ to a cell (Fig.10) with volume $\delta V_{i,i+1,i+2} = \delta V_m$, which is formed during time interval $\delta t_{i,i+1,i+2} = \delta t_m$, we get $M_m^{\delta} \delta t_m = 3\dot{\alpha}_m \delta t_m / \delta V_m = 3\Delta\alpha_m / \delta V_m$, where $\Delta\alpha_m$ is increment of information speed during $\delta t_m$.

The related increment of quantity information at the same $\delta t_m$ is $\Delta\alpha_m \delta t_m = a_m^{\Delta} = \dot{\alpha}_m \delta t_m^2$, where

$$a_m^{\Delta} = (\mathbf{a}_o\mathbf{a} - \mathbf{a}_o^2 \exp(2\mathbf{a}_o))\delta t_m^2 / t_{i+2,t}^2, \delta t_m^2 / t_{m,t}^2 = \varepsilon_m^2, t_{m,t}^2 = t_{i+2,t}. \tag{2.5}$$

Each $3a_m^{\Delta}$ invariant measures of the quantity of information produced during interaction of three equal eigenvalues within area $\varepsilon_m^2$. Increment of entropy (in 2.1) determines the related volume and $M_m^{\delta}$, measured by equivalent quantity information, related to a cell volume $\delta V_m$, during the time $\delta t_m$:

$$M_m^{\delta} \delta t_m^2 = M_m^{\Delta} = 3(\mathbf{a}_o\mathbf{a} - \mathbf{a}_o^2 \exp(2\mathbf{a}_o))\varepsilon_m^2 / \delta V_m. \tag{2.5a}$$

Information $3a_m^{\Delta}$ binds the three segments in $\varepsilon_m^2$ prior the two impulse attractions assemble them.

By the moment of interactive assembling $\tau_k^{i+2}$, the three equal eigenvalues have signs $\alpha_{it}(\tau_k^{i+2})sign\alpha_{it}(\tau_k^{i+2}) = \alpha_{i+1t}(\tau_k^{i+2})sign\alpha_{i+1t}(\tau_k^{i+2}) = -\alpha_{i+2t}(\tau_k^{i+2})sign\alpha_{i+2t}(\tau_k^{i+2})$.

The negative eigenvalues $\alpha_{it}(\tau_k^{i+2})sign\alpha_{it}(\tau_k^{i+2}) = \alpha_{i+1t}(\tau_k^{i+2})sign\alpha_{i+1t}(\tau_k^{i+2})$ are stable, and positive eigenvalue $-\alpha_{i+2t}(\tau_k^{i+2})sign\alpha_{i+2t}(\tau_k^{i+2})$ is unstable.

Their interaction associates with a choatic attraction, which leads to local instability localized within zone $\varepsilon_m^2$.

The attraction, delivering information $2\mathbf{a}_o^2$, *cooperates* within $\varepsilon_m^2$ the three segments by joining them into a single node. Thus, (2.5a) measures c*ooperative complexity* of the *interactive* three segments, forming *a single node* of *m*-th triplet.

The cooperative node forms its cell within volume $\delta V_m$, where both eigenvalues' interaction and cooperation takes place. Since quantity information $2\mathbf{a}_o^2 \cong 1bit$ of joint segment from *m*-th triplet's node is transferred to a first segment of the following $m+1$-th triplet, the quantity of *binding* information $3a_m^\Delta$ (in (2.5)), being spent on holding the *m*-th triplet, is concentrated in the volume $\delta V_m$.

Let $M_{cm}^\Delta = 3a_m^\Delta(\gamma)/[\delta V_m / \varepsilon_m^2]$ evaluates the quantity of information per cell volume $\delta V_m$ related to a cell size area $\varepsilon_m^2$. Then using $M_{cm}^\delta = 3\Delta\alpha_m / \Delta V_m$, $M_{cm}^\delta \delta t_m = 3\Delta\alpha_m \delta t_m / \Delta V_m$, we have $M_{cm}^\Delta = 3a_m^\Delta / \Delta V_m$ which for each $\Delta V_m$ evaluates $M_{cm}^{\Delta V} = 3a_m^\Delta$. At $\gamma = 0.5, \mathbf{a}_o \cong -0.75, \mathbf{a} \cong 0.25$. We get $M_{cmN}^{\Delta V} = 3a_m^\Delta(\gamma = 0.5) \cong -0.897 Nat$ per cell, or $M_{cmb}^{\Delta V}(\gamma = 0.5) \cong -1.29 bit$ per cell-volume that each $m$-th node *conserves* during it formation.

This invariant, produced during the considered interaction (that primary binds these segments), measures a *cooperative* effect of the interactions holding the node's *inner cooperative complexity*.

This a relative cooperative complexity does not depend on the IN actual cell volume and the number of nodes that the cell enfolds, since the $M_{cm}^{\Delta V}$ invariant quantity is not transferred along the IN nodes' hierarchy.

Actually, in function $\delta V_m / \varepsilon_m^2 = V_c t_m^2$, at a fixed invariant $\varepsilon_m^2$ and volume $V_c$, increment $\delta V_m$ grows with assembling more nodes. Since that, complexity $M_{cm}^{\Delta V} = inv(\gamma)$ for any fixed cell's volume (according to 2.5)) decreases with assembling more cooperating nodes within this volume.

With growing the size of a cooperative nodes, the cooperative complexity per its volume decreases in the ratio $M_{m+1}^\Delta / M_m^\Delta = t_m^2 / t_{m+1}^2 = (\gamma_m^\alpha)^{-2}$ while each following $M_{m+1}^\Delta$ enfolds complexity of the previous $M_m^\Delta$.

Absolute value of interval $\delta t_m = t_m \varepsilon$ grows with increasing $t_{m+1}/t_m = \gamma_2^\alpha$, which leads to $\delta t_{m+1}/\delta t_m = \gamma_2^\alpha$ and $M_m^\Delta = 3a_m^\Delta / \delta t_m \delta V_m$, $\delta V_m = 3V_c \delta t_m^2$, $M_m^\Delta = a_m^\Delta / V_c \delta t_m^3 = a_m^\Delta / V_c \varepsilon_m^3 t_m^3$, while $M_m^\delta = a_m^\Delta / V_c \delta t_m^4 = a_m^\Delta / V_c \varepsilon_m^4 t_m^4$.

This confirms the previous relations.

The ratio of the nearest triplet's complexities (2.4) is
$$M_{m+1}^\delta / M_m^\delta = t_m^4 / t_{m+1}^4 \text{ at } (\mathbf{a}_o \mathbf{a} - \mathbf{a}_o^2 \exp(2\mathbf{a}_o))/V_c \varepsilon(\gamma)^2 = A_M(inv(\gamma)). \tag{2.5b}$$

At $t_m^4 / t_{m+1}^4 = (\alpha_{m+1}/\alpha_m)^4 = (\gamma_{m+1})^{-4}$, and satisfaction of (2.5b) with $\gamma_{m+1} = \gamma_2(\gamma) = inv_o(\gamma)$, we get
$$M_{m+1}^\delta / M_m^\delta = \gamma_{m+1}^{-4}, \tag{2.6}$$

which for $\gamma_2(\gamma = 0.5) = 3.89$ takes values $M_{m+1}^\delta / M_m^\delta \cong 0.00437$.

Complexity $M_{m+1}^\delta$, measuring $m+1$ node, also enfolds and condenses the complexity of a previous node.

By moment $\tau_m$ of the $m$-th triplet's cooperation, its three eigenvalues equalize: $\alpha_{3\tau}^m = \alpha_{2\tau}^m = \alpha_{1\tau}^m$, and, at the *moment of* triplet's formation $\tau_m + o$, the cooperative eigenvalues $\alpha_m$ encloses the joint triplet eigenvalues:
$$\alpha_3^m(\tau_m + o) = 3\alpha_{3\tau}^m = \alpha_m, \tag{2.7}$$

In the IN, $m$-th triplet's first eigenvalue $\alpha_{1\tau1}^m$ *equals* to last eigenvalue of $(m-1)$-th triplet $\alpha_{m-1}$: $\alpha_{1\tau1}^m = \alpha_{m-1}$, where $\alpha_{1\tau1}^m$ enfolds all three eigenvalues of previous $(m-1)$- triplets. The $m$-triplet holds ratio of these eigenvalues

$$\alpha_{3\tau}^m / \alpha_{1\tau 1}^m = (\gamma_m^\alpha)^{-1}. \tag{2.7a}$$

Substituting (2.7a) to (2.7) we have $\alpha_m = 3\alpha_{1\tau 1}^m (\gamma_m^\alpha)^{-1}$, and with $\alpha_{m-1}$ we get get ratio

$$\alpha_m / \alpha_{m-1}^m = 3(\gamma_m^\alpha)^{-1}. \tag{2.7b}$$

The sustained cooperation of the IN eigenvalues requires $\gamma_m^\alpha (\gamma = 0.5) \cong 3.9$, which brings ratio (2.7b) to $\alpha_m / \alpha_{m-1}^m \cong (1.3)^{-1}$. Decreasing the eigenvalues of the triplets, cooperating along the IN, encloses the increased information density, which condenses more $MC_{ik}^\delta$ complexity.

Specifically, at $M_{m+1}^\delta / M_m^\delta = (\alpha_{m+1}^4 / \alpha_m^4) / \dot{V}_{m+1} / \dot{V}_m$ and $\dot{V}_{m+1} / \dot{V}_m = \alpha_m^2 / \alpha_{m+1}^2$, $\alpha_{m+1} / \alpha_m = (1/3\gamma_{m+1})^{-1}$, ratio $M_{m+1}^\delta / M_m^\delta = (\alpha_{m+1}^4 / \alpha_m^4) / \dot{V}_{m+1} / \dot{V}_m = (\alpha_{m+1}^6 / \alpha_m^6) = (1/3\gamma_{m+1})^{-6}$ brings decreasing $M_{m+1}^\delta / M_m^\delta \cong 0.203$.

Comparing (2.6) to the differrence of relative complexities:

$$\Delta M_m^\delta / M_m^\delta = (M_m^\delta - M_{m+1}^\delta) / M_m^\delta = (1 - \gamma_2^4),$$

we get $\Delta M_m^\delta / M_m^\delta \cong |0.996|$ at $\gamma_2(\gamma = 0.5) = 3.89$, indicating that the difference decreases insignificantly.

Relative sum of these complexities:

$$\Delta M_{m\Sigma}^\delta / M_m^\delta = (M_m^\delta + M_{m+1}^\delta) / M_m^\delta = (1 + \gamma_2^4), \Delta M_{m\Sigma}^\delta / M_m^\delta \cong 1.0044 \text{ also grows insignificantly.}$$

Comparing these complexities to the complexities of double cooperation within a triplet, we have

$M_{12}^\delta / M_1^\delta = (\alpha_{12} \delta t_{12} / \delta V_{12}) / (\alpha_1 \delta t_1 / \delta V_1) \cong 2(\alpha_2 / \delta V_{12}) / (\alpha_1 / \delta V_1)$, which at $\delta t_{12} \cong \delta t_1$,

$\alpha_2 / \alpha_1 = (\gamma_2^\alpha)^{-1}, \delta V_{12} / \delta V_1 = (\gamma_2^\alpha)^{-3}$, leads to $M_{12}^\delta / M_1^\delta \cong 2(\gamma_2^\alpha)^{-4}$. For a triplet we have

$$M_{123}^\delta / M_1^\delta \cong 3(\gamma_3^\alpha)^{-4}, \tag{2.8}$$

which at $\gamma_1^\alpha = 2.215$, $\gamma_2(\gamma = 0.5) = 3.89$ brings $M_{12}^\delta / M_1^\delta \cong 0.083$, $M_{123}^\delta / M_1^\delta \cong 0.013$. (2.8a)

During a triple cooperation, the complexity decreases more than that in douple cooperation within a triplet.

The $MC_{ik}^\delta$ (2.6) between the nearest triplets decreases much faster then that occurs in the cooperation within a triplet.

At the cooperative cooperation, each following nodes' complexity wraps and absorbs complexity of previous node, binding these node units and conserving the bound information.

The decrease of the IN cooperative complexity indicates that more cooperations have occurred, while, at negative eigenvalues jump, the complexity grows with decoupling nodes and rasing the choitic movement.

The MC for each extremal segment and their ratios determine the following relations:

$M_i^d = \alpha_{it} / \Delta V_{it}, M_{i+1}^d = \alpha_{i+1,t} / \Delta V_{i+1,t}, M_{i+2}^d = \alpha_{i+2,t} / \Delta V_{i+2,t}$, $M_{i+1} / M_i = (\alpha_{i+1,t} / \alpha_{it}) / (\Delta V_{i+1,t} / \Delta V_{it})$,

$M_{i+2} / M_i = (\alpha_{i+2,t} / \alpha_{it}) / (\Delta V_{i+2,t} / \Delta V_{it})$ at $(\alpha_{i+1,t} / \alpha_{it}) = \gamma_1^{-1}, (\alpha_{i+2,t} / \alpha_{it}) = \gamma_2^{-1}, (\Delta V_{i+1,t} / \Delta V_{it}) = (1 - \gamma_1^3)$,

$(\Delta V_{i+2,t} / \Delta V_{it}) = (1 - \gamma_2^3)$, we get $M_{i+1}^d / M_i^d = \gamma_1^{-1}(1 - \gamma_1^3)^{-1}$ and $M_{i+1}^d / M_i^d = \gamma_1^{-1}(1 - \gamma_1^3)^{-1}$. For $m$ th triple we get:

$$(M_i^d + M_{i+1}^d + M_{i+2}^d) / M_i^d = \Delta M_{m\Sigma}^d / M_m^d = 1 + \gamma_1^{-1}(1 - \gamma_1^3)^{-1} + \gamma_2^{-1}(1 - \gamma_2^3)^{-1}, \tag{2.8b}$$

and $\Delta M_{m\Sigma}^d / M_m^d \cong 1.05$ at $\gamma_2(\gamma = 0.5) = 3.89, \gamma_1(\gamma = 0.5) = 2.215$.

Invariants relaltions (2.3a) bring invariant forms for each $i, i+1, i+2$ triple of these complexities:

$$M_i^d = \mathbf{a} / t_i \Delta V_{it}, M_{i+1}^d = \mathbf{a} / t_{i+1} \Delta V_{i+1,t}, M_{i+2}^d = \mathbf{a} / t_{i+2} \Delta V_{i+2,t}, \tag{2.9}$$

where each of the volume encloses its invariant information.

Comparing two ratios of cooperative complexities in (2.8a) with the related sum of dynamic complexities(2.8b):

$$1 + \gamma_1^{-1}(1-\gamma_1^3)^{-1} + \gamma_2^{-1}(1-\gamma_2^3)^{-1} \gg 3(\gamma_2)^{-4}, \tag{2.9a}$$

for which, at $\gamma_2(\gamma = 0.5) = 3.89, \gamma_1(\gamma = 0.5) = 2.215$, we obtain inequality $1.05 > 0.013$.

The results indicate the essential differerence of both types of complexities, where $M_i^\delta$ measures the unit's information intensity, is determined by quantity of information intendent to spend on cooperation with other units.

When the cooperation occurs, the intensity is deminished, being compensated by information that binds these units and concerves bound information.

A collective unit holds a less information intensity than it was prior to cooperation, measured by a summary of each of unit complexities.

With more units in the collective, each complexity of attached unit $M_{i,i+1,i+2}^\delta, i = 1,....,m$ tends to decrease.

The growing cooperatives intend to spend less information for attracting other units, accepting assembled units with decreasing the information speeds under the minimax.

A total (integral) *relative MC*-complexity for entire IN with $m$ triplets approximates sum of (2.8b): $MC_m^\Sigma \cong m$, which grows with adding each new triplet. The IN integral relative differential *cooperative* complexity

$$MC_m^{\delta\Sigma} \cong \sum_1^m [3(\gamma_2^{-4})]^m = (1-[3(\gamma_2^{-4})])^m / [1-[3(\gamma_2^{-4})]] \tag{2.10}$$

decreases with adding each new triplet, and at $m \to \infty, \gamma_2(\gamma = 0.5) = 3.89, \gamma_1(\gamma = 0.5) = 2.215$ it holds $MC_m^{\delta\Sigma} \cong 1.013$.

As total $MC_m^{\delta\Sigma}$ grows, the complexity of each following cooperation diminishes the contribution to the IN cooperative complexity, and with growing number of the IN inits, the sum of the contribution approaches zero.

The $MC_m^\Sigma$, defined for a non-cooperating triplet's segment in $m$, is higher than the IN's cooperative complexy $MC_m^{\delta\Sigma}$ as the triplet's number grows.

The evolution dynamics with adaptive self-controls keep the $MC_m^{\delta\Sigma}$ decreasing at each evolving IN in the observation.

Therefore, the MC complexity decreases in the observation with reduction of the uncertainty of randomness.

This complexity emerges from the natural interactive process being observed.

The logic of the observing process encloses the entropy functional integral (EF).

The IN complexity decreases during formation of the nested structure where each following knot-node enfolds complexities of the previous formed node, and the ending IN node encloses and integrates complexities all IN.

### 6.3. The cooperative information mass and information space curvature

Information speed of triplet's eigenvectors $\alpha_m = 3\alpha_{i+2}$ cooperating in ajoint volume increment $v_m$, holds form $M_{vm} = \alpha_m v_m$, which we call *information cooperative mass* per this volume.

The triplet's eigenvalue Hamiltotian $\alpha_m = H_m$ and differential volume $v_m = \delta V_m / \delta t = \dot{V}_m$ define information mass of a diferential volume

$$M_{vm} = H_m \dot{V}_m. \tag{3.1}$$

The connection of entropy derivation $\partial \Delta S_m / \partial t = -H_m$ with entropy's divergence $\partial \Delta S_m / \partial t = c_m div \Delta S_m$ for the same volume $v_m$, linear speed $c_m$, at cooperation of $m$-the triplet, measures the information mass (3.1) in form

$$M_{vm} = -(c_m div \Delta S_m)\dot{V}_m. \tag{3.1a}$$

The ratio of the information masses for nearest triplets:

$$M_{vm}/M_{vm+1} = \alpha_m/\alpha_{m+1}(v_m/v_{m+1}), \quad v_m/v_{m+1} = \alpha_{m+1}^2/\alpha_m^2 = (1/3\gamma_{m+1}^\alpha)^2, \tag{3.1b}$$

$$M_{vm}/M_{vm+1} = 1/3\gamma^{\alpha}_{m+1}, \gamma^{\alpha}_{m+1} \cong 3.9 \qquad (3.1c)$$

grows in 1.3 times with adding each following IN's triplet.

The triplet cooperative complexity (2.3) for the same cooperative volume measure differential Hamiltonian:

$$M^{\delta}_m = 3\dot{H}_m/\dot{V}_m, \dot{H}_m/\dot{V}_m = \dot{H}^V_m, \qquad (3.2)$$

where the derivation applies on the EF trajectory' segment between emerging new information where $\dot{\alpha}_m$ is finite.

Multiplication information mass $M_{vm}$ on the complexity (3.2) leads to

$$M_{vm}M^{\delta}_m = 3H_m\dot{H}_m \text{, which for } H_m = \alpha_m, \dot{H}_m = \dot{\alpha}_m, \text{ brings } M^{\delta}_m M_{vm} = 3\alpha_m\dot{\alpha}_m. \qquad (3.2a)$$

Information curvature $K^{\alpha}_m$ defines the increment of information speed on an instant of geodesic line $ds$ in Riemann space, where the increment is relative to the information speed on this instant.

This curvature orginates in the curved impulse interactive cooperation.

Information curvature $K^m_{\alpha}$ at cooperation of three triplet's eigenvectors describes a curving phase space at locality within the cooperative volume $v_m$. This curvature connects to classical Gaussian curvature in a Riemann space [102], defined via fundamental metric tensor $\sqrt{g}$ and the phase space metric $ds = v_m dt$:

$$K^{\alpha}_m = (\sqrt{g})^{-1}d(\sqrt{g})/ds = (\sqrt{g})^{-1}d(\sqrt{g})/v_m dt, \qquad (3.3)$$

where $g$ describes a closeness of the space vectors in the cooperative curving interactions.

For the eigenvectors in information phase space, metrical tensor $\sqrt{g}$ is expressed [43] through information of triplet eigenvectors $\alpha_m$ localized in a space, which generates an increment of tensor $\sqrt{g}$ for the triple cooperation:

$$\sqrt{g} = (\alpha_m)^{-3}. \qquad (3.3a)$$

Substitution to (3.3) determines the triplet curvature

$$K^m_{\alpha} = -3\dot{\alpha}_m/\alpha_m v_m, \qquad (3.4)$$

which connects relations (3. 2) with the curvature:

$$K^m_{\alpha} = -M^{\delta}_m \dot{H}_m. \qquad (3.4a)$$

According to [102], multiplication of physical mass on $\sqrt{g}$ determines the mass density.

Multpuling information mass $M_{vm}$ on $\sqrt{g}$, expressed through the eigenvector (3.3a), leads to mass *density* $M^*_{vm}$:

$$M^*_{vm} = (\alpha_m)^{-3}\alpha_m v_m = (\alpha_m)^{-2}v_m. \qquad (3.5)$$

In simulated IN hierarchy (Fig.7), the cooperating eigenvalues $\alpha_m$ decrease with growing number of triplets $m \to n/2$, which increases $M^*_{vm}(m)$. That mass estimates information speed of the creating-encoding information units, whose energy $e_{ev}$ covers entropy cost of conversion to information $s_{ev} = i_{ev}$.

Expressing (3.4), at $MC_m = \alpha_m/v_m$, in form $K^m_{\alpha} = -3\dot{\alpha}_m\alpha_m v_m v_m/v_m\alpha_m\alpha_m v_m v_m$ leads to

$$K^m_{\alpha} = -M^*_{vm}, M^{\delta}_m MC_m = -M^*_{vm} = MC^{\delta e}_m = M^{\delta}_m MC_m, MC^{\delta e}_m = \dot{H}^V_m MC_m \qquad (3.6)$$

where $MC^{\delta e}_m$ is a triplet effective complexity and $\dot{H}^V_m$ is density of information Hamiltonian per volume.

The information curvature is a result of the joint cooperation and memorizing information mass- as the cooperative information mass, which generates the effective complexity.

The cooperation decreases uncertainty and increases information mass at forming each triplet.

The cooperation, accompanied by the decreases of the triplet's eigenvalues and complexity, declines the curvalure of the IN cooperating structure.

The negative curvature (3.5) characterizes a topology of the space area where the cooperation occurs.

*The information mass emerges as a curved information space per cooperative information complexity.*

Multiplication information mass density of the curvature: $M*_{vm} K_\alpha^m = -3\dot{H}_m H_m^{-1} = -3d^* H_m^*$ determines relative decrease of increment of energy $d^* H_m^*$ which the forming this mass requires.

Simple estimation the curvature by inverse radious $r_m$ of cooperating triplet node:
$$|K_{\alpha E}^m| = (r_m)^{-1}, r_m = \varepsilon(\gamma) = [(\gamma_1^\alpha / \gamma_2^\alpha)^2 - (\gamma_2^\alpha)^{-1}]^{1/2} \qquad (3.7)$$
connects the estimated curvature with the triplet invariants, and at $\varepsilon_m(\gamma^*) \cong 0.33$, it estimates the growing curvature of the IN node knot for each cooperating triplet.

To evaluate maximal speed $c_{mo}$ in elementary single cooperation, measured by information invariant **a**, we use relation $\partial \Delta S_m / \partial t = |\mathbf{a}|/t_m = c_m div \Delta S_m$ which leads to $c_{mo} = [t_{mo} div \Delta S_m / |\mathbf{a}|]^{-1}$,
where $t_{mo}$ estimates minimal admissible time interval $t_{mo} \cong 1.33 \times 10^{-15}$ sec of light wavelenght $l_{mo} = 4 \times 10^{-7} m$.

Structural invariant of minimal uncertainty estimates normalized divirgence: $div^* \Delta S_m = div \Delta S_m / |\mathbf{a}| \approx 1/137$.

Resulting maximal information speed up to $c_{mo} \approx 1.03 \times 10^{17}$ Nat/sec restricts the cooperative speed and minimal information curvature at other equal conditions. Information mass (3.1a):
$$M_{vm} = -|\mathbf{a}|(c_m div \Delta S_m / |\mathbf{a}|) v_m = -|\mathbf{a}| v_m h_\alpha^o \qquad (3.7a)$$
encloses the impulse information **a**, volume $v_m$, and triplet structural invariants $h_\alpha^o \cong 1/137$.

The elementary cooperation binds *space* information $div^* \Delta S_i$ that limits maximal speed of incoming $i$-information units, imposing *information* connection on the time and space.

Unbound information code's unit has not such limtation.

Ratio of speed $c_{mo}$ to speed of light $c_o$:
$$c_{mo} / c_o \approx 0.343 \times 10^9 \, Nat/m = 0.343 gigaNat/m \qquad (3.8)$$
(measured in a light's wavelength meters) limits a maximal information space speed.

In this case, each wavelength of speed of ligth delivers $\cong 137$ Nats during $t_{mo} \cong 1.33 \times 10^{-15}$ sec.

The material mass-energy that satisfies the law of preservation energy (following the known Einstein equation), distinguishes from the information mass (3.1), (3.5), defined for cooperating triplet's units.

The law satisfies when the triplet acquires energy - as the mass-energy while carying the information mass.

This triplet tinformation curvature and the effective cooperative complexity also hold physical energy.

### 6.4. The relative information observer

Ratio of the impulse space $h_k$ and time $o_k$ units:
$$h_k / o_k = c_k$$
defines the impulse linear speed $c_k$.

For the invariant impulse measure $|1|_M = [l_k \times \tau_k]_M$, this speed determines ratio
$$c_k = |1|_M / (o_k)^2.$$

More Bits concentrating in impulse time unit leads to $o_k \to 0$ and to $c_k \to \infty$, which is limited by the speed of light.

The persisting increase of information density grows the linear speed of the natural encoding, which associates with rising of the impulse curvature [91] at the observing interactions.

The curvature encloses the information density and enfolds the related information mass.

The information observer progressively increases both its linear speed and the speed of natural encoding combined with growing curvature of its information geometry.

The IPF integrates this density in observer' geometrical structure (Fig.10) whose rotating speed grows with increasing the linear speed, increasing the impulse space intervals and shortening time intervals of invariant impulse.

The growing density increases number of the enclosed events in both time and space intervals.

Comparing any current informational observer with speed $c_o$ relatively to the observer with maximal at $c_k > c_o$ leads to a wider impulse' time interval of observer' $c_o$ for getting the invariant information, related to that for observer $c_k$.

The IPF integrates less total information for observer $c_o$, if both of them start the movement instantaneously.

Assuming each observer total time movement, memorizing the natural encoding information, determines its life span indicates that for observer $c_o$ is less than for the observer $c_k$ which naturally encodes more information and its density.

At $c_o/c_k \to 1$ both observers approach the maximal encoding.

Hence, the relative moving information observer possesses both relative encoding and attractive curved gravitation emerging in the microprocess in addition to relative time and space.

The discussed approach adds information *versions* to Einstein's theory of relativity applied to moving information observer.

### 6.5. The Observer's IN self-replication and conditions of self-generation the observer new information quality

Each IN node' maximal admissible $m_M$-th level ends with a single dimensional process, which in attempt to attach new triplet over constrained $\gamma_{k1} \to 1$, loses the ability to enfold new attracting information.

Such IN stops satisfying the minimax information law, which leads to its instability.

That violation is *natural intention* growing with each additional IN, but is constrained by limitations (Secs.6.6,6.7).
Specifically, after completion of the IN cooperation, the last IN ending node initiates one dimensional process

$$x_n(t_n) = x_n(t_{n-1})(2 - \exp(\alpha^t_{n-1} t_{n-1})), \qquad (4.1)$$

which at $t_{n-1} = \ln 2 / \alpha^t_{n-1}$ approaches final state $x_n(t_n) = x_n(T) = 0$ with a potential infinite relative phase speed

$$\dot{x}_n / x_n(t_n) = \alpha^t_n = -\alpha^t_{n-1} \exp(\alpha^t_{n-1} t_n)(2 - \exp(\alpha^t_{n-1} t_n))^{-1} \to \infty. \qquad (4.2)$$

Since this node' process cannot reach zero final state $x_n(t_n) = 0$ with $\dot{x}_n(t_n) = 0$, a periodical process arises as result of alternating the macro movements with the opposite values of each *two* relative phase speeds:

$$\dot{x}_{n+k-1} / x_{n+k-1}(t_{n+k-1}) = \alpha^t_{n+k-1}, \dot{x}_{n+k} / x_{n+k}(t_{n+k}) = -\alpha^t_{n+k}. \qquad (4.3)$$

That leads to instable speeds' fluctuations at each $t = (t_{n+k-1}, t_{n+k})$ starting the alternations with eigenvalue

$$\dot{x}_n / x_n(t_n) = \alpha^t_n, \ k = 1, 2, .. \text{ at } \gamma \geq 1. \qquad (4.3a)$$

The instable fluctuations in three-dimensional process involves oscillation of interacting three ending nodes from other IN's approaching $\gamma \geq 1$. That generate frequency spectrum of the information model' eigenvalues $\lambda^*_i(t_{n+k})$ in each it space dimension $i = 1, 2, 3$.

Formal analysis of this instability associates with nonlinear fluctuations [103], which describes a superposition of linear fluctuations with frequency spectrum ($f_1, ..., f_m$) proportional to imaginary components of the spectrum eigenvalues ($\beta^*_1, ..., \beta^*_m$), where $f_1 = f_{\min}$ and $f_m = f_{\max}$ are the minimal and maximal frequencies of the spectrum accordingly.

In the model, the oscillations under the interactive actions generate imaginary eigenvalues $\beta^*_i(t)$:

$$\text{Im}\,\lambda^i_{n+k}(t) = \lambda^i_{n+k-1}[2 - \exp(\lambda^i_{n+k-1} t)]^{-1} \qquad (4.4)$$

at each $t = (t_{n+k-1}, t_{n+k})$ for these $i$-dimensions. This leads to relation

$$\operatorname{Im}\lambda_n^i(t_{n+k}) = j\beta_n^i(t_{n+k}) = -j\beta_{n+k-1}^i \frac{\cos(\beta_{n+k-1}^i t) - j\sin(\beta_{n+k-1}^i t)}{2 - \cos(\beta_{n+k-1}^i t) + j\sin(\beta_{n+k-1}^i t)},  \quad (4.5)$$

at $\beta_i^* = \beta_{n+k}^i, \beta_i^* \neq 0 \pm \pi k$, where $\beta_n^i(t_{n+k})$ includes real component:

$$\alpha_n^i(t_{n+k}) = -\beta_{n+k-1}^i \frac{2\sin(\beta_{n+k-1}^i t)}{(2 - \cos(\beta_{n+k-1}^i t))^2 + \sin^2(\beta_{n+k-1}^i t)}, \quad (4.5a)$$

which, at $\alpha_i^* = \alpha_i^*(t_{n+k}) \neq 0$, determines the relative parameter of dynamics

$$\gamma_i^* = \frac{\beta_n^i(t_{n+k})}{\alpha_n^i(t_{n+k})} = \frac{2\cos(\beta_{n+k-1}^i t) - 1}{2\sin(\beta_{n+k-1}^i t)}. \quad (4.6)$$

At $\gamma_i^* = 1$, it brings $(\beta_{n+k-1}^i t) \approx 0.423 rad(24.267^o) = 0.134645\pi$.

The fluctuations may couple the nearest dimensions by an interactive double cooperation overcoming *minimal elementary* uncertainty UR separating the model's dimensions' measure via invariant $h_\alpha^o$, which may border the IN maximal stable level $m_M$.

Suppose, the $k$-th interaction, needed for creation a single element in the double cooperation, conceals information $s_c(\gamma) = \mathbf{a}_{ok}^2(\gamma)$, which should compensate for increment of minimal uncertainty of invariant $h_\alpha^o \mathbf{a}_o(\gamma = 0) = 0.76805/137 = 0.0056$ Nat by information equals $\delta\mathbf{a}_k = \mathbf{a}_o(\gamma = 0) - \mathbf{a}_o(\gamma^*)$.

This invariant evaluates the UR information *border* through the increment of the segment's entropy concentrated in UR, while equality $h_\alpha^o \mathbf{a}_o(\gamma = 0) = \mathbf{a}_{ok}^2(\gamma)$ evaluates minimal interactive increment

$$\delta\mathbf{a}_k = \mathbf{a}_{ok}^2(\gamma) = 0.0056 \quad (4.6a)$$

with minimal information

$$\mathbf{a}_{ok}(\gamma) = 0.074833148, \quad (4.6b)$$

needed for a single interaction in each dimension.

Each $k$-th interaction changes initial $\gamma \geq 1$ on $-\Delta\gamma = -\gamma^*$ bringing minimal information increment in each dimension (4.6b). The minimal information attraction, enables cooperate a couple, needs three these increments, generating information

$$3\mathbf{a}_{ok}(\Delta\gamma) = 0.2244499443 \cong 0.23 = \mathbf{a}_k. \quad (4.6c)$$

Information attraction $\mathbf{a}_k$, generated in each dimensional interaction, can cooperate that interactive information in invariant $\mathbf{a}_{ok}(\gamma) \cong 0.7$ which binds three dimensions in single bit through total nine interactions.

The question is: how the interactive fluctuations enable creating a triplet which self-replicates new IN?
Dynamic invariant $\mathbf{a}(\gamma) = \mathbf{a}_k$ of information attraction determines ratios of starting information speeds $\gamma_1^\alpha = \alpha_{io}/\alpha_{i+1o}$ and $\gamma_2^\alpha = \alpha_{i+1o}/\alpha_{i+2o}$ needed to satisfy invariant relations (4.5.5.2).

To create new triplet's IN with ratio $\gamma_1^\alpha = \alpha_{io}/\alpha_{i+1o}$, relation (4.6) requires such ratio $\frac{\beta_i^*(t_{n+k})}{\beta_{n-1,o}(t_{n-1,o})} = l_{n-1}^m$ which deliver imaginary invariant $(\beta_{n+k-1}^i t) \to (\pi/3 \pm \pi k), k = 1, 2, ...$ at each $k$ with ending information frequency $\beta_{lo}(t_o) = \beta_i^*(t_{n+k})$ that would generate needed $\alpha_n^i(t_{n+k}) = \alpha_{lo}^m(t_o)$.

In case $\Delta\gamma \to 0 \to \gamma^*$, it can be achieved in (4.6) at $2\cos(\beta_{n+k-1}^i t) \to 1$, or at

$$(\beta_{n+k-1}^i t) \to (\pi/3 \pm \pi k), k = 1, 2, ..., \text{ with } \alpha_n^i(t_{n+k}) = \alpha_{lo}^m(t_o) = \lambda_{lo}^m \cong -0.577\beta_{n+k-1}^i. \quad (4.7)$$

That determines maximal *ratio* of *frequencies:*

$$l_{n-1}^m = \beta_{n+k-1}^i / \beta_{n-1,k=o}^i \tag{4.7a}$$

which at $\gamma = 1$, $\beta_{n-1,o}(t_{n-1o}) = \alpha_{n-1,o}(t_{n-1,o})$ and $\beta_{n+k-1}^i = \alpha_{l,k=3}^i / 0.577$ holds

$$l_{n-1}^m = \alpha_{l,k=3}^i / 0.577 / \alpha_{n-1,k=o}^i, \quad \alpha_{l,k=3}^i / \alpha_{n-1k=o}^i = \gamma_1^\alpha. \tag{4.7b}$$

Ratios (4.7a) and (4.7b) identifies the triplet invariant

$$l_{n-1}^m = \gamma_1^\alpha / 0.577, \tag{4.8}$$

generated by the initial $(n-1)$-dimensional spectrum with an imaginary eigenvalue $\beta_{n-1,o}(t_{n-1,o})$ by the end of the interactive movement. Invariants (4.7a),(4.8) lead to $\gamma_1^\alpha = \alpha_{l,k=3}^i / \alpha_{n-1,k=o}^i = 3.89$ and to ratio of initial frequencies $l_{n-1}^{m=1} \cong 6.74$. Next nearest $\alpha_{l+1,k=6}^i / \alpha_{lo,k=3}^i = \gamma_2^\alpha$ needs increasing first ratio in $l_{n-1}^{m=2} \cong 3.8$, and the following $\alpha_{l+2,k=9}^i / \alpha_{l+1,k=6}^i = \gamma_2^\alpha$ needs $l_{n-1}^{m=3} \cong 3.8$. The multiplied ratios

$$l_{n-1}^{m=1-3} = l_{n-1}^{m=1} \times l_{n-1}^{m=2} \times l_{n-1}^{m=3} \cong 97.3256 \tag{4.8a}$$

is needed to build new triplet, which at $\gamma = 1$ is not ending with a stable segment. While $l_{n-1}^{m=1-3}$ identifies maximal ratio of spectrum frequencies generated by the instable fluctuations.

Each three information $3\mathbf{a}_{ok}(\Delta\gamma)$ binds pair of nearest spectrum frequencies, starting with pair $\beta_{n-1,o}^i, \beta_{n,k=1}^i$, which sequentially grows with each $k$ interactive information (4.6a-c), intensifying the increase of frequency. First three pairs bind three dimensions in single Bit $\mathbf{a}_{ok}(\gamma) \cong 0.7$ through nine interactions of frequencies requiring ratio $l_{n-1}^{m=1} \cong 6.74$; next pair ratio grows in ~2.24 times, each next in 1.26 times.

Thus, a natural source to produce the very first triplet is nonlinear fluctuation of an initial dynamics, involving, as minimum, three such primary dynamics dimensions that enclose some memorized information by analogy with ending IN node.

That at $\mathbf{a}_o(\gamma = 1) = 0.58767$, $\mathbf{a}(\gamma = 1) = 0.29$ brings $\gamma_1^\alpha \cong 2.95$ and needs ratio $l_{n-1}^{m=1} \cong 5.1126$.

Natural source of maximal speed frequency is light wavelength whose time interval $t_{lo} \approx 1.33 \times 10^{-15}$ sec determines maximal frequency $f_{\max} \cong 0.7518 \times 10^{15}$ sec$^{-1}$ that at each interaction brings information

$$h_\alpha^o \mathbf{a}_o(\gamma = 0) = 0.005 \, \text{Nat} \tag{4.8b}$$

changing the initial frequency.

The required frequency ratio $l_{n-1}^{m=1-3}$ identifies minimal frequency $f_{\min} = 0.7724586 \times 10^{13}$ sec$^{-1}$.

The triplet information invariant allows finding the equivalent energy invariant for creating such triplet.

Invariants $h_\alpha^o \cong 1/137$ coincides with the Fine Structural constant in Physics:

$$\alpha^o = 2\pi \frac{e^2}{4\pi\varepsilon^o hc}, \tag{4.9}$$

where $e$ is the electron charge magnitude's constant, $\varepsilon^o$ is the permittivity of free space constant, $c$ is the speed of light, $h$ is the Plank constant.

The equality $h^o \cong \alpha^o$ between the model's and physical constants allows evaluate the model's structural parameter energy through energy of the Plank constant and other constants in (4.9):

$$h = \frac{e^2}{2\varepsilon^o ch^o} = C_h \alpha_h, \alpha_h = (h^o)^{-1} = inv, \quad \frac{e^2}{2\varepsilon^o c} = C_h = h/(h^o)^{-1} = 9.0831 \times 10^{32} J \cdot \text{sec} \tag{4.9a}$$

where $C_h$ is the energy's constant (in [J.s]), which transforms invariant $\alpha_h$ to $h$.

In this information approach, (4.9a) evaluates energy that conceals the IN bordered the stable level $m_M$. The triplet creation needs nine such interacting increments, which evaluate the triplet energy's equivalent

$$e_{tr} = 8.1748 \times 10^{-29} J \cdot \text{sec}. \tag{4.9b}$$

Invariant conditions (4.6), (4.9a) enable modeling *cyclic* renovation [104], initiated by the two mutual attractive processes, which do not consolidate by the moment of starting the interactive fluctuation.

After the model disintegration, the process can renew itself with the state integration and transformation of the imaginary intormation to the real information during the dissipative fluctuations which bring energy for a triplet.

Initial interactive process may belong to different IN macromodels (as "parents") generating new IN macrosystem (as a "daughter") at end of the "parents" process and beginning of the "daughters".

The macrosystem, which is able to continue its life process by renewing the cycle, has to transfer its coding life program into the new generated macrosystems and provide their secured mutual functioning.

A direct source is a joint information of three different IN nodes $\mathbf{a}_{o1}(\gamma_1), \mathbf{a}_{o2}(\gamma_2), \mathbf{a}_{o3}(\gamma_3)$ enable initiate attracting information with three information speeds where one has opposite sign of the two, whereas the information values cooperating an initial triplet will satisfy the above invariant relations.

Creating a triplet with specific parameters depends on the starting conditions initiating the needed attracting information. To achieve information balance, satisfying the VP and the invariants, each elementary $\mathbf{a}_{oi}(\gamma_i)$ searches for partners for the needed consumption of information.

A double cooperation conceals information $s_c(\gamma) = \mathbf{a}_o^2(\gamma)$, while a triple cooperation conceals information $s_{cm}(\gamma) = 2\mathbf{a}_o^2(\gamma)$. It could produce a less free information than needed for cooperation, while each of both above values depend on $\gamma$.

With more triplets, cooperating IN, the cooperative information grows, spending free information on joining each following triplet.

Minimal relative invariant $h_\alpha^o = 0.00729927 \cong 1/137$ evaluates a maximal *increment* of the model's dimensions $m_M \cong 14$, and the quantity of the hidden invariant information (4.8b) that produces an elementary triple code, enclosed into the cellular geometry of hyperbolic structure Fig.10.

This hidden *a non-removable uncertainly also enfolds a potential DSS information code.*

Results (4.6-4.9a) impose important *restrictions* on both maximal frequency generating new starting IN and maximal IN dimension which limits a single IN. That IN's ending node may initiate this frequency.

For example, $\gamma = 1$ corresponds to
$(\beta_{n+k-1}^i t) \approx 0.423 rad (24.267^o)$, with $\beta_i^*(t_{n+k}) \cong -0.6\beta_{n+k-1}^i$. (4.10)

Here $\beta_i^*(t_{n+k}) \cong \alpha_{lo}^m(t_o)$ determines maximal frequency $\omega_m^*$ of fluctuation by the end of optimal movement: $\alpha_{lo}^m(t_o) \cong -0.6\beta_{n+k-1}^i$, where $\alpha_{lo}^m(t_o) = \alpha_{1o}(t_o)(\gamma_{12}^\alpha)^{N_m}$.

Starting speed $\alpha_{1o}(t_o) = 0.002 \sec^{-1}$, at $m = 14$, determines $\alpha_{14o}(t_o) = 0.00414 \sec^{-1}$, $\beta_{14} \cong 0.0069 \sec^{-1}$. (4.10a)

The new macro-movement starts with that initial frequency. This newborn macromodel might continue the consolidation process of its eigenvalues, satisfying the considering restrictions on invariants and cooperative dynamics up to the ending consolidations and arising the periodical movements.

This leads to cyclic *micro-macro functioning* when the state integration alternates with state disintegration and the system decays with possible transformation of observable virtual process to the evolving information-certain process [105].

## 7. The Observer Information Cognition and Intelligence

### 7.1. Emerging the observer's cognition and intelligence

The observer logic emerges on the path from the collected observing interactive probabilistic logic, the curving impulse' interactive certain logic, and the IN nested information logics.

The ending triplet logic in every network enfolds all the IN logic with the *attracting triplet's* logic.

The multiple INs rotating speeds of the discrete nodes' attractive logic generate frequencies.

The attracting logic, equalizing the speeds, synchronizes the frequencies of the nodes' logic in resonance, which assemble a logical loop at each observer hierarchical level.

The starting logical loop at lower level involves in logical connection with other loops of the observer hierarchy.

The multiple logical loops form logical chain which encloses all hierarchy of the cooperating local loops with their hierarchical frequencies up to the observer highest level. This hierarchical logic consists of the mutual attracting free logic which self-organizes the cooperative logical rotating loops in a chain enclosing all observing logic.

Observer cognition

The observing time–space *logical* structure conserves the cooperative logical chains which we call the observer *cognitive logic*, or cognition. Such logical structure possesses both virtual probabilistic causality and real information causality and complexity, which measure the cognitive intentional actions.

The observer's cognition assembles the common units through the multiple resonances at forming the IN-triplet hierarchy, which accept only units that each IN node recognizes.

The cognitive rotating movement, at forming each node and level, processes a *temporary loop* (Fig.6) which might disappear after the new formed IN triplet is memorized.

The observer cognition emerges from the evolution process, as evolving intentional ability of requesting, integrating, and predicting the observer needed information that builds the observer growing networks.

The local logical resonances with self-equalizing the free logic perform the *cognitive functions,* which are distributed along hierarchy of assembling logical units: triplets, IN nested nodes, and the IN ending nodes.

These local functions self-organize the observer cognition up to the cognitive function at the upper level level synchronizing frequency. Along the IN hierarchy runs the distributed resonance frequencies spreading a chain of the loops.

The chain rotates the thermodynamic process (cognitive thermodynamics) with minimal energy which holds the chain resonance logic. This term is proposed my UCLA colleague M.Dean.

Arising the observer intelligence

Since each assembling logical unit possesses the free logic, its topological opposite curved interaction (Sec. 2.6) with an external impulse brings the asymmetry of interaction. The interactive Yes action can open access the external Landauer's energy starting erasure entropy and memorizing the information up to encoding a bit with its free information whose No action stops delivering this energy. Such bit encodes the assembling logical unit with its free information.

The multiple local bits, encoding both self-organized cognitive logic at all hierarchical levels and the levels free information, self-organize the observer hierarchical code.

Since such code holds energy of memorizing and encoding, it physically organizes the multiple IN, self-encoding their local codes in coding information structure of information Observer.

The code information structure, self-organizing the multiple local codes, we call *observer intelligence.*

The logical switching of the free information performs the *intelligence functions*, which generate each local code.

Therefore, these functions are also distributed hierarchically along those assembling units.

The observer intelligence cooperative code self-organizes these local functions.

The question is what initiates the switching to memorize the bit and its encoding?

Evidently, it is the necessity of external energy, which requires opening its access to each local logical unit.

It needs a coordinating connection of the observer inner and external times. As it had shown (Sec.2.6), such coordination takes place exactly at the moment ending interval of the free logic at each unit level. According to Sec.5.5, the switching time interval $\Delta t_{o1}$ (Sec.2.6) runs different units with specific time intervals $\delta T_{cm}$ and related frequencies of switching $f_{cm} = 1/\delta T_{cm}$. The $\delta T_{cm}$ changes from $\delta T_{co} = 12$ –for an elementary objective observer (with IN' two triplets units)-up to $\delta T_{cs} = 69242.359$ - for a subjective observer (with 9 triplets IN). If the unit measure is one hour, then the objective observer opens to get external energy with frequency $f_{co} = 1/12$, or each 12 hours. The subjective observer gets external energy with frequency $f_{cs} = 1/69242.359$, or ~1/30 min, which is equivalent to one opening for two seconds, or 30 times in minute. This means, each observer has own time clock with its time course (Sec.5.5) which commands the hierarchical switching. The clock interactive switches command encoding each cognitive function to intelligence function starting the observer intelligence. (The details of coordination of the times and frequencies of switching are in Secs.7.1.2, 7.14.).

Within the sequential segments of the observing dynamics (Fig.3a), each switch links a bridge between two segments when the second is encoding.

Since the cognitive rotating process, at forming each node and level, holds the coherent loop of harmonized speeds-information frequencies at different levels, these frequencies determine the clock time units at different levels.

At lower unit level with longer $\delta T_{cm}$, its frequencies are lower, and vice versa up to the highest frequencies of the subjective and an intelligence observer. The subjective observer, self-assembling hierarchy of logical structures, owns hierarchy of the frequencies and clock courses.

The *harmonized speeds-information frequencies* will self–setup the switching times and the frequencies when the cognitive loop at each level self-establishes.

Finally, the coherent cognitive dynamics, assembling the units cognitive functions, self–organizes the hierarchy of intelligence functions encoding the bits. Or the local cognitive units, involved in resonance movement, self-organize itself in local cognitive function which self-forms the observer cognition.

The hierarchical cognition schedules intelligence functions encoding the observer hierarchy in the observer cooperative code. Since each assembled unit encodes the triplet code, the observer cooperative code integrates the structural units of triplet code at each level. These local codes have increasing densities of encoded impulses according to their hierarchal locations. The cooperative code, which the clock synchronizes, has rhythmical sequence of time intervals when each observer logical structural unit gets the needed external energy. The clock time course assigns the frequency through the repeating time intervals, which determine each local resonance frequency of assembling the structural unit.

These frequencies-local rhythms identify the moments of ending interval of the free information at each unit level, or the interacting cognitive and intelligence local actions. Each stable observer conserves its switching time intervals.

Therefore, the observer code enhances multiple rhythms of the local structural units. That's why an external melody' rhythms, resonating with observer code rhythms, support the cognitive functions, intelligence actions, and generation of both the cooperative observer logic and the code encoding this logic [43, p.226].

Recently, it experimentally confirms the music influence on neural encoding [129].

The structure of observer code

The curving interactive movement, starting with observing the curving impulses, rotates the observing process trajectories forming spirals locating on a cone surfaces (Fig.3). During probabilistic observations, these spiral trajectories occur in a random periodic sequence. With emerging space and the conjugates entropy increments, the rotating trajectories' shape the conjugated space-time spirals (Fig.3a). The emerging information process continues rotating its trajectories in form of double spirals and assembling them (Fig.6). The path from each interacting entropy impulse to whole observing process' entropy integrates the EF. The minimax principle, preserving each interacting impulse measure, leads to minimax variation principle (VP) for the EF. The VP identifies both conjugating trajectories of the observing process, described as the EF extremals, and the information path integral (IPF) emerging with converging the entropy extremals to information processes conjugated information trajectories. These trajectories satisfy the VP Hamilton equations for the EF-IPF extremals as the observer dynamic process. The trajectories integrate all observer logic including the logic enclosed in the each ending IN node code. The multiple IN' ending triple codes integrates double space spiral structure (DSS) (Fig.9). Each encoding occurs when each double curving interaction memorizes its Bit, or two qubits, during the minimal Landauer's energy thermodynamic process. The encoding locates the between interacting actions of opposite directional conjugated spirals making bridges between the spirals (Fig.3a). The attracting free information of the encoding bit connects the bit, first, in the triplets, second, in the INs nested nodes, and then, in each IN ending triplet code. Each bit, memorized in the conjugated interactive bridge, divides the trajectory on reversible process section which not includes the bit' bridge and irreversible bridge between the reversible sections. Thus, the observer irreversible dynamic trajectory includes the reversible sections ending with each bridge, where each irreversible bit emerges from the current observation. The conjugated trajectories describe the EF extremals, while the emerging encoding bits on the bridge integrates the IPF enclosing the integral information in its final encoding bit.

Therefore, all observer' integral information identifies the IPF final code, which the observation predicts through the VP minimax optimal information law. This IPF code has increasing densities, which triple with each following bit (Sec.2.12).

We assume that EF prediction, based on the integration of all observing process with its both probabilistic and information logic, establishes *artificial designed observer cognition*.

The evolving logic self-organizes the specific time–space information logical structure at each structural unit level that assembles the triplets, building the IN and the domains, but preempts memorizing each of its assembled information.

The free information of this logic encloses each of the multiple IN ending node. When the logic, ending each assembling, switches to open an external energy, this information logic is memorized in the node codes. The EF logic predicts it conversion to the IPF and the observer cooperative encoding in the artificial designed observer intelligence. It starts with the persisting information speeds-frequencies of the attracting observing impulses on the EF-IPF trajectories.

The attracting information sequentially equalizes the information speeds-frequencies in the attracting resonances which assemble the cooperative (cognitive) logic. The assembling resonance frequencies identify the clock commanding the logical switching, which coordinates cognitive and intelligence actions.

The cognitive functions perform the impulse switches delivering Landauer's energy for memorizing each impulse logic.

The intelligence functions perform encoding the memorized bit. The observer EF-IPF integrates the multiple encoded information, coordinating and unifying the total observing information in the IPF code.

In the artificial designed observer, where the EF integrates cognitive logic and IPF integrates its encoding information, the automatic conversion the EF in the IPF implements the automatic conversion the EF in the IPF code.

The IPF frequencies determine the clock time course (which the EF additive functional (Ch.1) initiates ).

The time courses integrates the EF-IPF optimal observing process and information dynamics in the optimal observer DSS double spiral structure which finally encloses the predicting observer code.

The observer logical structure self-connects the local codes in the observer code, which encodes all these structures in the space-time information structure of information observer.

The observer triplet code memorizes the observer cooperative information structure and enhances multiple rhythms of the local structural units. The DSS coding structure memorizes total collected observer quantity and quality information, which determine the observer cooperative complexity.

This coding structure, which self-organizes all assembled information, integrates function of cognition and intelligences.

The EF-IPF observing process and information dynamics artificially design the DSS.

The artificial designed DSS measures total information IQ of this observer. The code of each natural observer measures its IQ. The difference of these observers IQs measures distinctness of their intelligence. The maximal information, obtained in the observation, allows designating the DSS with maximal achievable the IQ measure of the optimal AI observer.

The space-time information structure, enclosing the encoded the EF-IPF, integrates the observed information (Fig.10) in the analytically designed AI information observer.

The observing information of a particular observer is limited by the considered constraints of each observation.

The constraints also limit the conversion of observing process to the information process.

The thresholds between the evolving stages of the observation limit the stages' evolution.

All these limit the integral cognitive information and the following intellective actions, which also limits the amount of free information that reduces ability of making intelligent IN's connections.

### 7.1.1. Specific of the Information Cognition

The Observer logical structure possesses both virtual probabilistic and real information causality and complexity [26, 27].

A virtual observer, forming the rotational space-time displacement of the impulse' opposite actions during virtual observation, starts accumulating virtual information through temporal memorizing it in probabilistic logic, which initiates cognitive movement.

The rotating cognitive movement connects the impulse microprocess with the bits in macroprocess, composing triple macrounits through the created free information, which assembles each evolving IN. The INs ending triplets integrate multiple nested IN's information logic in information domains with the evolving grows the quality information.

The observer's cognitive dynamic movement models the observer hierarchical rotation mechanism, which enables transferring the observer through the evolution stages to overcome the stage thresholds.

The mechanism rotating movement characterizes potential of power $P_{in}(i)$ which measures multiplication the current $(i)$ rotating moment $M(i)$ on angular speed $\omega(i)$:

$$P_{in}(i) = M(i) \times \omega(i). \tag{7.1.1}$$

This power compensates the resonance movement along each loop of the observer cognition. The loop rotates the thermodynamic process (Sec.2.6.6) with minimal energy-cognitive thermodynamics.

The cognitive movement, at forming each nodes and level, processes a temporary loop (Fig.6) which might disappear after the new formed IN triplet is memorized. After emerging the memorized bit during the observation, the rotation movement develops information form of double helix movement (Figs.3, 3a).

The observer's cognition assembles the multiple resonances loops, forming in the IN-triplet hierarchy, which accepts only the common information units which the cognitive loop recognizes.

The rotating process in the *coherent* loop *harmonizes speeds-information frequencies* at different levels analogously to Efimoff's scenario. The loop includes Borromini knot and ring.

The cognitive functions model the correlated inter-actions and feed-backs between the IN levels, which the highest domain level' feedback controls. Both cognitive process and cognitive functions emerge from the evolving interactive observations, which maintain their emerging properties and provide discrete information language for the cognitive logic.

The rotating double spirals (DSS) compose the evolving information logic from the running the macrodynamic process segments' logic. The process' reversible segments memorize the DSS sequential knots encoding the cognitive process.

In the rotating DSS, the cognition merges with the natural memorizing of each bit on all evolution levels.

The cognition emerges in two forms: a virtual rotating movement processing temporal probabilistic logic following real information process' logic rotating in double helix structure. The DSS concurrently organizes the observing information bits in the IN nodes, whose sequential knots enable memorizing information causality and logic.

These processes start with the elementary virtual observer and emerging bit at microlevel, which holds the prehistory and participates in evolving information observer.

The starting cognitive thermodynamic have no actual physical cost.

Results [106] confirm that cognition arises at quantum level as "a kind of entanglement in time"…"in process of measurement", where…"cognitive variables are represented in such a way that they don't really have values (only potentialities) until you measure them and memorize", even "without the need to invoke neurophysiologic variables", while "perfect knowledge of a cognitive variable at one point in time requires there to be some uncertainty about it at other times". Moreover, this analysis shows that both cognition and intelligence have information nature.

Recent work in cognitive maps [107] confirms" large-scale are internal representations of navigable spaces", reveals how cognitive maps are encoded, anchored to environmental landmarks and used to plan routes.

Similar neural mechanisms might be used to form 'maps' of *nonphysical spaces*, and "applied to nonspatial domains to provide the building blocks for many core elements of human thought".

## 7.1.2. Self-forming hierarchical distributed logical structure of cognition

Multiple moving INs, sequentially equalizing the nodes speeds-frequencies' attracting information logic in a resonance, assembles total observer logic. The mutual attracting free information logic, sequentially interacting, self-organizes the cooperative logical rotating spiral loops in a chain with encloses the observing logic.

Each curved impulse invariant time-space measure $\pi$ enfolds information measure $1Nat$ which includes bit, free information, and information needed for encoding between a nearest impulse bits.

This free information attracts the impulse with intensity $1/3$ bit per impulse.

The IPF integrates the bits with free information connecting the bits sequence.

The minimax principle, applied along the IPFextreme trajectory, maximizes information enclosed in each current impulse, squeezes its time interval, while its growing attracting free information minimizes the time interval between the nearest impulses proportionally to $1/3$bit. For each third impulse, that interval of information distance becomes proportional to 1 bit, preserving the invariant information impulse' time-space measure $\pi$.

Hence, each invariant impulse along the IPF extreme trajectory squeezes a proceeding impulses distance to $h_i = \pi$.

The IPF extreme trajectory sequentially condenses the increasing information of each third impulse to a following impulse that consecutively grows the information density of the invariant impulse. Free information of such impulse increases that intensifies information attraction between the invariant impulses in above proportion. By the end of the IPF integration, all integrated information is concentrated in a last impulse, whose information density approach maximal limit. Since free information encloses information logic, the multiple bits with triple growing density raise the process information logic.

The IPF integral information with its logic is condensed in the last integrated impulse time–space interval volume.

For multiple information impulses, each third curved impulse having invariant measure $\pi$ appears in the information process with time frequency $f_i = k_i, k_i = 3, 5, 7, 9, ....$ which indicates the appearance of the triplets and their specific sequence.

The invariant impulse time coordinate $\tau_i = \pi/\sqrt{2}$, the flat surface space coordinate measure $l_i = \sqrt{2}$ and orthogonal to them space coordinate space coordinate measure $h_i = \pi$ determine impulse volume $v_i^s = \tau_i \times l_i \times h_i = \pi^2$.

Information time-space density $D_i^I = k_i Nat / v_i^s$, concentrating $k_i Nat$ for each third impulse, increases with growing $k_i$ while the time-space volume $v_i^s$ holds invariant measure in form $D_i^I = k_i Nat / \pi^2$.

So, the current impulse time–space geometry encloses information, density and frequency, concentrating information logic and information of all previous impulses along the information path which the IPF integrates.

The EFextreme trajectories, starting from the multi-dimensional observing process, the EF-IPF transformation converts to the multi-dimensional orthogonal processes (Chs.1-2) whose curved impulses hold the above information measures.

The EF-IPF space-time extremal trajectories rotates forming spirals located on conic surfaces Fig.3, which starts from virtual (entropy) process and continues as the information process.

Since each bit of this trajectory creates the cutting entropy in the impulse observation, the trajectory consists of segments of information process dynamics and the between segments intervals delivering each bit to the following segment.

On Fig.3 each segment starts on the cone vertex-point D and ends on point D4 which connects to a vertex of the following cone. The observing bit is delivering at each cone vertex. The segment includes the impulse process with its logical bit, intervals of free logic, and correlation connecting the nearest segment and temporary memorizing the segment logic.

The logical and information dynamics describes the process of sequential logical interaction of the multiple impulses, rotating with information speed determined by the impulse density.

The dynamics between the cone vertexes is reversible and symmetrical analogously to Hamiltonian dynamics.

The logical anti-symmetry brings the anti-symmetrical logical bit prior the interaction with an external impulse which starts delivering the external energy. This bit is supplying at each cone vertex.

After the external energy generates physical multiple bits, the physical information process starts.

According to Sec. 2.2.6., the moments of appearance the interactive logical bit from begging of impulse is $t_{11} = 0.2452$, which defines relative interval $\Delta t_1 = 0.2452/1.44 \cong 0.17$.

Time interval for memorizing the bit $\Delta t_B$ identifies the bit information measure $\ln 2$ which is equivalent of invariant impulse relative part $\Delta t_B = \ln 2/1.44 = 0.481352$. For impulse $1 Nat$, difference $1-(0.17+0.481352) = 0.348648$ includes interval of supplying external energy erasing asymmetric logical bit and memorizing, interval of encoding, and fee information logic $\Delta t_{fo} = 0.23/1.44 \cong 0.1597$. The deduction brings interval $0.188948$ which include interval of the following opposite asymmetrical interaction $0.01847$ (Sec.2.26) that after encoding transfers the free information to other infulse. When encoding on interval $\Delta t_{eno} = 0.188948$ ends, interval $0.01847$ has been already spent.

Therefore, interval of encoding is $\Delta t_{en} = 0.188948 - 0.01847 = 0.17039$ which equals to time interval of appearance asymmetrical logical bit $t_1$. The needed external impulse, erasing asymmetric logical bit takes interval $\Delta t_B$ and ends with interval of encoding $\Delta t_{en}$ for each invariant impulse.

On the IPF path on the trajectory, this moment follows interval $\Delta t_B$ of creation logical bit, ending with the emergence of the knot that binds the free logic on. The interval of memorizing physical bit requires the same interval $\Delta t_B$ during which the entropy of logical bit erases. The external energy, supplied on time interval $t_{\Sigma b} = 0.481352 + 0.18948 = 0.67083$, includes both erasure the logical bit and its encoding. Since external impulse interactive part is $0.025$, it brings total $t_{\Sigma bo} = 0.67083 + 0.025 = 0.69583 \cong \ln 2$ for the interval of the external bit.

Since each impulse' curved measure $\pi$ with its relative time interval $\Delta t_1$ appears in the information process with frequency $f_{10} = \pi$, this frequency is $f_1 = 0.17/\pi \times \pi = 0.17$.

The frequency of spectrum $\omega_1 = 2\pi f_1 = 1.068$ identifies the time of opening supply of the external energy, memorizing the bit. Time interval of memorizing the bit $\Delta t_B$ identifies frequency $f_B$ of appearance that interval within the impulse with frequency $\omega_2 = 2\pi f_B = 2\pi \ln 2/1.44 = 3.02 < \pi$

The time interval of impulse encoding $\Delta t_{en}$ determine the spectrum frequency $\omega_3 = \omega_1$

Therefore, the frequency spectrum, initiating the encoding, equals $\omega_1$ in sequence $\{\omega_1 \omega_2 \omega_1\}$. This triple sequence identifies the segments alternating on the trajectory with the repeating ratio of bridge-middle part-starting next bridge.

The ratio holds on the bridge relative interval $0.00653$ according to (2.6.3.55) in Sec.2.2.6.

Spectrum $\{\omega_1, \omega_2, \omega_1\} = \omega_o, \omega_o = (1.068, 3.0.2, 1.068)$ delivers logical bit, energy memorizing it and encoding the bit. Or these frequencies present spectrum $\omega_o = \{1, 28277, 1\} \times 1.068$.

After supplying the external energy during these time intervals, whose sum equals to the invariant impulse time interval, the whole impulse becomes the segment of physical information process.

Therefore, physical dynamics describe the IPF extremal trajectory rotating on sequential cones (Fig.3). Each cone vertex encodes the bit memorized in a previous impulse-segment with frequency $\omega_1$. Each current segment also repeats frequency $\omega_2$, and transfers to next cone its vertex with frequency $\omega_1$ of encoding the current impulse bit.

Hence, each physical information impulse carries spectrum $\{\omega_1\omega_2\omega_1\} = \omega_o$, while their sequential pair on the trajectory carries the impulses spectrum $\{\omega_1, \omega_2, [\omega_1 = \omega_1], \omega_2, [\omega_1 = ...]\} = \omega_\Sigma$. where $[\omega_1 = \omega_1]$ is the resonance frequency for two impulses whose distance is shortening on 1/3. That allows closely connecting the impulses in the resonance. Along the trajectory, each of these pairs appears with the growing frequency of the impulses appearance $f_{io} = 1/3k_i$, $k_i = 3, 5, 7,...$

Since these fixed time intervals $\Delta t_1, \Delta t_B, \Delta t_{en}$ are relative to the invariant impulse measure, they are repeating for each invariant impulse with the increasing information density and with growing frequency.

Thus, along the extreme trajectory, each third impulse will deliver triple frequency of spectrum $\{\omega_1, \omega_2, [\omega_1 = \omega_1]_{\Delta t_{10}}, \omega_2, [\omega_1 = \omega_1]_{\Delta t_{20}}, \omega_2, [\omega_1 = \omega_1]_{\Delta t_{30}},\} = \omega_{\Sigma 10}$ with related time intervals $|\Delta t_{10}, \Delta t_{20}, \Delta t_{30}|$.

These time intervals, being sequentially proportional to the impulse distance measuring in 1/3 bit proportion: $1/k_i$.

The invariant time measure of the impulse are shortening accordingly $\Delta t_{10} \sim 1/3 \sim \pi/3\sqrt{2}, \Delta t_{20} \sim 1/5 \sim \pi/5\sqrt{2}, \Delta t_{30} \sim 1/7 \sim \pi/7\sqrt{2}$. It sequentially shortens the distance between the impulses on the extreme trajectory assembling each such three impulses in a triple of the resonance frequencies.

The trajectory assembles these triple resonances in a collective resonance.

### 7.1.3. Self-forming triplet logical structures and their self-cooperation in the IN hierarchical logic

In the multi-dimensional observing process, minimum of three logical bits with free logics can appear, which, attracting each other, would cooperate in a logical triple.

Multiple probabilities of interacting impulses in this multi-dimensional process produce the numerous frequencies.

Some of those, minimum of three, can generate the attractive resonance cooperating the triple.

This triple logic starts temporary memorizing two sequential pair cross-correlations during in their time of correlation.

A locally asymmetric cross-correlation memorizes the asymmetrical logic during the time of this correlation process.

When this process is ending, the triple correlations temporary memorize the triple logical bits.

According to [91], Sec.2.26, the minimal entropy of cross correlation ln2 can be memorized at cost of the equivalent minimal energy of logical bit. This is information cost of memorizing the triple logical bit which includes the free logic.

The attracting free logic of the emerging three logical bits starts the bits self-cooperation in the following sequence.

The free logic of the emerging logical bit holding frequency $\omega_2$ attracts next logical bits of toward a resonance with the equal frequency of next bit's free logic, assembling the two in joint resonance.

This resonance process links these bits in duplets.

The free logic from one bit out of the pair gets spent on the binding of the duplet. The free logic from the duplet' bit attracts the third bit and binds all three in a knot bit creating the triplet logical structure.

The knot bit still has free information and it is used to attract a different bound pair of emerging bits, creating two bound triplets. This process continues creating nested layers of bound triplets, three triplets and more (Figs. 4-6).

Hence the triplet logical structure creates the resonance frequencies of the attracting logic joining the triple bits.

The free logic attraction toward the triple resonance of their equal frequencies is core information mechanism structuring an elementary triplet.

The trajectory of forming triplet describes the rotating segments of their cones (Fig.5), whose vertexes join the triplet knot starting the base of the following cone. The knot frequency joins the cone vertexes in resonance along the cone base when the next rotating segment starts. It connects next triplet in the resonance and so on, creating the nested layers of logical space-time information network (IN), where the layers' knots hierarchy identifies the nested nodes of the IN hierarchy.

Triplets are the basic structure elements that form a nested informational time-space network with a hierarchical structure.

Each triplet unit generates three symbols from three segments of information dynamics and one when the segment attracting triple logic which binds three in the logical triplet knot. These symbols can produce triplet code, while the knot logic symbol binds the triple code for a potential encoding all triple. Encoding the knot will release its free information logic which transfers this triple code to next triplet node. Thus, the nodes logically organize themselves in IN code.

The external energy, encoding the IN triple code sustains the spectrum above.

The IN emerging logical structure carries the triple code on each node space time-hierarchy, and the last triplet in the network collects and encloses the entire network's information code.

The network, built through the resonance, has limited stability and therefore each IN encloses a finite structure.

That's why the observing process self-builds multiple limited INs through free information of its ending nods.

The final triplet in every network contains the maximum amount of the enclosed free information.

The limited networks develop self-connection through attraction of their *ended triplets.*

Even after each IN potentially loses stability evolving in a chaos, it possesses ability of self-restoration.

The multiple INs develop self-cooperation in hierarchical domain, starting with the cooperation of each tree ended triplets' free information in a knot which joining this INs' triples in resonance. This IN ending knot's free information resonates with other three INs' ending free information, forming triplet structure analogous to the elemental triplet.

This high level triplet joins these three INs structuring in a next IN of the domain hierarchy.

The hierarchical logical trajectory describes the space-time spiral structure (Figs.7, 9), evolving in observations.

This hierarchy enables generating sequential triple code locating on the rotating trajectory of the cone vertexes, which are space distributed at the different hierarchical levels of the multiple IN and the domain hierarchy.

Such space-time code integrates the observing process in space-time information geometry self-organizing an observer.

### 7.1.4. The observer wave function self-forming hierarchical distributed logical structure of cognition

The self-creating units of the hierarchy generate frequency delivering spectrum $\{\omega_1, \omega_2, [\omega_1 = \omega_1]_{\Delta t_{10}}, \omega_2, [\omega_1 = \omega_1]_{\Delta t_{20}}, \omega_2, [\omega_1 = \omega_1]_{\Delta t_{30}},\} = \omega_{\Sigma 10}$ which is growing in the triple sequentially shortening intervals $|\Delta t_{10}, \Delta t_{20}, \Delta t_{30}|$ for each $i$ trajectory segment.

We specify more details in the following propositions at following conditions.

The space-time spiral trajectory of the EF extremal (Fig.3) describes sequence of multi-dimensional curving rotating segments, representing interacting impulses of the observing process, which integrates the observing process' logic.

Each segment' impulse has invariant entropy measure $1Nat$ moving along the trajectory and rotating the curved impulse invariant measure $\pi$, which includes time coordinate measure $\tau_i = \pi/\sqrt{2}$, flat surface' space coordinate measure $l_i = \sqrt{2}$, space coordinate measure $h_i = \pi$ orthogonal to both of them; measure $1Nat$ includes the impulse logical bit $\ln 2$ and free asymmetric logic of the $f_{li} = 1 - \ln 2 \cong 0.3 Nat$ on each segment. The logic density per each third segment volume $v_i^S = \tau_i \times l_i \times h_i = \pi^2$ increases according to $D_i^I = k_i Nat / v_i^s, k_i = 3, 5, 7, ...$

The asymmetric logic divides the sequential segments by barriers which transfer the between segments anti-symmetrical interaction' logic interval $\Delta t_1$ following interval $\Delta t_B$ of memorizing bit, and interval $\Delta t_{en}$ of the free information ending the segment. Along the space-time trajectory, each sequential segment repeats this triple with invariant frequency spectrum $\{\omega_1, \omega_2, \omega_1\} = \omega_o, \omega_o \cong (1.068, 3.0.2, 1.068)$. The ratios of alternating sequences of the bridge-middle part-the bridge along the segments on the trajectory identify frequencies $f_{io} = 1/3k_i$ of this spectrum.

Propositions

1. Along each $i$-dimensional space-time segment rotates *three dimensional space wave functions*, spinning like a top (Fig.A), with rotating speeds: around each spiral cross-section $\alpha_i^s = 1[square/radian]$, or $\alpha_i^{s_o} = \pi/radian$, and orthogonal to this rotation space speed $\alpha_i^h = 1[volume/radian]$, or $\alpha_i^{h_o} = \pi/radian$.

The related frequencies of the orthogonal rotations are $\omega_i^s = \alpha_i^s/2\pi, \omega_i^{s_o} = 1/2$ and $\omega_i^h = \alpha_i^h/2\pi, \omega_i^{h_o} = 1/2$ accordingly.

Each $i$-dimensional segment cross-sectional rotation spans the space rotation on space interval $\pi$ of the segments invariant measure. The three-dimensional wave function distributes the space rotation along the segments trajectory with the above invariant speeds delivering the invariant spectrum $\{\omega_1, \omega_2, \omega_1\} = \omega_o, \omega_o \cong (1.068, 3.0.2, 1.068)$. The wave spectrum frequencies identify the alternating ratio of bridge-middle part-bridge along each sequential segment.

*Prof* 1. We apply equation of a wave $u = F(vt - x)$ depending on velocity of movement $v$ and distance $x$ to the moving segment on the spiral trajectory. Function $u(u_s, u_h)$ describes rotation of components: $u_s$ running along its cross-section $s_i^w = \tau_i \times l_i = \pi$ and $u_h$ running along the rotating space impulse coordinate $h_i = \pi$. The wave function $\arg(F) = f = f_{ws} \times f_{wh}$ represents orthogonal functions of rotating impulse along the segments. So, the wave describes two-dimensional rotation $u_s = F(f_{ws}), f_{iws} = \alpha_i^s \rho_i^s - s_i^w$ of the trajectory' rotating $i$-cone radius $\rho_i^s$ of $s_i^w$ with speed $\alpha_i^s$, were $\rho_i^s$ is analog of the distance, to reach the impulse cross-section square $s_i^w = \pi$. From the geometry of rotating movement (Sec.5.5.1), the segment is rotating on the trajectory on the cone square- basis with angle $\varphi_i^s = k_{io}\pi, k_{io} = 1, 2, 3, ..$, Fig.3, 8. This rotation reaches the distance at $f_{iws} = \alpha_i^s \rho_i^s - s_i^w = 0$ when length of the radius $\rho_i^s$, rotating along the cross-section square with speed $\alpha_i^s$ traces the equal sizes of the impulse $\rho_i^s = \pi$ at $s_i^w = \pi$. It determines $\alpha_i^s = 1$.

Since the wave function argument $f_{iws}$, decreasing along the segment, reaches $f_{iws} = 0$ by the end of each $i$-segment, with period equals impulse measure $\pi$, the wave function $u_s$ is periodical with period $\pi$.

The wave function $u_s$, moving along its cross section with entropy speed $\alpha_i^s = 1[square/radian]$, or $\alpha_i^{s_o} = \pi/radian$, holds related frequency $\omega_i^s = \alpha_i^s/2\pi$.

The wave function $u_h = F(f_{wh})$ argument $f_{wh}$ moves along impulse length $h_i^w$ with space rotating speed $\alpha_i^h$ to reach the impulse volume $v_{ih}^t$ according to Eq. $f_{iwh} = \alpha_i^h h_i^w - v_i^w$, where at reaching volume $v_i^w = \pi^2$ holds $f_{iwh} = \alpha_i^h h_i^w - v_i^w = 0$ with speed $\alpha_i^h = \pi^2/\pi = 1[volume/radian]$ and related frequency $\omega_i^{h_o} = 1/2$. Thus, the wave carries along the spiral trajectory frequency $\omega_i^{h_o} = 1/2$ and along its cross section equal frequency $\omega_i^{s_o} = 1/2$. Or each $i$-rotation with frequency $\omega_i^s$ equal to frequency $\omega_i^h$ brings the space rotation interval during the cross-sectional rotation, or vise-versa. •

2. Let consider $i, i+1, i+2$ three dimensional segments along the multi-dimensional rotating segments on the extreme trajectory, where each of these segment delivers the invariant spectrum $\{\omega_1, \omega_2, \omega_1\} = \omega_o, \omega_o \cong (1.068, 3.0.2, 1.068)$ through the cross-section rotation and the speeding space rotation distributes the spectrum along each of three-dimensional space dimensional segments $i, i+1, i+2$. Then, the wave function' frequencies synchronize the segments triplet logic in collective resonance. The sequentially forming triple barrier-knots are squeezing the initial observing multi-dimensional process first to three-dimensional rotation and then to a single dimensional information process encoding the bits of all multiple knots. •

*Prof.*2. Along the extreme trajectory, each segment of equal measure $\pi$ has increasing density, which is proportional to the segments shortening intervals $|\Delta t_{10}, \Delta t_{20}, \Delta t_{30}|$ in their locations along the trajectory.

The wave consecutive three-dimensional space movements picks segments $i, i+1, i+2$ sequentially from each of these trajectory specific locations in these dimensions and simultaneously starts rotating each of them during interval $|\Delta t_{10}, \Delta t_{20}, \Delta t_{30}|$ placing the interavls between segments $i, i+1, i+2$ accordingly.

The densities increase proportionally to the squeezing time interval measures along each of these dimensions trajectory.

The first of the wave three-dimensional rotation moves $i$ segment rotating during interval $\Delta t_{10} = 1$ (equivalent to space interval $\pi$ with density proportional $k_i = 1$). The second of wave three-dimensional rotation moves $i+1$ segment during interval $\Delta t_{20} = 1/2 \Delta t_{10}$ (equivalent to space interval $\pi$ with density proportional to $k_i = 2$). The third of wave three-dimensional rotation moves segment $i+2$ during time interval $\Delta t_{30} = 1/3$ (equivalent to space interval $\pi$ with density proportional $k_i = 3$). These three-dimensional movements repeat shortening these intervals for each triple segment with increasing frequency $f_i = k_i, k_i = 3,5,7,...$ of growing information density along the trajectory.

Since each of the segments deliver the equivalent spectrums, the equal frequencies of the sequential segment's spectrum $\{\omega_1, \omega_2, [\omega_1 = \omega_1]_{\Delta t_{10}}, \omega_2, [\omega_1 = \omega_1]_{\Delta t_{20}}, \omega_2, [\omega_1 = \omega_1]_{\Delta t_{30}},\} = \omega_{\Sigma 10}$ can be synchronized during these time intervals sequence.

According to the proposition' condition, the invariant spectrum frequency $\omega_1$ repeats time interval $\Delta t_1$ of the logical anti-symmetrical interaction on a bridge separating $i-1$ and $i$ segments on the trajectory, and the interval end indicates beginning of time interval $\Delta t_B$ on a middle of $i$ segment repeating with frequency $\omega_2$. During time $\Delta t_B$ the segment bit is memorized. The end of $\Delta t_B$ end indicates beginning of time interval $\Delta t_{en}$ of free information logic, which identifies the beginning of bridge separated $i$ and $i+1$ segments.

The free information attracts the separated segments.

The sequentially squeezing segments' time intervals allows performing first, double synchronization during interval $\Delta t_{20} = 1/2 \Delta t_{10}$, and next, double synchronizes during interval $\Delta t_{30} = 1/3 \Delta t_{23} = \Delta t_{20} - \Delta t_{30} = 1/2 - 1/3 = 1/6$.

Sum $\Delta t_{33} = \Delta t_{20} + \Delta t_{23} + \Delta t_{30} = 1/2 + 1/6 + 1/3 = 1$ equals to the first interval $\Delta t_{10}$, during which all two doublets are forming. Three segments finally deliver three memorized bits with their three free information intervals, which sequentially attract the synchronizing doublets during the rotation movement.

The information attraction on these time intervals adjoins the synchronized intervals information in a triple during the $i$ dimensional interval $\Delta t_{10} = 1$. Forming the triplet completes free information which delivers each $i+2$ segment with triple frequency while holding the invariant spectrum.

The wave function' frequencies synchronize the triplet logic in collective resonance. The delivering three ending free information join the tree memorized bits in a triple's knot where during additional interval $0.01847$ the bits encode in the triple.

The frequencies of the shortening time intervals distribute the orthogonal space rotations along the segments of the multiple dimensional observing trajectory which is moving *three dimensional space wave function* for each of the this trajectory multiple dimensions. Each of the three dimensions' shortening time intervals, which the three-dimensional rotation moves, bring the triplet knot that joins that three-dimensions to one.

The sequentially forming triple knots are squeezing the initial observing multi-dimensional process first to three-dimensional rotation and then to a single dimensional information process encoding the bits of all multiple knots.

Finally the periodical wave function includes the sequence of repeating arguments along both orthogonal rotations: $u_{sh} = u_s \times u_h, f_{ws} = \{f_{iws}\}, f_{wh}\{f_{iws}\}, i = 1,...n$, which performs the multiple three-dimensional movement with *three dimensional space wave functions* the like a top (Fig.A).

The shape of the multiple wave functions describes the extreme multi-dimensional trajectory formalizing the minimax observation process which models rotating segments on cones (Fig.3). •

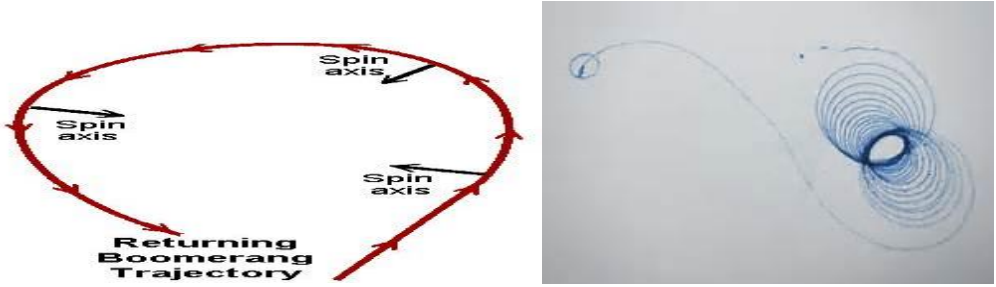

**Fig.A. Illustrative schematic of spinning top (from Google' top trajectory), where schematic bellow illustrates a rotating segment with bridge on the EF-IPF trajectory distributing the wave space –time movement.**

The wave function frequencies and properties

1).The wave function with above speeds and frequencies emerges during the observation process when a space interval appears within the impulse microprocess during reversible time interval of $\varepsilon_{ok} = 0.015625$ of the impulse invariant measure $\pi$ equivalent to $1 Nat$. Before that, the observing trajectory has described the probabilistic time function whose probability $P_\Delta^* \cong 0.821214$ indicates the appearance of a space-time probabilistic wave. During the probabilistic time observation, entropy of the Bayes priori-posteriori probabilities measures probabilistic symmetric logic of a sequence of these probabilities. Thus, wave function starts emerging in probabilistic observation as a probability wave in probability field.
At beginning of the microprocess, the probabilistic wave measures only time of its propagation.

2). The asymmetrical logic emerges with appearance of free logic interval $\Delta t_{fo} \cong 0.1597 \cong 1/2\pi$ which, repeating with equal wave frequency $\omega_i^s$, indicates beginning of the interactive rotating asymmetry on a segment' bridge.

From that, the observation logic on the trajectory becomes the asymmetric part of total free logic $f_{li} = 1 - \ln 2 \cong 0.3 Nat$.

The asymmetric *logical wave* emerges. The asymmetric logic probability approaching $p_{\pm a} = \exp(-2h_\alpha^{o*1}) \cong 0.9866617771$ appears with certainty-reality of observing the previously hidden asymmetrical bit.
Such logic temporary memorizes correlation with the probability which carries a logical bit of the certain logic.
Such certain logical bit may carry energy in real interactive process, which had covered by the entropy of the multiple random interactions of the Markov diffusion process. The path to creation of the certain bit includes an increment of the probability $\Delta P_{ie} = 0.9855507502 - 0.981699525437 = 0.1118$ starting injection of energy from an interacting impulse of the Markov process, Sec.3.5. Thus, the certain free impulse logic carries the certain logical attraction.
The wave function in the microprocess is probabilistic until the certain logical information bit appears.
The certain asymmetrical logical bit become physical bit through erasure the entropy of this logic, which allows replacing the logic by memorizing its bit.

3). The wave function starts on the observing process which the EF extreme trajectory prognosis, carrying the probabilistic wave which transforms the observing process to certainty of real observation. The spinning movement of the space –time trajectory describes the invariant speed around the cross-section of its rotating impulses-segments, which spreads the invariant rotation space speed along the segments trajectory. The segments' invariant spectrum $\{\omega_1, \omega_2, [\omega_1 = \omega_1]_{\Delta t_{10}}, \omega_2, [\omega_1 = \omega_1]_{\Delta t_{20}}, \omega_2, [\omega_1 = \omega_1]_{\Delta t_{30}},\} = \omega_{\Sigma 10}$ repeats the triple frequencies of these three time intervals between them. That shortens distance of the equal spectrum frequencies and assembles them in resonance creating joint logical structures-triplets units up to the IN hierarchies and domains. The frequency absolute maximum indicates a finite end of its creation. A minimal energy of the resonance supports the forming logical loop. •

Distribution of the space-time hierarchy

1).The hierarchy of self-cooperating triplet's units distributes the space rotation emerging along the EF segments of the time-space extreme trajectory, where each third impulse progressively increases the information density measure of its bit in triple.

The time-space hierarchy of the units starts with emerging in observation the symmetrical logic at appearance of space interval in the microprocess. This logic self–forms a hierarchy of the logical unit structures through the impulse' mutual attracting free logic which, sequentially attracting the moving unit' speeds, equalizes their frequencies in resonance that assembles the observer logic along the units hierarchy.

2).The hierarchy of the logical cooperating units becomes asymmetrical with appearance of certain logical bit on the extreme trajectory. The repeating free logic interval indicates the wave frequency $\omega_1 = f_i^s = 1/2\pi$.

The EF rotating trajectory of three segments equalizes their information speeds joining in the resonance frequency during the space rotation, which cooperates each third logical bit's segment on the trajectory and logically composes each triplet structure in the unit space hierarchy.

3).The appearance of the asymmetrical logical bit on the extreme trajectory indicates entrance the IPF information measure on its path to forming logical bit-ln2. The path starts on relative time interval $\Delta t_{fo} = 0.23/1.44 \cong 0.1597$ of the logical asymmetry, which identifies the segment bridge. During the triple impulses, the third time interval $\Delta t_{3r} = 3\Delta t_{1r} = 3/2\pi \cong 0.4775152$ indicates the end of the triple cooperative logic, starting build the triplet knot. Forming the triplet knot needs a time interval during which the triple free logic binds in the triplet bit. The time interval of creating the bit approaches $\Delta t_B = \ln 2/1.44 \cong 0.481352$. Difference $\Delta t_B - 3\Delta t_{3r} \cong 0.004$ evaluates the time of binding the triplet. Thus, the wave space interval delivers the logical bit with wave spectrum frequency $\omega_2 = 2\pi\Delta t_B = 2\pi \ln 2/1.44 = 3.02 < \pi$, while the triplet knot repeats with spectrum frequency $\omega_{20} = 2\pi 3/2\pi = 3$.

4). Delivering external energy for memorizing the logical bit identifies relative moment $t_1 = 0.2452/1.44 \cong 0.17$ ending the interval of the asymmetry. The resonance frequencies of the asymmetrical logic by this moment have already created.

Along the IPF path on the trajectory, this moment follows interval $\Delta t_B$ of creation logical bit, ending the emergence of the knot that binds the free logic. The interval of memorizing physical bit requires the same interval $\Delta t_B$ during which the entropy of logical bit erases. The needed external impulse, erasing asymmetric logical bit, starts with interval $\Delta t_B$ and ends with interval of encoding the bit $\Delta t_{en} = 0.17$. The external energy, supplied on time interval $t_{\Sigma b} = 0.481352 + 0.19 = 0.671352$, includes both erasure the logical bit and its encoding. Whereas interval of information free logic $\Delta t_{fo} = 0.23/1.44 \cong 0.1597$ is left for attracting a new bit, interaction. Since external impulse interactive part is $0.025$, it brings total $t_{\Sigma bo} = 0.67083 + 0.025 = 0.69583 \cong \ln 2$ for the interval of the external bit.

Therefore, the frequency spectrum, initiating the encoding, equals $\omega_1$ in sequence $\{\omega_1\omega_2\omega_1\}$. This triple sequence identifies the segments alternating on the trajectory with the repeating ratio of bridge-middle part-starting next bridge.
The ratio holds on the bridge relative interval $\Delta t_{en} = 0.17$.

Thus, the sequence of segments on the EF-IPF extreme trajectory carries its wave function' frequencies which self-structure the space-time unit of logical bits' hierarchy that self-assembles the observer logic. The logic controls memorizing and encoding physical bits and the hierarchical structure of these units space-time information geometry.

5).The segments-impulses on the EF spiral trajectory sequentially interact through the frequencies repeating on the bridges time–space locations, which connect the segments in trajectory. The segments' sequence on the EF-IPF extreme trajectory (Figs.3, 4) carries its wave function' frequencies which self-structure of the unit logical bit hierarchy that self-assembles total observer logic.This logic controls memorizing and encoding both physical bits and their units hierarchical structure. •

The observer cognitive logic, which encloses both, probabilistic and information causalities distributed along all observer hierarchy. The logical functions of the self-equalizing free information in the resonance perform the cognitive functions, which are distributed along hierarchy of assembling units: triplets, IN nested nodes, and the IN ending nodes.

These local functions self-organize the observer cognition.

Assembling runs the resonance frequencies $[\omega_1 = \omega_1]$ spreading along this hierarchy. Since each unit, ending high level structure encloses all its information logic, the high unit' impulse invariant time-space interval, containing this information, increases more information density than the unit of lower level hierarchy.

The resonance frequencies of spectrum $\{\omega_1, \omega_2, \omega_1\} = \omega_o$ holding the cognitive logic loop-self creates the unit hierarchy.

The observer logical structure

1). The observer logical structure self–forms the attracting free information, which self –organizes hierarchy of the logical triplet units assembling in resonance frequencies. Each triplet logical structure models Borromean ring consisting of three topological circles linking by Brunnian link-loop. The spinning top space trajectory, FigA, as well as EF-IPF trajectory, includes a Borromean rings' chain modeling a distributed hierarchical logic.

2). The observer logical structure carries the wave along the trajectory' segments, where each third segment delivers triple logic of information spectrum $\{\omega_1, \omega_2, [\omega_1 = \omega_1]_{\Delta t_{10}}, \omega_2, [\omega_1 = \omega_1]_{\Delta t_{20}}, \omega_2, [\omega_1 = \omega_1]_{\Delta t_{30}},\} = \omega_{\Sigma 10}$ with sequentially shortening intervals $|\Delta t_{10}, \Delta t_{20}, \Delta t_{30}|$, at $\Delta t_{io} / k_{io}, k_{io} = 3, 5, 7, 9$ and the increasing segment's information density.

Two sequential segments synchronize resonance frequencies $[\omega_1 = \omega_1]_{\Delta t_{10}}$ and $[\omega_1 = \omega_1]_{\Delta t_{20}}$ while the triplet synchronizes resonance frequency $[\omega_1 = \omega_1]_{\Delta t_{30}}$. This triple logic holds one bit in each observer's triplet logical structure unit.

The sequential triplets' attracting free logic conveys that resonance spectrum with progressively shortening time intervals and growing their frequencies, which cooperate the logical units in IN nested hierarchy. The needed spectrum with the increasing frequencies automatically carries each consecutive segment along the EF-IPF trajectory.

The emanating wave function delivers the frequencies cooperating a growing hierarchy of the logical units.

The self-built hierarchy of the logical structures self-integrates the observed logic which the structure encloses.

3). The hierarchy of distributed logical loops self-connects logical chain. The logical chain wide determines the invariant impulse' relative interval $0.17$ enclosing the assembled logical code. The growing density of consecutive impulses along the trajectory sequentially squeezes absolute value of this interval whose ratio preserves the invariant impulse.

The absolute time-space sizes of the logical chain are squeezing through the multi-level distributed hierarchy.

4). The logical chain rotation, carrying the frequencies of synchronized spectrum, requires a minimal energy to support the chain. This energy is equivalent to the logical bits code. The integrated chain logic holds this code.

The observer logic chain synchronizes the triple rhythms along the EF-IPF trajectory in a melody. Or the hierarchical space-time chain harmonizes the melody of the rhythms. •

Therefore, the wave function frequencies, initiating self-forming the observer cognition, emerges along the EF-IPF extreme trajectory in form of probabilistic time wave in probability field. The probabilistic impulse observation starts the microprocess, where entangled space rotation code develops the rotating space-time probability wave. The emerging opposite asymmetrical topological interaction shapes the space–time wave function, becoming certain, as well as the observer' cognitive logic. *These results conclusively and numerically determine the structure and functions of cognition.*

### 7.1.5. Specific of Information Intelligence and estimation its information values

The causal probabilities, following from Kolmogorov-Bayes probabilities' link, start the Markovian correlation connection with minimum of tree probabilistic events. An observer integrates the observing events in the information networks, which accumulate the nested triple connections, depending on the IN information invariant properties.

Each IN has invariant information geometrical structure and maximal number of nodes-triple bits, whose ability of cooperating more triplet nodes limits a possibility of the IN self-destruction by arising a chaotic movement (Sec. 6.5).

The intelligence measures the *memorized ending node of the IN highest levels*, while cognitive process at each triplet level preempts its memorizing. This means each memorizing node involves the cognition. The information measure of intelligence is *objective for each particular observer* while the IQ is an *empiric subjective* measure.

This theory shows that an observer, during current observation, can build each IN with maximum 24-26 nodes with average $3^{26}$ bits and enfold maximum of 26 such IN's.

Since each IN following level integrates information from all the IN previous levels, it measures the relative information quality of this level, which exposes information relationships between the levels in the triplet forms.

Because the subsequent relationships have been enclosed by the cognitive rotating mechanism, they formalize a causal comparative information meaning getting for the observing substances.

The *Observer Intelligence* has ability to uncover causal relationships enclosed in the evaluated observer $N_{o1} = 3^{26} \times 26 bits$ networks bits. That requires not only to build each of $N_{1I} = 26$ INs but also sequentially enfolds them in a final node whose single bit accumulating $N_{oI}$ bits:

$$N_{oI} = (3^{26}) \times 26 = 2,541.865.828329 \times 26 = 66,088.511.536.554 \cong 6.61 \times 10^9 \text{ bits.} \qquad (7.1)$$

However, since each IN node holds single triplet's information, the final IN node' bit keeps the triple causal information relationship with density $D_{oI} = N_{oI} / bit$ -per bit.

To support the IN node impulse feedback communication with the requested attracting information, this node requires information density:

$$i_{md} \cong 1.8 \times 10^{14} Nat/\sec = 1.44 \times 1.8 \times 10^{14} bit/\sec, \qquad (7.2)$$

where each such bit accumulates $N_{oI}$. Thus, total information density of the observer final IN bit:

$$i_{do} \cong 1.44 \times 1.8 \times 10^{14} \times (3^{26}) \times 26 bit/\sec \qquad (7.3)$$

evaluates the intelligent observer's information density.

With this density, the intelligent observer can obtain maximal information from the EF through the impulse interaction with entropy random process during time observation $T$.

Let us evaluate the EF according to (Sec.4.4)):

$$I_e = 1/8 \ln[r(T)/r(t_s)] \approx 1/8 \ln(T/t_s), T = m_N t_s.$$

Here $m_N$ is a total number of the IN nodes needed to build intelligent observer, $t_s$ is time interval of invariant impulse which is also invariant. At $m_N = 26 \times 26$ it allows estimate $I_e = 1/8 \ln 26^2 = 11.729 Nat$.

Thus, the intelligent observer needs $N_i \cong 12$ invariant impulses to build its total IN during time interval of observation $T$.

<u>Comments.</u> The human brain consists of about 86 billion neurons [109], which approximately in 14 times exceeds $N_{oI}$ (7.1), if each single bit of the cognition commands each neuron?

Nonetheless it agree with this estimation, if each neuron builds own IN with about five-six triplets (with levels $3 + 2^4 = 11, or 3 + 2^5 = 13$), while ending triplet bit condenses this $N_{oI}$.

Ability of a neuron building a net concurs with [110] and [109]. If its thrue, then

$N_{oI}$ measures information memory of human being. •

According to estimation [67, others] maximal information in Universe approximates

$$I_U \cong 3 \times 10^{29} Nat = 4.328 \times 10^{29} bits, \qquad (7.4)$$

from which each invariant intelligent observer can get $I_{ob} \cong 6.61 \times 10^9 bits$.

To obtain all $I_U$ information, number $M_{ob} \cong 1,527 \times 10^{16}$ of such intelligent observers is needed.

Each IN triplet node may request $I_m \cong (3.45 - 2.45) bits$, which measures this IN level of quality information that memorizes the node bit. Depending on each IN's levels $N_{1I} = 26$, such node's level accumulates average information between $I_m bits$ and $N_{om} = 3^{26} bits$.

Quantity $N_{oI}$ (8.1) measures invariant transformation to build the extreme IN nodes' structure during the observation, which transforms a probable observing process to information process in emerging information observer with intelligence.

The initial probability field of random processes, evaluated by entropy functional, contains potential information which an intelligent observer can obtain through the invariant transformation.

Information threshold $N_{oI}$ limits level of intelligence the intelligent observer satisfying the minimax variation principle.

The intelligent (human) observer can overcome this threshold requiring highest information up to $I_U$.

Such an observer that conquers the threshold possess a supper intellect, which can control not only own intellect, but control other observers.

Multiple joint supper intellectual observers can form a super-intellectual system (with $I_U$) controlling Universe, or would destroy themselves and others. In an intelligent machine, collecting the observing information, the emerging invariant regularities of the mimax law limits the AI observer actions.

### 7.2. Interacting intelligent observers through communication

Since any information intelligent observer emerges during the evolving observation, important issue is interaction of such observers in a common observation, which preserves the invariant information properties.

Suppose an intelligent observer sends a message, containing information encoding its meaning.

Another intelligent observer, receiving this information, would be able to *read the message, recognize its meaning*, *select and accept* it, if this information *satisfies the observer needed information quality* being memorized through its DSS code. Fulfillment of these five issues is considered subsequently.

### 7.2.1. How the interacting intelligence observers can understand meaning in each communication

When an intelligent observer sends a message, containing its information, which emanates from this intelligent observer's IN node, another intelligent observer, receiving that information, enables recognize its meaning if its information is equivalent to this observer IN nodes information quality satisfying coherence its cognitive logic.

Since the DSS code is invariant for all information observers, each observer encodes its message in that coding language, whose logic and length depend on sending information, possibly collected from the observer–sender's different INs nodes.

The observer's request for growing quality of needed information measures by the specific qualities of free information emanating from the IN distinctive nodes that need the compensation.

The observer request initiates recognition of the needed information, which includes understanding of its meaning.

That process comprises the following steps.

1). The IN node is requesting the needed quality by the Bayes high posteriori probability correlation (closed to certainty) which memorizes the message logical information making its temporary copy.

2). Copying logical information builds a temporary logical IN.

The number of the IN nodes triplets enable adjoining and cohere in the resonance, constraining them.

3).The forming temporary resonance structure-as a temporal cognition, initiates the requested IN nodes' high level probabilistic free logic, which allows involving the incoming copy in the observer cognitive logic.

The copies mirror the transitive impulses providing asymmetrical free logic with $\Delta t_1$ intervals.

4).Each logic interval allows access an external impulse' interval $\Delta t_B$, which, erasing the copy temporal logic, reveals its information bit by starting process of memorizing bit and its decoding. Decoding of this memorized bit holds interval $\Delta t_{en}$.

5).The coherence of observer cognitive logic actually allows starting the decoding from a low level of the observer hierarchical structure, if such structure needs updating information using its part of the frequency spectrum.

The message information delivers the wave function frequencies along the observer space –time hierarchy.

6).Decoding finalizes requesting IN nodes whose acceptance of the message comparative qualities indicates its ability to cohere, cooperating the message quality with the quality of the IN node, enclosed in the observer-receiver IN structure.

7). Since the acceptance of the message quality changes the existing observer logic encoded in the INs hierarchy, understanding the meaning of the message through its logic requires high level observer intelligent logic.

The observer logic' coherence with the message logic allows memorizing and encoding the decoded message information.

Delivering the message logic to the IN related observer logic needs a high frequency of the wave function spectrum, which generates the cognitive loop recognizing the message logic.

The message recognition allows its memorizing and encoding in the IN hierarchy up to the observer coding structure. Accepting the message quality, the intelligent observer recognizes its logic and encodes its copying digital images in space codes, being self-reflective in understanding the message meaning.

Thus, the intelligent observer uncovers a meaning of communicating message in the self-reflecting process, using the common message information language, temporary memorized logic, the cognitive acceptance, and logic of the memorized decoding whose coherence with intelligence observer cognitive logic allows memorizing and encoding the accepted message. Understanding the meaning of an observing process includes the coherence of its information with the observer current coding structure, which creates and evolves all previous observations, interactions, and communications.

### 7.2.2. How a biological observer's neurons accept a message?

Understanding the message describes the information formalism, which includes copying the accepted message on the cognitive moving helix which temporary memorizes it as triplets' entropies in a virtual IN structure.

This converting mechanism includes a compression of observing image in virtual impulse ending the virtual IN.

The virtual impulse, holding the entropy equivalent of the image information, moves the cognition scanning helix along the observer's INs until its negative curved step-up action, carrying the entropy equivalent of energy, will attract a positive curvature of the IN node bit's step-down action. The forming Bit encloses the equivalent energy's quality measured by its entropy value. When the IN bit's step-down action interacts with the moving image's step-up action, it injects energy capturing the entropy of impulse' ending step-up action. This inter-action models 0-1 bit (Fig.2A, B).

The opposite curved interaction provides a time–space difference (an asymmetrical barrier) between 0 and 1 actions, necessary for creating the Bit. The interactive impulse' step-down ending state memorizes the Bit when the observer interactive process provides Landauer's energy with maximal probability (up to a certainty).

Such energy drives the cognitive helix movement having minimal entropy production to overcome the bridge to the intellectual action memorizing and encoding the bit. This is energy of cognitive thermodynamic process (Sec.2.6.6), spending minimal quantity equal to binding the triplet structure by Landauer's energy ln 2.

At cooperating a triplet, this energy can be spent on memorizing the joint triple bit in the knot after third bit gets asymmetrical structure needed for memorizing. Such triplets carry a message.

The erasure and then memorizing each observing bit can run equal neuron information bits.

If the incoming information coheres with the cognitive loop, the message can be accepted and memorized in the receiver's IN. The forming IN bit encloses the equivalent energy's quality measured by its entropy value.

Therefore, the cognitive thermodynamic process practically has not thermodynamic cost, which models of a cognitive software with the minimal algorithmic complexity. The important coordination of an observer external time-space scale with its internal time-space scale happens when an external step-down jump action interacts with observer inner cognitive thermodynamics' time-space interval, which, in the curved interaction measures the difference of these intervals. Understanding the receiving information includes classifying and selecting such information that concurs with this observer's memorized meaning of other comparative images. Thus, cognitive movement, beginning in virtual observation, holds its imaginary form, composing an entropy microprocess, until the memorized bit transfers it to an information macromovement.

That brings two forms for cognitive helix process: imaginary reversible with temporal memory, and real-information which moves irreversible cognitive thermodynamics and ends memorizing the incoming information.

Explaining the mechanism modeling the message acceptance and understanding requires admitting, first, that the developed math-information formalism is considering as software controlling a brain structures–a hardware.

Connecting them requires a converting mechanism, which copies an observation and starts action on intelligence hardware. Those perform different sensors bending neurons, which make a mirror virtual copy of the observing image-message (analogous to transition impulse (Sec.2.6, Figs.1A, B)) *on* the cognitive moving helix.

For example, eyes scanning a TV screen, integrate the screen picks in a reflected image, accompany with the ears accumulating sound of the seeing image. That first composes cognition which allows intelligent memory and encoding.

The observer may not need to memorize each currently observed virtual image, which is reflected temporally in some sequence. Accordingly, such multiple virtual copies are formed by temporary triplets units composing a temporal collective IN whose ending node encloses a virtual impulse entropy' bit.

Such virtual IN with temporal memory is forming in a reversible logical process without permanent memory, which comprises a part of observation process (when Kolmogorov-Bayes probabilities link the triple events).

This converting mechanism includes a virtual compression of observing image in virtual impulse ending a virtual IN with the following coordination internal and external time course. Specifically, the time–space difference between particular 0 and 1 actions determines the clock coordinating the observer external and internal time course.

When the image bits memorize the observer IN's specific node, this node information quality and its precise position allow the observer to *recognize* this image information among other distinctive information qualities. The nodes positions already contain the observer INs' memorized information qualities. Recognizing a collective information image associates with understanding it by that observer enclosed information. Understanding implies that the observer can classify and select such information according to this observer's memorized *meaning* among other comparative images.

The information model of observer's understanding a message includes:

1. Sensor conversion of the observing message-image with building the virtual IN of the message, as a virtual mirror copy of the image collective information, which the IN compresses it in the virtual impulse.

2. Copying on the moving cognitive helix, which scans the observer' INs information enclosed from all IN domain levels.

3. Interaction of a sensor' neuron impulse, initiating yes-no actions, with the virtual impulse of the image through its yes-action, which injects an energy capturing the virtual entropy of the impulse' ending step-up action when the scanning helix cognitive movement contacts the observer IN node that provides this energy.

4. Memorizing the yes-no interactive Bit by the neuron interactive impulse' step-down no-action through the cognitive dynamic interactive process which provides Landauer's energy for erasure the observing image. That builds a mirror's bits memory, which decodes the message-image. In this neuron-message communication, the neuron yes-action, capturing the virtual impulse's ending step-up action, connects it with this neuron' no-action, which provides step-down action memorizing the message through the cognitive dynamic energy. Thus, the neuron curving interaction connects virtual and real actions, which actually binds the cognitive software with the brain hardware structure.

5. The memorized information bit stops the scanning cognitive mechanism on such IN level, where this information is understood through its observer' IN recognition. That ends the process of understanding of a current message.

Scanning the observer understood meanings allows recovering the message semantic, and then encoding the required one in a sending message. The virtual impulse of cognitive interaction provides logical Maxwell demon, while it transformation to the memorized IN information runs physical IMD. It presumes that neuron's yes-action starts its impulse entropy microprocess until the neuron' no-action, interacting with the observer macroprocess via the IN node bit by the jumping no-action, memorizes the incoming image in the observer IN structure. Thus, cognitive movement, beginning during virtual observation, holds its imaginary form, composing entropy microprocess, until the memorized IN bit transfers it to an information macro movement. That brings *two forms* for the cognitive helix process: *imaginary reversible without memory,* *and real-information moving by the irreversible cognitive thermodynamics memorizing incoming information.* The imaginary starts with the neuron yes-action and ends with the neuron no-action at the ending state of the neuron impulse, while the real starts with the IN bit yes-action memorizing the accepted bit, which processes the cognitive thermodynamics continuing moving the cognitive helix irreversibly. The threshold between the imaginary and real cognition holds memory and energy of the cognitive thermodynamics. This is how the observing quantity and quality of interacting information emerge in observer as the memorized quality encoding the observer cognition.

### 7.2.3. Analysis of some experimental studies

Study [111] provides explicit quantities for the energetic cost of processing sensory information.

"The findings in blowfly visual sensory system revealed that for visual sensory data, the cost of one bit of information is around $5 \times 10^{-14}$ Joules, or equivalently $10^4$ ATP molecules. That amount of information was delivered to the "blowfly retina' photoreceptors in the form of fluctuations of light intensity. This neural processing efficiency is still far from Landauer's limit of K Tln (2) J" and its bit's minimum *ln2,* "but it is still much more efficient than modern computers."

This limit evaluates minimal cost of neuron yes-action, which starts capturing virtual observation upward to real observing action. "A number of studies conclusively demonstrate that the large monopolar cell (LMC), the second-order retinal neuron, is optimized to maximize bit rate". That unique single cell holds "photoreceptors and an LMC of the blowfly retina code light level in a single pixel of the compound eye. Six photoreceptors carrying the same signal converge on a single LMC and drive it via multiple parallel synapses. The signals are intracellular recordings of the graded changes of membrane potential induced by a randomly modulated light source. Analysis of these analog responses yielded the rate at which photoreceptors and LMCs transmit information. The *oval* inset shows a photoreceptor-to-LMC synapse. The presynaptic site on the photoreceptor axon terminal (PR), contains synaptic vesicles, grouped around a prominent presynaptic ribbon. This *release site* faces four postsynaptic elements, containing cisternae.

The central pair of elements is invariably the dendrites of the two parallel LMCs, as captured in this tracing of an electron microscope section"."The low capacity (55 bits per second) synapse transmits at a much lower cost per bit than the high capacity (1600 bits per second) interneuron, the LMC capacity 1,500bits per second."

That result limits the yes-no neuron transmission rate in our cognitive model.

The brain neurons communicate [112] when presynaptic dopamine terminals demand neuronal activity for neurotransmission; in a response to depolarization [112a], dopamine vesicles utilize a cascade of vesicular transporters to dynamically increase the vesicular pH gradient, thereby increasing dopamine vesicle content.

In [113], neurogenesis provides fresh fields of interaction at the cellular (neuronal) level. Prior responses are generalized and the stage is set for future responses to assess probabilities and store temporary and intermediate data.

Study [114] found that brain computes Bayes probability distribution which generates a current observation, and this "belief distribution" representing the (log-transformed) posterior distributions encoded in pattern of brain activity reinforcing learning and decision making".

Cortical networks exhibit different modes of activity such as oscillations, synchrony and neural avalanches [115].

Since brain cerebral cortex covers the outer layer, its last activity phenomena is especially interesting to support moving this network with the enclosed cognition through a cognitive thermodynamic process.

Study [116] shows that dopamine modulates the brain dynamics boosting cognitive performance large–scale cortical networks, specifically, "enhanced dopaminergic signaling modulates the two potentially interrelated aspects of large-scale cortical dynamics during cognitive performance". Thus "dopamine enhances information-processing capacity in the human cortex during cognitive performance". That confirms the communication of interacting bits modeling the neurons.

Results [117] confirms that structure of thought arises from the author hidden Markov model providing the cognitive model. The suggested pyramidal model supports the pyramidal space structure of the information observer' domain intelligence. Consciousness is really "physics from the inside", whose more supporting results follow from the author image of conscience [118]:

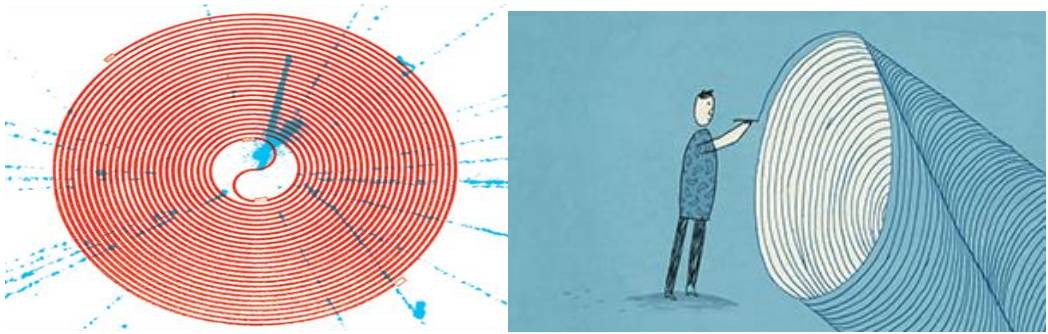

Where the first represents a plane projection of Fig.3, and the second replicates Fig.10.

The hypothesis of connection brain activity and cognition following from thermodynamics and information theory discusses [118].

The idea of *predictive information* (as the mutual information between the past and the future of a time series) unifies connections between learning and complexity [119].

According to [120] "Intelligence measures general cognitive functioning capturing a wide variety of different cognitive functions. It has been hypothesized that the brain works to minimize the resources allocated toward higher cognitive functioning, and working memory performance is associated with the excitation–inhibition balance in the brain".

"If consciousness is inherent to all mentation [121], it may be fundamental in nature...".That requires a physical account, a detailed description of the physics of *attention*, self-reflection which is *interaction*.

Results [122] experimentally confirm the coherent dynamic cognition and conceptual retrieval of semantic cognition.

According to experimental study [120], the brain energy supply regulates astrocytes which adjust blood flow forming an anatomical bridge between the vasculature and neuronal synapses. That molecular mechanism confirms our information model of supplying external energy for the observer intelligence through cognitive bridge rising the intelligence.

## 7.3. The Observer individuality specifics:

-The probability field' triad observing specific set of probabilistic events that arise in emerging particular information observer. The asymmetry of final bit will conserve this specific;

-Time of observation, measuring quantity and density of information of the delivered bits;

-Cooperation the observing information in a limited number of the IN-triplet nodes and limited number of the observer's IN, which depends on individual observer selective actions [34].

-The selec*t*ive actions define the cooperative information forces, which depends on the number of the IN nodes. The minimal cooperative force, forming very first triplet, defines minimal selective observer.

The individual ability for selection classifies information observers by levels of the IN hierarchy, time-space geometrical structure, and inner time scale whose feedback holds *admissible* information spectrum of observation. The individual observers INs determine its explicit ability of self-creation.

-These specifics classify the observers also by level of cognition and intelligence.

-The observer goal unifies its multiple DSS codes, which measure integral quality of information events that a primary observer accumulates during the observing process, starting from the probability field triad.

The information *mechanism* of building all observers is invariant, which describes the invariant equation of information dynamics following from minimax variation principle.

## 7.4. How the intelligence code self-controls the observer physical irreversible processes

The distributed intelligence coding actions at each hierarchical level control entrance the needed external physical processes. The DSS encodes the triplet dynamics in information macrodynamic process which implements the observer encoding logic.

We assume the observer requests the needed energy to implement own actions from such levels of his hierarchical structure, which enclose the requested code.

The request follows the same steps that perform communication, except encoding the levels' information macrodynamic in related physical irreversible thermodynamic.

After the request approval from the observer cognition, the request interactive action attracts impulses with the needed external energy bringing entropy gradient $dS/dx$ between the interactive actions' states $dx$.

The gradient provides equivalent information force $X = dS/dx$. The impulse correlation determines` diffusion $b$.

The force acting on the diffusion initiates thermodynamic flow (speed) $I = bX$ of the needed external thermodynamic process. The thermodynamic process' forces and flows-speeds determine power through the process Hamiltonian $H = X \times I$ to physically implement the requested actions.

The following observation of performance of these actions provides feedback to observer self-controlling the performance.

### 7.5. Metaphysical aspects of rising information observer

According to [123], God initiates randomness creating an initial random process' field.

The field of random waves, interacting with an observer, expose doted (discreet) impulses acting as a random control (Sec.1.4). An analogy is light wave which brings multiple quanta at interaction (measurement).

The linking Kolmogorov-Bayes probabilities' objective measures generate observable process of interacting random impulses where information observer emerges.

Quantity $N_{oI}$ measures the invariant transformation built during the observation, which transforms a probable observing process to information process of the emerging information observer (comparable to human being).

That shows that randomness finally can produce a real information observer whose multiplicities approach the God created Universe' information.

The question is how a natural intelligent observer, a human being, can self-transform the God initiated randomness in God's information? We assume that such transformer is a human faith.

According the Bible, Hebrews 11:1: "Now faith is the assurance of things hoped for, the conviction of things not seen."

And following [124]:..'That definition of faith contains two aspects: intellectual assent and trust. Intellectual assent is *recognizing* the true and *agreeing* that it supports a person'...'Trust is actually relying on the fact that the something is true. '

Information cognitive model of intellectual observer is recognizing information through human observer faith, agreeing with the observer accepted information, which the observer had collected. The information observer relies on the fact–information obtained during the evolving observation path from random uncertainty to information certainty with maximal probability approaching 1. Since faith, arising intuitively in a person-observer, encloses the described mathematical information formalism, it connects faith and science, as we believe. From that point of view, the cognitive movement, both virtual and real, models the faith, enables the human transformation of the God's created randomness to the human information, generated from the randomness. The transformation starts with the virtual logical cognitive process up to cognitive information and thermodynamics as a receptor of intellectual information. According to contemporary astronomy, the Universe is limited by a boundary-edge or horizon. Assuming that this boundary generates a huge amount of Information which produces the random field inside the Universe, I suppose that the generator of this Information is God located on the boundary horizon.

The curved horizon may generate **the resonance frequencies of spectrum** $\{\omega_1, \omega_2, \omega_1\} = \omega_o$ holding the cognitive logic loop-self-creating multiple information. Humans, and other possible observers, select and accept portions of the field of Information, following a path from uncertainty to certainty. These Observers are created by God and perform His Laws. Moreover, the observers enable communicating with God through possibility of selecting information from the random field.

It follows that God creates all observers inside Himself; God physically builds Himself though the information-structured Observer; the Universe is physically built by God, as an Information generator; Observer functions and destiny result from performing the emergent laws and information assignments.

**7.3. Math Summary**

1. Probabilities and conditional entropies of random events.

A *priori* $P_{s,x}^a(d\omega)$ and *a posteriori* $P_{s,x}^p(d\omega)$ probabilities observe Markov diffusion process $\tilde{x}_t$ distributions for random variable $\omega$.

For each $i,k$ random event $A_i, B_k$ along the observing process, each conditional a priori probability $P(A_i/B_k)$ follows conditional a posteriori probability $P(B_k/A_{i+1})$.

Conditional Kolmogorov probability

$$P(A_i/B_k) = [P(A_i)P(B_k/A_i)]/P(B_k) \tag{3.1}$$

after substituting an average probability

$$P(B_k) = \sum_{i=1}^{n} P(B_k/A_i)P(A_i)$$

defines Bayes probability by averaging this finite sum or integrating [19].

Conditional entropy

$$S[A_i/B_k)] = E[-\ln P(A_i/B_k))] = [-\ln \sum_{i,k=1}^{n} P(A_i/B_k)]P(B_k) \tag{3.1a}$$

averages the conditional Kolmogorov-Bayes probability for multiple events along the observing process.

Conditional probability satisfies Kolmogorov's 1-0 law [19] for function $f(x)|\xi$ of $\xi, x$ infinite sequence of independent random variables:

$$P_\delta(f(x)|\xi) = \begin{cases} 1, f(x)|\xi) \geq 0 \\ 0, f(x)|\xi) < 0 \end{cases}. \tag{3.1b}$$

This probability measure has applied for the impulse probing of an observable random process, which holds opposite Yes-No probabilities – as the unit of probability impulse step-function.

Sequence of Bayes probabilities for each three ratio of the impulse a posteriri –a priori probabilities satisfy

$$P(A_{i=1.23} \cup B) = \max. \tag{3.1c}$$

Random current conditional entropy is

$$\tilde{S}_{ik} = -\ln P(A_i/B_k)P(B_k). \tag{3.1d}$$

Probability density measure on the process trajectories:

$$p(\omega) = \frac{\tilde{P}_{s,x}(d\omega)}{P_{s,x}(d\omega)} = \exp\{-\varphi_s^t(\omega)\}, \tag{3.1e}$$

is connected with the process additive functional

$$\varphi_s^T = 1/2 \int_s^T a^u(t,\tilde{x}_t)^T (2b(t,\tilde{x}_t))^{-1} a^u(t,\tilde{x}_t) dt + \int_s^T (\sigma(t,\tilde{x}_t))^{-1} a^u(t,\tilde{x}_t) d\xi(t), \tag{3.1f}$$

defined through controllable functions drift $a^u(t,\tilde{x}_t)$ and diffusion of the process, where (3.1f) also describes transformation of the Markov processes' random time traversing the various sections of a trajectory.

2. The *integral measure* of the observing *process* trajectories are formalized by an *Entropy Functional* (EF), which is expressed through the regular and stochastic components of Markov diffusion process $\tilde{x}_t$:

$$\Delta S[\tilde{x}_t]|_s^T = 1/2 E_{s,x}\{\int_s^T a^u(t,\tilde{x}_t)^T (2b(t,\tilde{x}_t))^{-1} a^u(t,\tilde{x}_t) dt\} = \int_{\tilde{x}(t)\in B} -\ln[p(\omega)]P_{s,x}(d\omega) = -E_{s,x}[\ln p(\omega)], \tag{3.2}$$

3. Cutting the EF by impulse delta-function determines the increments of information for each impulse:

$$\Delta I[\tilde{x}_t]\vert_{t=\tau_k^{-o}}^{t=\tau_k^{+o}} = \begin{cases} 0, t < \tau_k^{-o} \\ 1/4 Nat, t = \tau_k^{-o} \\ 1/4 Nat, t = \tau_k^{+o} \\ 1/2 Nat, t = \tau_k, \tau_k^{-o} < \tau_k < \tau_k^{+o} \end{cases} \quad (3.3) \text{ with total } \sum_{t=\tau_k^{-o}}^{t=\tau_k^{+o}} \Delta I[\tilde{x}_t]_{\delta t} = 1 Nat. \quad (3.3a)$$

**4. Information path functional** (IPF) unites the information cutoff contributions $\Delta I[\tilde{x}_t / \varsigma_t]_{\delta_k}$ taking along $n$ dimensional Markov process impulses during its total time interval $(T-s)$:

$$I[\tilde{x}_t]\vert_s^{t \to T} = \lim_{k=n \to \infty} \sum_{k=1}^{k=n} \Delta I[\tilde{x}_t / \varsigma_t]_{\delta_k} \to S[\tilde{x}_t] \quad (3.4)$$

which in the limit approach the EF.

The IPF along the cutting time correlations on optimal trajectory $x_t$, in the limit, determines Eq

$$I[\tilde{x}_t / \varsigma_t]_{x_t} = -1/8 \int_s^T Tr[(r_s \dot{r}_t^{-1}] dt = -1/8 Tr[\ln r(T)/r(s)]. \quad (3.4a)$$

**5.** The equation of the EF for a microprocess:

$$\partial S(t^*)/\delta t^* = u_{\pm}^{t1} S(t^*), u_{\pm}^{t1} = [u_+ = \uparrow_{\tau_k^{+o}} (j-1), u_- = \downarrow_{\tau_k^{+o}} (j+1)] \quad (3.5)$$

under inverse actions of function $u_{\pm}^{t1}$, starting the impulse opposite time $t_{\pm}^* = \pm \pi/2t^i$ measured in space rotating angle relative to the impulse inner time $t^i$, determine the solutions-conjugated entropies $S_+(t_+^*)$, $S_-(t_-^*)$:

$S_+(t_+^*) = [exp(-t_+^*)(Cos(t_+^*) + jSin(t_+^*))], S_-(t_-^*) = [exp(-t_-^*)(Cos(t_-^*) + jSin(t_-^*))]$ at

$$S_{\pm}(t_{\pm}^*) = 1/2 S_+(t_+^*) \times S_-(t_-^*) = 1/2[exp(-2t_+^*)(Cos^2(t_+^*) + Sin^2(t_+^*) - 2Sin^2(t_+^*))] =$$
$$1/2[exp(-2t_+^*)((+1 - 2(1/2 - Cos(2t_+^*))))] = 1/2 exp(-2t_+^*) Cos(2t_+^*) \quad (3.5a)$$

The interactive entropy $S_{\pm}(t_{\pm})$ becomes a minimal *which begins the space* during reversible relative time interval of $0.015625\pi$ part of the impulse invariant measure $\pi$.

Overcoming entropy-information gap starts the information bit and Observer.

**6.** The information macrodynamic equations

$$\partial I / \partial x_t = X_t, a_x = \dot{x}_t = I_f, I_f = b_t X_t, \quad (3.6)$$

where $X_t$ is gradient (force) of information path functional $I$ (3.4) on macroprocess' trajectories $x_t$, $I_f$ is information flow defined through speed $\dot{x}_t$ of the macroprocess; the flow emerges from drift $a^u(t, \tilde{x}_t)$ being averaged by $a_x$ along all microprocesses; as well as the averaged diffusion $b_t \to b$ for the macroprocess force. Information Hamiltonian:

$$-\frac{\partial \tilde{S}}{\partial t} = (a^u)^T X + b \frac{\partial X}{\partial x} + 1/2 a^u (2b)^{-1} a^u = -\frac{\partial S}{\partial t} = H. \quad (3.7)$$

determines the macro equations from the minimax variation principle.

Equations (2.6) are information form of the equations of irreversible thermodynamics, which the *information macrodynamic process* generalizes.

The discretely changed information Hamiltonian divides irreversible dynamic trajectory on the partial reversible segments, predicting next emerging information unit. The flows and forces determine the macroprocess Hamiltonian $H = X \times I$.

Information curvature $K_\alpha^m$, density of information mass $M_{vm}^*$, and effective complexity $MC_m^{\delta e}$ connect Eqs

$$K_\alpha^m = -M_{vm}^* = MC_m^{\delta e}, \quad (3.8)$$

where $MC_m^{\delta e} = 3 \dot{H}_m^V MC_m \quad (3.9)$

includes differential of Hamiltianian per volume $\dot{H}_m^V$ and the IN cooperative complexity $MC_m$.

Single Eq.(3.8) unifies all previous math formalism.

**The significance of main finding**:

The composite structure of observer's generated information process, including:

**1**. *Reduction the process entropy under probing impulse, observing by Kolmogorov-Bayesian probabilities link, increases each posterior correlation; the impulse cutoff correlation sequentially converts the cutting entropy to information that memorizes the probes logic in Bit, participating in next probe-conversions; finding the curved interactive creation of Bit.*

**2**. *Creation wave function emerging in probabilistic observation whose frequencies self-forming the observer cognition, and the wave space distributes multiple bits hierarchy, becoming certain along with the observer' cognitive logic. The interacting observation self-creates its hierarchical cognition which self-encodes the observer intelligence with multilevel encoding.*

**3**. *Identifying this process stages at the information micro-and macrolevels, which govern the minimax information law. The intelligence code self-controls the observer evolution.*

**4**. *Finding self-organizing information triplet as structure macrounit self-forming information time-space cooperative distributed network enables self-scaling, self-renovation, adaptive self-organization, and cognitive and intelligent actions. The Observers, communicating by message covering a meaning, are self-reflective, whose intelligence enables understanding the message meaning.*

**5**. *Finding how the Observers, communicating by message covering a meaning, are self-reflective, whose intelligence enables understanding the message meaning and implement needed actions. The Observer encodes the triplet units' information macrodynamic process in related physical irreversible thermodynamic process which implements the observer encoding logic. The thermodynamic process' forces and flows-speeds determine power to physically implement the encoding actions. The following observation of performance of these actions provides feedback to observer self-controlling the performance. The results validate analytical and computer simulations and experimental applications from physics to biology* [43-62,131-139].